\documentclass[aps,prd,onecolumn,superscriptaddress,preprintnumbers,%
    floatfix,nofootinbib,longbibliography,eqsecnum]{revtex4-2}

\usepackage[T1]{fontenc}
\usepackage[utf8]{inputenc}
\usepackage[english]{babel}
\usepackage{lmodern}

\usepackage{amsmath,amssymb,amsfonts}
\usepackage{bm}
\usepackage{longtable,boldline}
\usepackage{cellspace} 
\usepackage[dvipsnames]{xcolor}
\usepackage{xfrac}
\usepackage{multirow}
\usepackage{pgf}

\usepackage{placeins}

\usepackage{titlesec}
\titleformat{\paragraph}{\centering\normalfont\small\itshape}{\theparagraph}{1em}{}
\titlespacing*{\paragraph}{0pt}{3.25ex plus 1ex minus .2ex}{1.5ex plus .2ex}



\usepackage{hyperref}
\hypersetup{colorlinks=true,citecolor=OliveGreen,linkcolor=BrickRed,urlcolor=MidnightBlue}

\newcommand{\Nstates}{\ensuremath{N_{\text{st}}}}
\newcommand{\tmin}{\ensuremath{\tau_{\text{min}}}}
\newcommand{\tmax}{\ensuremath{\tau_{\text{max}}}}
\newcommand{\ttmin}{\ensuremath{t_{\text{min}}}}
\newcommand{\ttmax}{\ensuremath{t_{\text{max}}}}
\newcommand{\ensemble}[2]{\mbox{\texttt{$\beta$ #1 M #2}}}

\newcommand{\Tmin}{\ensuremath{T_{\text{min}}}}
\newcommand{\Tmax}{\ensuremath{T_{\text{max}}}}

\newcommand{\rmax}{\ensuremath{r_{\text{max}}}}

\newcommand{\mrow}[2]{\multirow{#1}{*}{#2}}

\renewcommand{\d}{\text{d}}
\renewcommand{\i}{\text{i}}
\newcommand{\tr}{\mathop{\mathrm{tr}}}

\newcommand{\e}[1]{\text{e}^{#1}}
\newcommand{\order}{\mathrm{O}}
\newcommand{\gammaE}{\gamma_{\text{E}}}
\newcommand{\Ei}{\text{Ei}}
\renewcommand{\case}[2]{{\textstyle\frac{#1}{#2}}}
\renewcommand{\sfrac}[2]{\ensuremath{#1/#2}}

\newcommand{\half}{\ensuremath{\case{1}{2}}}

\newcommand{\CF}{C_{\text{F}}}

\newcommand{\CA}{C_{\text{A}}}
\newcommand{\Nf}{N_{\text{f}}}
\newcommand{\Nc}{N_{\text{c}}}
\newcommand{\Ns}{N_{\sigma}}
\newcommand{\Nt}{N_{\tau}}
\newcommand{\MS}{\ensuremath{\overline{\text{MS}}}}
\newcommand{\LQCD}{\Lambda_{\text{QCD}}}

\newcommand{\als}{\alpha_{\text{s}}}

\newcommand{\alsr}{\als(\sfrac{1}{r})}
\newcommand{\alsNf}{\alpha_{\text{s}}^{(\Nf)}}
\newcommand{\alsNfm}{\alpha_{\text{s}}^{(\Nf-1)}}
\newcommand{\alsNfp}{\alpha_{\text{s}}^{(\Nf+1)}}
\newcommand{\alsNfr}{\alpha_{\text{s}}^{(\Nf)}(\sfrac{1}{r})}
\newcommand{\muus}{\mu_{\text{us}}}

\newcommand{\ml}{m_{\text{l}}}
\newcommand{\ms}{m_{\text{s}}}
\newcommand{\mc}{m_{\text{c}}}

\newcommand{\mtot}{m_{\text{tot}}}

\newcommand{\FIGDIR}{Figures}

\newcommand{\HISQlats}{MILC:2010pul, MILC:2012znn, Bazavov:2017lyh}

\newcommand{\new}[1]{{\color{Black}#1}} 
\newcommand{\str}[1]{} 

\allowdisplaybreaks
\begin{document}

\title{Static Energy in (\texorpdfstring{\boldmath{$2+1+1$}}{2+1+1})-Flavor Lattice QCD: %
    Scale Setting and Charm Effects}
\author{Nora Brambilla}
\email{nora.brambilla@ph.tum.de}
\affiliation{Physik Department, Technische Universität München, James-Franck-Straße~1, \\
    D-85748 Garching b.\ München, Germany}
\affiliation{Institute for Advanced Study, Technische Universität München, Lichtenbergstraße~2a, \\
    D-85748 Garching b.\ München, Germany}
\affiliation{Munich Data Science Institute, Technische Universität München, Walther-von-Dyck-Straße~10, \\
    D-85748 Garching b.\ München, Germany}
\author{Rafael L. Delgado}
\email{rafael.delgado@upm.es}
\affiliation{Universidad Politécnica de Madrid, Nikola Tesla, s/n, 28031-Madrid, Spain}
\author{Andreas S. Kronfeld}
\email{ask@fnal.gov}
\affiliation{Particle Theory Department, Theory Division, Fermi National Accelerator Laboratory, %
    Batavia, Illinois 60510-5011, USA}
\affiliation{Institute for Advanced Study, Technische Universität München, Lichtenbergstraße~2a, \\
    D-85748 Garching b.\ München, Germany}
\author{Viljami Leino}
\email{viljami.leino@tum.de}
\affiliation{Physik Department, Technische Universität München, James-Franck-Straße~1, \\
    D-85748 Garching b.\ München, Germany}
\author{\\ Peter Petreczky}
\email{petreczk@bnl.gov}
\affiliation{Physics Department, Brookhaven National Laboratory, Upton, New York 11973-5000, USA}
\author{Sebastian Steinbeißer}
\email{sebastian.steinbeisser@tum.de}
\affiliation{Physik Department, Technische Universität München, James-Franck-Straße~1, \\
    D-85748 Garching b.\ München, Germany}
\affiliation{Leibniz-Rechenzentrum der Bayerischen Akademie der Wissenschaften, Boltzmannstraße~1, \\
    D-85748 Garching b.\ München, Germany}
\author{Antonio Vairo}
\email{antonio.vairo@ph.tum.de}
\affiliation{Physik Department, Technische Universität München, James-Franck-Straße~1, \\
    D-85748 Garching b.\ München, Germany}
\author{Johannes H. Weber}
\email{johannes.weber@physik.hu-berlin.de}
\affiliation{Physik Department, Technische Universität München, James-Franck-Straße~1, \\
    D-85748 Garching b.\ München, Germany}
\affiliation{Institut für Physik \& IRIS Adlershof, Humboldt-Universität zu Berlin, Zum Großen Windkanal~6, \\
    D-12489 Berlin, Germany}
\collaboration{TUMQCD}
\noaffiliation
\date{\today}
\preprint{TUM-EFT 154/21}
\preprint{HU-EP-22/19-RTG}
\preprint{FERMILAB-PUB-22-438-T}

\begin{abstract}
We present results for the static energy in ($2+1+1$)-flavor QCD over a wide range of lattice spacings and several
quark masses, including the physical quark mass, with ensembles of lattice-gauge-field configurations made available by
the MILC Collaboration.
We obtain results for the static energy out to distances of nearly $1$~fm, allowing us to perform a simultaneous
determination of the scales $r_{1}$ and $r_{0}$, as well as the string tension $\sigma$.
For the smallest three lattice spacings we also determine the scale $r_{2}$.
Our results for $\sfrac{r_{0}}{r_{1}}$ and $r_{0}\sqrt{\sigma}$ agree with published ($2+1$)-flavor results.
However, our result for $\sfrac{r_{1}}{r_{2}}$ differs significantly from the value obtained in the ($2+1$)-flavor
case, which is most likely due to the effect of the charm quark.
We also report results for $r_{0}$, $r_{1}$, and $r_{2}$ in fm, with the former two being slightly lower than published
($2+1$)-flavor results.
We study in detail the effect of the charm quark on the static energy by comparing our results on the finest two
lattices with the previously published ($2+1$)-flavor QCD results at similar lattice spacing.
We find that for $r > 0.2$~fm our results on the static energy agree with the ($2+1$)-flavor result, implying the
decoupling of the charm quark for these distances.
For smaller distances, on the other hand, we find that the effect of the dynamical charm quark is noticeable.
The lattice results agree well with the two-loop perturbative expression of the static energy incorporating finite
charm mass effects.
This is the first time that the decoupling of the charm quark is observed and quantitatively analyzed on lattice data
of the static energy.
\end{abstract}

\maketitle

\newcounter{response}

\clearpage
\tableofcontents

\clearpage
\section{Introduction}
\label{sec:intro}

The energy of a static quark-antiquark pair separated by a distance $r$, $E_{0}(r)$ has played an important role in
QCD since early days~\cite{Wilson:1974sk}.
Nonperturbative calculations with lattice gauge theory~\cite{Bali:2000gf} were important in establishing confinement in
QCD and in understanding its interplay with asymptotic freedom.
Confinement manifests itself in the linear rise of $E_{0}(r)$ at large $r$; the corresponding slope is known as the
string tension.
In the literature, $E_{0}(r)$ is sometimes also called the static potential.
The term ``static energy'' is, however, preferable because in the context of nonrelativistic effective field theories
of QCD the term ``static potential'' is understood to be the contribution to $E_{0}(r)$ coming solely from soft gluons,
i.e., gluons of energy or momentum of order $\sfrac{1}{r}$.
The static potential is infrared divergent~\cite{Appelquist:1977es}.
Up to a constant shift, the energy is a physical quantity not affected by infrared divergences.
In particular, the infrared divergence of the static potential cancels in the static energy against an ultraviolet
divergence coming from ultrasoft gluons, i.e., gluons of energy and momentum of order
$\sfrac{\als}{r}$~\cite{Brambilla:1999qa, Brambilla:1999xf}.

In lattice QCD, the static energy plays also an important role in setting the lattice scale, i.e., in the conversion
from lattice to physical units.
In quenched lattice QCD calculations, the scale has been set using the string tension, but in full QCD the string
breaks at the pair-production threshold, making a precise definition difficult.
Instead of the static energy, one can also use the force
\begin{equation}
    F(r) \equiv \frac{\d E_{0}(r)}{\d r},
    \label{eq:force}
\end{equation}
which is easier to manage in dimensional regularization as it is free of the order
$\Lambda_\text{QCD}$ renormalon~\cite{Hoang:1998nz, Pineda:1998id, Necco:2001xg} and in lattice gauge theory because it
is free of the self-energy linear divergence.
The dimensionless product $r^{2}F(r)$ can be used to set the scale~\cite{Sommer:1993ce}, especially at distances where
statistical and systematic uncertainties are under good control.
Examples of such a scale setting are the scales $r_{0}$, $r_{1}$, and $r_{2}$ defined by
\begin{equation}
    r_{i}^{2} F(r_{i})= c_{i}, \quad i=0,1,2,
    \label{eq:scales}
\end{equation}
with $c_{0}=1.65$~\cite{Sommer:1993ce}, $c_{1}=1$~\cite{Bernard:2000gd}, and
$c_{2}=\sfrac{1}{2}$~\cite{Bazavov:2017dsy}.

The static energy has been extensively studied in QCD with two light quarks and a (physical) strange quark, referred to
as $(2+1)$-flavor QCD~\cite{Bernard:2000gd, Aubin:2004wf, Cheng:2006qk, Cheng:2007jq, Bazavov:2011nk, HotQCD:2014kol,
Bazavov:2017dsy, Bazavov:2019qoo}, and the scales $r_{0}$ and $r_{1}$ have been determined for a wide range of lattice
spacing.
The study of the static energy in ($2+1+1$)-flavor QCD, i.e., in QCD with two light quarks, a (physical) strange quark,
and a (physical) charm quark, is less established.
The MILC Collaboration~\cite{MILC:2010pul, MILC:2012znn} calculated the static energy in a narrow region of distances
and obtained the $r_{1}$ scale using the highly improved staggered quark (HISQ) action~\cite{Follana:2006rc} for sea
quarks and one-loop tadpole-improved Symanzik gauge action~\cite{Symanzik:1983dc, Symanzik:1983gh, Luscher:1984xn,
Luscher:1985zq, Luscher:1985wf, Hao:2007iz, Hart:2008sq}.
In MILC's work, four lattice spacings were used, $a\approx0.06$~fm, $0.09$~fm, $0.12$~fm, and $0.15$~fm, and the three
light quark masses, $\ml=\sfrac{\ms}{27}$, $\sfrac{\ms}{10}$, and $\sfrac{\ms}{5}$, the first corresponding to the
physical light quark mass.
Here, $\ms$ is the physical strange quark mass.
The ETM Collaboration studied the static energy using the twisted-mass formulation in the quark sector and tree-level
Symanzik gauge action~\cite{EuropeanTwistedMass:2014osg}.
The calculations were performed at three lattice spacings, $a\approx0.065$~fm, $0.082$~fm, and $0.089$~fm, and several
values of the light quark masses corresponding to pion mass in the range
210--450~MeV~\cite{EuropeanTwistedMass:2014osg}.
In that work, the static energy was calculated in a narrow distance range around $r\sim r_{0}$ and the scale $r_{0}$
was determined.

In this paper, our aim is to extend the studies of the static energy in ($2+1+1$)-flavor QCD to smaller lattice
spacing, namely $a\approx0.032$~fm and $0.043$~fm, and a large range of distances on MILC's ($2+1+1$)-flavor HISQ
ensembles.
We perform a simultaneous determination of the scales $\sfrac{r_{0}}{a}$, $\sfrac{r_{1}}{a}$ and the string tension on
11--12 ensembles.
We proceed to take the continuum limit of these scales and the combinations $\sfrac{r_{0}}{r_{1}}$ and $\sqrt{\sigma
r_{0}^{2}}$.
In addition, we also determine the scale $\sfrac{r_{2}}{a}$ and the ratio $\sfrac{r_{1}}{r_{2}}$ on the six
ensembles at the three smallest lattice spacings.
Finally, we determine the continuum limits in fm of $r_{0}$, $r_{1}$, and $r_{2}$ as well.

The ($2+1+1$)-flavor HISQ ensembles are described in Refs.~\cite{MILC:2010pul, MILC:2012znn, Bazavov:2017lyh}.
Taken together these ensembles have yielded impressive results for a wide range of observables.
The observables cover spectroscopy~\cite{Gregory:2010gm, Gregory:2011sg, Briceno:2012wt, Dowdall:2012ab,
Hughes:2017xie, Lin:2019pia}, the decay constant ratio $\sfrac{f_{K^{+}}}{f_{\pi^{+}}}$ ~\cite{Bazavov:2013vwa,
Dowdall:2013rya}, the $B$-, $D$-, and $J/\psi$-meson decay constants~\cite{Dowdall:2013tga, Bazavov:2014wgs,
Colquhoun:2015oha, Bazavov:2017lyh, Hughes:2017spc, Hatton:2020vzp}, quark condensates~\cite{McNeile:2012xh}, the
hadronic vacuum polarization for the anomalous magnetic moment of the muon~\cite{Chakraborty:2014mwa,
Chakraborty:2016mwy, FermilabLattice:2017wgj, Davies:2019efs}, quark masses and $\als$~\cite{Chakraborty:2014aca,
Bazavov:2018omf, Lytle:2018evc}, the hindered M1 transition $\Upsilon(2S) \to
\eta_{b}(1S)\gamma$~\cite{Hughes:2015dba}, the electromagnetic form factor of the pion~\cite{Koponen:2015tkr,
Koponen:2017fvm}, the Cabibbo–Kobayashi–Maskawa (CKM) element $|V_{us}|$ from $K \to \pi l \nu$~\cite{Bazavov:2013maa,
Bazavov:2018kjg}, $B_{s} \to D_{s}^{(\ast)}$ form factors~\cite{Harrison:2017fmw, McLean:2019sds, McLean:2019qcx},
neutral $B^{0}_{d,s}$ mixing matrix elements~\cite{Dowdall:2019bea, Davies:2019gnp}, and $B_{c} \to B_{d,s}, J/\psi$
form factors~\cite{Cooper:2020wnj, Harrison:2020gvo}.
Significant, though less extensive work has been carried out on ensembles with ($2+1+1$)-flavor of twisted-mass Wilson
fermions~\cite{Baron:2010bv, EuropeanTwistedMass:2010voq, Ottnad:2012fv}, for example, quark
masses~\cite{EuropeanTwistedMass:2014osg}.

The static energy is also an important way to determine the strong coupling $\als$ or, equivalently, $\Lambda_{\MS}$;
see~Ref.~\cite{dEnterria:2022hzv} for a recent review.
Such studies started with quenched QCD~\cite{Necco:2001gh, Pineda:2002se,Brambilla:2010pp}.
Thereafter, the static energy~in ($2+1$)-flavor QCD has been used to determine~$\als$ in several lattice
setups~\cite{Bazavov:2012ka, Bazavov:2014soa, Takaura:2018lpw, Bazavov:2019qoo, Ayala:2020odx}.
These works have showed that perturbative QCD describes well the lattice results up to distances
$r\approx$~0.15--0.2~fm.
These distances include the inverse charm quark mass, so the charm quark can neither be considered massless nor
infinitely heavy.
It is important to account for finite charm quark mass effects when analyzing the static energy in $(2+1+1)$-flavor
QCD, particularly when determining~$\als$.
In this paper, we show the impact of finite charm quark mass effects on the static energy by comparing our new lattice
QCD results for the static energy in ($2+1+1$)-flavor QCD with published results in ($2+1$)-flavor QCD at similar
lattice spacings.
The comparison also demonstrates for the first time how the charm quark decouples from the static energy when
going from short to large distances.\footnote{%
Decoupling of $\Nf=2$ charmlike heavy quarks at very large distances has already been observed for the force, and
was published in conference proceedings~\cite{Cali:2017brl}.
Decoupling of heavy quarks in a similar setup has been proposed as a scheme for determining
$\als$~\cite{DallaBrida:2019mqg, dEnterria:2022hzv}.}
Further, we compare the ($2+1+1$)-flavor lattice data with the two-loop expression of the static energy, including charm
mass effects.

The rest of the paper is organized as follows.
Our numerical calculation of the static energy on the HISQ ensembles is described in Sec.~\ref{sec:sim}.
We then take these results and analyze them in Sec.~\ref{sec:scales} to obtain the scales $\sfrac{r_{i}}{a}$ and string
tension $a^{2}\sigma$.
Section~\ref{sec:cont-limit} forms several universal ratios or products of these quantities among each other and
combined with $a_{f_{p4s}}$ (the lattice spacing defined via the decay constant of a fictitious meson with quark and
antiquark having mass $0.4\ms$) from Ref.~\cite{Bazavov:2017lyh}.
We then turn in Sec.~\ref{sec:charm} to the comparison of the static energy with perturbation theory, in particular,
studying the effect of the massive charm quark sea.
Section~\ref{sec:conclusions} offers some outlook and conclusions.
Several technical appendices follow.
We found some inconsistencies in the gauge fixing of the publicly available and widely used HISQ ensembles, which we
document in Appendix~\ref{app:gauge-fixing}.
Additional plots and tables in support of Secs.~\ref{sec:sim}, \ref{sec:scales}, and~\ref{sec:cont-limit} can be found
in Appendix~\ref{app:data}.
Formulas from perturbative QCD needed for our study of charm quark loops are collected in Appendix~\ref{app:pQCD}.
Preliminary results based on these data have been published in conference proceedings~\cite{Steinbeisser:2021jgc}; we
have refined that analysis to permit quantitative studies of the impact on the various uncertainties.

\begin{figure}[b]
    \centering
    \vspace*{-6pt}
    \includegraphics[width=0.5\textwidth]{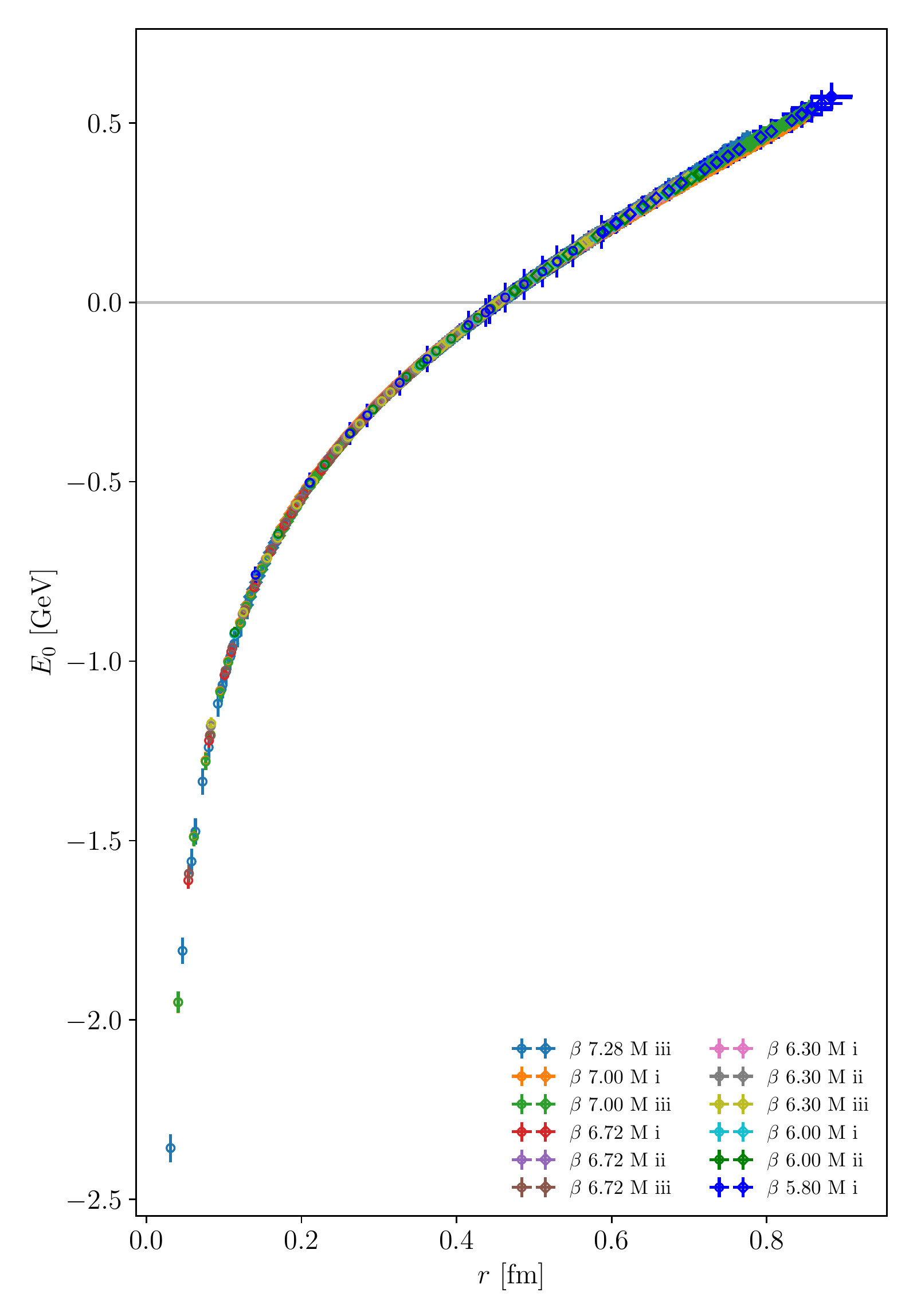}
    \caption{\label{fig:curve_collapse}%
    Results for the static energy in physical units from the calculations described in this paper.
    The data are from twelve ensembles of varying lattice spacing (keyed by $\beta$) and three choices of light quark
    mass (denoted ``M~i'', ``M~ii'', ``M~iii'').
    Lattice units are eliminated via $\sfrac{r_{0}}{a}$, and the unphysical constant is eliminated by setting
    $E_{0}(r_{0})=0$.
    See Sec.~\ref{sec:summary_plot} for details.}
\end{figure}

To conclude this introduction, Fig.~\ref{fig:curve_collapse} shows
the ($2+1+1$)-flavor QCD static energy obtained from our calculations for all ensembles in this work.
As detailed below, we compute the static energy with both ``bare'' and ``smeared'' links, and in the figure we show
only the bare (smeared) data for $\sfrac{r}{a} \le 4$ ($\sfrac{r}{a} > 4$).
(For details, see Sec.~\ref{sec:sim}).
On the scale of Fig.~\ref{fig:curve_collapse}, it is possible to see light quark mass dependence only at the larger $r$,
but it is very difficult to spot lattice-spacing dependence.
The data are, however, precise enough for both to be (statistically) significant, requiring the painstaking analysis of
the rest of the paper.
Figure~\ref{fig:curve_collapse} demonstrates for the first time the progression of the static energy in
($2+1+1$)-flavor QCD from the Coulombic to the confining region.

\section{Simulations}
\label{sec:sim}

In this section we give an overview of the simulation details, i.e., the gauge and fermion action, and further ensemble
details.
After that, we describe the operators used and how we extract the static energy, i.e., the ground state of the
underlying correlation function.

\subsection{HISQ ensembles and lattice setup}
\label{sec:setup}

{\refstepcounter{response}\label{resp:topology}}
We employ ensembles of lattice gauge fields with ($2+1+1$)-flavors of sea quark, generated by the MILC
Collaboration~\cite{\HISQlats}.
The subset used in this paper is listed in Table~\ref{tab:ensembles}.\footnote{%
\new{While the \ensemble{7.00}{i,iii} or \ensemble{7.28}{iii} ensembles are affected by insufficient sampling of
topological sectors, this does not lead to statistically significant effects for heavy-light
mesons~\cite{Bazavov:2017lyh}.
There are indications in ($2+1$)-flavor QCD that insufficient sampling of topological sectors does not affect the
static energy at a statistically significant level either~\cite{Weber:2018bam}.
Hence, we have disregarded quantitative effects of topological freezing in our calculations.}}
The sea quarks, namely two isospin-symmetric light quarks and physical strange and charm quarks, are simulated with the
(rooted) determinant of the HISQ action~\cite{Follana:2006rc}.
In most figures, we denote the ensembles by their respective $\beta$ values and their light quark mass labeled with
roman numerals i, ii, or iii, indicating $\sfrac{\ml}{\ms}$ at the physical value $\sfrac{1}{10}$ or $\sfrac{1}{5}$,
respectively.
The gluon action is the on-shell Symanzik-improved action~\cite{Symanzik:1983dc, *Symanzik:1983gh, Luscher:1984xn} with
the couplings determined at the one-loop level~\cite{Luscher:1985zq, *Luscher:1985wf, Hao:2007iz, *Hart:2008sq} with
tadpole improvement~\cite{Lepage:1992xa}.
Thus, the gluon action has leading discretization effects of order $\als^{2} a^{2}$ and~$a^{4}$.
The sea quark action eliminates discretization effects of order $\als^{0} a^{2}$, as well as those from staggered
taste-symmetry violation of order $\als^{1} a^{2}$, but does not realize full $\order(\als a^{2})$ improvement.
In short-distance quantities, the sea quarks contribute in loops, so the quark-action discretization artifacts in the
static energy are of order $\als^{2}a^{2}$ and~$\als a^{4}$.
The three-link improvement term for the charm quark is adjusted to eliminate higher-dimension discretization effects
with powers of $(a\mc)^{2}$ at the tree level.
In the characterization of these ensembles, we use the lattice scale $a_{f_{p4s}}$, which was introduced
in~\cite{MILC:2012znn} as an extension of the $f_{ps}$ scale~\cite{Davies:2009tsa}, determined via the decay constant
of a pseudoscalar meson made up from two quarks at the mass of $0.4\,\ms$~\cite{Bazavov:2010pi, MILC:2012znn}, which is
a compromise between good chiral behavior and only modest staggered taste-symmetry violation.

\begin{table}[b]
    \newcommand{\cB}{\cite{Bazavov:2017lyh}}
    \newcommand{\cpM}{\cite{private_milc}}
    \centering
    \caption{\label{tab:ensembles}%
    MILC gauge ensembles used in this study.
    The ensembles in the four upper rows have successive configurations separated by 5 time units (TU); the other
    ensembles use a separation of 6 TU.\footnote{%
    There are two exceptions from this rule, one stream of \ensemble{6.30}{i} with a separation of 4 TU, and one stream
    of \ensemble{6.72}{i} with a separation of 8 TU.}}
    \begin{tabular}{lccccccccccc}
    \hline \hline
\multirow{2}{*}{Our naming} & \multirow{2}{*}{$\Ns^{3} \times \Nt$} & \multirow{2}{*}{$\beta$} & $a_{f_{p4s}}$ (fm) & $u_{0}$ & \multirow{2}{*}{$a\ml$} & \multirow{2}{*}{$a\ms$} & \multirow{2}{*}{$a\mc$} & \multirow{2}{*}{$\sfrac{\ml}{\ms}$} & $(a\ms)_{\text{tuned}}$ & $M_\pi$ (MeV) & $N_{\text{conf}}$ \\  
&&& \cB & \cpM &&&&& \cB & \cB & \cB \\
\hline
\ensemble{5.80}{i}   & $32^{3} \times 48$   & 5.80           & 0.15294                & 0.85535            & 0.00235  & 0.0647           & 0.831           & physical               & 0.06852                     & 131               & 1041        \\[0.5em] 
\ensemble{6.00}{ii}  & $32^{3} \times 64$   & \mrow{2}{6.00} & \mrow{2}{0.12224}      & \mrow{2}{0.86372}  & 0.00507  & \mrow{2}{0.0507} & \mrow{2}{0.628} & \sfrac{1}{10}      & \mrow{2}{0.05296}           & 217               & 1000        \\        
\ensemble{6.00}{i}   & $48^{3} \times 64$   &                &                        &                    & 0.00184  &                  &                 & physical               &                             & 132               & 709         \\[0.5em] 
\ensemble{6.30}{iii} & $32^{3} \times 96$   & \mrow{3}{6.30} & \mrow{3}{0.08786}      & \mrow{3}{0.874164} & 0.0074   & 0.037            & 0.44            & \sfrac{1}{5}       & \mrow{3}{0.03627}           & 316               & 1008        \\        
\ensemble{6.30}{ii}  & $48^{3} \times 96$   &                &                        &                    & 0.00363  & \mrow{2}{0.0363} & 0.43            & \sfrac{1}{10}      &                             & 221               & 1031        \\        
\ensemble{6.30}{i}   & $64^{3} \times 96$   &                &                        &                    & 0.0012   &                  & 0.432           & physical               &                             & 129               & 1074        \\[0.5em] 
\ensemble{6.72}{iii} & $48^{3} \times 144$  & \mrow{3}{6.72} & \mrow{3}{0.05662}      & \mrow{3}{0.885773} & 0.0048   & \mrow{2}{0.024}  & \mrow{2}{0.286} & \sfrac{1}{5}       & \mrow{3}{0.02176}           & 329               & 1017        \\        
\ensemble{6.72}{ii}  & $64^{3} \times 144$  &                &                        &                    & 0.0024   &                  &                 & \sfrac{1}{10}      &                             & 234               & 1103        \\        
\ensemble{6.72}{i}   & $96^{3} \times 192$  &                &                        &                    & 0.0008   & 0.022            & 0.26            & physical               &                             & 135               & 1268        \\[0.5em] 
\ensemble{7.00}{iii} & $64^{3} \times 192$  & \mrow{2}{7.00} & \mrow{2}{0.0426}       & \mrow{2}{0.892186} & 0.00316  & 0.0158           & 0.188           & \sfrac{1}{5}       & \mrow{2}{0.01564}           & 315               & 1165        \\        
\ensemble{7.00}{i}   & $144^{3} \times 288$ &                &                        &                    & 0.000569 & 0.01555          & 0.1827          & physical               &                             & 134               & 478         \\[0.5em] 
\ensemble{7.28}{iii} & $96^{3} \times 288$  & 7.28           & 0.03216                & 0.89779            & 0.00223  & 0.01115          & 0.1316          & \sfrac{1}{5}       & 0.01129                     & 309               & 821         \\        
\hline
\hline

    \end{tabular}
\end{table}

For reference, we also employ ($2+1$)-flavor ensembles from the HotQCD Collaboration~\cite{HotQCD:2014kol,
Bazavov:2017dsy}, again with the (rooted) HISQ determinant for the sea quarks but now with a \emph{tree}-level
Symanzik-improved gauge action.
These ensembles correspond to continuum pion masses of $M_{\pi}\approx160$ or $320$~MeV, respectively, while the
strange quark is physical; in their characterization we use the lattice scale $a_{r_{1}}$ determined from the static
energy~\cite{Bernard:2000gd}, which had been obtained through, e.g., continuum extrapolation of
$\sfrac{r_{0}}{r_{1}}$~\cite{Aubin:2004wf}, or chiral-continuum extrapolation of
$\Upsilon$-splittings~\cite{Davies:2009tsa}, or of $r_{1}f_{\pi}$~\cite{Bazavov:2010pi}.
Since $r_{1}$ is derived from a gluonic operator, it is rather insensitive to the light quark masses in the sea or to
staggered taste-symmetry violation; our analysis confirms this well-known fact, see
Sec.~\ref{sec:quark_mass_dependence}.

The gauge configurations have been fixed to Coulomb gauge.
Due to miscommunication, we accidentally employed two different schemes, fixed tolerance and fixed iteration count.
Subsets of the \ensemble{6.72}{i} ensemble had each in turn; for details, see Appendix~\ref{app:gauge-fixing}.
As a consequence, we analyzed the two subsets with different gauge-fixing schemes separately and confirmed the
independence of the energy levels; see Figs.~\ref{fig:gaugefix_effm} and~\ref{fig:gaugefix_fits} in
Appendix~\ref{app:gauge-fixing}.
In the further analysis, we restricted ourselves to the subset with the fixed-tolerance scheme due to having better
statistics.

\subsection{Correlation functions and fitting}
\label{sec:corrfit}

The static energy is obtained from the time dependence of the Wilson-line correlation function
$C\left(\bm{r},\tau,a\right)$ at separation $\sfrac{\bm{r}}{a} \in \mathbb{Z}^{3}$ computed after fixing to a
Coulomb~gauge (see Sec.~\ref{sec:setup}):
\begin{align}
    W\left(\bm{r},\tau,a\right) &= \prod_{u=0}^{\sfrac{\tau}{a}-1} U_{4}\left(\bm{r},ua,a\right),
    \label{eq:wiline} \\
    C\left(\bm{r},\tau,a\right) &= \left\langle \frac{1}{\Ns^{3}} \sum\limits_{\bm{x}} \sum\limits_{\bm{y}=R(\bm{r})}
        \frac{1}{\Nc\,N_{\bm{r}}} \tr\left[W^{\dagger}\left(\bm{x}+\bm{y},\tau,a\right)
            W\left(\bm{x},\tau,a\right)\right] \right\rangle,
    \label{eq:corr} \\
        &= \sum_{n=0}^{\infty} C_{n}\left(\bm{r},a\right) \left(\e{-\tau E_{n}\left(\bm{r},a\right)} +
            \e{-(a\Nt-\tau)E_{n}\left(\bm{r},a\right)} \right),
    \label{eq:tower}
\end{align}
where, on the first line, $U_{4}$ is a temporal link.
On the second line, one sum is over all spatial sites $\bm{x}$ with $\Ns$ the isotropic spatial extent of the lattice
in each direction, and $\langle\dots\rangle$ denotes the average over all dynamical quark and gauge field
configurations.
The other sum is over all distances $\bm{y}$ that are either a cubic rotation reflection of $\bm{r}$, or that correspond
to the same geometric distance $|\bm{r}|$ with large enough $\sfrac{|\bm{r}|}{a}$;\footnote{%
For paths of length $\sfrac{|\bm{r}|}{a} > 6$ that are inequivalent under the hypercubic group, we average over the
path-dependent correlation function and neglect non-smooth discretization artifacts; see Sec.~\ref{sec:artifacts}.}
$N_{\bm{r}}$ is the total number of distances included in this sum.
On the same line, $\Nc=3$ is the number of colors.
Finally, on the last line, $\Nt$ is the temporal extent of the lattice, and this spectral decomposition holds---with
improved gauge action---only for $\sfrac{\tau}{a}\ge2$.
Because, in our notation, $\sfrac{\tau}{a}$ are dimensionless integers, fits to the $\sfrac{\tau}{a}$ dependence yield
dimensionless energies $aE_{n}$.
For each ensemble, we have also constructed a Wilson-line correlation function replacing the bare links~$U_{4}$ with
links after one iteration of four-dimensional hypercubic (HYP) smearing~\cite{Hasenfratz:2001hp} with standard smearing
parameters ($\alpha_{1}=0.75$, $\alpha_{2}=0.6$, $\alpha_{3}=0.3$).
Smearing improves greatly the signal-to-noise ratio even at large distances, as discussed in Secs.~\ref{sec:artifacts}
and~\ref{sec:quark_mass_dependence}, but the exponents in Eq.~\eqref{eq:tower} can be interpreted as static energies
only when at least one component of $\sfrac{\bm{r}}{a}$ is greater than 2.

In this work, we are interested only in the lowest-lying state, namely $aE_{0}(\bm{r},a)$.
We want to combine data for a huge range of $\bm{r}$ from $\sfrac{\bm{r}}{a} = (1,0,0)$ out to $|\bm{r}|\approx1$~fm
and from a wide range of lattice spacing~$a$ from $0.03$~fm up to $0.15$~fm.
The former (short distances, fine lattices) are indispensable for the comparison to weak-coupling calculations in
Sec.~\ref{sec:charm}, while the latter (large distances, coarser lattices) are indispensable for a determination of
some lattice scales and the string tension in Sec.~\ref{sec:scales}.
The data written to disk are limited to within a sphere $|\bm{r}|\le R_{\text{max}}$ and to a maximum time $\tau \le
T_{\text{max}}$, which are collected in Table~\ref{tab:data-intervals}.
\new{In particular, our data are restricted to distances smaller than those where string breaking occurs.}\footnote{%
{\refstepcounter{response}\label{resp:string_breaking}}
\new{Since Wilson line correlators in Eq.\eqref{eq:corr} do not overlap with two static-light mesons, our correlators
are expected to be insensitive to string breaking.}}
\begin{table}
    \caption{\label{tab:data-intervals}%
    Time and distance intervals in the full data set.
    The minimum distance is always $(1,0,0)$.}
    \newcommand{\h}{\phantom{2}}
    \begin{tabular}{ccccccc}
    \hline\hline
    $a_{f_{p4s}}$ (fm) & $\beta$ & $\sfrac{\Tmin}{a}$ & $\sfrac{\Tmax}{a}$ & \Tmax~(fm) & $\sfrac{\rmax}{a}$ & $\rmax$~(fm) \\
    \hline
    0.15294      &  5.8\h  &     1     &   \h9     &   1.35     &    \h6    &   0.92     \\
    0.12224      &  6.0\h  &     1     &   \h6     &   0.73     &    \h6    &   0.73     \\
    0.08786      &  6.3\h  &     1     &   \h8     &   0.70     &    \h8    &   0.70     \\
    0.05662      &  6.72   &     1     &    10     &   0.57     &     12    &   0.68     \\
    0.0426\h     &  7.0\h  &     1     &    20     &   0.85     &     20    &   0.85     \\
    0.03216      &  7.28   &     1     &    28     &   0.91     &     24    &   0.78     \\
    \hline\hline
    \end{tabular}
\end{table}

Because of the limited time range, and because of the exponential degradation of the signal-to-noise ratio,\footnote{%
This degradation is much alleviated in correlation functions obtained with smeared links since the ultraviolet noise,
due to short-distance fluctuations, is diminished and since the divergent contribution to static energy is decreased
dramatically.} we can safely neglect the backwards-propagating states in Eq.~\eqref{eq:tower} and are, in practice,
limited to extractions of the ground state energy from multiexponential fits with a finite number of states.
We reparametrize $C\left(\bm{r},\tau,a\right)$ using energy differences
$a\Delta_{n}(\bm{r},a)=aE_{n}(\bm{r},a)-aE_{(n-1)}(\bm{r},a)>0$, $n\ge1$ instead of the equivalent\footnote{%
We denote the collective set of fit parameters as $\{C_{n}, E_{n}\}$, $n \ge 0$, even though it means, in practice,
$\{C_{0}, \{C_{n}\}, E_{0}, \{\Delta_{n}\}\}$, $n \ge 1$, which contains the same information.} full excited state
energies $aE_{n}(\bm{r},a)$, $n \ge 1$,
\begin{align}
    C\left(\bm{r},\tau,a\right) &= \e{-\tau E_{0}\left(\bm{r},a\right)} \left( C_{0}\left(\bm{r},a\right) +
        \sum\limits_{n=1}^{\Nstates-1} C_{n}\left(\bm{r},a\right)
            \prod\limits_{m=1}^{n} \e{-\tau \Delta_{m}\left(\bm{r},a\right)}
        \right) + \ldots,
    \label{eq:forward_tower}
\end{align}
and choose $\Nstates=1$, $2$, or~$3$, such that the highest state is labeled by $(\Nstates-1)$.
The spectrum depends strongly on~$|\bm{r}|$, so the time interval $\tau\in[\tmin,\tmax]$ in the fit must be chosen to
depend on $|\bm{r}|$.
Whereas, the ground state energy is essentially an attractive Coulomb interaction for small $\bm{r}$, the low-lying
excited states correspond to a repulsive Coulomb interaction instead.
At larger distances, all energy levels are controlled by the QCD string tension, with much smaller energy differences
of order $\LQCD$.
Hence, as the excited states survive longer at larger $\bm{r}$, we choose standard values of $\tmin$ for each
$\Nstates$, depending on the distance $|\bm{r}|$, i.e.,
\begin{equation}
\begin{aligned}
    &\phantom{\case{2}{3}}|\bm{r}| + 0.2~\text{fm} \le \tau_{\text{min},1} \le 0.3~\text{fm}
        & ~\text{for}~\Nstates=1, \\
    &\case{2}{3}|\bm{r}| + 0.1~\text{fm} \le \tau_{\mathrm{min},2} \le \tau_{\mathrm{min},1}-2a
        & ~\text{for}~\Nstates=2, \\
    &\case{1}{3}|\bm{r}|\phantom{\;+ 0.2~\text{fm}} \le \tau_{\mathrm{min},3} \le \tau_{\mathrm{min},2}-2a
        & ~\text{for}~\Nstates=3,
\end{aligned}
\label{eq:time-ranges}
\end{equation}
and round it to the next larger integer multiple of the lattice spacing $a_{f_{p4s}}$.
Since we cannot always follow these criteria, we amend the fit ranges as necessary.
The two main reasons for doing so are the limited number of available data due to finite $\sfrac{\Tmax}{a}$ (see
Table~\ref{tab:data-intervals}) and the inability to constrain the two extra parameters for each further state, if less
than two data could be added.
We test the robustness of the fits by varying $\sfrac{\tmin}{a}$ by $\pm 1$ wherever possible.
Occasionally, the reduction by $-1$ sets $\sfrac{\tmin}{a} = 1$, which with the Symanzik-improved action is marred by a
contact term, so we do not use such fits further.
Hence, we arrive at a two-dimensional table, labeled by $(\sfrac{\tmin}{a},\Nstates)$, of results $\{C_{n},E_{n}\}$ for
each $(\bm{r},a)$.
A complete account of the time ranges is given in Table~\ref{tab:fit-intervals} in Appendix~\ref{app:correlator_fits}.

{\refstepcounter{response}\label{resp:autocorrelations}}
For a few representative pairs of $(\bm{r},\tau)$, we find autocorrelation times of $C\left(\bm{r},\tau,a\right)$ in the
range of 1 or 2 separations of successive configurations \new{on the \ensemble{7.00}{i} ensemble, which is the worst
case due to the interplay of small ensemble size, fine lattice spacing, and physical light-quark mass,
see Table~\ref{tab:ensembles}.
Hence, in this case a block size of $2\tau_\text{int} \approx 4$ is justifiable, which permits up to $100$ blocks.
Autocorrelation times on other ensembles are, if anything, smaller than this.}
Hence, we assemble for each ensemble $N_{J}=100$ jackknife pseudoensembles of the correlation function data for each
$(\bm{r},\tau)$.
From these $N_{J}=100$ jackknife pseudoensembles, we estimate the correlation matrix, which obviously has non-zero
off-diagonal entries in both directions of the $(\bm{r},\tau)$ space.
The available data span an $(\bm{r},\tau)$-space of $\order(10^{2})$ to $\order(10^{4})$ points.
While proximity in the $\tau$-direction certainly provides a hint on the actual strength of the correlations, such a
naive expectation is not justified at all towards proximity in~$|\bm{r}|$.
Given $N_{J}=100$ jackknife pseudoensembles, we may expect to be able to obtain good estimates for $\sqrt{N_{J}}=10$
eigenvalues.
In order to avoid or reduce eigenvalue smoothing\footnote{%
We follow standard procedures~\cite{Michael:1994sz} with adaptations spelled out in the text.} as much as possible, we
have to slice the data and reduce them to a subset of about $\sqrt{N_{J}}=10$ points in $(\bm{r},\tau)$-space and
estimate the correlation matrix for that subset.
In order to propagate the statistical correlations of the correlation function into the analysis of $\bm{r}$-dependence
of the static energy $E_{0}(\bm{r},a)$, we repeat the analysis on the original sample and on all $N_{J}=100$ jackknife
pseudoensembles.

In the correlation function fits discussed in this section, we slice $(\bm{r},\tau)$-space in the $\bm{r}$ direction,
i.e., we consider the correlation matrix only between data at different $\tau$ for the same $\bm{r}$.
If the correlated fit\footnote{%
We use R statistical package~\cite{Rpackage} with the NLME library~\cite{nlme} for these fits.} (on the original
sample) does not converge, then we thin out the set of $\tau$ values---potentially going down to zero degrees of
freedom---by iteratively eliminating one datum in a randomized manner (keeping at least two data in each of the first
$\sfrac{2}{3}$ and last $\sfrac{1}{3}$ of the fit interval) and repeating the fit attempt.
If we include $N_{D} \le \sqrt{N_{J}}=10$ data, we do not smooth eigenvalues of the correlation matrix.
Otherwise, if we include $N_{D} > 2\sqrt{N_{J}}=20$ data, we smooth the $N_{D}-\sqrt{N_{J}}$ lowest eigenvalues; or
else, we apply smoothing to the $\sfrac{N_{D}}{2}$ lowest eigenvalues.
In some cases we have one large and copiously many very small eigenvalues\footnote{%
Such cases occur typically at small $\sfrac{\bm{r}}{a}$ and even more so with smeared links.
As some of these fits failed altogether, we have missing entries in the $(\sfrac{\tmin}{a},\Nstates)$-table of results
for some $aE_{0}(\bm{r},a)$.} leading to a condition number of $\order(10^{6})$ even after the smoothing; correlated
fits for these cases could only succeed by means of the randomized thinning out of the fit interval.
We also perform uncorrelated fits by neglecting the off-diagonal elements of the correlation matrix, which never
require the randomized thinning out of the fit interval, and thus provide additional cross-checks on the results.
Below, in the study of the $\bm{r}$ dependence, $\tau$ is not a variable anymore, and we consider the correlation
matrix between data at different $\sfrac{|\bm{r}|}{a}$; see Sec.~\ref{sec:interpolation}.

For $0 < n \le \Nstates$, we use Bayesian priors for $C_{n}(\bm{r},a)$, $aE_{0}(\bm{r},a)$, and $a\Delta_{n}(\bm{r},a)$.
The prior distributions in $\chi^{2}_{\text{prior}}(\{C_{n},aE_{n}\})$ are of Gaussian form for each parameter, i.e.,
\begin{equation}
    \chi^{2}_{\text{prior}}(\{C_{n},aE_{n}\}) = \frac{(aE_{0} - a\tilde{E}_{0})^{2}}{\sigma^{2}_{a\tilde{E}_{0}}} +
        \sum\limits_{n=0}^{\Nstates-1} \frac{(C_{n} - \tilde{C}_{n})^{2}}{\sigma^{2}_{\tilde{C}_{n}}} +
        \sum\limits_{n=1}^{\Nstates-1} \frac{\left[a\Delta_{n} - a\tilde{\Delta}_{n}\right]^{2}}
            {\sigma^{2}_{a\widetilde{\Delta}_{n}}},
    \label{eq:chi2prior}
\end{equation}
where the reasoning behind central values and width is spelled out in detail in the following.
Since we are interested only in the ground state energy, $aE_{0}(\bm{r},a)$, we marginalize over the parameters related
to the excited states; therefore, priors may be chosen to aid the stability of the fits, particularly during resampling.
To set the prior central values and widths for fits with $\Nstates=1$, we automate procedures based on the effective
mass and scaled correlator.
Automation is necessary because of the large (1000s) of Wilson-line correlators in this work.

In practice, we use the results of the $\Nstates=1$ fits as the starting guess for the $\Nstates=2$ fits and similarly
for the $\Nstates=3$ fits.
On each resample we follow the exact same procedure that was used on the mean to choose the Bayesian priors to determine
starting values for the fit parameters.

For fits with $\Nstates>1$, the choice of priors faces several challenges.
Since the values of the overlap factors $C_{n}(\bm{r},a)$ change by an order of magnitude across the available $\bm{r}$
range, we cannot use a simple functional form that works over a wide $\bm{r}$ range.
A further challenge is the decrease of the ground state overlap factor $C_{0}(\bm{r},a)$ and the increase of the ground
state energy $aE_{0}(\bm{r},a)$ for larger $|\bm{r}|$, which gets compounded with an increase of the excited state
overlap factors $C_{n}(\bm{r},a)$ and the decrease of the excited state energy differences $a\Delta_{n}(\bm{r},a)$.
These features require the priors to become narrower for larger $|\bm{r}|$.
Further, we require priors on the ground state parameters to avoid an outcome where the parameter $C_{0}(\bm{r},a)$
approaches zero with poorly constrained $aE_{0}(\bm{r},a)$, while $aE_{1}(\bm{r},a)$ approaches the true ground state
energy.
Thus, we use multiple stages of simpler fits for each $\bm{r}$ to gain information for use as prior knowledge in fits
with larger $\Nstates$.
We ensure for all ground state parameters, i.e., $(aE_{0}(\bm{r},a),C_{0}(\bm{r},a))$, loose priors with a width of at
least 10\%, which is orders of magnitude wider than the respective statistical uncertainties.
For the excited state energy differences $(a\Delta_{1}(\bm{r},a),a\Delta_{2}(\bm{r},a))$, we use loose priors with
widths of 10\% or more.
Lastly, for the excited state overlap factors $(C_{1}(\bm{r},a),C_{2}(\bm{r},a))$, we determine very loose priors in
terms of small positive values with widths of at least 100\%.
Due to their large widths, the individual priors do not rely on unacceptable examination of the data and could be
modified without significant changes of the fit results.

In more detail, our procedures are as follows:
\begin{enumerate}
\renewcommand{\labelenumi}{(\roman{enumi})}
\item
For fits with $\Nstates=1$, we estimate the initial parameters, central values and widths of the priors via linear
regression.
For fits with any $\Nstates$, we assign $10\%$ of the respective central value or $100\%$ of the previous error
(estimate)---whichever is larger---to the widths of the two priors related to the ground state.
The main purpose of the fits with $\Nstates=1$ in our analysis is to suggest suitable central values of the priors for
the ground state parameters $C_{0}(\bm{r},a)$ and $aE_{0}(\bm{r},a)$ in the ensuing fits with $\Nstates=2$.
\item
The fits with $\Nstates=2$ serve as our main result, as we are interested only in the ground state energy, i.e.,
$aE_{0}(\bm{r},a)$.
We use the (uncorrelated) fits with $\Nstates=1$ to obtain prior central values for the ground-state parameters.
We assign $10\%$ of this central value or $100\%$ or the $\Nstates=1$ error (estimate)---whichever is larger---to
the widths of the two priors related to the ground state.
For the energy difference $a\Delta_{1} = aE_{1} - aE_{0}$, we take a calculation in $\text{SU}(3)$ pure gauge
theory~\cite{Juge:2002br, *Morningstar:2002br} fit to a Cornell parametrization,
\begin{equation}
    a\Delta_{1} =
         -\frac{A}{R} + a_{f_{p4s}}B + a_{f_{p4s}}^{2} \sigma R,
    \label{eq:Cornell_delta1}
\end{equation}
with $A=-0.09364$~GeV~fm, $B=1.11218$~GeV, and $\sigma=-0.309585~\text{GeV\,fm}^{-1}$; here $R$ is a dimensionless
measure of distance defined in Sec.~\ref{sec:scales}, and we employ $a_{f_{p4s}}$ from Table~\ref{tab:ensembles} to
convert the right-hand side to lattice units.
As we do not have robust prior information about the overlap factor $C_{1}(\bm{r},a)$ in ($2+1+1$)-flavor QCD, we
choose a fairly loose prior $C_{1}(\bm{r},a) = 0.10(0.10)$, which coincides with the usual sign and order of magnitude
seen in earlier stages of the analysis.
To err on the side of caution, we assign $20\%$ of the respective central value to the width of the prior related to
$a\Delta_{1}$.
\item
For fits with $\Nstates=3$, we use the (uncorrelated) fits with $\Nstates=2$ to obtain prior central values for the
ground-state and first excited-state parameters.
We retain the assignment of $10\%$ of the respective central value or $100\%$ of the previous error
(estimate)---whichever is larger---to the widths of the priors related to these states.
However, we choose a width of $0.10$ or 100\%---whichever is larger---for the overlap factor $C_{1}(\bm{r},a)$ of
the first excited state since we anticipate that we may have been incapable of separating it from the second excited
state in the fit with $\Nstates=2$.
For $a\Delta_{2}$, we choose $2a\Delta_{1}$ and $\sfrac{a\Delta_{1}}{2}$ as the prior central value and width,
respectively.
As we have even less prior information about the overlap factor $C_{2}(\bm{r},a)$, and since it is known that the
correlation functions with Symanzik action contain negative spectral weights for small $\sfrac{|\bm{r}|}{a}$, see,
e.g., Refs.~\cite{Bazavov:2019qoo, Bala:2021fkm}, we choose a very loose prior $C_{2}(\bm{r},a) = 0.02(0.20)$ since
this coincides with magnitude seen in earlier stages of the analysis.
The main purpose of the fits with $\Nstates=3$ in our analysis is to serve as cross-checks that confirm that neither of
the two lowest states would be modified significantly if another state were added.
\end{enumerate}

With these priors in hand, we define an augmented $\chi^{2}$ function for each $(\bm{r},a)$:\footnote{%
The label $(\bm{r},a)$ for various quantities is suppressed to reduce clutter.}
\begin{align}
    \chi^{2}_{\text{aug}} (\{C_{n},aE_{n}\}) &= \chi^{2}_{\text{data}}(\{C_{n},aE_{n}\}) +
        \chi^{2}_{\text{prior}}(\{C_{n},aE_{n}\}),
    \label{eq:chi2aug} \\
    \chi^{2}_{\text{data}} (\{C_{n},aE_{n}\}) &= \sum\limits_{u,w \in \sfrac{[\tmin,\tmax]}{a}}
        \Delta(u;\Nstates|\{C_{n},aE_{n}\}) [\sigma^{-2}]_{uw} \Delta(w;\Nstates|\{C_{n},aE_{n}\}),
    \label{eq:chi2data} \\
        \Delta(u;\Nstates|\{C_{n},aE_{n}\}) &= C(u) - F(u;\Nstates|\{C_{n},aE_{n}\}),
    \label{eq:chi2-residual}
\end{align}
where $C(u)$ denotes a Monte Carlo estimate of the correlator $C(\bm{r},ua,a)$, $\sigma^{2}$ their covariance in the
sample, and $F$ the right-hand side of Eq.~\eqref{eq:tower} truncated to $\Nstates$ states and considered to be a
function of the $C_{n}$ and $aE_{n}$ and parametrized by the lattice time $u$ (or $w$).
The prior term $\chi^{2}_{\text{prior}}$ is given in Eq.~\eqref{eq:chi2prior} above.
For each $(\bm{r},a)$, we minimize $\chi^{2}_{\text{aug}}$ to obtain the best-fit values of $(\{C_{n},aE_{n}\})$,
$0 \le n<\Nstates$.

We show representative plots of $p$-value distribution (across the $N_{J}=100$ jackknife pseudoensembles) for the
physical \ensemble{7.00}{i} ensemble in Fig.~\ref{fig:p-values} in the Appendix~\ref{app:correlator_fits}.
Here, $p$ is defined as described in Appendix~B of Ref.~\cite{FermilabLattice:2016ipl}.
The energy levels and overlap factors in Fig.~\ref{fig:gaugefix_fits} (concerning another ensemble, namely the physical
\ensemble{6.72}{i} ensemble), suggest that constraining excited states is challenging at small distances, hence the
presence of a few outliers for small $\sfrac{|\bm{r}|}{a}$.
At large enough $\sfrac{|\bm{r}|}{a}$ the distribution is quite flat and close to the ideal case, suggesting that the
fit functions are good descriptions of the data.

We estimate the statistical errors of the fit parameters from either the Hessian matrix of the fit or from the
distribution of the jackknife resamples.
We find these two estimates to be similar in magnitude without obvious trends of one being usually larger or smaller
than the other.
In practice, we keep the jackknife error estimate and propagate statistical correlations in terms of the resamples.
Moreover, given the stability analysis for the physical \ensemble{7.0}{i} ensemble---see
Fig.~\ref{fig:stability_analysis} in Appendix~\ref{app:correlator_fits}---we find that the influence of reasonable
variation of $\Nstates$ or $\sfrac{\tmin}{a}$ is covered by this statistical error estimate, so we do not modify the
error of the ground state energy $aE_{0}(\bm{r},a)$ further.
Similarly, including or neglecting the off-diagonal entries of the correlation matrix does not lead to a statistically
significant or systematic trend in the results.
Because the uncorrelated fits do not require the randomized thinning in~$\tau$, described above, we carry these results
to the next step as a cross-check.
The final result of this analysis consists of the $(\sfrac{\tmin}{a},\Nstates)$ table of $aE_{0}(\bm{r},a)$ and the
respective (statistical) error estimate, each on the mean and on the $N_{J}=100$ jackknife pseudoensembles, and each
with or without including the off-diagonal elements of the correlation matrix.
This analysis permits quantitative studies of the impact of the various uncertainties on the physical results obtained
in Secs.~\ref{sec:scales}, \ref{sec:cont-limit}, and~\ref{sec:charm}.
The results for $aE_{0}$ for the $\Nstates=2$ fits are contained in Supplemental Material~\cite{SupplMat}.

\section{Fits of the static energy}
\label{sec:scales}

In this section, we take the results from the $\Nstates=2$ fits described in Sec.~\ref{sec:sim} to determine the
``potential'' scales $\sfrac{r_{i}}{a}$, $i=0,1,2$ and the string tension $a^{2}\sigma$.
The scales $r_{i}$ are defined in Eq.~\eqref{eq:scales} via the force in Eq.~\eqref{eq:force}.
Earlier calculations in ($2+1$)-flavor QCD~\cite{MILC:2010hzw, HotQCD:2014kol, Bazavov:2017dsy} find the scales to be
\begin{equation}
    r_{0} \approx 0.475~\text{fm}, \quad r_{1} \approx 0.3106~\text{fm}, \quad r_{2} \approx 0.145~\text{fm},
    \label{eq:scale_estimates}
\end{equation}
corresponding to distinct physical regimes.
On the one hand, $r_{2} \sim \sfrac{1}{\mc}$ is similar to the inverse charm quark mass and, being right at the edge of
the perturbative regime, expected to be insensitive to the light sea quarks.
On the other hand, $r_{0} \sim \sfrac{1}{\LQCD}$ is in the non-perturbative regime and, hence, is known to be sensitive
to the pion mass, but is expected to be insensitive to charm sea quarks.
As $r_{1}$ is in between these two, it might be sensitive to both the light and the charm quarks in the sea.
At distances beyond $r_{0}$, but before string breaking, the force is a constant, namely the ``string tension''
$\sigma=-F(r)$, $r_{0} \lesssim r \lesssim 1~\text{fm}$.
As discussed in Sec.~\ref{sec:sim}, our data set is intended to obtain accurate results for the scales $r_{i}$ (and
$\als$), rather than the string tension, which we obtain from data with $r\ge0.58$~fm.

At non-zero lattice spacing, the static energy is available only at discrete distances, requiring some sort of
numerical derivative in place of Eq.~\eqref{eq:scales}.
Moreover, $aE_{0}(\bm{r},a)$ depends on the direction of $\bm{r}$, so it is not a smooth function of the usual spatial
Euclidean distance~$r=|\bm{r}|=\sqrt{x_{1}^{2}+x_{2}^{2}+x_{3}^{2}}$.
A good alternative is a measure of distance defined via the tree-level gluon propagator, known as the tree-level
corrected distance~\cite{Bazavov:2019qoo}.

In the following, we begin with the explicit definition of the tree-level corrected distance in
Sec.~\ref{sec:artifacts}.
We then proceed in Sec.~\ref{sec:interpolation} to the fits of the static energy, which yield the force and, thus,
values for the scales $\sfrac{r_{i}}{a}$ and string tension $a^{2}\sigma$ in lattice units and at fixed lattice spacing.
This subsection includes discussion of the results, including the quark mass dependence and a comparison to earlier
work.
For relative scale setting in future work, it is convenient to combine the data in a fit to a smooth
curve~\cite{Allton:1996kr, MILC:2009mpl}, which we do in Sec.~\ref{sec:quark_mass_dependence}.

\subsection{Discretization artifacts and tree-level correction}
\label{sec:artifacts}

On the lattice, the static energy is given at the tree level of perturbation theory by one-gluon exchange, just like in
the continuum.
In Coulomb gauge, its temporal component reads
\begin{equation}
    D_{44}(k) = a^{2} \left[4\sum\limits_{j=1}^{3} \sin^{2}\left(\half ak_{j}\right)
        + c_{w} \sin^{4}\left(\half ak_{j}\right)\right]^{-1},
    \label{eq:gluon_propagator}
\end{equation}
where $c_{w}=0$ for the (unimproved) Wilson gauge action and $c_{w} = \sfrac{1}{3}$ for the (improved) Lüscher-Weisz
action~\cite{Luscher:1985zq}.
As in the continuum, this component is independent of $k_{4}$ (in Coulomb gauge).
For bare links, one simply takes the Fourier transform,
\begin{equation}
    E_{0}^{\text{tree}}(\bm{r},a) =
        -\CF g_{0}^{2} \int \frac{\d^{3} k}{(2\pi)^{3}} \e{\i \bm{k} \cdot \bm{r}} D_{44}(k)
        \equiv -\frac{\CF g_{0}^{2}}{4\pi}\frac{1}{r_{I}},
    \label{eq:tree_level_E}
\end{equation}
where $g_{0}$ is the bare gauge coupling, $\CF = \sfrac{(\Nc^{2}-1)}{(2\Nc)}$ is a color factor, and the last
expression defines $r_{I}$, which is discussed further below.
Because the gluon propagator is a direction-dependent function of $\bm{k}$, the static energy $E_{0}(\bm{r},a)$ is a
non-smooth function of the Euclidean distance~$r$.
Even beyond the tree level, one finds that the static energy is much smoother in $r_{I}$, which we refer to below as
the tree-level improved or tree-level corrected distance.
For example, $r=3a$ for both $\bm{r}=(3,0,0)a$ and $(2,2,1)a$, but $r_{I}(3,0,0)=2.979a$ while $r_{I}(2,2,1)=3.013a$.
Even beyond the tree level, $E_{0}(3,0,0)<E_{0}(2,2,1)$.
We have computed the tree-level corrected distances $\sfrac{r_{I}}{a}$ in the infinite-volume limit for each vector
$\sfrac{\bm{r}}{a}$ with $\sfrac{|\bm{r}|}{a} \le 6$ both for bare links or for links after one step of HYP
smearing~\cite{Hasenfratz:2001hp} using the \texttt{HiPPy} software package and the \texttt{HPsrc} software
framework~\cite{Hart:2004bd, Hart:2009nr}.
HYP smearing introduces a nontrivial vertex on either side of $D_{44}$ in Eq.~(\ref{eq:tree_level_E}), thus modifying
$r_{I}$.
For example, in this case $r_{I}(3,0,0)=3.020a$ while $r_{I}(2,2,1)=2.997a$.
The results are in Table~\ref{tab:tree-level_corrections_both} in Appendix~\ref{app:tree-level corrections} for bare
and HYP-smeared links.\footnote{%
These results are part of an ongoing project aiming at a full one-loop calculation in lattice perturbation
theory~\cite{lpt_paper}.
To our knowledge, the improved distance with HYP smearing appears in Table~\ref{tab:tree-level_corrections_both} for
the first time.}
The former are consistent with previous results~\cite{Bazavov:2014soa, Bazavov:2018wmo} up to very small finite-volume
effects.

For the rest of this section, it is convenient to switch to lattice units.
We introduce $E(R,a)=aE_{0}(\bm{r},a)$ and $R=\sfrac{r_{I}}{a}$.
The tree-level correction reduces the size of non-smooth discretization artifacts considerably but not completely.
\begin{figure}
    \centering
    \includegraphics[width=0.49\textwidth]{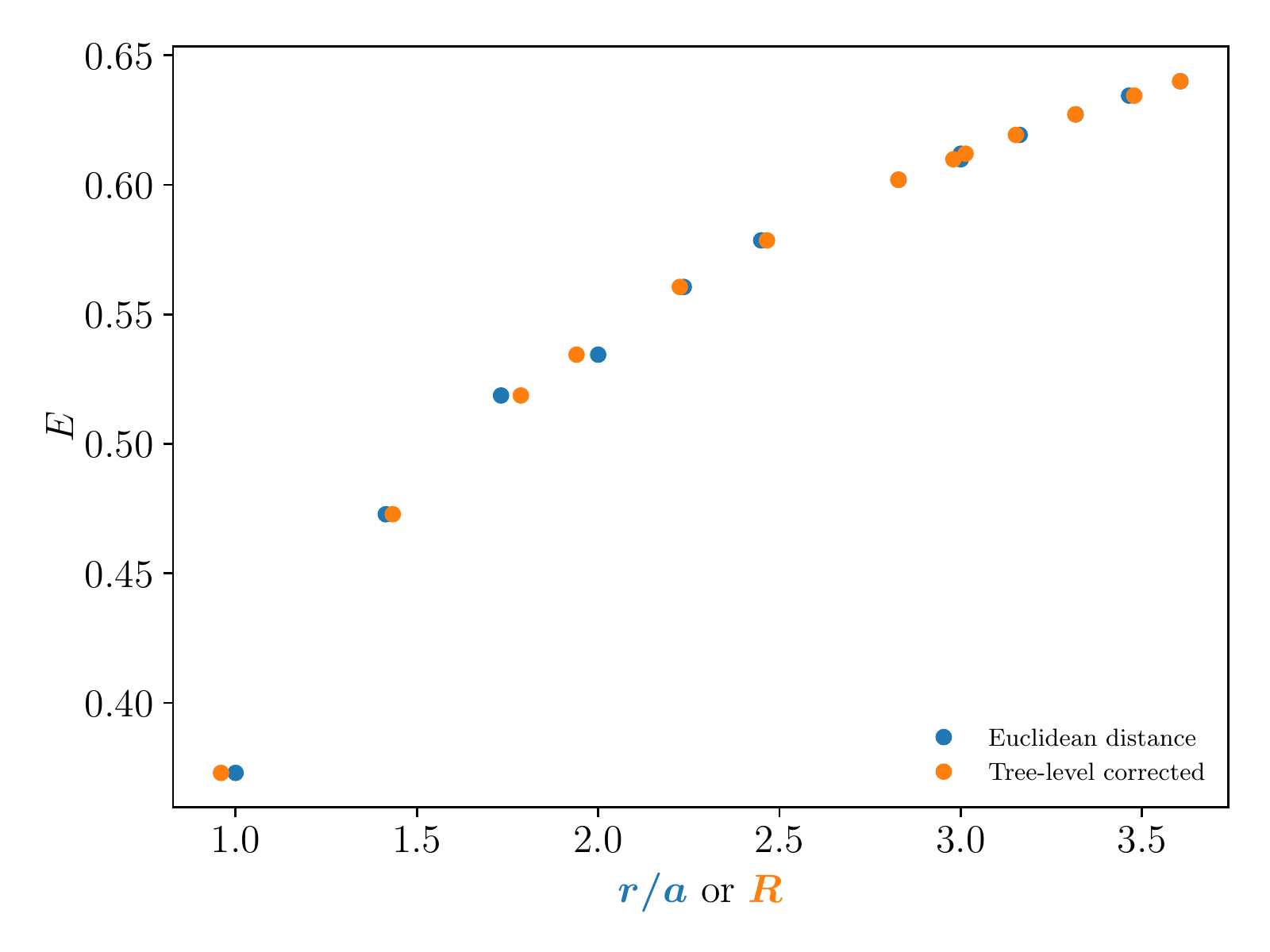}%
    \hfill%
    \includegraphics[width=0.49\textwidth]{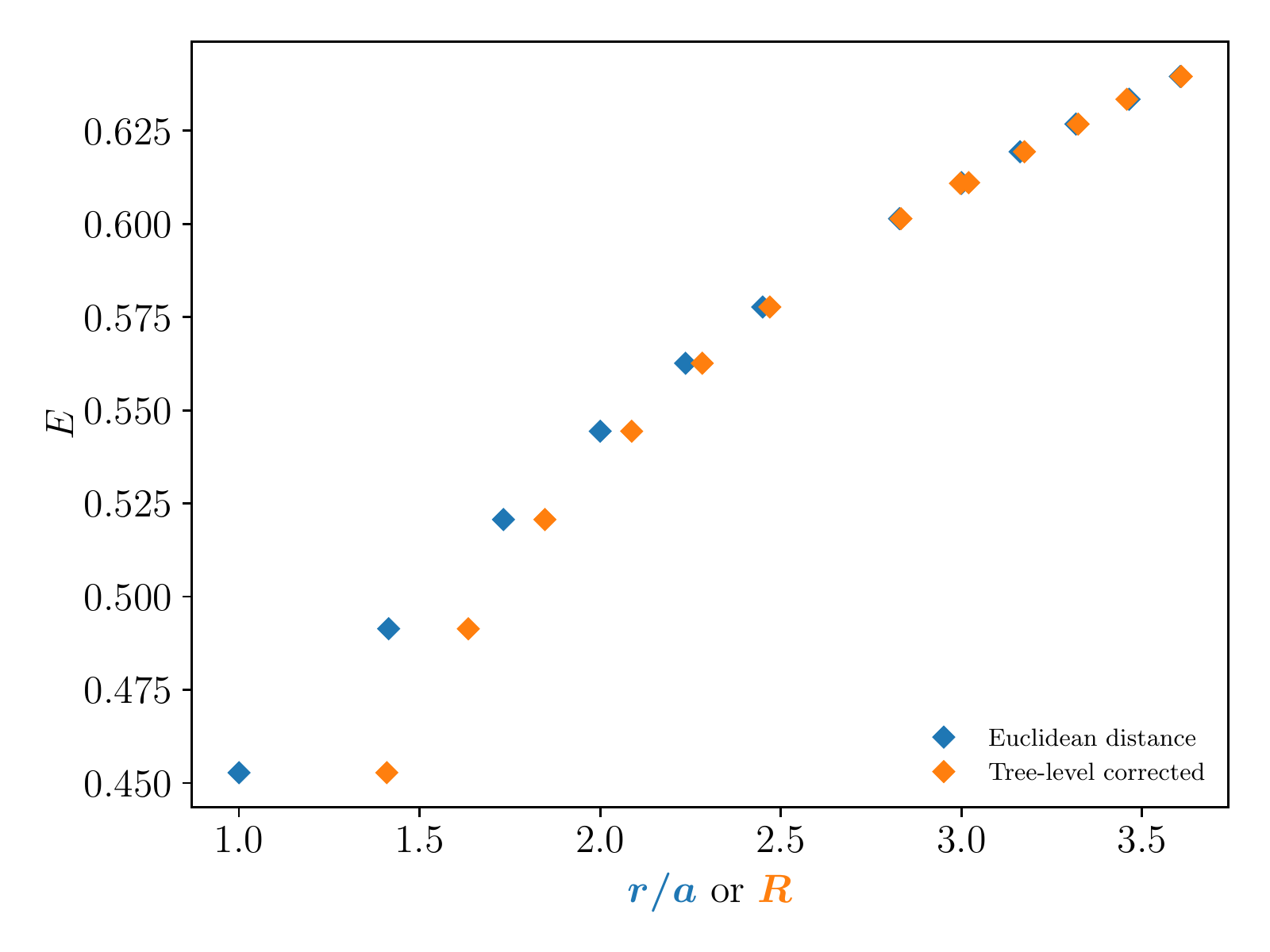}%
    \caption{\label{fig:tree_level_correction}%
        The static energy $E(R,a)$ from the fits with $\Nstates=2$ with preferred $\sfrac{\tmin}{a}$ for the physical
        \ensemble{7.00}{i} ensemble, vs two measures of the distance.
        Left: data from bare links (right: HYP-smeared).
        The static energies plotted against the tree-level improved distance $R$ (Euclidean distance $\sfrac{r}{a}$)
        are colored orange (blue).
        The static energies with bare links roughly follow a $\sfrac{1}{r}$ in terms of both distances measures up to
        small non-smooth discretization artifacts; cf., Fig.~\ref{fig:tree_level_correction-2}.
        The static energy with HYP-smeared links is far from Coulomb-like when plotted against the Euclidean distance
        $\sfrac{r}{a}$, when $\sfrac{r}{a}\lesssim2.5$, but using the improved distance $R$ removes this distortion.
        Serious non-smooth discretization artifacts remain; cf., Fig.~\ref{fig:tree_level_correction-2}.}
\end{figure}
Figure~\ref{fig:tree_level_correction} shows how the results on the \ensemble{7.00}{i} ensemble change (apparent) shape
when switching from the Euclidean distance $\sfrac{r}{a}$ to the improved distance~$R$.
The behavior is similar to previous calculations in ($2+1$)-flavor QCD~\cite{Bazavov:2019qoo}; see the side-by-side
comparison of Figs.~12 and 13 of Ref.~\cite{Bazavov:2019qoo}.
The improvement, especially for HYP-smeared data, is readily apparent.
That said, a closer look---dividing the data by a Cornell fit over the range $2.7\le R\le4.7$ as in
\begin{figure}
    \centering
    \begin{minipage}[t]{0.49\textwidth}\vspace{0pt}%
        \includegraphics[width=\textwidth]{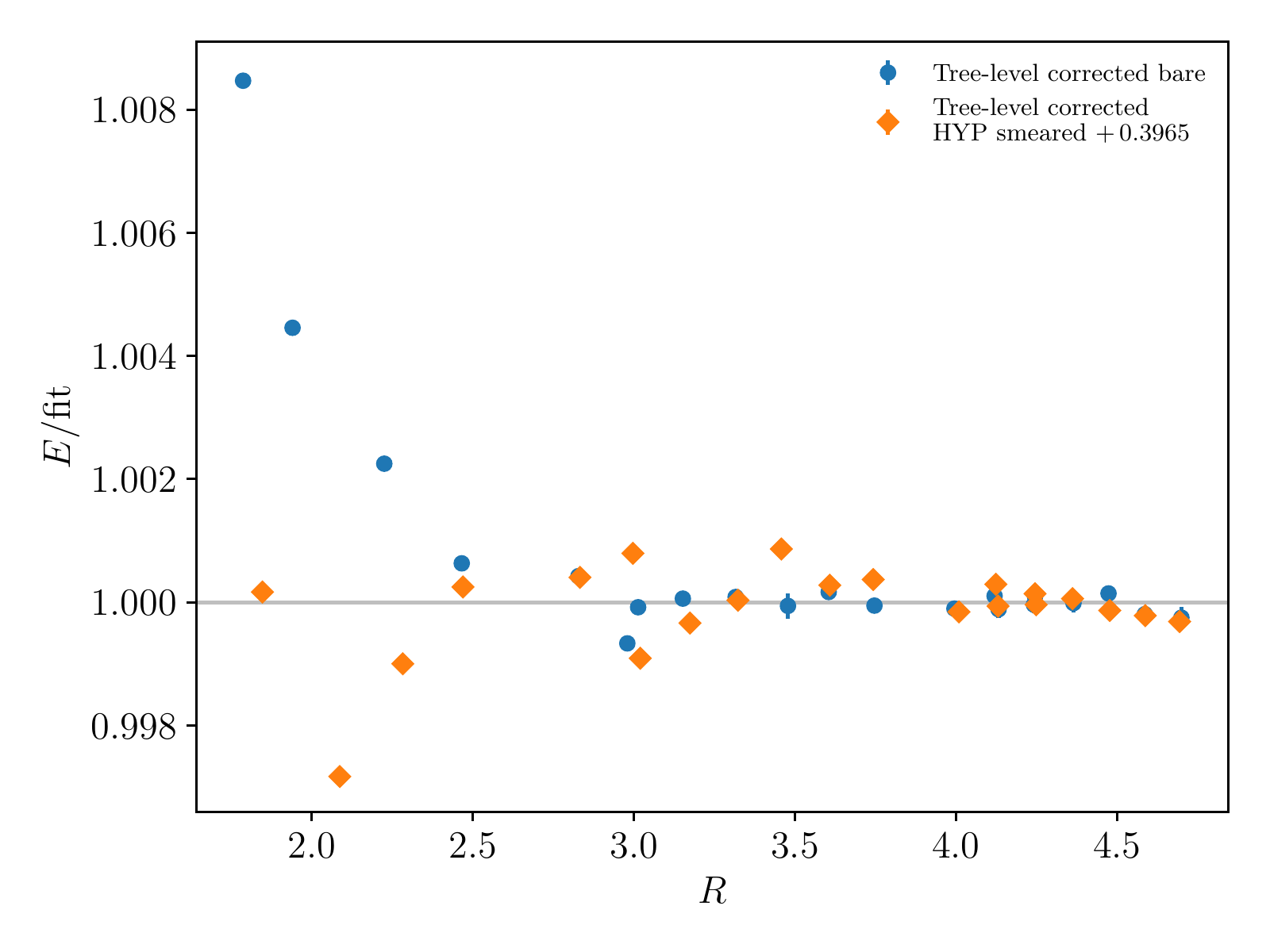}%
    \end{minipage}%
    \hfill%
    \begin{minipage}[t]{0.49\textwidth}\vspace{0pt}%
        \caption{\label{fig:tree_level_correction-2}%
        The static energy $E(R,a)$ again on the \ensemble{7.00}{i} ensemble, divided by a Cornell-fit performed in the
        range $2.7\le R\le4.7$, for bare-link (blue circles) and HYP-smeared (orange diamonds) data.
        Even after using the tree-level improved distance $R$, residual non-smooth discretization artifacts remain:
        the bare-link data are not smooth at $R=3$ and $R=\sqrt{17}$, for example, while the HYP-smeared data are not
        smooth until (at least) $R>4.5$.}
    \end{minipage}
\end{figure}
Fig.~\ref{fig:tree_level_correction-2}---shows the tree-level correction is insufficient to produce a result for
$E(R,a)$ that is smooth at the level of its statistical errors.
In previous calculations in $(2+1)$-flavor QCD with a much denser set of lattice spacings~\cite{Bazavov:2014soa,
Bazavov:2019qoo}, the residual discretization artifacts were taken care of through a heuristic non-perturbative
correction procedure~\cite{Bazavov:2014soa, Bazavov:2019qoo, Komijani:2020kst}, but here we do not pursue such an
approach.

Correlation functions are distorted at small $\sfrac{|\bm{r}|}{a}$ by contact-term interactions between overlapping
``fat links'' from which the temporal Wilson lines are constructed.
With one iteration of HYP smearing applied to each temporal Wilson line, the distance vectors up to
$\sfrac{\bm{r}}{a}\le(2,2,2)$ are, in principle, affected by such contact terms.
The contribution along the cubic diagonal is suppressed (for the standard choice of parameters, $\alpha_{1}=0.75$,
$\alpha_{2}=0.6$, and $\alpha_{3}=0.3$~\cite{Hasenfratz:2001hp}) by $(0.135)^{2} \approx 2\%$ against a corresponding
``thin link'' contribution, giving rise to effects commensurate with the differences between Euclidean or improved
distances with bare links.
The contact-term contributions remain quantitatively significant even at the maximal range
$\sfrac{\bm{r}}{a}\le(2,2,2)$.
The intermittent ordering of $\sfrac{r}{a}$ for vectors with largest component $2a$ or $3a$ leads to discontinuous
changes in the HYP-smeared result much larger than the tiny statistical errors, see
Fig.~\ref{fig:tree_level_correction-2}, in particular, between $\sfrac{\bm{r}}{a}=(3,0,0)$ and
$\sfrac{\bm{r}}{a}=(2,2,1)$ or between $\sfrac{\bm{r}}{a}=(2,2,2)$ and its neighbors.
To reduce the impact of these discontinuities, we omit $\sfrac{\bm{r}}{a}=(3,0,0)$ and $\sfrac{\bm{r}}{a}=(2,2,2)$ from
our data set with smeared links.

\subsection{Determination of the scales and the string tension from the static energy}
\label{sec:interpolation}

Even as a function of $R$, the lattice result for the static energy $E(R,a)$ contains non-smooth residual
discretization artifacts larger than its statistical errors, yet we require a smooth interpolation to define its
derivative.
We choose the Cornell potential,
\begin{equation}
    E(R,\sfrac{r}{a},a) = -\frac{A}{R} + B + \Sigma R
    \label{eq:Cornell_potential}
\end{equation}
as a functional form because it encodes the main features of the static energy.
In practice, we adjust the constant term~$B$ by adding a shift such that $E((3,0,0),a)+E((2,2,1),a)=0$, i.e.,
$B=\sfrac{A}{R_{\ast}}-\Sigma R_{\ast}$, where $R_{\ast}\equiv\half[R(3,0,0)+R(2,2,1)]$.
We consider $RE(R,a)$ in order to get rid of the leading Coulomb behavior, which results in the functional form
\begin{equation}
    RE(R,a) = -A + BR + \Sigma R^{2} =
        -A \left(1-\frac{R}{R_{\ast}}\right) + \Sigma \left(R^{2}-RR_{\ast}\right).
    \label{eq:fit_function}
\end{equation}
On each ensemble, we fit the data to the right-hand side of Eq.~(\ref{eq:fit_function}) to obtain $A$ and $\Sigma$,
from which we solve $c_{i}=A+\Sigma(\sfrac{r_{i}}{a})^{2}$ to obtain the scale $\sfrac{r_{i}}{a}$ (for each
$i = 0,1,2$).
For fits at large distances, we identify $\Sigma$ with the string tension (in lattice units, i.e.,
$\Sigma=a^{2}\sigma$).
\begin{figure}[b]
    \centering
    \includegraphics[width=0.49\textwidth]{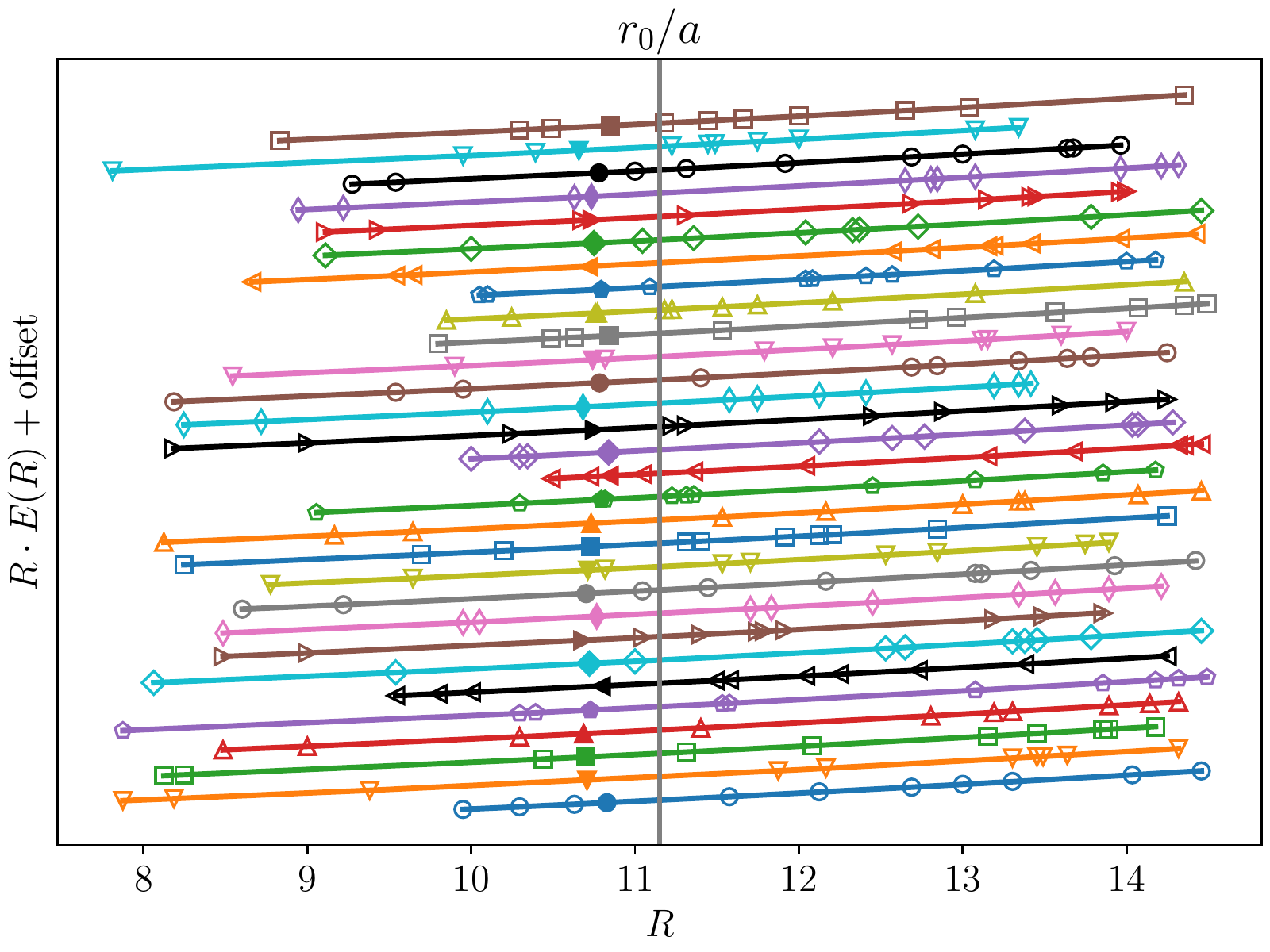} \hfill
    \includegraphics[width=0.49\textwidth]{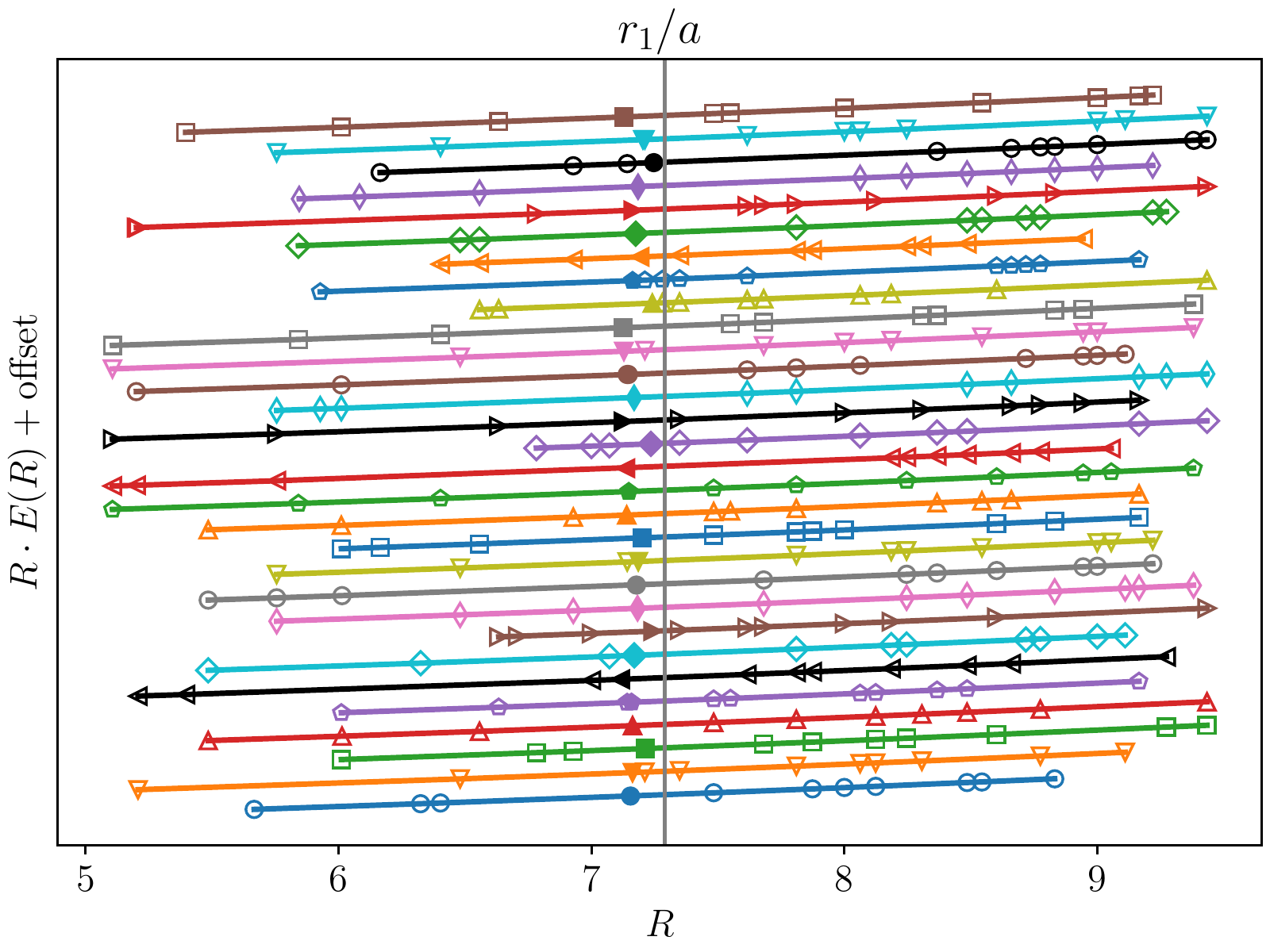} \\
    \begin{minipage}[t]{0.49\textwidth}\vspace{0pt}%
        \includegraphics[width=\textwidth]{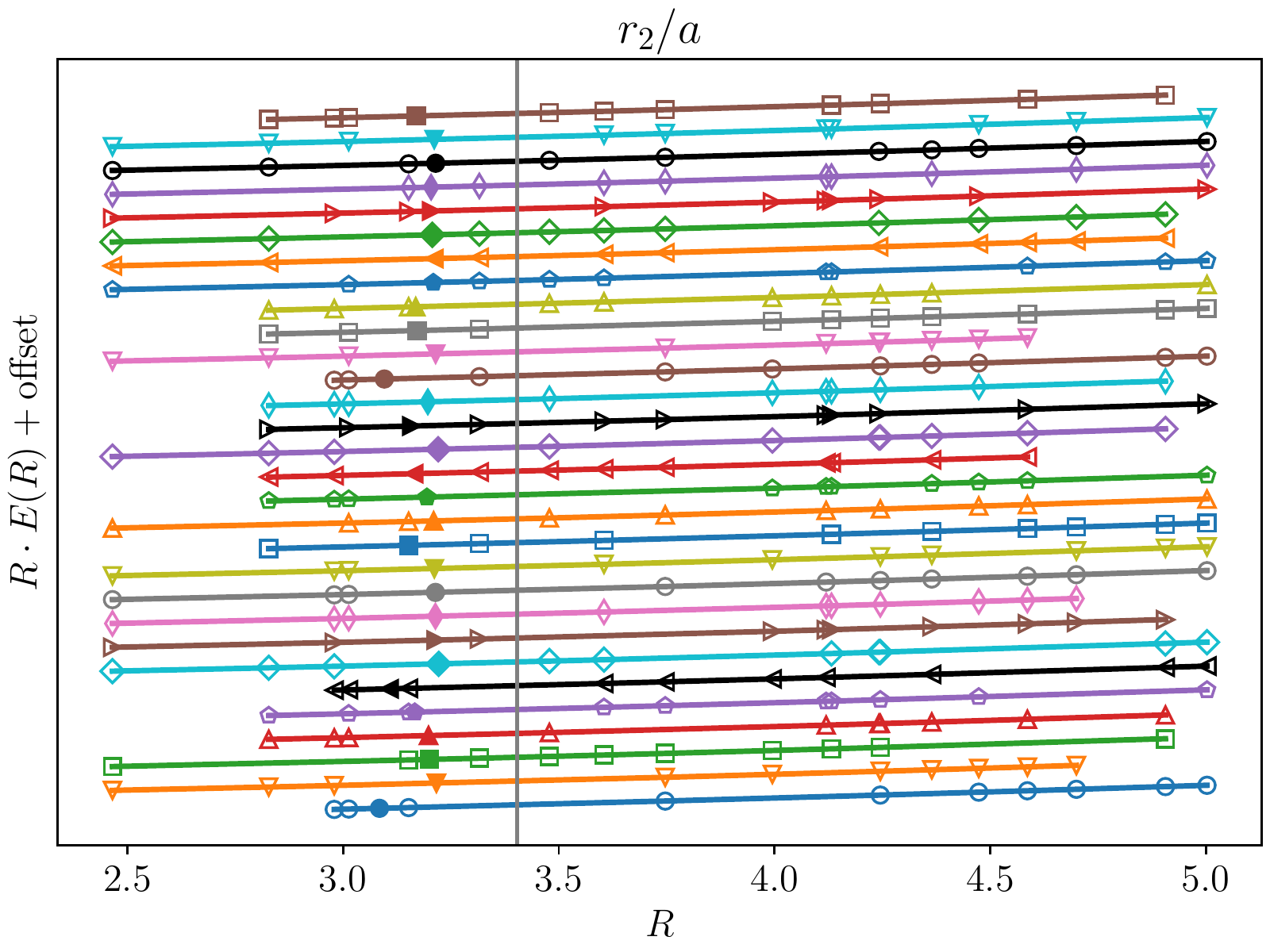}
    \end{minipage}
    \hfill
    \begin{minipage}[t]{0.49\textwidth}\vspace{0pt}%
        \caption{\label{fig:random_picks}%
        First 30 of the $N_{P}$ randomly selected data points (open symbols) and corresponding fit results
        (filled symbols) for the first jackknife pseudoensemble of the bare-link data on the physical
        \ensemble{7.00}{i} ensemble.
        A vertical offset is introduced for clarity, while the colors and symbol shapes are for visual distinction only.
        The separation between the lower and upper half of the interval is indicated by a gray vertical line, based on
        Eq.~\eqref{eq:scale_estimates} and $a_{f_{p4s}}$ as in Table~\ref{tab:ensembles}.}
    \end{minipage}
\end{figure}

{\refstepcounter{response}\label{resp:direction_dependence}}
\new{For tests, we try adding to the right-hand side of Eq.~(\ref{eq:Cornell_potential}) direction-dependent terms
$\kappa_{\text{p}}\Delta_{\text{p}}(\sfrac{r}{a})$ or $\kappa_{\text{LW}}\Delta_{\text{LW}}(\sfrac{r}{a})$, which are
defined via Eq.~\eqref{eq:tree_level_E} in terms of the gluon propagator for the plaquette or Lüscher-Weisz action:
\begin{equation}
    \begin{split}
        \Delta_{\text{p}}(\sfrac{r}{a}) &\equiv \left(\frac{1}{R_{\text{p}}} - \frac{a}{r}\right)=\order(a^{2}), \\
        \Delta_{\text{LW}}(\sfrac{r}{a}) &\equiv \left(\frac{1}{R} - \frac{a}{r}\right)=\order(a^{4}).
    \end{split}
    \label{eq:direction-dependence}
\end{equation}
Here, $R_{\text{p}}$ is the same as $R$ but for the plaquette-action gluon propagator.
The coefficients $\kappa_{\text{P}}$ or $\kappa_{\text{LW}}$ are expected to be numbers of order 1 times leading powers
of $\als^{3}$ or $\als^{2}$, respectively, expected from power counting arguments in the Symanzik effective theory.
The non-smooth contributions from both terms of Eq.~\eqref{eq:direction-dependence} is dropped in the determination of
the force as the derivative of a smooth function.}

These fits entail several challenges.
The Cornell potential is too simple to describe the full range of distances, so for each scale $r_{i}$ we fit to a
narrow interval around $\sfrac{r_{i}}{a_{f_{p4s}}}$ with $r_{i}$ as in Eq.~(\ref{eq:scale_estimates}) and $a_{f_{p4s}}$
as in Table~\ref{tab:ensembles}.
For the ensemble \ensemble{7.28}{iii}, we choose the interval to be $\pm 35\%$ and $\pm{30}\%$, otherwise.
We also require six (fourteen) or more points below (above) $\sfrac{r_{i}}{a_{f_{p4s}}}$ and expand the interval
towards smaller (larger) $\sfrac{r}{a}$ if needed (as happens on the coarser ensembles).
If the latter criterion cannot be met, then we relax it to five (eleven) instead.
The coarsest ensembles cannot provide enough points, particularly below $\sfrac{r_{i}}{a_{f_{p4s}}}$, so then we do not
attempt fits.
In practice, this means we quote results for $\sfrac{r_{1}}{a}$ ($\sfrac{r_{2}}{a}$) only for $\beta>5.80$
($\beta>6.30$).
For the string tension, we fit the range $0.58~\text{fm}\le r_{I}<\rmax$, with \rmax\ from
Table~\ref{tab:data-intervals}.

The next challenge is the correlations among the $aE_{0}$ data in each fit.
As discussed in Sec.~\ref{sec:sim}, we use $N_{J}=100$ jackknife pseudoensembles to estimate the covariance matrix,
permitting good control of up to $\sim\sqrt{N_{J}}=10$ eigenvalues (in practice, of the correlation matrix).
Unless we restrict the fit to only $\sim10$ distinct $R$, we encounter many unphysically small eigenvalues.
We address this issue by thinning the data, choosing ten (fifteen) points at random in the intervals specified above
for the scales $r_{i}$ (string tension).
We pick three $R$ values in the lower half of the interval and seven $R$ values in the upper half for fits of the
scales as illustrated in Fig.~\ref{fig:random_picks}.
This asymmetric selection is motivated by three general properties, namely the decrease of the slope of the static
energy at larger $R$, the increase of the noise at larger $R$, and the higher density of data at larger $R$ due to the
larger number of Euclidean spatial vectors with integer components.
Without the random picks the first two properties would be counterbalanced by the latter in terms of the constraining
power attributable to data at smaller or larger $R$.
In our procedure that uses a fixed number of data, we have to make a somewhat arbitrary choice how to skew the
selection procedure to mimic these properties.
At very large $R$, the data are similarly noisy across the available $R$ range, the slope does not change visibly, and
the density is always fairly high for all ensembles.
Hence, these considerations do not apply, and we pick five $R$ values out of each third of the interval for fits of the
string tension.

Now, the very restrictive fit form, Eq.~\eqref{eq:fit_function}, may underestimate the uncertainties in the derivative,
and a thinned-out set of $R$ values may exaggerate the influence of non-smooth discretization artifacts.
For this reason, we repeat the random picks $N_{P}$ times.
For the finest \ensemble{7.28}{iii} ensemble we use $N_{P}=200$, and for the others $N_{P}=100$.
The same $N_{P}$ sets of random picks are used on each of the $N_{J}$ jackknife pseudoensembles.
The procedure is illustrated in Fig.~\ref{fig:random_picks}, which shows the first 30 fits on the first of the
jackknife pseudoensemble of the physical \ensemble{7.00}{i} ensemble.
Sometimes there are fewer than three points to the left of the fit result, which happens because the separator is set
by $r_{i}$ from ($2+1$)-flavor QCD, Eq.~\eqref{eq:scale_estimates}, and the ($2+1+1$)-flavor QCD scale $a_{f_{p4s}}$ in
Table~\ref{tab:ensembles}.
This effect is found to happen most often for $r_{2}$, which is the distance most sensitive to the charm quark sea.

\begin{figure}
    \centering
    \includegraphics[width=0.49\textwidth]{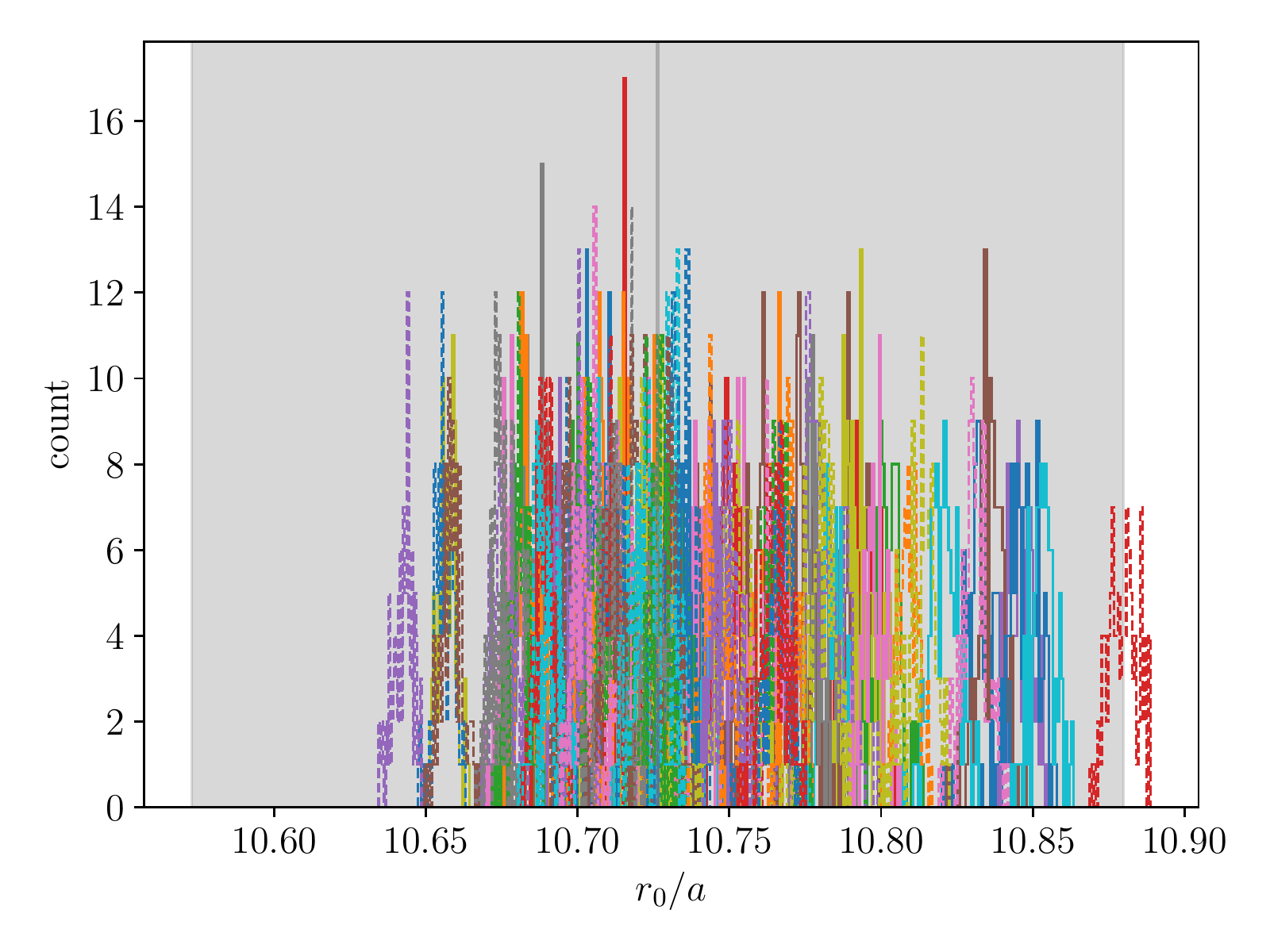} \hfill
    \includegraphics[width=0.49\textwidth]{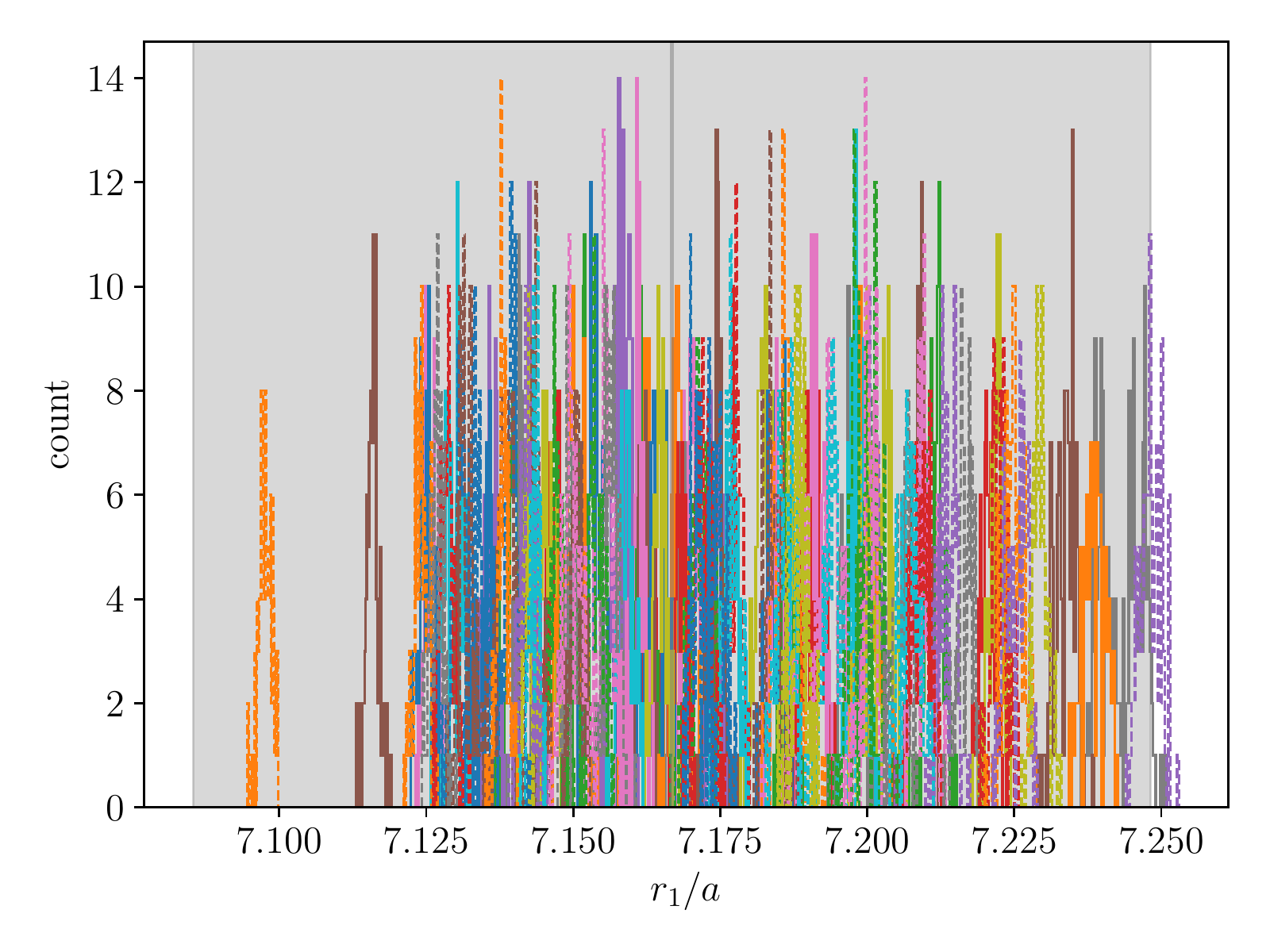} \\
    \begin{minipage}[t]{0.49\textwidth}\vspace{0pt}%
        \includegraphics[width=\textwidth]{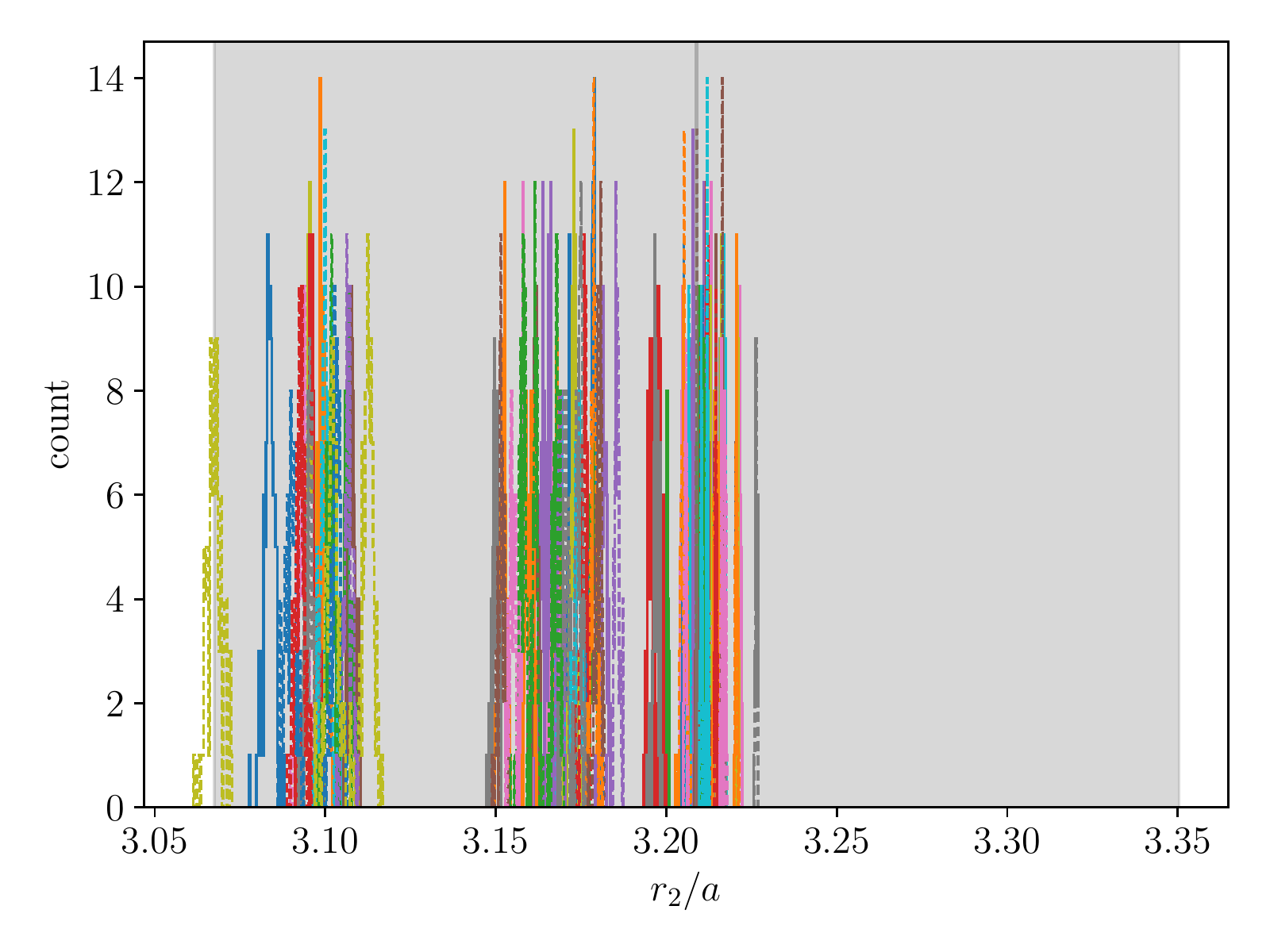}
    \end{minipage}
    \hfill
    \begin{minipage}[t]{0.49\textwidth}\vspace{0pt}%
        \caption{\label{fig:systematical_distribution}%
        Jackknife histograms of the $\sfrac{r_{i}}{a}$ for each of $N_{P}$ random picks, distinguished by color for the
        bare-link data on the \ensemble{7.00}{i} ensemble.
        The gray vertical lines and bands represent the corresponding mean value and error estimate, described in the
        text and collected in Table~\ref{tab:r_i_over_a}. Similar plots for $a^{2}\sigma$ are shown in
        Fig.~\ref{fig:systematical_distribution_sigma} in Appendix~\ref{app:scales and string tension}.}
    \end{minipage}
\end{figure}

While repeating the fit on all jackknife pseudoensembles takes care of the correlated statistical fluctuations, using
different random picks accounts for the systematic uncertainties that arise from non-smooth discretization artifacts
and thinning the data.
For each of the $N_{P}$ sets of random picks, we obtain the mean and statistical error from the variation over the
$N_{J}$ jackknife pseudoensembles.
Figure~\ref{fig:systematical_distribution} shows the jackknife histograms of the $N_{P}$ picks for each of the three
$\sfrac{r_{i}}{a}$.
The variation over the random picks is much larger than the statistical variance of each individual pick.
(Bear in mind that the width of a jackknife histogram has to be multiplied by $\sqrt{N_{J}}$ to get the statistical
error.)
{\refstepcounter{response}\label{resp:direction_dependence_cont}}
\new{Under the natural assumption of some uncorrelated component in the statistical fluctuations across different $R$,
the random picks partially account for statistical errors, too.
It is not a priori obvious whether such statistical or systematic effects dominate the spread of the distribution of
the histograms associated with the random picks.
Statistically significant variation of the (weighted) mean in Fig.~\ref{fig:systematical_distribution} upon including
some direction-dependent parametrization of discretization artifacts via $\kappa_{\text{P}}\Delta_{\text{P}}$ or
$\kappa_{\text{LW}}\Delta_{\text{LW}}$ as in Eq.~\eqref{eq:Cornell_potential} would indicate dominance of the latter,
while a small variation of the (weighted) mean would suggest dominance of the former.
We observe a small variation of the weighted mean, covered by the spread without direction-dependent terms, which
increases slowly toward smaller distances $R$.
We conclude that statistical effects are dominant for the distances considered.}

There are systematic dependencies between the extracted scale $\sfrac{r_{i}}{a}$ and the details of the $N_{P}$ random
picks, which can be visualized if the $N_{P}$ random picks are projected to a more simple measure such as the (randomly
chosen) minimum distance $R_{\text{min}}$.
For example, Fig.~\ref{fig:random_picks_rmin} in Appendix~\ref{app:scales and string tension} shows that the extracted
$\sfrac{r_{i}}{a}$ sometimes is, and sometimes is not, correlated with $R_{\text{min}}$.
For some other cases, these dependencies may be clearly monotonic, rather flat, or clearly non-monotonic.
Lacking clear patterns, we account for the variation by taking for the central value an average of the $N_{P}$
different mean values, weighted by the statistical (jackknife) errors and estimating the systematic uncertainty by
considering the maximal absolute difference between this weighted mean and any of the $N_{P}$ random picks.
This systematic uncertainty estimate is much larger than the (statistical) sample standard deviation for small
$\sfrac{r_{i}}{a}$, but smaller than it for large enough $\sfrac{r_{i}}{a}$.
Nevertheless, the statistical error of the mean---proportional to the sample standard deviation---is practically
always smaller than this systematic uncertainty estimate.
Although the statistical error of the mean is further reduced for the smeared-link result, the systematic error remains
similar.
As a consequence, the benefits of smearing are astonishingly small for the determination of the lattice scale.

Turning to the string tension, our data are insufficient to constrain the coefficient $A$ when fitting the static
energy over the range $r\ge0.58$~fm.
{\refstepcounter{response}\label{resp:string tension}}
This range lies between the Coulomb and (asymptotic) string regime, where a $\sfrac{1}{R}$ behavior is also expected
albeit on very different physical grounds~\cite{Luscher:1980ac}.
With no obvious physical origin for a $\sfrac{1}{R}$ term in this range, we choose fits fixing $A$ to either
$A_{r_{0}}$, the fit results from the $r_{0}$ fit, or $\sfrac{\pi}{12}$~\cite{Luscher:1980ac}.
In fact, $A_{r_{0}}$ turns out to be within a factor of 2 of $\sfrac{\pi}{12}$\new{, and it is natural to expect a
coefficient of an effective $\sfrac{1}{R}$ term within this range}.
As the string tension is not the main objective of this work, we simply present both choices in
Appendix~\ref{app:scales and string tension}.

\begin{figure}
    \centering
    \includegraphics[width=0.49\textwidth]{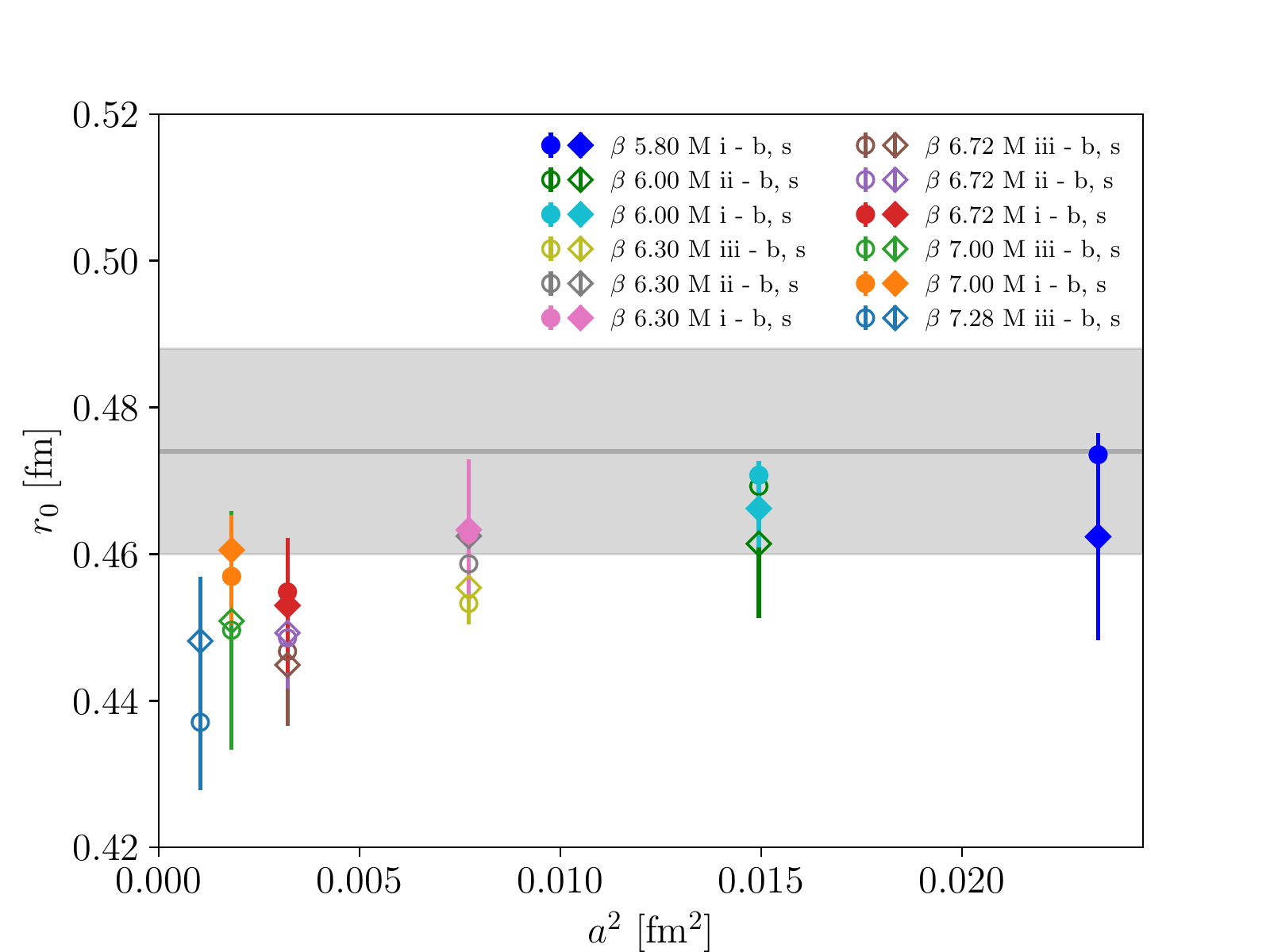} \hfill
    \includegraphics[width=0.49\textwidth]{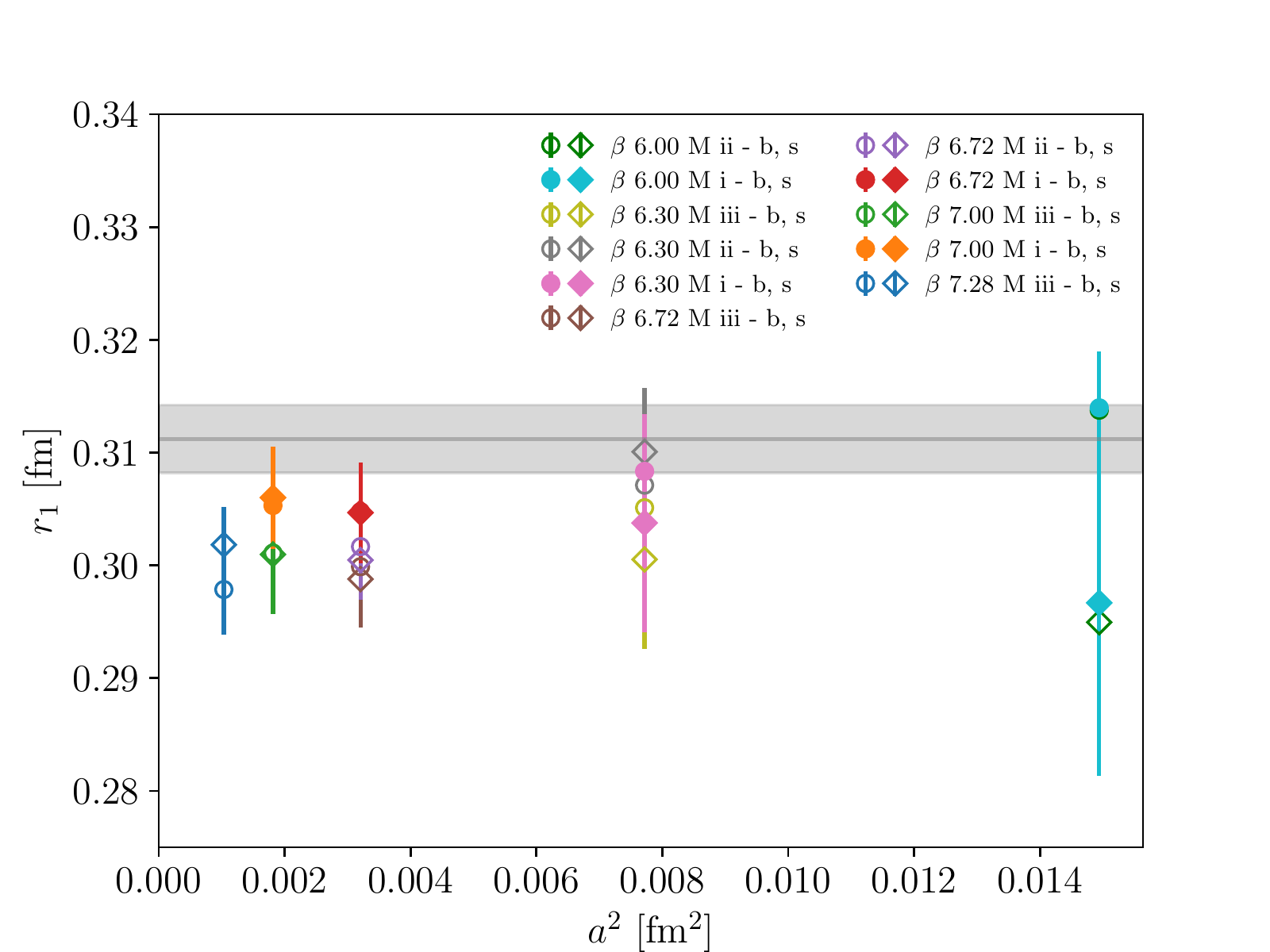} \\
    \begin{minipage}[t]{0.49\textwidth}\vspace{0pt}%
        \includegraphics[width=\textwidth]{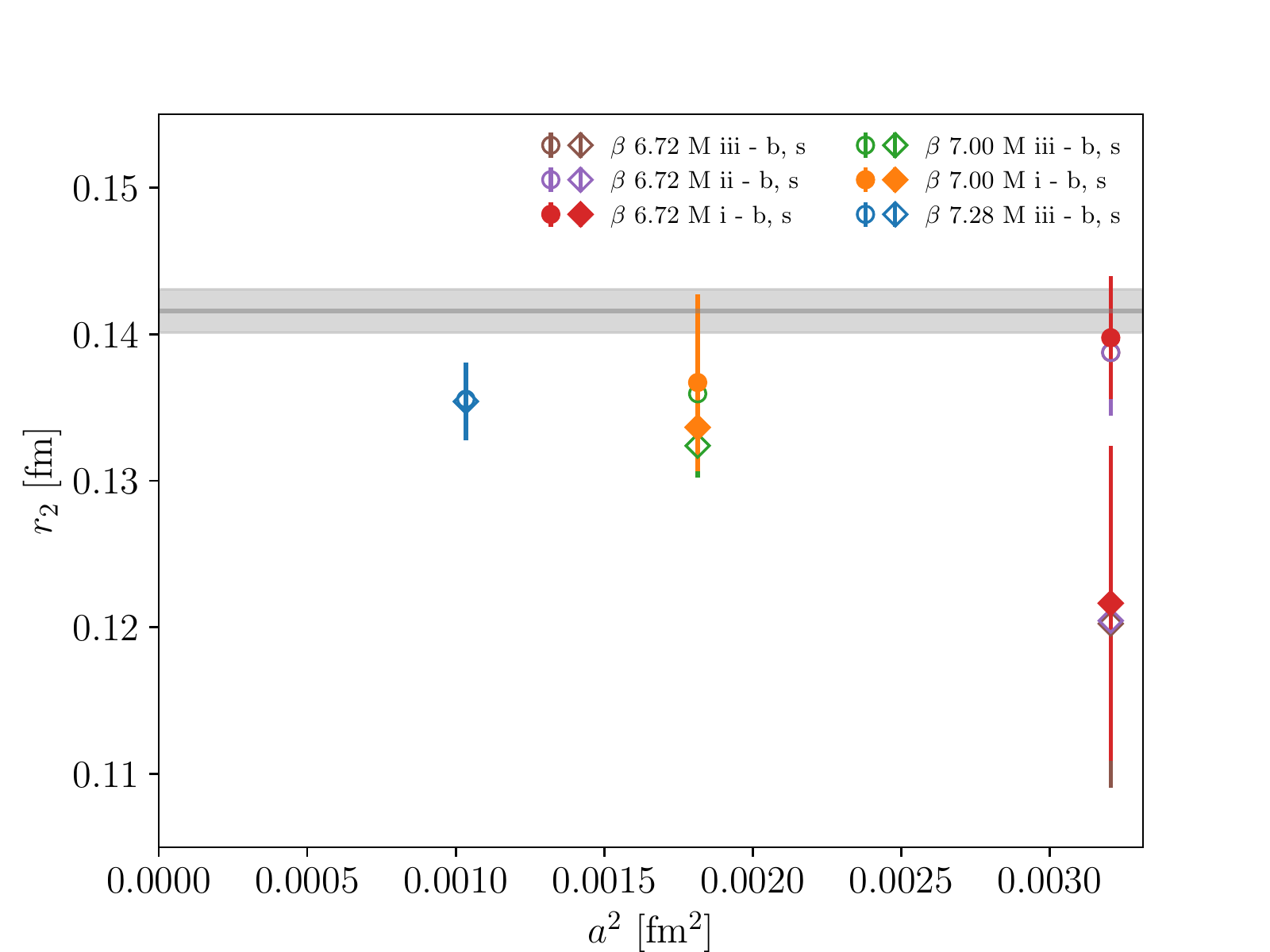}
    \end{minipage}
    \hfill
    \begin{minipage}[t]{0.49\textwidth}\vspace{0pt}%
        \vspace*{1.5em}
        \caption{The potential scales $\sfrac{r_{i}}{a}$, $i=0,1,2$ for all ensembles (indicated by colors) and bare
        ($\circ$) and smeared ($\diamond$) gauge links.
        We use the lattice scale $a_{f_{p4s}}$ to convert our $\sfrac{r_{i}}{a}$ results to physical units and
        $a_{f_{p4s}}^{2}$ for the $x$-coordinate.
        Filled symbols correspond to physical light quark mass ensembles, while open symbols represent larger than
        physical quark masses.
        The gray band indicates the ($2+1$)-flavor value from Flavour Lattice Averaging Group (FLAG)
        2021~\cite{Aoki:2021kgd} for $r_{0}$ and $r_{1}$, and from Ref.~\cite{Bazavov:2017dsy} for $r_{2}$; see
        those references for details on the conversion to physical units.
        Similar plots for $\sqrt{\sigma}$ are shown in Fig.~\ref{fig:sigma_over_a} of
        Appendix~\ref{app:scales and string tension}.}
        \label{fig:r_i_over_a}
    \end{minipage}
\end{figure}

The resulting values and errors for the scales~$\sfrac{r_{i}}{a}$ and the string tension~$a^{2}\sigma$ are given in
Table~\ref{tab:r_i_over_a} of Appendix~\ref{app:scales and string tension}.
We observe a strikingly non-trivial quark mass dependence for all scales $\sfrac{r_{i}}{a}$.
First, as naively expected and observed in previous calculations in ($2+1$)-flavor QCD~\cite{Bazavov:2017dsy}, we
obtain larger values of $\sfrac{r_{i}}{a}$ at smaller light quark masses,\footnote{%
For unclear reasons, the smeared result for $\sfrac{r_{1}}{a}$ with the intermediate mass
$\sfrac{\ml}{\ms}=\sfrac{1}{10}$ (\ensemble{6.30}{ii}) does not follow a consistent mass ordering and is a clear
outlier from many other trends, too.} which is clearly visible in Fig.~\ref{fig:r_i_over_a}.
However, this effect seems to have a very peculiar lattice spacing dependence.
On the one hand, the physical $\sfrac{\ml}{\ms}$ or $\sfrac{\ml}{\ms}=\sfrac{1}{10}$ results are very close at
$\beta=6.00$ or $\beta=6.30$, while the $\sfrac{\ml}{\ms}=\sfrac{1}{5}$ is somewhat off at $\beta=6.30$.
On the other hand, at $\beta=6.72$, the $\sfrac{\ml}{\ms}=\sfrac{1}{10}$ or $\sfrac{\ml}{\ms}=\sfrac{1}{5}$ results are
very close, while the physical $\sfrac{\ml}{\ms}$ is somewhat off.
Instead of being due to a statistical fluke, this effect may be caused by the variation of the charm or strange quark
masses, which are highly correlated across the ensembles; see Table~\ref{tab:ensembles}.
At $\beta=6.72$, the $\sfrac{\ml}{\ms}=\sfrac{1}{10}$ or $\sfrac{\ml}{\ms}=\sfrac{1}{5}$ ensembles have a charm quark
mass that is 10\% larger than for the physical $\sfrac{\ml}{\ms}$ ensemble.
However, at $\beta=6.30$ the physical $\sfrac{\ml}{\ms}$ or $\sfrac{\ml}{\ms}=\sfrac{1}{10}$ ensembles have almost the
same charm quark mass, which is about 2\% smaller than for the $\sfrac{\ml}{\ms}=\sfrac{1}{5}$ ensemble.
And at $\beta=6.00$ the physical $\sfrac{\ml}{\ms}$ or $\sfrac{\ml}{\ms}=\sfrac{1}{10}$ ensembles have identical charm
masses.
This observation suggests that the dependence of the potential scales on the charm or strange quark masses may be
significantly larger than previously anticipated.
In Sec.~\ref{sec:quark_mass_dependence}, we study this quark mass dependence quantitatively when fitting the
$\sfrac{r_{i}}{a}$ data to a smooth curve in $g_{0}^{2}$ and quark masses.
The light quark mass dependence becomes insignificant for $\sfrac{r_{2}}{a}$, in line with results in ($2+1$)-flavor
QCD~\cite{Bazavov:2017dsy}.

Since the correlators with bare- or smeared-link variables represent different discretizations, different values of the
scales $\sfrac{r_{i}}{a}$ with bare or smeared links are to be expected.
This effect needs to be distinguished from the distortions of the smeared-link correlators at small distances due to
the unphysical contact-term interactions.
While the former is not a problem, the latter needs to be avoided.
The smeared-link data yield substantially smaller values when $\sfrac{r_{i}}{a} < 3$, namely for $\sfrac{r_{2}}{a}$ at
$\beta=6.72$ or $\sfrac{r_{1}}{a}$ at $\beta=6.0$, which are clearly inconsistent with the bare-link results.
It is suggestive to attribute this discrepancy to the contact-term interactions seen in
Fig.~\ref{fig:tree_level_correction}.
Hence, we discard these smeared results (enclosed by square brackets in Table~\ref{tab:r_i_over_a} in
Appendix~\ref{app:scales and string tension}) and use the bare results in their place when smoothening the scales and,
in Sec.~\ref{sec:cont-limit}, when extrapolating to the continuum limit.
In the range $3 \lesssim \sfrac{r_{i}}{a} \lesssim 4$, which includes the maximal $R$ where the contact-term
interactions between the smeared link variables distort the correlation function, this underestimation of
$\sfrac{r_{i}}{a}$ with smeared links becomes mild and usually consistent within errors.
However, the shift between $\sfrac{r_{1}}{a}$ with bare and smeared links in the \ensemble{6.30}{ii} ensemble clearly
deviates from the pattern exhibited by the other two masses at this (or any larger)~$\beta$.
Since the result with bare links is consistent with the expected pattern of a monotonic light quark mass dependence, we
conclude that this smeared-link result is not reliable.
To be consistent, we discard the $\sfrac{r_{1}}{a}$ results with smeared links for all sea quark masses at $\beta=6.30$
(enclosed by square brackets in Table~\ref{tab:r_i_over_a} in Appendix~\ref{app:scales and string tension}) and use the
bare results in their place.
We opt, however, to keep the smeared-link results for $\sfrac{r_{0}}{a}$ at $\beta \le 6.0$ and for $\sfrac{r_{2}}{a}$
at $\beta = 7.0$ since there is nothing obviously wrong with these.
With smeared links we find compatible or slightly larger $\sfrac{r_{i}}{a}$ for $\sfrac{r_{i}}{a} > 5$, where
small-distance distortions can be ruled out; see Sec.~\ref{sec:artifacts}.

\begin{figure}
    \centering
    \begin{minipage}[t]{0.49\textwidth}\vspace{0pt}%
    \includegraphics[width=\textwidth]{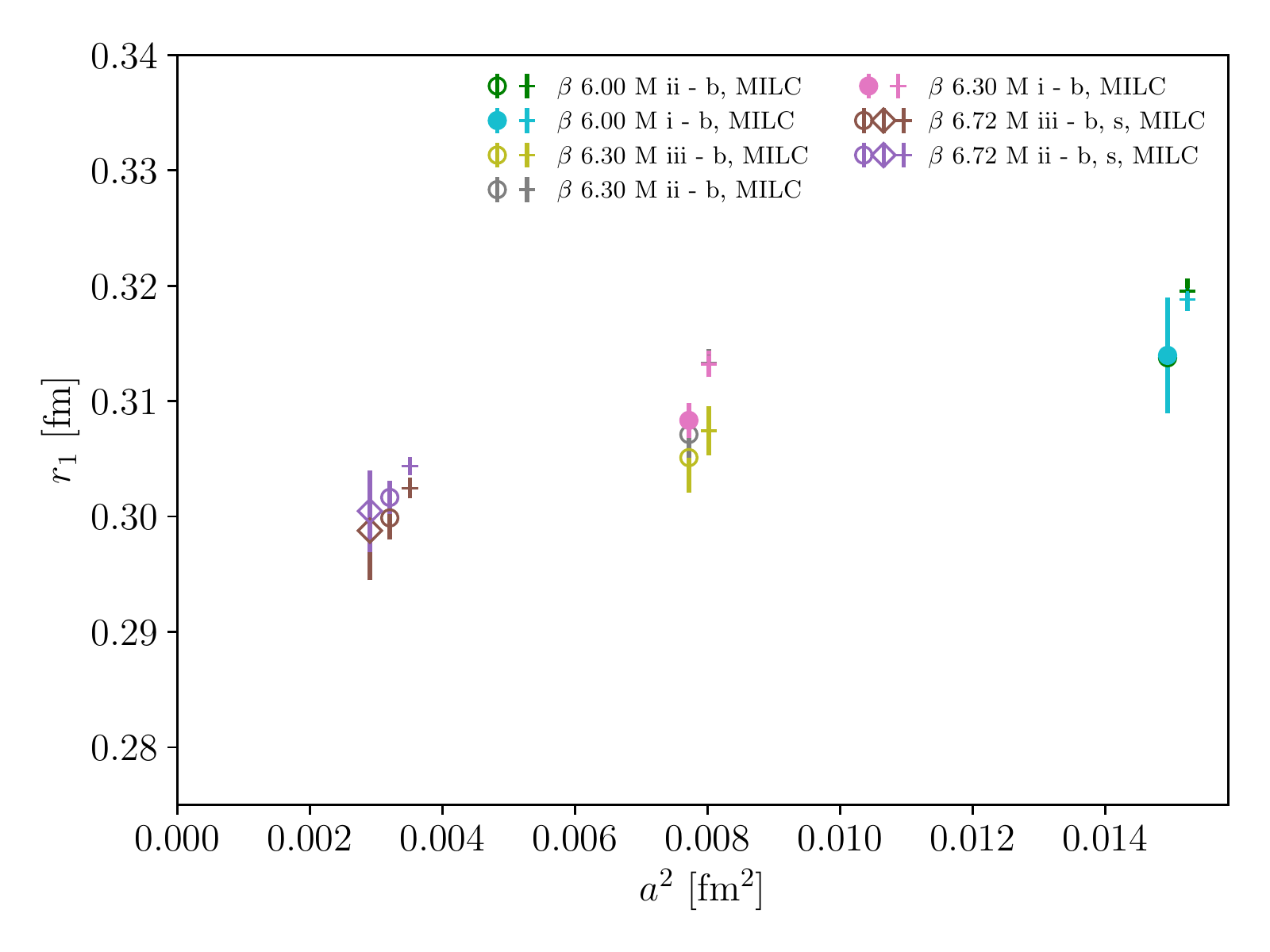}%
    \end{minipage}
    \hfill%
    \begin{minipage}[t]{0.49\textwidth}\vspace{0pt}%
    \caption{Comparison of our direct determinations of $\sfrac{r_{1}}{a}$,
    Table~\ref{tab:r_i_over_a}, to previous results on some on the ensembles from the MILC
    Collaboration~\cite{MILC:2010pul, MILC:2012znn}.
    As in Fig.~\ref{fig:r_i_over_a}, we convert all results to physical units via $a_{f_{p4s}}$ and use
    $a_{f_{p4s}}^{2}$ for the $x$-coordinate.}
    \label{fig:r_1_data_vs_MILC}
    \end{minipage}
\end{figure}

Another striking feature of Fig.~\ref{fig:r_i_over_a} is that our data lie consistently lower than the $(2+1)$-flavor
results (shown as gray bands).
In Fig.~\ref{fig:r_1_data_vs_MILC}, we compare our $\sfrac{r_{1}}{a}$ results to those from earlier calculations using
common subsets of the ensembles obtained by the MILC Collaboration~\cite{MILC:2010pul, MILC:2012znn}.
Our results are systematically lower than MILC's, significantly so at $\beta=6.3$.
Since the two sets of results are based on different fit procedures, they can differ by discretization effects.
Our larger errors originate in the systematic error estimate from the full spread to the randomized variation of the
$R$ values in Sec.~\ref{sec:interpolation}.
Even so, the trend of both data sets is toward a lower value of $r_{1}$ (in~fm) than that from the FLAG compilation of
$(2+1)$-flavor results; cf.\ Fig.~\ref{fig:r_i_over_a}.
This trend is corroborated by our data at smaller lattice spacings, as discussed further in Sec.~\ref{sec:cont-limit}.

\subsection{Smoothening}
\label{sec:quark_mass_dependence}

For relative scale setting in future work on the HISQ ensembles, it is useful to summarize the results for
$\sfrac{r_{i}}{a}$ as functions of the squared bare gauge coupling $g_{0}^{2}$ and the bare quark masses
$am_{q}$.\footnote{%
Here, it is not possible to do so for $\sfrac{r_{2}}{a}$ because we have data at only three lattice spacings.}
In this work, we use an Allton \emph{Ansatz}~\cite{Allton:1996kr}, in particular, the very generic form found, for
example, in Ref.~\cite{MILC:2009mpl}, adapted to include the charm quark mass dependence,
\begin{equation}
    \frac{a}{r_{i}} = \frac{C_{0} f_{\beta} + C_{2} g_{0}^{2} f_{\beta}^{3} + C_{4} g_{0}^{4} f_{\beta}^{3}}
        {1 + D_{2} g_{0}^{2} f_{\beta}^{2}}. 
    \label{eq:Allton_fit}
\end{equation}
Here,
\begin{equation}
    f_{\beta} = (b_{0} g_{0}^{2})^{-\sfrac{b_{1}}{(2 b_{0}^{2})}} \e{-\sfrac{1}{(2 b_{0} g_{0}^{2})}}, \quad
    b_{0} = \frac{\beta_{0}^{(\Nf)}}{(4\pi)^{2}}, \quad
    b_{1} = \frac{\beta_{1}^{(\Nf)}}{(4\pi)^{4}},
    \label{eq:two_loop_beta_function}
\end{equation}
is the integrated $\beta$ function to two loops, which scales asymptotically as $f_{\beta}\propto a$, and
$\beta_{0,1}^{(\Nf)}$ are the first two coefficients of the $\beta$ function; see Appendix~\ref{app:coefficients}.
In the present case, $\Nf=4$.
Further,
\begin{equation}
\begin{aligned}
    C_{0} &= C_{00} + C_{01l} \frac{a\ml}{f_{\beta}} + C_{01s} \frac{a\ms}{f_{\beta}} +
        C_{01} \frac{a\mtot}{f_{\beta}} + C_{02} \frac{(a\mtot)^{2}}{f_{\beta}}, \\
    C_{2} &= C_{20} + C_{21} \frac{a\mtot}{f_{\beta}}, \quad a\mtot = 2 a\ml + a\ms + a\mc,
    \end{aligned}
\label{eq:Allton_fit_coefficients}
\end{equation}
where $C_{00}$, $C_{01l}$, $C_{01s}$, $C_{01}$, $C_{02}$, $C_{20}$, $C_{21}$, $C_{4}$, and $D_{2}$ are parameters to be
determined from fits described below.
In $C_{00}$, the second through fourth terms parametrize continuum limit quark mass dependence, while the $C_{02}$ term
represents a discretization effect on the largest (i.e., fourth) term.
We find we cannot constrain $C_{4}$ and $C_{21}$, so in the following, they are set to zero.

\begin{figure}
    \centering
    \includegraphics[width=1.0\textwidth]{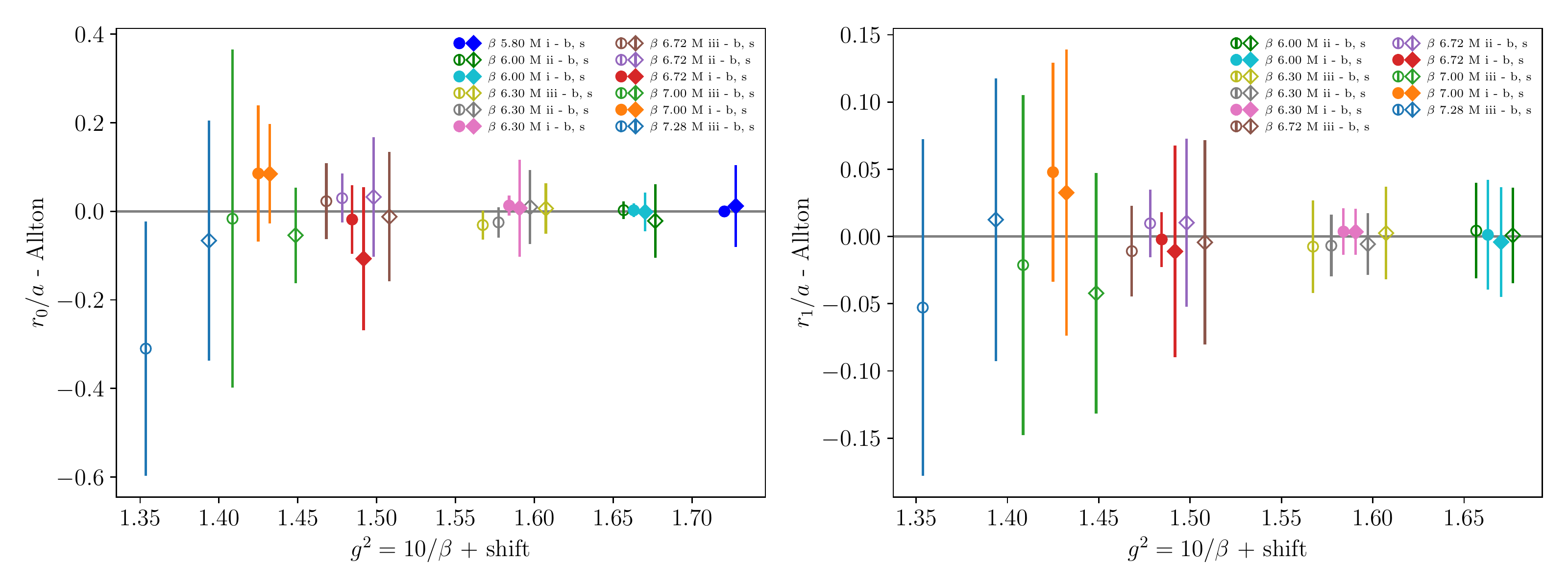}%
    \caption{\label{fig:r_i_over_a_Allton}%
    Residues of the Allton fits for $\sfrac{r_{i}}{a}$ using all ensembles (indicated by the color).
    Filled symbols correspond to physical light quark mass ensembles, while open symbols represent larger-than-physical
    quark masses; circles (diamonds) denote bare- (smeared)-link data.
    We use the squared bare gauge coupling $g_{0}^{2}$ for the $x$-coordinate, but shift bare- and smeared-link data
    horizontally by $\mp 0.1\sfrac{\ml}{\ms}$ to improve the visibility.}
\end{figure}

Further, we cannot reliably constrain the coefficients of multiple quark mass dependent terms.
Given that the charm quark mass is much larger than light or strange quark masses, $a\mtot$ is dominated by the
variation of $a\mc$; fits using only $a\mc$ (in place of $a\mtot$) typically have larger reduced $\chi^{2}$ than those
incorporating the light quark mass dependence as well through $a\mtot$.
Since the strange quark mass usually varies quite similarly to the charm quark mass, such that the physical value of
$\sfrac{\mc}{\ms}$ is realized to a fair approximation, using $a\mc + a\ms$ (in place of $a\mtot$) would not lead to
different conclusions.
Thus, parametrizations with some light quark mass dependence are preferred by the data.
The parametrization yielding smallest reduced $\chi^{2}$ (averaged over four fits for $\sfrac{r_{0}}{a}$ or
$\sfrac{r_{1}}{a}$ using both bare-link or smeared-link data) is quadratic in $a\mtot$ with only $C_{00}$, $C_{02}$,
$C_{20}$, and $D_{2}$ being allowed to vary.
For $\sfrac{r_{1}}{a}$, fits are similarly good with a parametrization linear in $a\mtot$ with only $C_{00}$,
$C_{01}$, $C_{20}$, and $D_{2}$ being allowed to vary.
Finally, for $\sfrac{r_{0}}{a}$, fits with a parametrization linear in $a\ml$ and $a\ms$ (neglecting $a\mc$) are
similarly good, too.
The coefficients for the total quark mass term are compatible between all linear or between all quadratic fits; see
Table~\ref{tab:Allton_coefficients} in Appendix~\ref{app:relative scale setting}.
The respective covariance matrices are supplemented as text files.
Moreover, the coefficients of fits for bare or smeared links are compatible.
We point out that the dominant light quark mass dependence in both the linear or the quadratic fit is actually linear,
as $(a\mtot)^{2} = (a\mc + a\ms)^{2} + 4(a\mc + a\ms)(a\ml) + \ldots$.
We use the quadratic fits including all ensembles as our main results.
Their residuals do not show a consistent pattern of the light quark mass dependence; see
Fig.~\ref{fig:r_i_over_a_Allton}.

\begin{figure}
    \centering
    \includegraphics[width=\textwidth]{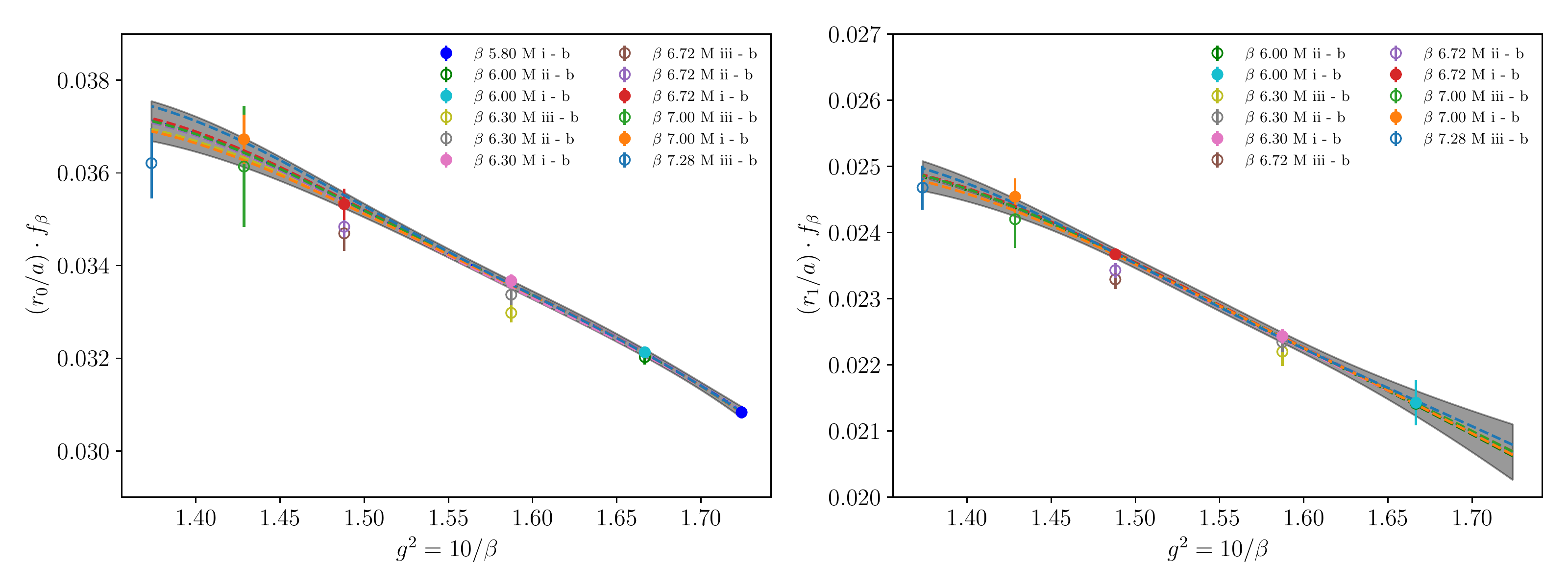}%
    \caption{\label{fig:r_i_over_a_Allton_summary}%
    The potential scales $\sfrac{r_{i}}{a}$, $i=0,1$ multiplied by the two-loop $\beta$-function, $f_{\beta}$ as in
    Eq.~\eqref{eq:two_loop_beta_function}, for all ensembles (indicated by colors) and bare links.
    Filled symbols correspond to physical light quark mass ensembles, while open symbols represent larger than physical
    quark masses.
    The curves correspond to the Allton fit, Eq.~\eqref{eq:Allton_fit}, evaluated at the masses of the physical mass
    ensembles using the parameters given in Table~\ref{tab:Allton_coefficients} in
    Appendix~\ref{app:relative scale setting}.
    The color of the lines indicates the ensemble that has been left out, while the black curve (hidden behind the
    other lines) is the one including all, with the band representing its regression error.
    We use the squared bare gauge coupling $g_{0}^{2}$ for the $x$-coordinate.
    A corresponding plot for smeared links is in Fig.~\ref{fig:r_i_over_a_Allton_summary_smeared} in
    Appendix~\ref{app:relative scale setting}.}
\end{figure}

The regression errors of the interpolated values $\sfrac{r_{i}}{a}$ obtained from the Allton fit coefficients are
reported in Table~\ref{tab:r_i_over_a_Allton} in Appendix~\ref{app:relative scale setting}.
They are similarly large as the errors from the direct determination, while some outliers among the errors have been
eliminated; cf.\ Table~\ref{tab:r_i_over_a} in Appendix~\ref{app:relative scale setting}.
In order to test whether the Allton fit might assign an undue, large weight to any ensemble, we repeated the same
Allton fit on each subset of the data leaving out one ensemble in each; all of these fits are covered by the regression
error of the Allton fit using the full data set, see Fig.~\ref{fig:r_i_over_a_Allton_summary}.
In the corresponding curves, evaluated at $\ml \ge \sfrac{\ms}{10}$ with the corresponding $a\ms$ and $a\mc$ values
(not shown in Fig.~\ref{fig:r_i_over_a_Allton_summary}), we see a wiggly structure between $\beta=7.28$ and
$\beta=6.30$, which is more pronounced in $\sfrac{r_{0}}{a}$ than in $\sfrac{r_{1}}{a}$; hints of such a trend were
already seen in Fig.~\ref{fig:r_i_over_a} and are interpreted as an effect due to the $10\%$ variation of the charm
mass between the different $\beta=6.72$ ensembles.
We also use the parametrization in terms of the Allton fit to obtain results in the chiral limit of the light quark
mass, where we use $\ms$ and $\mc$ of the physical mass ensembles, or $\ms$ and $\mc$ of the only existing $\beta=7.28$
ensemble.
In the latter case, we estimate the physical value of the light quark mass from the sea strange quark mass using
the physical $\sfrac{\ml}{\ms}$-ratio, i.e., $\sfrac{1}{27.3}$, to be $a\ml=0.000409$.

\section{Continuum limits}
\label{sec:cont-limit}

In Sec.~\ref{sec:quark_mass_dependence}, we have determined the individual results for the scales $\sfrac{r_{i}}{a}$
and the string tension $a^{2}\sigma$ on each ensemble.
Here, we form dimensionless combinations of the $\sfrac{r_{i}}{a}$ and $a^{2}\sigma$.
In particular, we compute $\sfrac{r_{0}}{r_{1}}$ and $\sfrac{r_{1}}{r_{2}}$ for which we use the smoothened values for
$\sfrac{r_{0,1}}{a}$ given in Table~\ref{tab:r_i_over_a_Allton} in Appendix~\ref{app:relative scale setting} and the
direct determination of $\sfrac{r_{2}}{a}$ given in Table~\ref{tab:r_i_over_a} in
Appendix~\ref{app:scales and string tension}.
The results for the string tension are conveniently multiplied by the smoothened $(\sfrac{r_{0}}{a})^{2}$; the square
root of this product is collected in Table~\ref{tab:r_i_over_a_Allton} in Appendix~\ref{app:relative scale setting}.
We then perform continuum extrapolations of these universal dimensionless quantities.
We also multiply the smoothened values for $\sfrac{r_{0,1}}{a}$ with $a_{f_{p4s}}$, cf.~the last few paragraphs of
Sec.~\ref{sec:interpolation}, in order to perform a continuum extrapolation of these two dimensionful quantities.

\subsection[Ratios \texorpdfstring{$\sfrac{r_{0}}{r_{1}}$}{r0/r1} and \texorpdfstring{$\sfrac{r_{1}}{r_{2}}$}{r1/r2}]%
{Ratios \texorpdfstring{\boldmath{$\sfrac{r_{0}}{r_{1}}$}}{r0/r1} and %
\texorpdfstring{\boldmath{$\sfrac{r_{1}}{r_{2}}$}}{r1/r2}}
\label{sec:ratios}

\begin{figure}
    \centering
    \includegraphics[width=1.0\textwidth]{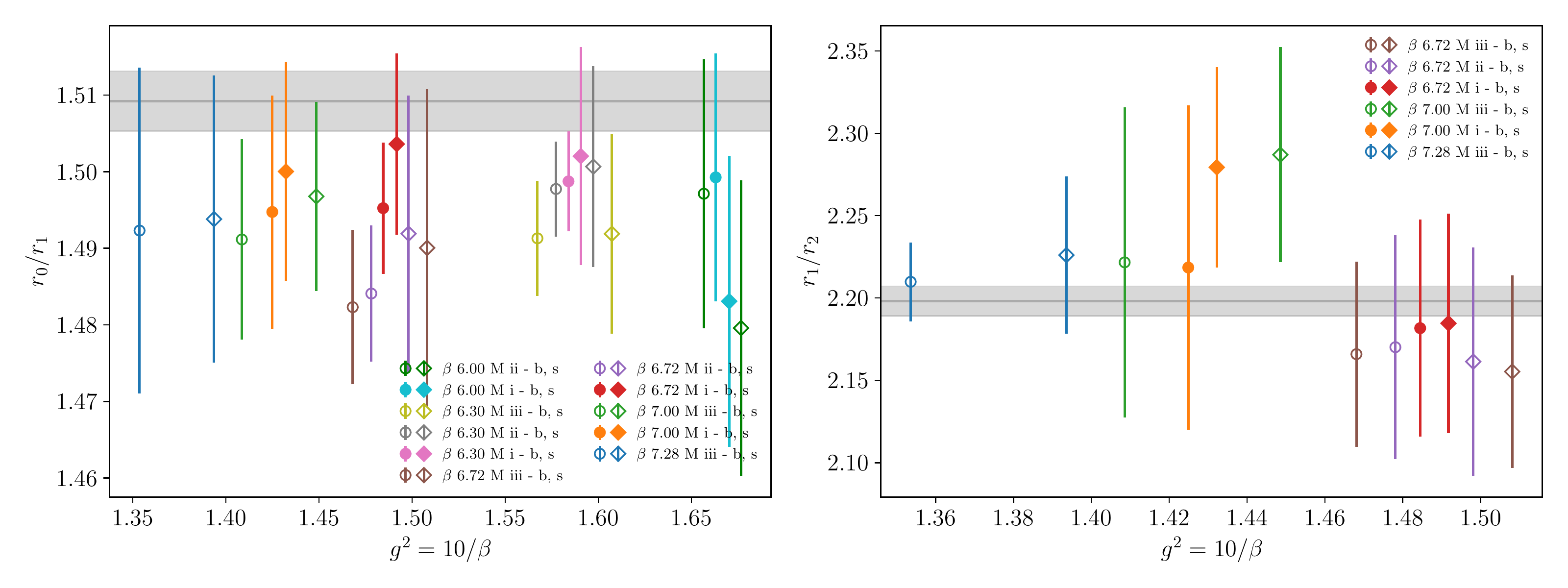} \\
    \caption{\label{fig:ratios}%
    Ratios of smoothened $\sfrac{r_{0}}{r_{1}}$ or $\sfrac{r_{1}}{r_{2}}$ for all ensembles.
    For visibility's sake the data are shifted horizontally by $\mp 0.1 \sfrac{\ml}{\ms}$ for the
    bare-/smeared-link data.
    The gray solid line and band show published ($2+1$)-flavor values~\cite{HotQCD:2014kol, Bazavov:2017dsy} for
    $\sfrac{r_{0}}{r_{1}}$ and $\sfrac{r_{1}}{r_{2}}$, respectively.}
\end{figure}
The errors of the individual $\sfrac{r_{i}}{a}$ contain our estimates of systematic uncertainties, dominated by the
variation of the independent randomly chosen sets of $R$ values.
These are considerably larger than the statistical errors, and as explained in the discussion around
Fig.~\ref{fig:systematical_distribution}, we add them and the statistical uncertainties in quadrature.
The regression errors of the smoothened $\sfrac{r_{i}}{a}$ therefore reflect the systematic errors.
We show the results for the ratios of the smoothened scales evaluated at the parameters of the individual ensembles as
a function of the squared bare gauge coupling in Fig.~\ref{fig:ratios}.
The gray bands indicate published ($2+1$)-flavor values~\cite{HotQCD:2014kol, Bazavov:2017dsy}.
Across all ensembles, our results for $\sfrac{r_{0}}{r_{1}}$ in ($2+1+1$)-flavor QCD are marginally lower than the
HotQCD result in ($2+1$)-flavor QCD~\cite{HotQCD:2014kol}, where the approach to the continuum limit has been found to
be flat within errors for similar lattice spacings with a variety of actions~\cite{Cheng:2007jq, HotQCD:2014kol}.
The ($2+1+1$)-flavor QCD result shows a fairly flat behavior, too, although there are hints of some curvature that
point to a mild decrease towards the continuum limit.
A systematic dependence on the light quark mass with larger $\sfrac{r_{0}}{r_{1}}$ for smaller pion mass is visible.
With the exception of the results on the corresponding coarsest lattices, $\beta=6.72$, our results for
$\sfrac{r_{1}}{r_{2}}$ in ($2+1+1$)-flavor QCD turn out to be systematically higher than the result in ($2+1$)-flavor
QCD, where the approach to the continuum limit has been found to be flat within errors for a similar range of lattice
spacings~\cite{Bazavov:2017dsy}.
The coarsest lattices for which $\sfrac{r_{2}}{a}$ has been obtained in either analysis have
$\sfrac{r_{2}}{a} < 3$.
Such distances are still affected by substantial non-smooth discretization artifacts after the tree-level correction,
see Sec.~\ref{sec:artifacts}.
Since the ($2+1$)-flavor QCD analysis had benefited from non-perturbative corrections, they may have not been affected
by a similar discretization artifact that impacts the ($2+1+1$)-flavor QCD result at $\beta=6.72$; this might explain
the somewhat lower value for $\sfrac{r_{1}}{r_{2}}$.
No systematic dependence on the light quark masses can be resolved.

We now describe our continuum extrapolations following the same procedures for all quantities.
For brevity and clarity, we denote each of these as $\xi$ in the following.
The leading discretization effects are of order $\als^{2}a^{2}$ and $a^{4}$, as discussed in Sec.~\ref{sec:setup}.
With the lattice spacing dependence represented by $x=(\sfrac{a}{r_{0}})^{2}$ or $(\sfrac{a}{r_{1}})^{2}$, and the
light quark mass dependence represented by $y=\sfrac{(a\ml)_{\text{sea}}}{(a\ms)_{\text{sea}}}$ or
$\sfrac{(a\ml)_{\text{sea}}}{(a\ms)_{\text{tuned}}}$, we consider the functional forms\footnote{%
The abbreviations shown below are also the ones used in the Supplemental Material~\cite{SupplMat} providing details
of the individual fit results.}
\begin{align}
    \xi &= \xi_{0} && \text{(weighted average)} , \label{eq:weighted_average} \\
    \xi &= \xi_{0} + \alpha^{2}  \xi_{1} x && \text{(lin)} , \label{eq:continuum_lin} \\
    \xi &= \xi_{0} + \alpha^{2}  \xi_{1} x + \xi_{2} x^{2} && \text{(quad)} , \label{eq:continuum_quad} \\
    \xi &= \xi_{0} + \alpha^{2} [\xi_{1} x + \xi_{2} x y] && \text{(l,lm)} , \label{eq:continuum_l_lm} \\
    \xi &= \xi_{0} + \alpha^{2} [\xi_{1} x + \xi_{2} x y] + \xi_{3} x^{2} && \text{(q,lm)} , \\
    \xi &= \xi_{0} + \alpha^{2} [\xi_{1} x + \xi_{2} x y^{2}] && \text{(l,qm)} , \\
    \xi &= \xi_{0} + \alpha^{2} [\xi_{1} x + \xi_{2} x y^{2}] + \xi_{3} x^{2} && \text{(q,qm)} , \\
    \xi &= \xi_{0} + \alpha^{2} [\xi_{1} x + \xi_{2} x y] + \xi_{3} y && \text{(l,lm,mc)} , \\
    \xi &= \xi_{0} + \alpha^{2} [\xi_{1} x + \xi_{2} x y] + \xi_{3} x^{2} + \xi_{4} y && \text{(q,lm,mc)} , \\
    \xi &= \xi_{0} + \alpha^{2} [\xi_{1} x + \xi_{2} x y^{2}] + \xi_{3} y && \text{(l,qm,mc)} , \\
    \xi &= \xi_{0} + \alpha^{2} [\xi_{1} x + \xi_{2} x y^{2}] + \xi_{3} x^{2} + \xi_{4} y && \text{(q,qm,mc)} ,
    \label{eq:continuum_q_qm_mc}
\end{align}
where we assume either $\alpha = \alpha_{b} \equiv \sfrac{g_{0}^{2}}{(4\pi u_{0}^{4})}$, including the tadpole factors
given in Table~\ref{tab:ensembles} (originally from Ref.~\cite{Bazavov:2017lyh}), or $\alpha = 1$, i.e., we either
incorporate or ignore the one-loop improvement of the $a^{2}$ dependence.

\begin{figure}
    \centering
    \includegraphics[width=0.49\textwidth]{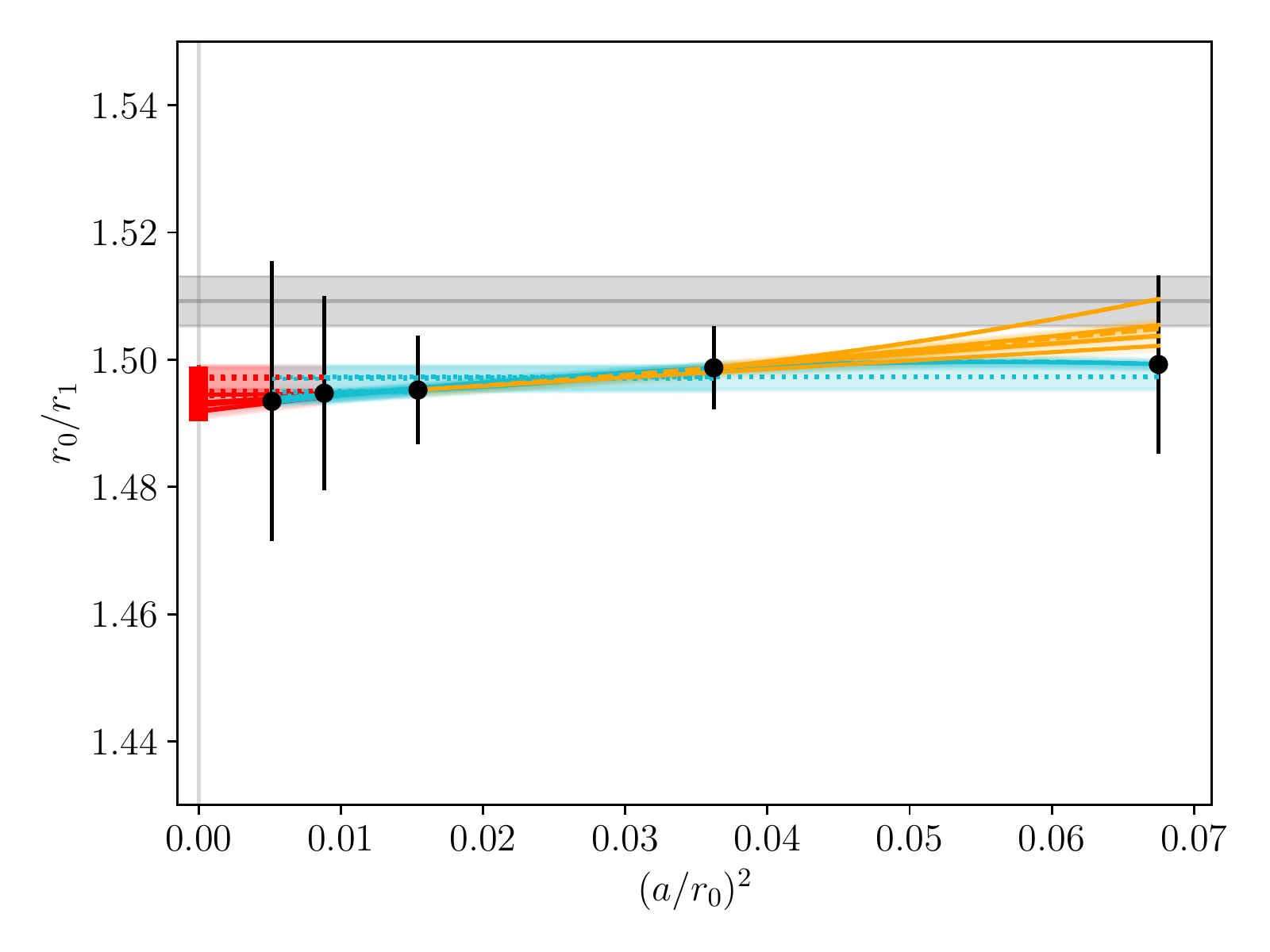}
    \includegraphics[width=0.49\textwidth]{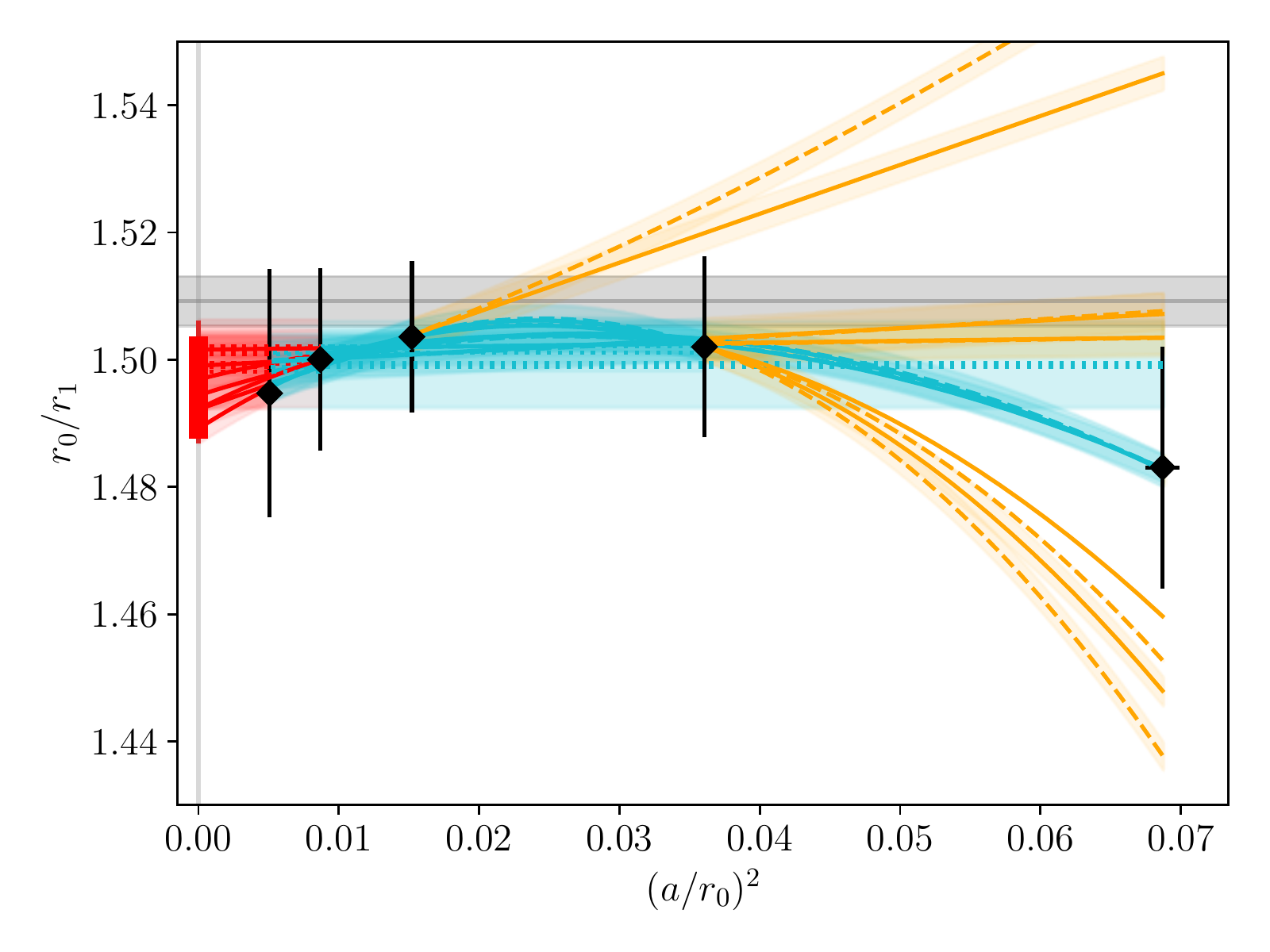}
    \caption{\label{fig:continuum_ri_rj_I}%
    Continuum extrapolation of the parametrization of $\sfrac{r_{0}}{r_{1}}$ evaluated at the physical
    $\sfrac{\ml}{\ms}$-ratio for $6.0\le\beta\le7.28$ as a function of $(\sfrac{a}{r_{0}})^{2}$.
    The black points show the bare-link data (left) and the smeared-link data (right) with the corresponding continuum
    results shown in red.
    The lines and bands show the fit curves and errors; within the fit range in cyan, as extrapolations towards the
    continuum or coarser lattices in red/orange, respectively.
    The gray solid line and band indicate the HotQCD result in ($2+1$)-flavor QCD~\cite{HotQCD:2014kol}.
    A corresponding plot as a function of $(\sfrac{a}{r_{1}})^{2}$ is shown in Fig.~\ref{fig:continuum_ri_rj_I_r1} in
    Appendix~\ref{app:continuum extrapolations}.}
\end{figure}

We fit the ratio $\sfrac{r_{0}}{r_{1}}$ using the parametrization evaluated at four fixed $\sfrac{\ml}{\ms}$-ratios,
namely, in the chiral limit of the light quark masses, or at the three sets of actual values present in the
simulations, see Table~\ref{tab:ensembles}.
Here we have five\footnote{%
Not six because we do not determine $\sfrac{r_{1}}{a}$ on the \ensemble{5.80}{i} ensemble.} data points available,
which allows us to vary $\beta_{\text{min}}$ and $\beta_{\text{max}}$.
We start with a weighted average, Eq.~\eqref{eq:weighted_average}, for $\beta_{\text{min}} \in \lbrace 7.0, 6.72, 6.3,
6.0 \rbrace$, and we use linear, Eq.~\eqref{eq:continuum_lin}, for $\beta_{\text{min}} \in \lbrace 6.72, 6.3 \rbrace$,
and quadratic, Eq.~\eqref{eq:continuum_quad}, for $\beta_{\text{min}} \in \lbrace 6.3, 6.0 \rbrace$ fits in
$(\sfrac{a}{r_{0}})^{2}$.
We use $\beta_{\text{max}} \in \lbrace 7.28, 7.0 \rbrace$.
We repeat these fits with the exception of the weighted average as a function of $(\sfrac{a}{r_{1}})^{2}$.
These constitute inequivalent extrapolations with different error budgets: on the one hand, due to the different error
in $x$,\footnote{%
All of the fits here are performed using orthogonal distance regression fits that, in contrast to ordinary least
squares minimization, also takes into account uncertainties in the independent variable~\cite{Boggs1990:ODR}.} and on
the other hand, due to the different cutoff and quark mass dependence of $\sfrac{r_{0}}{a}$ and $\sfrac{r_{1}}{a}$.
{\refstepcounter{response}\label{resp:parameterization}}
The smoothened data, continuum results, and fit curves of the parametrization evaluated at the physical
$\sfrac{\ml}{\ms}$-ratio as a function of $(\sfrac{a}{r_{0}})^{2}$ are shown in Fig.~\ref{fig:continuum_ri_rj_I}.
\new{The errors of the ratio are obtained by adding the errors derived from the full parameter covariance matrix of
each parametrization in quadrature.}

\begin{figure}
    \centering
    \begin{minipage}[t]{0.49\textwidth}\vspace{0pt}%
    \includegraphics[width=\textwidth]{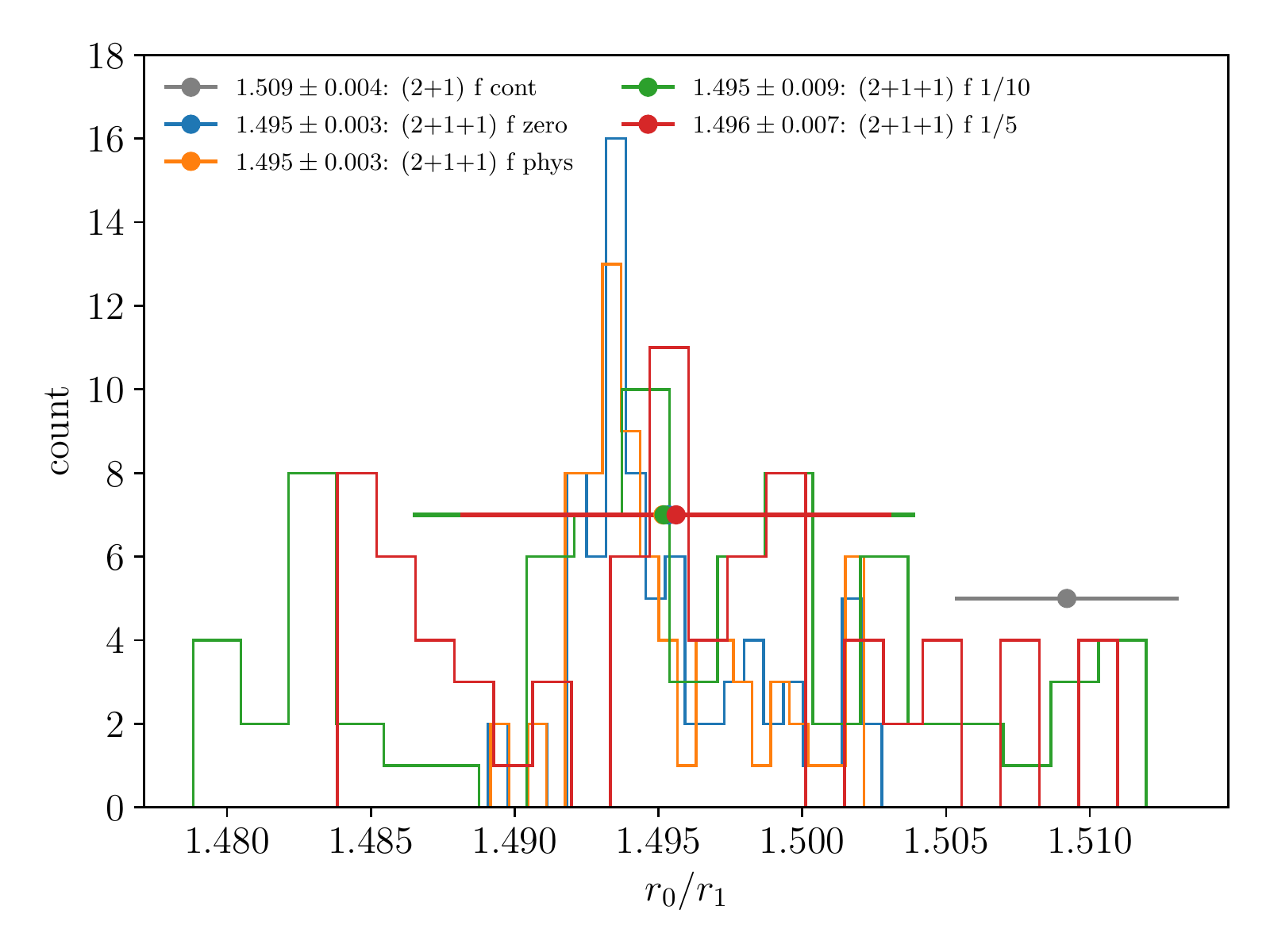}%
    \end{minipage}%
    \hfill%
    \begin{minipage}[t]{0.49\textwidth}\vspace{0pt}%
    \caption{\label{fig:continuum_ri_rj_II}%
    Continuum results for $\sfrac{r_{0}}{r_{1}}$ are obtained at four different fixed $\sfrac{\ml}{\ms}$-ratios
    (distinguished by color).
    The histograms show the distributions accumulated from fits like those shown in Fig.~\ref{fig:continuum_ri_rj_I}.
    The symbols of each color correspond to the respective mean and standard deviation.
    The gray symbol indicates the HotQCD result in ($2+1$)-flavor QCD~\cite{HotQCD:2014kol} for $\sfrac{r_{0}}{r_{1}}$.}
    \end{minipage}
\end{figure}

The distribution of the results for the four different light quark mass ratios (chiral limit, physical,
$\sfrac{1}{10}$, and $\sfrac{1}{5}$) is shown in Fig.~\ref{fig:continuum_ri_rj_II}.
While the central values are indistinguishable, the distributions are much broader for the two larger masses, a
consequence of the wiggles mentioned in Sec.~\ref{sec:quark_mass_dependence}.
We include all four of them for our final determination of $\sfrac{r_{0}}{r_{1}}$.

\begin{figure}
    \centering
    \includegraphics[width=1.0\textwidth]{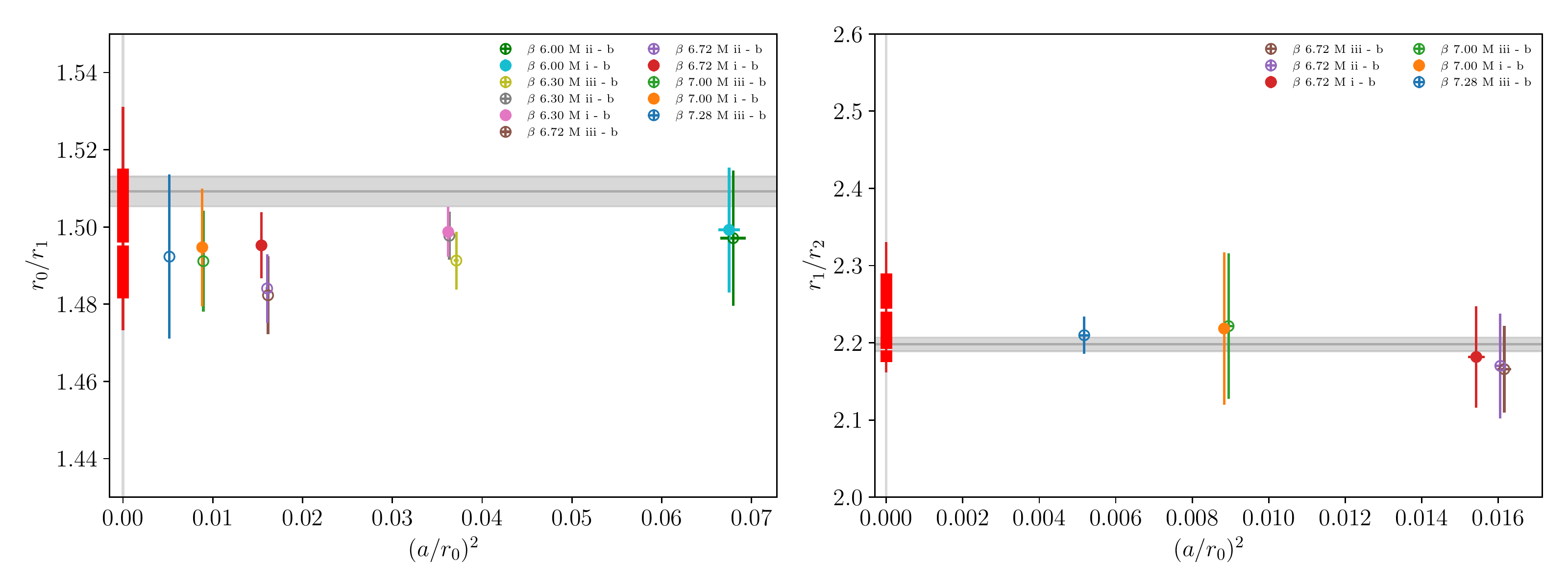}
    \caption{\label{fig:continuum_ri_rj_III}%
    Continuum results (red) and smoothened data (other colors) for the ratios $\sfrac{r_{0}}{r_{1}}$ (left) or
    $\sfrac{r_{1}}{r_{2}}$ (right) as functions of $(\sfrac{a}{r_{0}})^{2}$, using bare links.
    The gray solid line and band show the published ($2+1$)-flavor QCD values~\cite{HotQCD:2014kol, Bazavov:2017dsy}
    for $\sfrac{r_{0}}{r_{1}}$ and $\sfrac{r_{1}}{r_{2}}$, respectively.
    The corresponding plots for smeared links are shown in Fig.~\ref{fig:continuum_ri_rj_III_smeared} of
    Appendix~\ref{app:continuum extrapolations}.}
\end{figure}

{\newcounter{dummy}\refstepcounter{dummy}\label{expl:joint_fits}}
Furthermore, we perform joint fits combining different light quark masses.
Namely, we use the parametrization of $\sfrac{r_{0}}{r_{1}}$ evaluated at the actual ensemble parameters in our study,
or at subsets thereof.
For $\sfrac{r_{1}}{r_{2}}$ we exclusively use joint fits and combine the smoothened $\sfrac{r_{1}}{a}$ data with the
direct $\sfrac{r_{2}}{a}$ data.
We start the joint fits with weighted averages as above and employ fits linear and quadratic in
$(\sfrac{a}{r_{0}})^{2}$, where we neglect explicit light quark mass dependence.
Again, for $\sfrac{r_{0}}{r_{1}}$ and the linear fits in $(\sfrac{a}{r_{0}})^{2}$, we vary $\beta_{\text{min}} \in
\lbrace 6.72, 6.3 \rbrace$ and for the quadratic fits in $(\sfrac{a}{r_{0}})^{2}$, we vary $\beta_{\text{min}} \in
\lbrace6.3,6.0\rbrace$.
For $\sfrac{r_{1}}{r_{2}}$, we use $\beta_{\text{min}} = 6.72$, using only linear fits in $(\sfrac{a}{r_{0}})^{2}$.
For either, we use $\beta_{\text{max}} \in \lbrace 7.28, 7.0 \rbrace$.
We additionally supplement the fits with terms linear and quadratic in the $\sfrac{\ml}{\ms}$-ratio, and furthermore, we
also repeat these fits, adding a term proportional to the $\sfrac{\ml}{\ms}$-ratio that survives in the continuum limit.
At this stage, we perform the fits with either the sea strange quark mass of the ensemble or with the tuned
strange quark mass given in Table~\ref{tab:ensembles}, doubling the amount of fits.
On top, we choose either $\alpha = \alpha_{b} \equiv \sfrac{g_{0}^{2}}{(4\pi u_{0}^{4})}$ or $\alpha = 1$, again
doubling the amount of fits.
The fit functions are given in Eqs.~\eqref{eq:continuum_l_lm} to~\eqref{eq:continuum_q_qm_mc}.
In order to get the continuum contribution due to the terms proportional to $\sfrac{\ml}{\ms}$ that survive in the
continuum, we substitute the values for $\sfrac{\ml}{\ms}$ by expressions using the neutral or charged pion masses and
the average squared kaon mass, respectively.
For this, we use that in the isospin limit, we have, using the Gell-Mann--Oakes--Renner (GMOR)
relation~\cite{Gell-Mann:1968hlm},
\begin{equation}
    M_{\pi}^{2} = 2\ml B_{0}, \quad M_{K}^{2} = (\ml + \ms)B_{0} ,
\end{equation}
where $B_{0}$ is a low-energy constant related to the chiral condensate in the chiral limit~\cite{Gell-Mann:1968hlm}
that cancels in the ratio.
We thus get from the GMOR-relation
\begin{equation}
    \sfrac{\ml}{\ms} = \sfrac{1}{(\sfrac{2M_{K}^{2}}{M_{\pi}^{2}} - 1)}.
\end{equation}
Inserting Particle Data Group (PDG)~\cite{ParticleDataGroup:2020ssz} values we can fix the $\sfrac{\ml}{\ms}$-ratio
in the continuum.
We use the average squared kaon mass, $2M_{K}^{2} = M_{K^{\pm}}^{2} + M_{K^{0}}^{2}$, and either the neutral or charged
pion mass squared, $M_{\pi^{\pm}}^{2}$ or $M_{\pi^{0}}^{2}$, yielding
\begin{align}
    \left.\sfrac{\ml}{\ms}\right|_{M_{\pi}^{2} = M_{\pi^{\pm}}^{2}} &= 0.04128, \\
    \left.\sfrac{\ml}{\ms}\right|_{M_{\pi}^{2} = M_{\pi^{0}}^{2}}   &= 0.03851.
\end{align}
We show the data together with the respective continuum results in Fig.~\ref{fig:continuum_ri_rj_III}.

\begin{figure}
    \centering
    \includegraphics[width=1.0\textwidth]{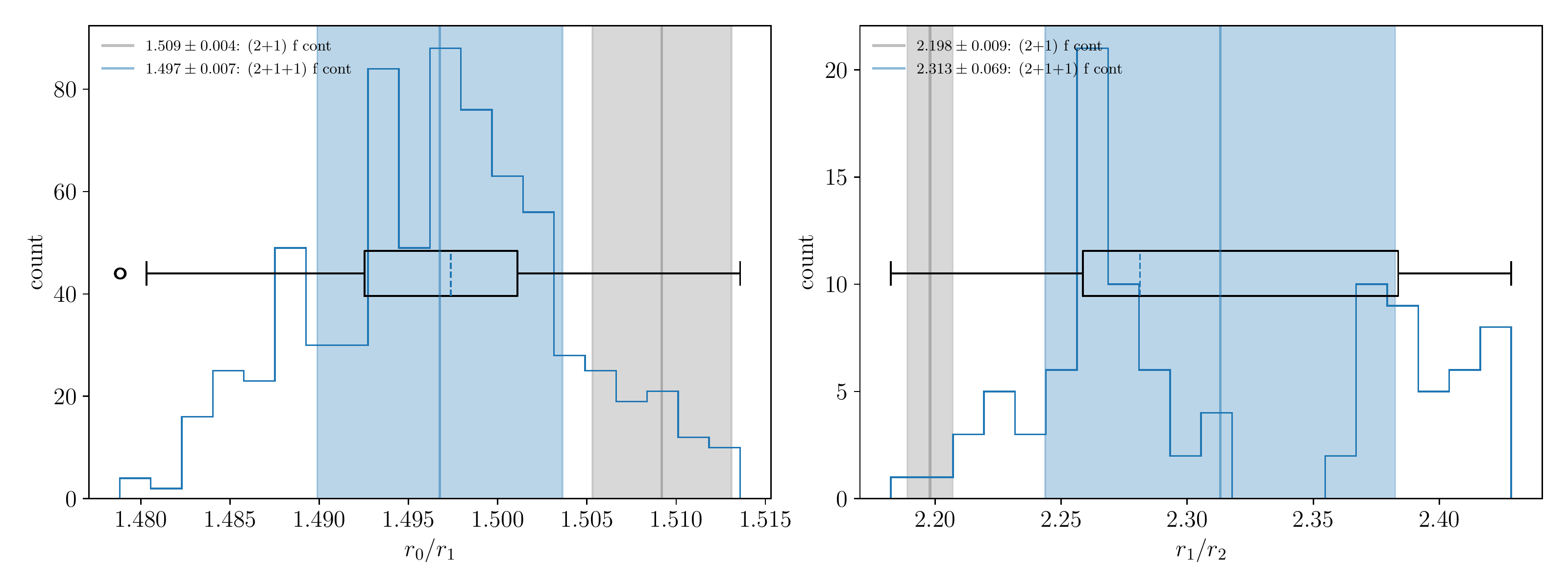}%
    \caption{\label{fig:continuum_ri_rj_hist}%
    Histogram of the continuum extrapolations for the ratios $\sfrac{r_{0}}{r_{1}}$ and $\sfrac{r_{1}}{r_{2}}$ using
    the \emph{Ansätze} discussed in the text.
    The box plots are explained in a footnote on page~\pageref{expl:BoxPlot}.
    We take the mean and the standard deviation of the respective distributions as our final value and uncertainty.
    The gray solid line and band show published ($2+1$)-flavor values~\cite{HotQCD:2014kol, Bazavov:2017dsy} for
    $\sfrac{r_{0}}{r_{1}}$ and $\sfrac{r_{1}}{r_{2}}$, respectively.
    The distribution of the errors is shown in Fig.~\ref{fig:errors_ratio} of
    Appendix~\ref{app:continuum extrapolations}.}
\end{figure}

All of these combinations lead to about 100 trial fits whose results, together with the ones previously discussed,
are shown in the histograms of Fig.~\ref{fig:continuum_ri_rj_hist}.
The blue lines and bands correspond to the mean and the standard deviation of the distributions.
We also show a box plot\footnote{%
{\refstepcounter{dummy}\label{expl:BoxPlot}}
The box plots use the standard, yet arbitrary, definition where the box extends from the first quartile (Q1) to the
third quartile (Q3) of the data, with the dashed line at the median.
The whiskers extend from the box by $1.5\times$~the interquartile range (IQR).
Flier points are those past the end of the whiskers.
They are sometimes considered outliers to be omitted; however, we do include them as regular data points.
In all of our results the mean (solid lines) and the median in the box lie close to one another, and the median lies
within the box supporting the decision to take the mean as our final result.
Furthermore, the standard deviation in our results (areas) and the IQR coincide very well supporting the decision to
take the former as our final error estimate.} together with the histograms.
The gray bands correspond to the published ($2+1$)-flavor values~\cite{HotQCD:2014kol, Bazavov:2017dsy} for
$\sfrac{r_{0}}{r_{1}}$ and $\sfrac{r_{1}}{r_{2}}$, respectively.
Because the distribution for $\sfrac{r_{0}}{r_{1}}$ is fairly similar to a Gaussian, the width appears to be an
appropriate estimate of the error; cf. Fig.~\ref{fig:errors_ratio} in Appendix~\ref{app:continuum extrapolations}.
However, the distribution for $\sfrac{r_{1}}{r_{2}}$ does not resemble a Gaussian and is, at least, bimodal.
{\refstepcounter{response}\label{resp:hist_cdf}}
\new{Even so, the confidence interval derived from the cumulative distribution function of the histogram is quite
similar to the one from a Gaussian interpretation.}
We obtain the best Akaike information criterion (AIC)~\cite{Akaike:1974, Cavanaugh:1997, Jay:2020jkz} with weighted
averages of all included ensembles, both for $\sfrac{r_{0}}{r_{1}}$ or $\sfrac{r_{1}}{r_{2}}$, both for bare- or
smeared-link data, and both for separate fits at any of the light quark masses---if applicable---or joint fits of
the different masses.
For $\sfrac{r_{0}}{r_{1}}$, these weighted averages scatter within the central $1\sigma$ interval of the distribution.
For $\sfrac{r_{1}}{r_{2}}$, however, they are very close to the ($2+1$)-flavor QCD result, while the distribution of
the fits yields a significantly larger central value.
Given that there are hints that $\sfrac{r_{1}}{r_{2}}$ for $\beta=6.72$ could be a bit on the low side due to
discretization artifacts, this situation could be cleared up once a correction beyond the tree level becomes available
in ($2+1+1$)-flavor QCD as well.
Our final continuum results for the ratios read
\begin{align}
    \sfrac{r_{0}}{r_{1}} &= 1.4968 \pm 0.0069, \label{eq:r0/r1} \\
    \sfrac{r_{1}}{r_{2}} &= 2.313 \pm 0.069. \label{eq:r1/r2}
\end{align}

\subsection[The scales \texorpdfstring{$r_{0}$}{r0} and \texorpdfstring{$r_{1}$}{r1} and the string tension]%
{The scales \texorpdfstring{\boldmath{$r_{0}$}}{r0} and \texorpdfstring{\boldmath{$r_{1}$}}{r1} and the string tension}
\label{sec:scales_and_string_tension}

\begin{figure}
\centering
    \includegraphics[width=1.0\textwidth]{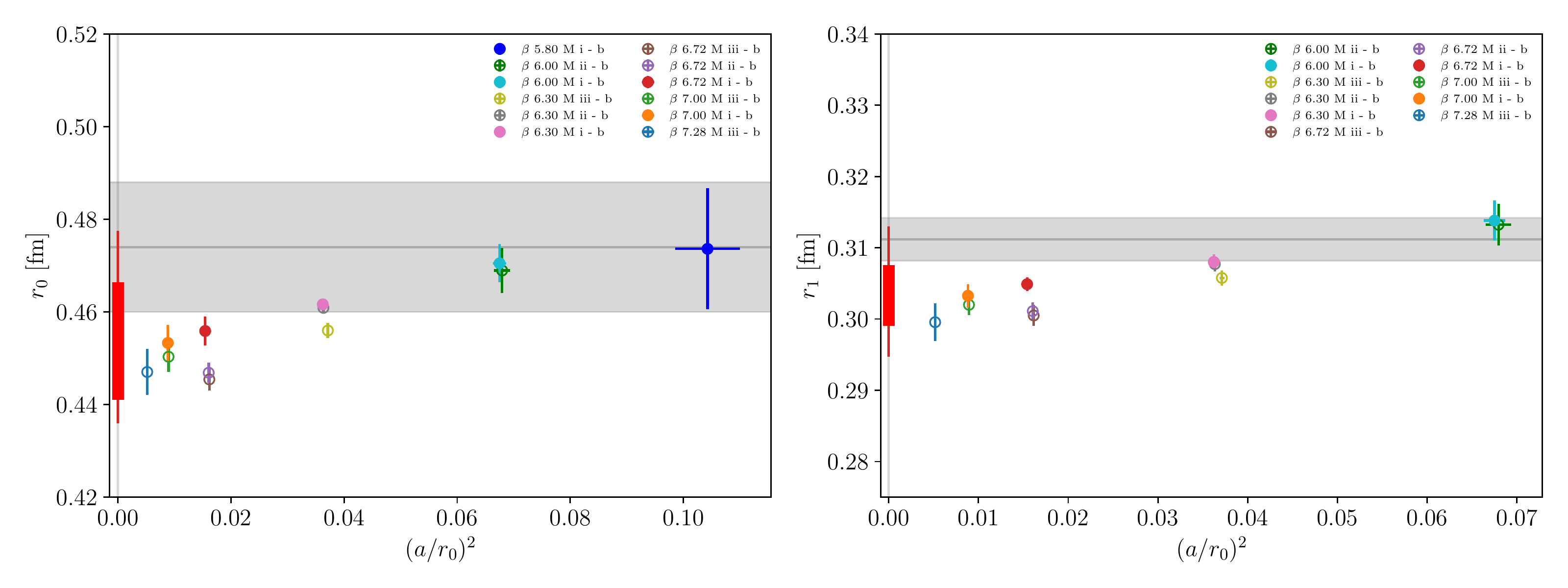}%
    \caption{\label{fig:continuum_r0_r1}%
    Continuum results (red) and smoothened data (other colors) for $a_{f_{p4s}}\sfrac{r_{0,1}}{a}$ are shown in the
    left or right columns, respectively, as functions of $(\sfrac{a}{r_{0}})^{2}$, using bare links.
    The gray solid line and band show the published ($2+1+1$)-flavor QCD values~\cite{EuropeanTwistedMass:2014osg,
    Dowdall:2013rya} for $r_{0}$ and $r_{1}$, respectively.
    The corresponding plots for smeared links are shown in Fig.~\ref{fig:continuum_r0_r1_smeared} of
    Appendix~\ref{app:continuum extrapolations}.}
\end{figure}

\begin{figure}
\centering
    \includegraphics[width=1.0\textwidth]{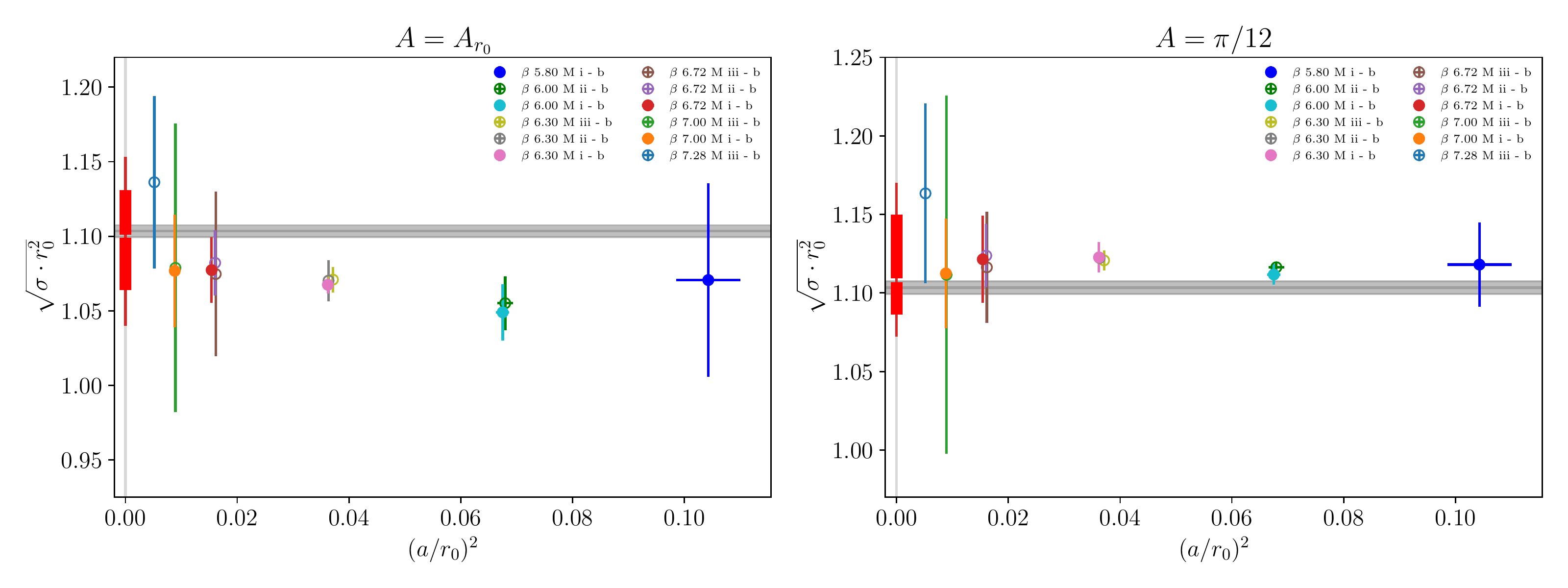}
    \caption{\label{fig:continuum_sqrt_sigma_r0_sq}%
    Continuum results (red) and smoothened data (other colors) for $\sqrt{\sigma r_{0}^{2}}$ assuming two different
    coefficients $A$ for the Coulomb term are shown in the left or right columns, respectively, as functions of
    $(\sfrac{a}{r_{0}})^{2}$, using bare links.
    The gray solid line and band show the published ($2+1$)-flavor QCD value~\cite{Cheng:2007jq}.
    The corresponding plots for smeared links are shown in Fig.~\ref{fig:continuum_sqrt_sigma_r0_sq_smeared} of
    Appendix~\ref{app:continuum extrapolations}.}
\end{figure}

We repeat the analysis via the joint fits described earlier on page~\pageref{expl:joint_fits} for the two scales
$r_{0,1}$, or for the string tension~$\sigma$.
To be more precise, we extrapolate $a_{f_{p4s}}(\sfrac{r_{0,1}}{a})$, as well as $\sqrt{\sigma r_{0}^{2}}$ for the two
choices of the coefficient $A$ of $\sfrac{1}{R}$, discussed in Sec.~\ref{sec:interpolation} as functions of
$(\sfrac{a}{r_{0}})^{2}$.
The parametrizations are evaluated at the actual ensemble parameters in our study, or at subsets of these.
We show the data together with the respective continuum results in Figs.~\ref{fig:continuum_r0_r1}
and~\ref{fig:continuum_sqrt_sigma_r0_sq}, respectively.

\begin{figure}
    \centering
    \includegraphics[width=\textwidth]{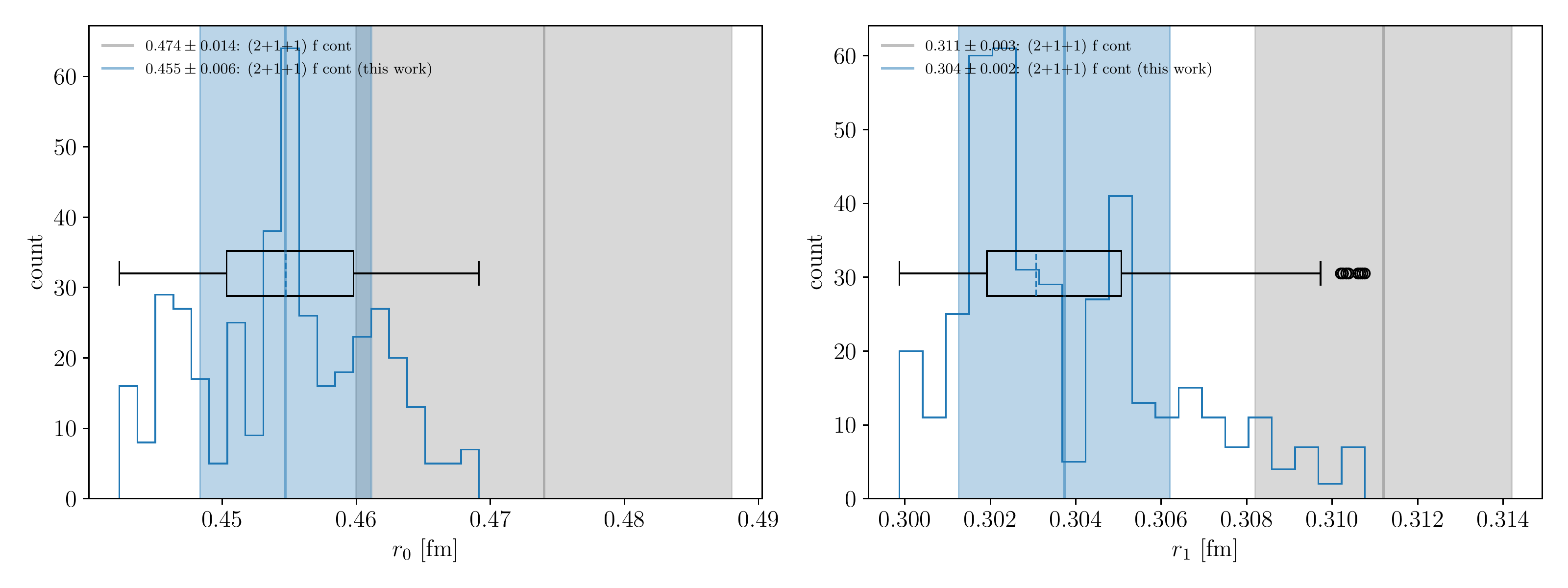}%
    \caption{\label{fig:continuum_r0_r1_hist}%
    Histogram of the continuum extrapolations for the individual scales $(\sfrac{r_{0,1}}{a}) a_{f_{p4s}}$ using the
    \emph{Ansätze} discussed in the text.
    The box plots are explained in the text on page~\pageref{expl:BoxPlot}.
    We take as our final value and uncertainty the mean and the standard deviation of the respective distribution.
    The gray bands corresponds to the literature values~\cite{EuropeanTwistedMass:2014osg, Dowdall:2013rya} for $r_{0}$
    and $r_{1}$, respectively.
    The distribution of the errors is shown in Fig.~\ref{fig:errors_ri} of Appendix~\ref{app:continuum extrapolations}.}
\end{figure}

\begin{figure}
    \centering
    \begin{minipage}[t]{0.49\textwidth}\vspace{0pt}%
    \includegraphics[width=\textwidth]{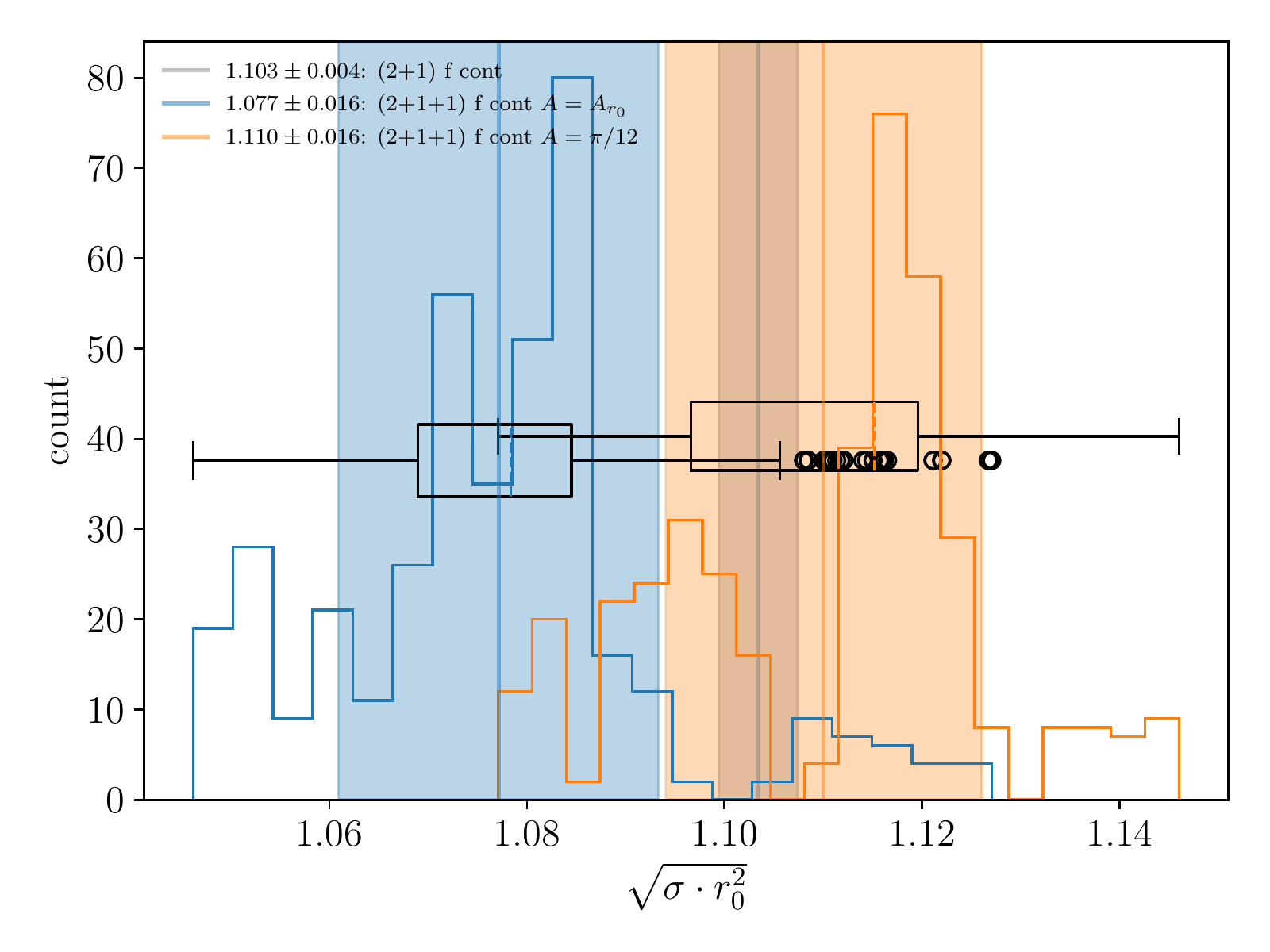}%
    \end{minipage}%
    \hfill%
    \begin{minipage}[t]{0.49\textwidth}\vspace{0pt}%
    \caption{\label{fig:continuum_sqrt_sigma_r0_sq_hist}%
    Histogram of the continuum extrapolations for the string tension using the \emph{Ansätze} discussed in the text.
    The box plots are explained in the text on page~\pageref{expl:BoxPlot}.
    The blue and orange lines and bands correspond to the mean and the standard deviation of the distributions using
    the two different $A$ values, respectively.
    The gray band corresponds to the ($2+1$)-flavor value~\cite{Cheng:2007jq} for $\sqrt{\sigma r_{0}^{2}}$.
    The distribution of the errors is shown in Fig.~\ref{fig:errors_string} of
    Appendix~\ref{app:continuum extrapolations}.}
\end{minipage}
\end{figure}

The histograms of the results are shown in Figs.~\ref{fig:continuum_r0_r1_hist}
and~\ref{fig:continuum_sqrt_sigma_r0_sq_hist}.
For the physical mass ensembles, the products $a_{f_{p4s}}(\sfrac{r_{0,1}}{a})$ approach their respective continuum
limits from above, with clearly monotonic behavior throughout the scaling window.
In the case of $r_{0}$, the best AIC is reached for the bare-link result with quadratic $x$-dependence, or for
smeared-link result with weighted averages, in both cases for the full $\beta$ range.
In the case of $r_{1}$, the best AIC is reached for the bare- or smeared-link results with quadratic $x$-dependence for
the full $\beta$ range.
Fits with quadratic $x$-dependence usually yield rather low continuum results in the first quartile, while fits in the
fourth quartile are obtained by omitting smaller $\beta$ values, and are substantially disfavored in terms of AIC.
On the other hand, for the string tension weighted averages are favored in terms of AIC and close to the center of each
distribution.
The distributions of errors suggest that the width of the histograms are good estimates of the uncertainty; cf.
Fig.~\ref{fig:errors_ri} in Appendix~\ref{app:continuum extrapolations}.

Both histograms of $\sqrt{\sigma r_{0}^{2}}$ have a more pronounced tail towards lower values.
The blue and orange lines and bands correspond to the mean and the standard deviation of the distributions using the
two different $A$ values, respectively.
The gray band corresponds to the published ($2+1$)-flavor QCD result~\cite{Cheng:2007jq} for $\sqrt{\sigma
r_{0}^{2}}$, which had been determined in simultaneous fits of $\sfrac{r_{0}}{a}$ and $a\sqrt{\sigma}$.
This result is bracketed by our two calculations and conceptually closer to our analysis with $A = A_{r_{0}}$;
after taking into account the lower value for $r_{0}$ in our analysis, see Fig.~\ref{fig:FLAG}, the results for
$\sigma$ are in perfect agreement.

Our final results for the scales themselves, and the string tension, are given by the mean and standard deviation of
the respective distributions of our fit results.
Finally, we may combine the continuum limits of $\sfrac{r_{1}}{r_{2}}$ and $r_{1}$ adding errors in quadrature, i.e.,
Eqs.~\eqref{eq:r1/r2} and~\eqref{eq:r1}, to obtain our final result for $r_{2}$.
The corresponding procedure, i.e., combining the continuum limits of $\sfrac{r_{0}}{r_{1}}$ and $r_{1}$, i.e.,
Eqs.~\eqref{eq:r0/r1} and~\eqref{eq:r1}, and adding the errors in quadrature, yields a consistent result for $r_{0}$
with smaller errors, namely $0.4546 \pm 0.0043$~fm.
Our final results read
\begin{align}
    r_{0}       &= 0.4547 \pm 0.0064~\text{fm}, \label{eq:r0}\\
    r_{1}       &= 0.3037 \pm 0.0025~\text{fm}, \label{eq:r1}\\
    r_{2}       &= 0.1313 \pm 0.0041~\text{fm}, \\
    \sqrt{\sigma r_{0}^{2}} &= 1.077 \pm 0.016 \quad (A = A_{r_{0}}), \label{eq:sigma_Ar0} \\
    \sqrt{\sigma r_{0}^{2}} &= 1.110 \pm 0.016 \quad (A = \sfrac{\pi}{12}). \label{eq:sigma_AL}
\end{align}

The decreasing trend in the scale $\sfrac{r_{1}}{a}$ in Fig.~\ref{fig:continuum_r0_r1} is similar to the one already
discussed in Fig.~\ref{fig:r_i_over_a} and is reflected in the continuum value of $r_{1}$ that is lower than the
published value, namely $r_{1} = 0.3112(30)$~fm~\cite{Dowdall:2013rya}.
A similar statement holds for $r_{0}$.
{\refstepcounter{response}\label{resp:sigmas}}
\new{The difference between the results in Eqs.~\eqref{eq:sigma_Ar0} and~\eqref{eq:sigma_AL} must be regarded as a
measure of the inherent uncertainty of defining the string tension in QCD in an intermediate regime between Coulomb
behavior at small distances and string breaking at large distances.}

\subsection{Summary plot}
\label{sec:summary_plot}

We finally have all building blocks available to achieve a curve collapse and generate Fig.~\ref{fig:curve_collapse} in
Sec.~\ref{sec:intro}.
For all ensembles we use the static energy results from the $\Nstates=2$ fits described in Sec.~\ref{sec:sim} as a
function of the tree-level corrected distance $R=\sfrac{r_{I}}{a}$ described in Sec.~\ref{sec:artifacts}.
We convert both the dimensionless static energy $E=aE_{0}(\bm{r},a)$ and the distance $R$ to $r_{0}$ units, i.e.,
$(\sfrac{r_{0}}{a})E$ as a function of $(\sfrac{a}{r_{0}})R$.
We replace $\sfrac{r_{0}}{a}$ by $(\sfrac{r_{1}}{a}) (\sfrac{r_{0}}{r_{1}})$ using Eq.~\eqref{eq:r0/r1} and
$\sfrac{r_{i}}{a}$ from Table~\ref{tab:r_i_over_a_Allton}, if the latter has a smaller error than the former.
We use bare-link rather than smeared-link results for the scales.
Next, we normalize all results to zero at $(\sfrac{a}{r_{0}})R=1$ (using Eq.~\eqref{eq:fit_function} to interpolate in
a $\pm 25\%$ interval).
Then, we combine our normalized continuum results from the bare-link data for $R \le 4$ and the smeared-link data for
$R > 4$, and convert the combined result $(\sfrac{r_{0}}{a})E$ as the ordinate and $(\sfrac{a}{r_{0}})R$ as the
abscissa to physical units using the continuum result for $r_{0}$ from Eq.~\eqref{eq:r0}.
This is the set of static-energy results shown in Fig.~\ref{fig:curve_collapse}.

\subsection{Comparison to published results}

In Fig.~\ref{fig:FLAG}, we compare our (2+1+1)-flavor QCD results for the ratio $\sfrac{r_{0}}{r_{1}}$ and for the
scales $r_{0,1}$ to earlier $(2+1)$- and $(2+1+1)$-flavor QCD results and the corresponding FLAG 2021
average~\cite{Aoki:2021kgd}.
\begin{figure}
    \centering
    \begin{minipage}[t]{0.49\textwidth}\vspace{0pt}%
        \nocite{Aubin:2004wf, Cheng:2007jq, RBC:2010qam, Bazavov:2011nk}
        \nocite{Aubin:2004wf}
        \nocite{Cheng:2007jq}
        \nocite{RBC:2010qam}
        \nocite{Bazavov:2011nk}
        \resizebox{\textwidth}{!}{\input{\FIGDIR/r0_r1_literature_comparison.pgf}}
    \end{minipage}
    \hfill
    \begin{minipage}[t]{0.49\textwidth}\vspace{4ex}%
    \caption{\label{fig:FLAG}%
    Comparison plots for $\sfrac{r_{0}}{r_{1}}$, $r_{0}$, and $r_{1}$ with the FLAG 2021 averages (gray
    bands)~\cite{Aoki:2021kgd}.
    Multiple errors on inputs are added in quadrature.
    References to results entering the FLAG averages are shown in the plots, and we also include ($2+1$)-flavor results
    (gray symbols) for $\sfrac{r_{0}}{r_{1}}$~\cite{Cheng:2007jq} and $r_{0}$~\cite{Aoki:2009sc} that are omitted from
    the FLAG report.
    The blue bands constitute our ``new'' averages explained in the text.}
    \end{minipage}
    \nocite{EuropeanTwistedMass:2014osg}
    \nocite{Aubin:2004wf}
    \nocite{Gray:2005ur}
    \nocite{PACS-CS:2008bkb}
    \nocite{Aoki:2009sc}
    \nocite{RBC:2010qam}
    \nocite{Bazavov:2011nk}
    \nocite{Yang:2014sea}
    \nocite{MILC:2010hzw}
    \nocite{HPQCD:2011qwj}
    \nocite{Dowdall:2013rya}
    \nocite{Davies:2009tsa}
    \nocite{MILC:2009ltw}
    \nocite{MILC:2009mpl}
    \newline
    \resizebox{0.49\textwidth}{!}{\input{\FIGDIR/r0_literature_comparison.pgf}} \hfill
    \resizebox{0.49\textwidth}{!}{\input{\FIGDIR/r1_literature_comparison.pgf}}
\end{figure}
For the ratio $\sfrac{r_{0}}{r_{1}}$, the ($2+1$)-flavor FLAG value has a $\sfrac{\chi^{2}}{\text{d.o.f.}} \approx
\sfrac{4.741}{2}$ [degrees of freedom (d.o.f.)].
Under the assumption of the decoupling of the charm quark, we compute a new weighted average of our result, the known
($2+1$)-flavor results~\cite{Aubin:2004wf, RBC:2010qam, Bazavov:2011nk} and the result~\cite{Cheng:2007jq} omitted in
the FLAG report that encompasses the ($2+1$)-flavor FLAG average and most of its uncertainty band; however, it comes
with a slightly larger uncertainty itself.
The $\sfrac{\chi^{2}}{\text{d.o.f.}} \approx \sfrac{14.800}{3}$ increases slightly but not significantly when including
our result and the one~\cite{Cheng:2007jq} omitted by FLAG.
The change in average is from $1.5049(74)$ to $1.490(20)$.

For the scales $r_{0,1}$, the ($2+1$)-flavor FLAG values, $r_{0} = 0.4701(36)$~fm and $r_{1} = 0.3127(30)$~fm, have a
$\sfrac{\chi^{2}}{\text{d.o.f.}} \approx \sfrac{3.790}{4} = 0.948$ and $\sfrac{\chi^{2}}{\text{d.o.f.}} \approx
\sfrac{7.281}{4} = 1.820$, respectively.
The ($2+1+1$)-flavor results consist of one determination each:
$r_{0} = 0.474(14)$~fm~\cite{EuropeanTwistedMass:2014osg} (with twisted-mass Wilson sea quarks) and
$r_{1}=0.3112(30)$~fm~\cite{Dowdall:2013rya} (with a subset of the ensembles used here).
Performing a weighted average of the respective determinations with our results yields $r_{0} = 0.4586(71)$~fm with
$\sfrac{\chi^{2}}{\text{d.o.f.}} \approx \sfrac{1.472}{1}$ and $r_{1} = 0.3076(37)$~fm with
$\sfrac{\chi^{2}}{\text{d.o.f.}} \approx \sfrac{3.021}{1}$.

\section{Charmed loops}
\label{sec:charm}

As anticipated in the Introduction, now that we have data for the static energy in ($2+1+1$)-flavor QCD, it is possible
to study the effect of the massive charm loops.
We review the weak-coupling result for $\Nf$ massless sea quarks in Sec.~\ref{sec:perturbation_theory} and discuss the
corrections from a massive sea quark in Sec.~\ref{subsec:charm_pert}.
We collect all relevant two-loop formulas in Appendix~\ref{app:pQCD}.
Finally, in Sec.~\ref{sec:comparison}, we conclude the discussion with the explicit, quantitative comparison of our
results with perturbation theory, either with finite-quark-mass effects at the two-loop level, or with ($2+1$)-flavor
QCD, i.e., without charm at all.
We see a clear difference in the lattice-QCD data with and without charm, and the comparison with perturbation theory
validates the expected decoupling.

\subsection{Static energy and force in perturbation theory}
\label{sec:perturbation_theory}

Similarly to Eqs.~\eqref{eq:corr} and~\eqref{eq:tower} in lattice gauge theory, the static energy is related to the
large time behavior of the real-time rectangular Wilson loop of spatial length $r$ and temporal length
$t$~\cite{Wilson:1974sk, Susskind:1976pi, Fischler:1977yf, Brown:1979ya},
\begin{equation}
    E_{0}(r) = \lim\limits_{t \to \infty} \frac{\i}{t} \ln\left\langle \tr \mathcal{P}
        \exp\left[ \i g \oint\limits_{r \times t} \d z^{\mu} A_{\mu}(z) \right] \right\rangle,
\end{equation}
where $\mathcal{P}$ stands for the path ordering of the color matrices, $g$ is the QCD gauge coupling ($\als =
\sfrac{g^{2}}{(4\pi)}$), and $A_{\mu}$ are the SU(3) gauge fields, which are time ordered.
In non-singular gauges, like the covariant gauges and the Coulomb gauge, the Wilson lines at equal initial and final
time do not contribute to the energy and may be ignored or replaced with any initial and final state that overlaps with
the ground state.
The static energy is, up to a constant shift, a physical observable, hence, gauge invariant and renormalization scheme
and scale independent.

At short distances, i.e., when $r \LQCD \ll 1$, it holds that $\alsr \ll 1$ and $E_{0}(r)$ may be expanded as a series
in~$\als$.
In the following of this section, we will restrict ourselves to the case of massless sea quarks.
The perturbative expansion of $E_{0}(r)$ has then the form
\begin{equation}
    \label{eq:statenergyI}
    E_{0}(r) = \Lambda - \frac{\CF \als}{r} \left( 1 + \# \als + \# \als^{2} + \# \als^{3} \ln\als + \# \als^{3} +
        \# \als^{4} \ln^{2}\als + \# \als^{4} \ln\als + \dots \right) ,
\end{equation}
where $\Lambda$ is a constant of mass dimension one and the $\#$ stand for the numerical coefficients that have been
analytically computed so far (some of them are given in Appendix~\ref{app:coefficients}).
It is precisely because the expansion~\eqref{eq:statenergyI} is known to high order, that fitting the static energy
computed with lattice QCD to~\eqref{eq:statenergyI} has the potential to provide an accurate determination of $\als$.

Up to two loops, the only scale that sets the running of the strong coupling constant is $\sfrac{1}{r}$.
Starting from three loops, however, another scale contributes to the static energy, it is the energy scale
$\sfrac{\als}{r}$~\cite{Appelquist:1977es}.
Because this scale is much smaller than $\sfrac{1}{r}$, it may be called the ultrasoft scale, and the latter, the soft
scale.
Ultrasoft gluons may be emitted by static quark-antiquark pairs when changing their color configuration from a color
singlet to a color octet.

Soft and ultrasoft effects are conveniently factorized in an effective field theory framework~\cite{Brambilla:1999qa,
Brambilla:1999xf},
\begin{equation}
\label{eq:statenergyII}
E_{0}(r) = \Lambda + V(r,\nu,\muus) + \delta_{\text{us}}(r,\nu,\muus) ,
\end{equation}
where $V(r,\nu,\muus)$ contains all soft contributions and can be identified with the color-singlet static potential,
and $\delta_{\text{us}}(r,\nu,\muus)$ encodes the ultrasoft contributions.
The scale $\nu$ is the renormalization scale of the strong coupling constant.
It is typically of the order of the soft scale $\sfrac{1}{r}$.
The energy scale $\sfrac{1}{r} \gtrsim \muus \gtrsim \sfrac{\als}{r}$ is a factorization scale separating soft from
ultrasoft modes.

While the static energy is up to a constant shift finite, the functions $V(r,\nu,\muus)$ and
$\delta_{\text{us}}(r,\nu,\muus)$ are not.
Indeed, the $\ln\als$ terms appearing in the expansion~\eqref{eq:statenergyI}, first at order $\als^{4}$, are remnants
of cancellations happening between infrared divergences affecting the potential $V(r,\nu,\muus)$ and ultraviolet
divergences affecting $\delta_{\text{us}}(r,\nu,\muus)$:
\begin{equation}
    \ln\als = \ln\frac{\muus}{\sfrac{1}{r}} + \ln\frac{\sfrac{\als}{r}}{\muus}.
\end{equation}
The potential satisfies renormalization group equations that have been determined and solved up to subleading
logarithmic accuracy~\cite{Pineda:2000gza, Brambilla:2009bi}.
This means that all logarithms of the form $\als^{3+n}\,\ln^{n} (\muus r)$ and $\als^{4+n}\,\ln^{n}(\muus r)$ entering
the potential have been computed.
The two-loop expression of the static potential (energy) supplemented by the logarithms $\als^{3+n}\,\ln^{n} (\muus r)$
($\als^{3+n}\,\ln^{n} \als$) is said to provide the static potential (energy) at next-to-next-to-leading logarithmic
accuracy (N$^{2}$LL).
The three-loop expression of the static potential (energy) supplemented by the logarithms $\als^{4+n}\,\ln^{n} (\muus
r)$ ($\als^{4+n}\,\ln^{n} \als$) is said to provide the static potential (energy) at next-to-next-to-next-to-leading
logarithmic accuracy (N$^{3}$LL).

In lattice regularization, the constant $\Lambda$ in Eq.~\eqref{eq:statenergyII} accounts for the linear divergence of
the self energy.
In dimensional regularization the linear divergence vanishes but the constant $\Lambda$ encodes a renormalon of order
$\LQCD$ that cancels against a renormalon of the same order in the color-singlet static potential~\cite{Pineda:1998id,
Hoang:1998nz}.
The renormalon in the static potential is responsible for the poor convergence of the perturbative expansion of
$E_{0}(r)$.
The poor convergence of the static energy may be treated by subtracting the renormalon of the static potential in a
suitable renormalon subtraction scheme and reabsorbing it into a redefinition of $\Lambda$~\cite{Pineda:2002se}.
Another way to enforce the renormalon cancellation in the perturbative expansion of the static energy is by computing
the force~\cite{Necco:2001gh} defined in Eq.~\eqref{eq:force}.
The force is free of renormalons and therefore well behaved as an expansion in $\als$.
One then recovers the static energy by integrating back over the quark-antiquark distance $r$,
\begin{equation}
    \label{eq:V_from_F}
    E_{0}(r) = \int\limits_{r^{\ast}}^{r} \d r' \; F(r') + \text{const}.
\end{equation}
The distance $r^{\ast} < r$ is arbitrary and contributes only with an additive constant.
This constant can be reabsorbed into an additive shift when comparing with lattice data. Equation~\eqref{eq:V_from_F}
effectively amounts to a rearrangement of the perturbative series enforcing the renormalon
cancellation~\cite{Pineda:2002se}.
The integral in Eq.~\eqref{eq:V_from_F} can be computed (numerically) while setting the renormalization scale $\nu$ at
$\sfrac{1}{r}$.
At two-loop accuracy the static force with $\Nf$ massless quarks reads~\cite{Bazavov:2014soa}
\begin{align}\label{eq:statfor}
    F^{(\Nf)}(r,\nu=\sfrac{1}{r}) &= \frac{\CF \alsNfr}{r^{2}} \Biggl\{1 + \frac{\alsNfr}{4\pi}\left[a_{1}^{(\Nf)}
            + 2\gammaE \beta_{0}^{(\Nf)} - 2\beta_{0}^{(\Nf)}\right] \nonumber \\
        &+ \left(\frac{\alsNfr}{4\pi}\right)^{2} \Biggl[a_{2}^{(\Nf)} + \left(\frac{\pi^{2}}{3} + 4\gammaE^{2}\right)
            \left(\beta_{0}^{(\Nf)}\right)^{2} + \gammaE \left(4 a_{1}^{(\Nf)} \beta_{0}^{(\Nf)}
                + 2 \beta_{1}^{(\Nf)}\right) \nonumber \\
        &- 4 \left(a_{1}^{(\Nf)} + 2\gammaE \beta_{0}^{(\Nf)}\right) \beta_{0}^{(\Nf)}
        - 2\beta_{1}^{(\Nf)}\Biggr]\Biggr\}.
\end{align}
Resumming the ultrasoft leading logarithms in the expression of the static potential yields the expression of the force
at N$^{2}$LL accuracy~\cite{Bazavov:2014soa},
\begin{align}\label{eq:statfor_resummed}
    F^{(\Nf)}(r,\nu=\sfrac{1}{r}) =& \frac{\CF \alsNfr}{r^{2}} \Biggl\{1 + \frac{\alsNfr}{4\pi}\left[a_{1}^{(\Nf)}
            + 2\gammaE \beta_{0}^{(\Nf)} - 2\beta_{0}^{(\Nf)}\right] \nonumber \\
        &+ \left(\frac{\alsNfr}{4\pi}\right)^{2} \Biggl[a_{2}^{(\Nf)}
            + \left(\frac{\pi^{2}}{3} + 4\gammaE^{2}\right) \left(\beta_{0}^{(\Nf)}\right)^{2}
            + \gammaE \left(4 a_{1}^{(\Nf)} \beta_{0}^{(\Nf)} + 2 \beta_{1}^{(\Nf)}\right) \nonumber \\
        &- 4 \left(a_{1}^{(\Nf)} + 2\gammaE \beta_{0}^{(\Nf)}\right) \beta_{0}^{(\Nf)} - 2\beta_{1}^{(\Nf)}\Bigg]
        + \left(\frac{\alsNfr}{4\pi}\right)^{2} \left[-\frac{a_{3}^{\text{L}}}{2\beta_{0}^{(\Nf)}}
            \ln\left(\frac{\alsNf(\muus)}{\alsNfr}\right)\right]\Biggr\},
\end{align}
where we have set the ultrasoft scale to be
\begin{equation}
    \muus = \frac{\CA \alsNfr}{2r},
\end{equation}
which is the difference between the Coulomb potential in the adjoint and in the fundamental representation of SU(3).
The coefficients $a_{1}^{(\Nf)}$, $a_{2}^{(\Nf)}$, $a_{3}^{\text{L}}$, $\beta_{0}^{(\Nf)}$, and $\beta_{1}^{(\Nf)}$ can
be found in Appendix~\ref{app:coefficients}; $\gammaE$ is the Euler--Mascheroni constant.

\subsection{Charm quark mass effects in perturbation theory}
\label{subsec:charm_pert}

Effects due to the finite mass of a heavy quark, while keeping $\Nf$ quarks massless, can be cast into a correction
$\delta V_{m}^{(\Nf)}(r)$ to be added to the static potential or energy.
This correction has been computed at $\order(\als^{2})$ in Ref.~\cite{Eiras:1999xx} and at $\order(\als^{3})$ in
Refs.~\cite{Melles:2000dq, Melles:2000ey, Hoang:2000fm}.
For a typo-free summary, see Ref.~\cite{Recksiegel:2001xq} and Appendix~\ref{app:charm}.
In our case of interest, the relevant massive quark is the charm quark.

The expression for the static energy that we use in this work for comparison to lattice simulations with $\Nf=3$ nearly
massless quarks and a charm quark of mass $m = \mc = 1.28$~GeV is
\begin{equation}
    \label{eq:full_statenergy}
    E^{(\Nf)}_{0,m}(r) = \int\limits_{r^{\ast}}^{r} \d r' \; F^{(\Nf)}(r') + \delta V_{m}^{(\Nf)}(r) + \text{const},
\end{equation}
where we have explicitly indicated for each quantity the number of massless quarks.
In particular, in the right-hand side all couplings are computed with $\Nf$ massless flavors.
The expression of $F^{(\Nf)}(r)$ at two loops is given in Eq.~\eqref{eq:statfor}, and the expression of
$F^{(\Nf)}(r)$ at N$^{2}$LL accuracy is given in Eq.~\eqref{eq:statfor_resummed}.
The expression of $\delta V_{m}^{(\Nf)}(r)$ up to two-loop accuracy is given by
\begin{equation}
    \delta V_{m}^{(\Nf)}(r) = \delta V_{m}^{(\Nf),[2]}(r,\nu) + \delta V_{m}^{(\Nf),[3]}(r,\nu),
\end{equation}
where $\nu$ is the renormalization scale, and $\delta V_{m}^{(\Nf),[2]}(r,\nu)$ and $\delta V_{m}^{(\Nf),[3]}(r,\nu)$
are the one- and two-loop corrections, given in Eqs.~\eqref{eq:deltaV2} and~\eqref{eq:deltaV3}, respectively.
The renormalization scale of the coupling is set to be $\sfrac{1}{r}$.
The integral over the force $F^{(\Nf)}(r)$ is performed numerically while keeping $\als$ running at three-loop
accuracy using the \texttt{RunDec} package~\cite{Chetyrkin:2000yt, Schmidt:2012az, Herren:2017osy}.

The static energy with a massive quark and $\Nf$ massless quarks reduces to the static energy with $\Nf$ massless
quarks, $E^{(\Nf)}_{0}(r)$, for $m \gg \sfrac{1}{r}$, and it reduces to the static energy with $\Nf+1$ massless quarks,
$E^{(\Nf+1)}_{0}(r)$, for $m \ll \sfrac{1}{r}$.
This is a consequence of the decoupling of the static potential discussed in Appendix~\ref{app:charm}.

Finally, we remark that since the finite mass corrections to the static potential, $\delta V_{m}^{(\Nf)}(r)$, are known
only up to two loops, the available three-loop information on the force, $F^{(\Nf)}(r)$, cannot be used in a
consistent manner.
In particular, adding the three-loop correction to $F^{(\Nf)}(r)$ without the three-loop correction to $\delta
V_{m}^{(\Nf)}(r)$ would lead to a violation of the decoupling theorem in the static energy at order
$\als^{4}$.

\subsection{Charm quark mass effects on the lattice}
\label{sec:comparison}

In this section, we study how a finite charm quark mass affects the determination of the static energy on lattices with
($2+1+1$) flavors.
In particular, we focus on the short distance behavior of the static energy and compare it with the expectation from
perturbation theory, i.e., Eq.~\eqref{eq:full_statenergy}.
In principle, we could use this comparison to extract $\als$, as this is the only free parameter (up to the constant
shift) in Eq.~\eqref{eq:full_statenergy}.
For the argument given at the end of the previous section, this would lead to a determination of $\als$ accurate at two
loops.
A two-loop determination of $\als$ would, however, not be competitive with respect to existing three-loop
determinations based on ($2+1$)-flavor lattices~\cite{Bazavov:2014soa, Takaura:2018vcy, Bazavov:2019qoo, Ayala:2020odx}.
Hence, we will refrain from a determination of $\als$ in this work, while we will limit ourselves to some observations
on the impact of finite charm quark mass effects on the static energy.
This is a first time study of this kind of effects.

\begin{figure}
    \centering
    \begin{minipage}[t]{0.49\textwidth}\vspace{0pt}%
        \includegraphics[width=\textwidth]{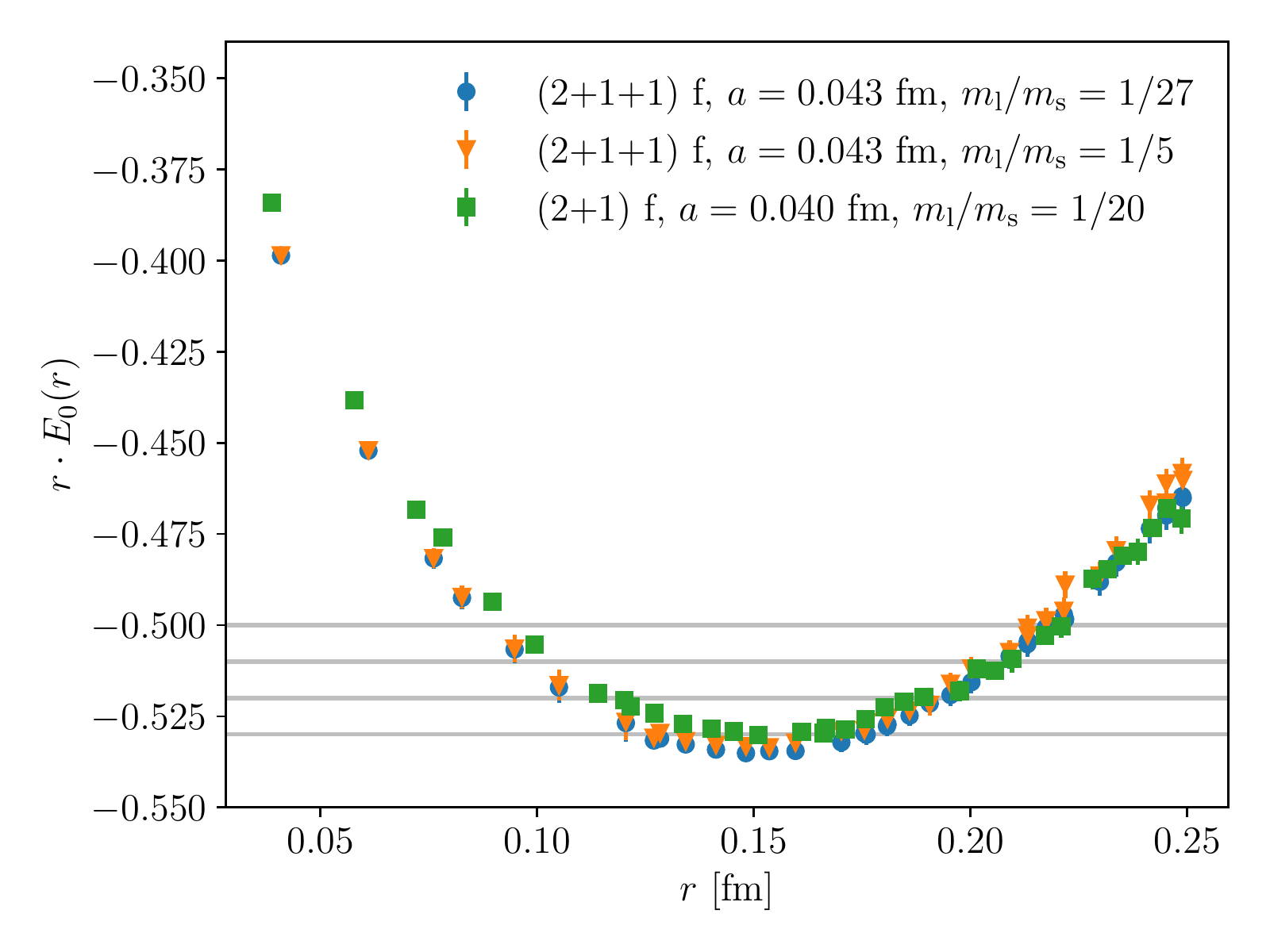}%
    \end{minipage}%
    \hfill%
    \begin{minipage}[t]{0.49\textwidth}\vspace{0pt}%
        \caption{\label{fig:charm1}%
        The dimensionless quantity $rE_{0}(r)$ for two different ($2+1+1$)-flavor ensembles using different light quark
        masses and one ($2+1$)-flavor ensemble of similar lattice spacing.
        The latter has been matched to the ($2+1+1$)-flavor ensemble of the similar light quark mass ratio at large
        distances.}
    \end{minipage}
\end{figure}

In Fig.~\ref{fig:charm1}, we show ($2+1$)-flavor and ($2+1+1$)-flavor lattice data for $rE_{0}(r)$,\footnote{%
We prefer to show $rE_{0}(r)$ rather than $E_{0}(r)$ because $rE_{0}(r)$ is a dimensionless quantity.
Moreover, it has no Coulomb singularity, which facilitates plotting and comparisons.} which correspond to different
discretizations and to light quark mass over strange quark mass ratios $\sfrac{\ml}{\ms} = \sfrac{1}{20}$ and
$\sfrac{\ml}{\ms} = \sfrac{1}{27}$, respectively.
For the ($2+1+1$)-flavor data we use the scale $a_{f_{p4s}}$ in Table~\ref{tab:ensembles} to convert the abscissa to
physical units; for the ($2+1$)-flavor data, we use the published value $\sfrac{r_{1}}{a}=7.690(58)$ combined with the
published value of $r_{1}$ in Eq.~\eqref{eq:scale_estimates}, both from Ref.~\cite{HotQCD:2014kol}.
We add a mass independent constant to the ($2+1+1$)-flavor $E_{0}(r)$ such that the shifted data at physical mass (in
blue color) are rather flat in the range of interest to facilitate the visualization of the small finite mass effects
that we are investigating.
We additionally show another ($2+1+1$)-flavor data set (in orange color) with larger light quark mass $\sfrac{\ml}{\ms}
= \sfrac{1}{5}$, whose data set has not been shifted relative to the physical one.
Therefore, the difference between the two ($2+1+1$)-flavor data sets is due to the different light quark masses.
We match the ($2+1$)-flavor data (in green color) to the ($2+1+1$)-flavor data of the similar $\sfrac{\ml}{\ms}$-ratio,
whose additive shift is different due to the difference in discretizations, at large distances, $r \gg \sfrac{1}{\mc}
\sim 0.15$~fm, where they must agree up to a constant due to the decoupling of the charm quark.
This matching of the ($2+1$)-flavor data to the ($2+1+1$)-flavor data is done by minimizing their difference over the
range $r \in [0.18,0.27]$~fm and by varying the range to estimate the matching error.
This corresponds to a relative shift of the ($2+1$)-flavor data compared to the ($2+1+1$)-flavor data by an amount of
$0.028 \pm 0.001$ at $r=0.15$~fm.
The difference in the light quark mass between the ($2+1$)-flavor data and the ($2+1+1$)-flavor data is smaller than
the one between the two sets of ($2+1+1$)-flavor data.
Since the latter are hardly distinguishable, we deduce that the light quark mass difference should be irrelevant in
this entire range and that the difference between the ($2+1$)-flavor data and the ($2+1+1$)-flavor data is due to the
dynamical charm quark in the sea.
The effect of the dynamical charm is therefore significant and visible in the data.
{\refstepcounter{response}\label{resp:charm}}
\new{Comparison between ($2+1$)-flavor and ($2+1+1$)-flavor lattice data using different ensembles with
$\sfrac{\ml}{\ms}=\sfrac{1}{5}$ (the \ensemble{7.28}{iii} ensemble compared to two ($2+1$)-flavor ensembles with
$\sfrac{r_1}{a}=10.653(60)$ or $8.905(60)$ from Ref.~\cite{Bazavov:2017dsy}) gives qualitatively similar results.}

\begin{figure}
    \centering
    \includegraphics[width=0.49\textwidth]{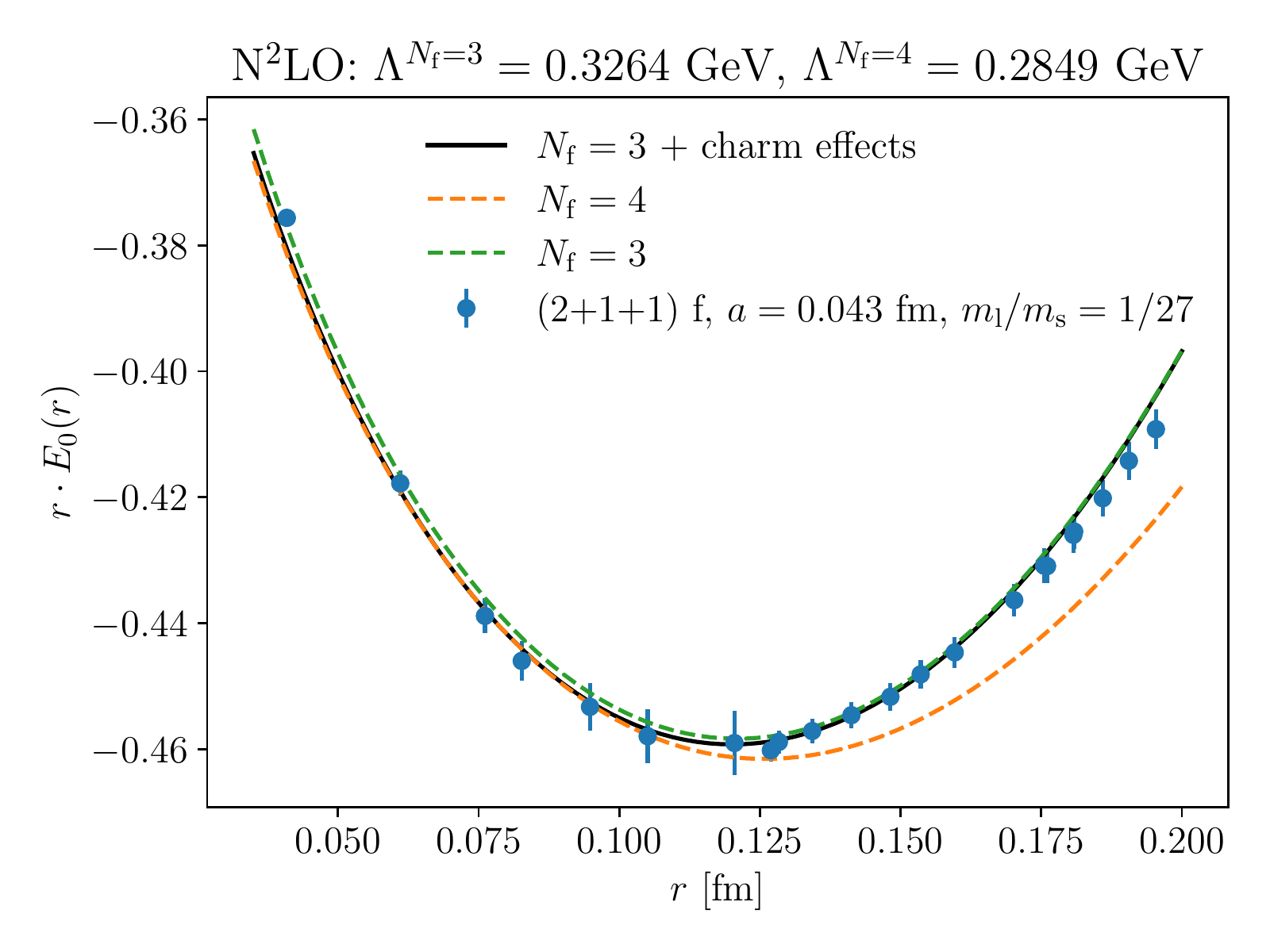}%
    \hfill%
    \includegraphics[width=0.49\textwidth]{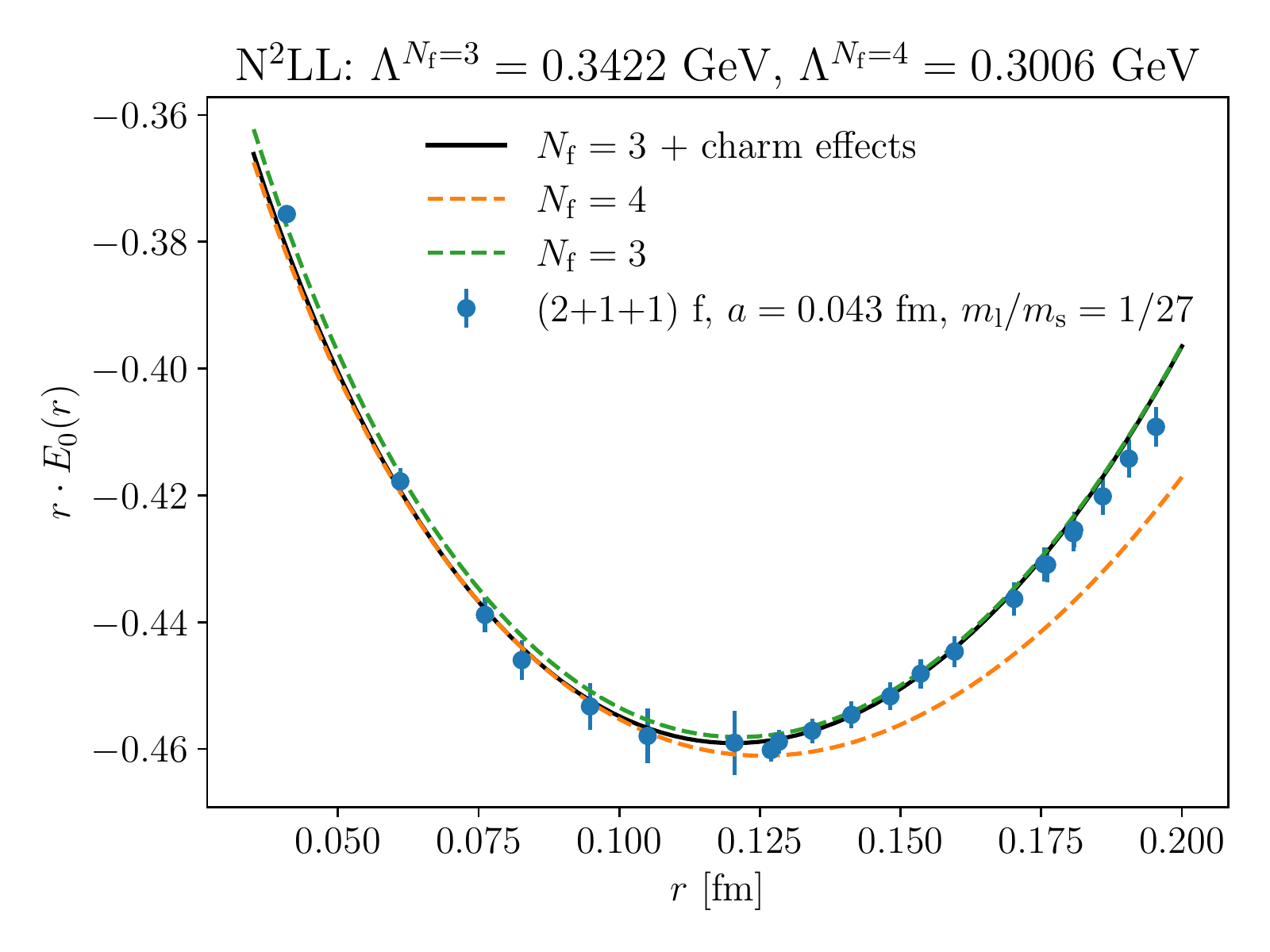}%
    \caption{\label{fig:charm2}%
    Comparison of the ($2+1+1$)-flavor data with curves obtained from different perturbative expressions of the static
    energy times the distance.
    Left: in black, green, and orange we show $r E^{(3)}_{0,m}(r)$, $r E^{(3)}_{0}(r)$, and $r E^{(4)}_{0}(r)$,
    respectively.
    The perturbative curves have been obtained from the static force at two loops [next-to-next-to-leading order
    (N$^{2}$LO)], Eq.~\eqref{eq:statfor}, using three-loop running of $\als$.
    Charm mass effects have been included in the black curve at two-loop accuracy using $\mc^{\MS}(\mc^{\MS}) =
    1.28$~GeV.
    Right: as in the left panel but at N$^{2}$LL accuracy, i.e., the force is given by Eq.~\eqref{eq:statfor_resummed}.}
\end{figure}

As discussed in Sec.~\ref{subsec:charm_pert} and Appendix~\ref{app:charm}, the effective number of active flavors that
enters the running of $\als$ and the static energy changes at different distances with $\mc$ fixed.
At large distance, $r \gg \sfrac{1}{\mc}$, the charm quark decouples, and in this region the static energy behaves
effectively as with three massless flavors.
At short distance, $r \ll \sfrac{1}{\mc}$, the charm quark contributes as an active massless flavor, and thus, in this
region the static energy behaves effectively as with four massless flavors.
One expects to see this behavior realized by the ($2+1+1$)-flavor lattice data of the static energy.
In the following, we will superimpose Fig.~\ref{fig:theory_curves} to the lattice data and verify that this is indeed
the case in the distance region for which we expect perturbation theory to work: $r < \sfrac{1}{\LQCD} \approx 0.2$~fm.

In order to compare with perturbation theory, we need first to determine $\Lambda_{\MS}$.\footnote{%
Although we do not attempt to give a precision extraction of $\als$ or $\Lambda_{\MS}$, we need to determine a
reference value and use it throughout the analysis.}
We determine $\Lambda_{\MS}^{(\Nf=3)}$ by fitting Eq.~\eqref{eq:full_statenergy} to the physical ($2+1+1$)-flavor
ensemble.
We leave out data at $\sfrac{r}{a}=1$ from all the fits and vary the fit range up to $r \approx 0.19$~fm using
$\mc^{\MS}(\mc^{\MS}) = 1.28$~GeV and the three-loop running of $\als$.
To account for the residual discretization artifacts, see Fig.~\ref{fig:tree_level_correction-2}, we enlarge the error
to $3$\textperthousand{} of the raw data at $\sfrac{r}{a} \le \sqrt{8}$, or to $1$\textperthousand{} of the raw data,
otherwise.
The numerical running of $\als$ and the conversion between the three-flavor and the four-flavor values of
$\Lambda_{\MS}$ is performed using the \texttt{RunDec} package~\cite{Chetyrkin:2000yt, Schmidt:2012az, Herren:2017osy}.
The value of $\Lambda_{\MS}^{(\Nf=3)}$ that we obtain using the N$^{2}$LO expression of the force,
Eq.~\eqref{eq:statfor}, is $\Lambda_{\MS}^{(\Nf=3)} \approx 326$~MeV.\footnote{%
This value is about $3.8\%$ higher than the ($2+1$)-flavor determination of Ref.~\cite{Bazavov:2019qoo}, yet still
covered within the perturbative truncation error.
Note that the determination in Ref.~\cite{Bazavov:2019qoo}, based on ($2+1$)-flavor lattice data, is accurate up to
three loops, although the central value is the same between two or three loops with leading ultrasoft resummation.
If we compare, instead, the values for $r_{1}\Lambda_{\MS}^{(\Nf=3)}$, then we see a partial compensation between the
smaller value of $r_{1}$ and the larger value of $\Lambda_{\MS}^{(\Nf=3)}$ in ($2+1+1$)-flavor QCD---the difference
shrinks to the level expected from combined lattice uncertainties.}

The static energy at N$^{2}$LO, including N$^{2}$LO massive charm loop effects, is shown by the black curve in the left
panel of Fig.~\ref{fig:charm2}.
Lattice data are the blue dots.
Omitting the data point at the smallest distance, we obtain $\chi^{2}_{\text{red}} \approx 0.5$.
We use $r \le 0.19$~fm.
The static energy, Eq.~\eqref{eq:V_from_F}, with four massless active flavors, $\Nf=4$, which is the orange dashed
curve, is matched to the black curve at 0.08~fm to compensate for truncation effects of order $\als^{4}$.
It begins to deviate from the lattice data at distances $r \gtrsim 0.12$~fm.
The static energy, Eq.~\eqref{eq:V_from_F}, with three massless active flavors, $\Nf=3$, is shown by the green dashed
curve.
In this case no shift is performed to match with the black curves, as the two overlap exactly by construction at large
distances.
The green dashed curve shows a systematic overshooting of the data at distances $r \lesssim 0.12$~fm (with the
exception of the first data point, corresponding to one lattice spacing, which is possibly affected by large
discretization artifacts).
The ($2+1+1$)-flavor lattice data behave therefore accordingly to the decoupling theorem.
At large distance they are well described by the perturbative static energy with three massless flavors and at short
distance by the perturbative static energy with four massless flavors.
The static energy with three massless flavors and one massive charm interpolates smoothly between these two curves and
on the overall describes well the data.
We have seen, indeed, in Fig.~\ref{fig:charm1} that the lattice data are sensitive to finite charm mass effects in the
intermediate region $r \sim \sfrac{1}{\mc}$.

A similar analysis can be done using the N$^{2}$LL expression of the force, Eq.~\eqref{eq:statfor_resummed}.
We get, in this case, $\Lambda_{\MS}^{(\Nf=3)} \approx 342$~MeV.\footnote{%
This value is only about $1.1\%$ away from, and therefore consistent inside uncertainties with, the N$^{3}$LL fit of
Ref.~\cite{Ayala:2020odx}, which found $\Lambda_{\MS}^{(\Nf=3)} \approx 338$~MeV by reanalyzing a subset of the
($2+1$)-flavor data.}
As before, the black curve, which includes the charm mass effects, reproduces the data with $\chi^{2}_{\text{red}}
\approx 0.5$, while the orange dashed curve with four massless flavors deviates significantly from the data at $r
\gtrsim 0.12$~fm, and the green dashed curve with three massless flavors overshoots the data at $r \lesssim 0.12$~fm,
with the possible exception of the first data point.
Again, we see that the lattice data reflect the expectations from the decoupling theorem, i.e., at large distance they
are well described by the perturbative static energy with three massless flavors and at short distance by the
perturbative static energy with four massless flavors, while the static energy with three massless flavors and one
massive charm interpolates smoothly between these two curves and describes well the data.

\section{Conclusions}
\label{sec:conclusions}

In this paper, we present results for the static energy in ($2+1+1$)-flavor QCD over a wide range of lattice spacings
and several quark masses, including the physical quark mass.
To gain better control of the statistical errors in the static energy at large distances, the calculations have been
performed using bare links, or links after one level of HYP smearing.
This enabled us to obtain reliable results for the static energy also at relatively large distances.
We perform a simultaneous determination of the scales $r_{1}$ and $r_{0}$, as well as the string tension $\sigma$, and
for the smallest three lattice spacings, we also determine the scale $r_{2}$.
For the scales, direction-dependent discretization uncertainties dominate over statistical errors.
Our values of $\sfrac{r_{1}}{a}$ on the coarser lattices are marginally lower than previous ones from the MILC
Collaboration~\cite{MILC:2010pul, MILC:2012znn} and have larger uncertainties due to the differences in the procedure
for obtaining $\sfrac{r_{1}}{a}$.
Our results on $\sfrac{r_{0}}{r_{1}}$ and $r_{0} \sqrt{\sigma}$ agree with published ($2+1$)-flavor results.
On the other hand, our result for $\sfrac{r_{1}}{r_{2}}$ differs significantly from the value obtained in the
($2+1$)-flavor case~\cite{Bazavov:2017dsy}, which is most likely due to the effect of the charm quark.

We study in detail the effect of the charm quark on the static energy by comparing our results on some of the finest
two lattices with previously published ($2+1$)-flavor QCD results at similar lattice spacing.
Significant influence of the different light quark masses can be ruled out.
We have found that for $r > 0.2$~fm our results on the static energy agree with the ($2+1$)-flavor results, implying
the decoupling of charm quark for these distances.
For smaller distances, on the other hand, we find that the effect of the dynamical charm quark is noticeable.
The behavior of the ($2+1+1$)-flavor lattice data for the static energy is well reproduced by the perturbative
expression of the static energy incorporating the charm mass effects at two loops.
This shows at a quantitative level how the ($2+1+1$)-flavor lattice data smoothly interpolate between the large
distance region, where the charm quark decouples, and the short distance region, where the charm quark may be treated
as massless.

A precision extraction of $\als$ from lattice QCD data of the static energy with ($2+1+1$) flavors is at the moment
problematic if data are included for distances around $\sfrac{1}{\mc}$.
At such distances, as we have seen, finite charm mass effects have to be included in the fitting perturbative
expression.
Since these are known up to two loops, this is also the maximal precision one may obtain at present for the strong
coupling from these data.
The computation of finite charm-mass corrections to the static energy at three loops is certainly challenging.

\begin{acknowledgments}
We thank the MILC Collaboration for allowing us to use of their ($2+1+1$)-flavor HISQ ensembles.
The simulations were carried out on the computing facilities of the Computational Center for Particle and Astrophysics
(C2PAP) in the project \emph{Calculation of finite $T$ QCD correlators} (pr83pu) and of the SuperMUC cluster at the
Leibniz-Rechenzentrum (LRZ) in the project \emph{The role of the charm-quark for the QCD coupling constant} (pn56bo),
both located in Munich (Germany).
This research was funded by the Deutsche Forschungsgemeinschaft (DFG, German Research Foundation) cluster of excellence
``ORIGINS'' (\href{www.origins-cluster.de}{www.origins-cluster.de}) under Germany's Excellence Strategy
EXC-2094-390783311.
This research is supported by the DFG and the NSFC through funds provided to the Sino-German CRC 110 ``Symmetries and
the Emergence of Structure in QCD''.

R.L.D.\ is supported by the Ramón Areces Foundation, the INFN postdoctoral fellowship AAOODGF-2019-0000329, and the
Spanish Grant MICINN: PID2019-108655GB-I00.
Fermilab is managed by Fermi Research Alliance, LLC, under Contract No.\ DE-AC02-07CH11359 with the U.S.\ Department of
Energy.
P.P.\ is supported by the U.S.\ Department of Energy under Contract No.\ DE-SC0012704.
A.V.\ is supported by the EU Horizon 2020 research and innovation programme, STRONG-2020 project, under Grant Agreement
No.~824093.
J.H.W.’s research is funded by the Deutsche Forschungsgemeinschaft (DFG, German Research
Foundation)---Projektnummer 417533893/GRK2575 ``Rethinking Quantum Field Theory''.

The lattice QCD calculations have been performed using the publicly available
\href{https://web.physics.utah.edu/~detar/milc/milcv7.html}{MILC code}.
The data analysis for the ground state extraction was performed using the \texttt{R-base} and \texttt{NLME}
packages~\cite{Rpackage, nlme}.
The data analysis for all further quantities was performed using \texttt{python3}~\cite{python3, python_II, python_III}
and the libraries \texttt{gvar}~\cite{gvar}, \texttt{matplotlib}~\cite{Hunter:2007},
\texttt{numpy}~\cite{oliphant2006guide, 5725236, harris2020array}, \texttt{pandas}~\cite{reback2020pandas,
mckinney-proc-scipy-2010}, and \texttt{scipy}~\cite{scipy}.

V.L.\ and S.S.\ would like to thank Xavier Garcia i Tormo for discussions.
S.S.\ would like to thank Florian M. Kaspar for discussions.
\end{acknowledgments}

\appendix
\section{Wilson line correlation function at different levels of gauge fixing}
\label{app:gauge-fixing}

\begin{figure}
    \centering
    \includegraphics[width=0.49\textwidth]{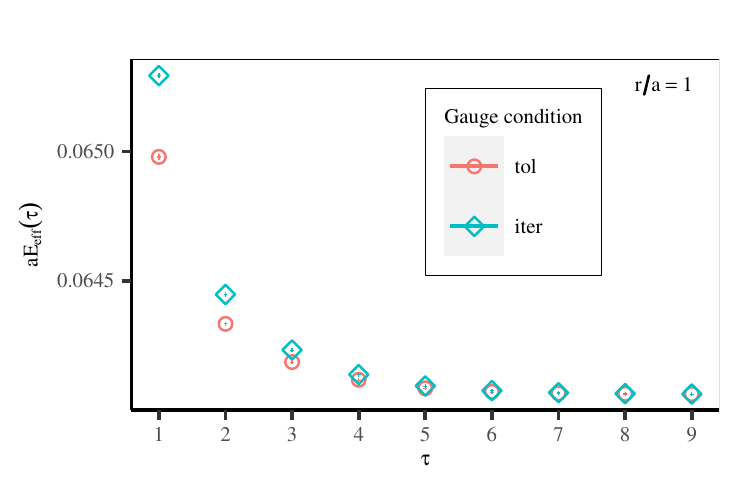}%
    \hfill%
    \includegraphics[width=0.49\textwidth]{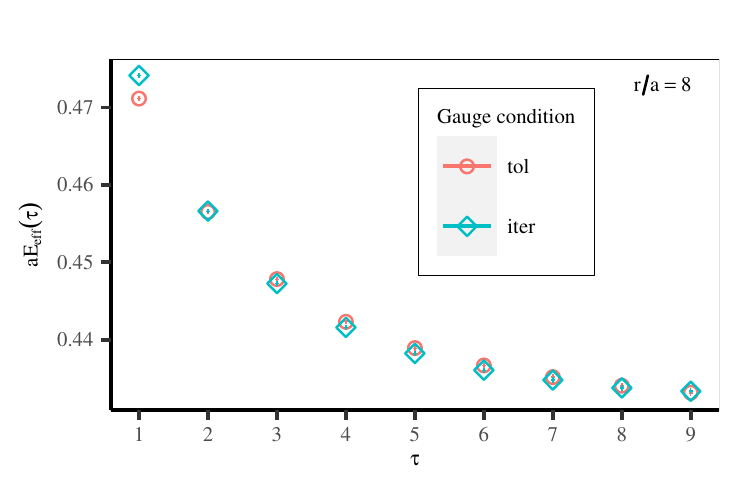}%
    \caption{\label{fig:gaugefix_effm}%
    The effective mass $aE_{\text{eff}}(\tau)$ has been calculated on the two subsets of the ensemble
    \ensemble{6.72}{i} with different gauge-fixing schemes, which are labeled \texttt{tol} for fixed tolerance (red)
    and \texttt{iter} for fixed number of iterations (teal).
    $aE_{\text{eff}}(\tau)$ differs at small $\tau$ but approaches the same plateau at large $\tau$.
    Note, that the ordering of the two gauge-fixing schemes changes in a statistically significant manner with $\tau$.
    At least two crossings occur for larger $\sfrac{r}{a}$; the first of these occurs at smaller $\tau$ for larger
    $\sfrac{r}{a}$.
    We show the effective mass for one iteration of HYP smearing since the errors and fluctuations are larger without
    smearing.
    Jackknife errors are obtained from the distribution of the resamples.}
\end{figure}

\begin{figure}
    \centering
    \includegraphics[width=0.49\textwidth]{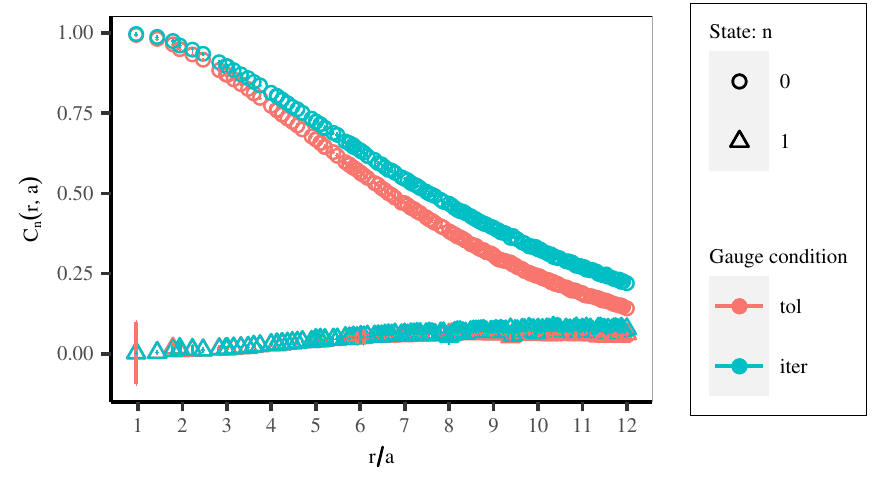}%
    \hfill%
    \includegraphics[width=0.49\textwidth]{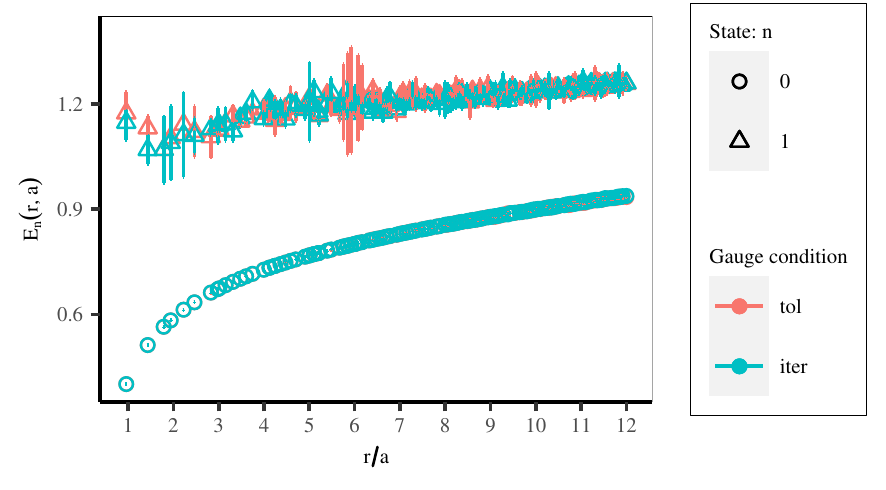}%
    \caption{\label{fig:gaugefix_fits}%
    The correlation function (without smearing) has been analyzed on the two subsets of the ensemble \ensemble{6.72}{i}
    with different gauge-fixing schemes via three-state fits (for the schemes and the color code, see
    Fig.~\ref{fig:gaugefix_effm} and text).
    Left: both overlap factors, $C_{0,1}$, of the ground state or first excited state, respectively, decrease as the
    volume-averaged final gauge-fixing functional is reduced towards a lower tolerance.
    The relative decrease of the ground state overlap factor $C_{0}(r,a)$ increases quite dramatically from $0.3\%$ at
    $\sfrac{r}{a}=1$ to $30\%$ at $\sfrac{r}{a}=12$.
    For the excited state, the overlap factor $C_{1}(r,a)$ changes mildly by about 40\% at $\sfrac{r}{a}=1$ to 25\% at
    $\sfrac{r}{a}=12$; however, since $C_{1}(r,a)$ increases for large $\sfrac{r}{a}$, this is statistically
    significant only for large enough $\sfrac{r}{a}$.
    Right: the change of the energy levels between the two smearing schemes is statistically insignificant.
    $E_{1}(r,a)$ represents in this plot the full first excited state energy.
    Thick error bars represent Hessian errors, while thin error bars represent jackknife errors obtained from the
    distribution of the resamples.}
\end{figure}

As stated in Sec.~\ref{sec:setup} in the body of this paper, different gauge-fixing schemes were unintentionally mixed
during the simulations.
On the one hand, we utilized the original scheme with a tolerance of $\epsilon = 2 \times 10^{-6}$ for the coarser
ensembles with $\beta \le 6.30$.
On the other hand, we have employed a fixed number (320) of steps for the finer ensembles with $\beta \ge 7.00$.
Lastly, we could use gauge-fixed ensembles for $\beta = 6.72$ with unphysical masses (\ensemble{6.72}{ii} or
\ensemble{6.72}{iii}) with a tolerance of $\epsilon = 2 \times 10^{-6}$.
However, for \ensemble{6.72}{i} we could use only a fraction of the ensemble gauge fixed with a tolerance of $\epsilon
= 2 \times 10^{-6}$ and had to gauge fix the rest ourselves.
Due to an initial misunderstanding, we used a prescription with a fixed number (320) of steps instead.
These lead to slight deviations in the final gauge-fixing precision, as the procedure usually stopped $\order(1)$ steps
before reaching the tolerance of $\epsilon = 2 \times 10^{-6}$ (usually less than $10\%$ level deviation between the
volume-averaged final gauge-fixing functional).
Since the time history mostly consisted of consecutive segments that used either tolerance or step number as criteria,
this led to unexpectedly large autocorrelations in the correlator size in some streams that were, however, practically
absent in the effective mass.

As a consequence, we analyzed the subsets of the \ensemble{6.72}{i} ensemble with different gauge-fixing schemes
separately and confirmed the independence of the energy levels.
We show the results of this analysis on the level of the effective mass in Fig.~\ref{fig:gaugefix_effm} and on the
level of the fit parameters in Fig.~\ref{fig:gaugefix_fits}.

\FloatBarrier
\section{Additional plots and tables of numerical data}
\label{app:data}

This Appendix contains additional material in which we discuss further details of the analysis.
In Sec.~\ref{app:correlator_fits} we provide further details on the correlator fits.
We tabulate the tree-level corrections in Sec~\ref{app:tree-level corrections}.
Section~\ref{app:scales and string tension} extends the discussion of the systematic uncertainty and the discretization
effects of the scales and string tension.

\subsection{Fit ranges, quality, and stability}
\label{app:correlator_fits}

Table~\ref{tab:fit-intervals} shows the actual time ranges used in the correlator fits.
The choices have been informed by keeping similar time ranges in physical units across all ensembles accounting for the
number of states used, see Eq.~\eqref{eq:time-ranges}, with an extra variation to check for systematic effects.

\clearpage


{%
\newcommand{\h}{\phantom{2}}
\renewcommand{\arraystretch}{1.1}
\LTcapwidth=\textwidth
\begin{longtable}[c]{ccc|cc|ccc|ccc|ccc}
    \caption{\label{tab:fit-intervals}%
    Time intervals used in the correlator fits, see Sec.~\ref{sec:corrfit}; $t=\sfrac{\tau}{a}$.
    An open-ended dash means ``until upper or lower end of available data'', respectively.
    Entries marked as ``\dots'' indicate that the fit was not possible.}\\
    \hline
\hline
$\approx a$~(fm)& $\beta$ & $\ttmax$ & Operator & $\sfrac{r_{I}}{a}$
& $\ttmin^{(1,-)}$ & $\ttmin^{(2,-)}$ & $\ttmin^{(3,-)}$
& $\ttmin^{(1,0)}$ & $\ttmin^{(2,0)}$ & $\ttmin^{(3,0)}$
& $\ttmin^{(1,+)}$ & $\ttmin^{(2,+)}$ & $\ttmin^{(3,+)}$ \\*

\hline
\endfirsthead
\hline
$\approx a$~(fm)& $\beta$ & $\ttmax$ & Operator & $\sfrac{r_{I}}{a}$
& $\ttmin^{(1,-)}$ & $\ttmin^{(2,-)}$ & $\ttmin^{(3,-)}$
& $\ttmin^{(1,0)}$ & $\ttmin^{(2,0)}$ & $\ttmin^{(3,0)}$
& $\ttmin^{(1,+)}$ & $\ttmin^{(2,+)}$ & $\ttmin^{(3,+)}$ \\*

\hline
\endhead
\hline
\endfoot
\endlastfoot
\mrow{3}{0.15\h} &  \mrow{3}{5.8\h}  &   \mrow{3}{9}
&  Bare             &  all
&     1             &     1             &   \dots
&     2             &     2             &   \dots
&     3             &     3             &   \dots     \\*
\cline{4-14}
&                   &
&  \mrow{2}{HYP}    &  \phantom{0.0}--3.5
&     1             &     1             &   2
&     2             &     2             &   2
&     3             &     3             &   2      \\*
&                   &
&                   &  3.5--\phantom{0.0}
&     1             &     1             &   1
&     2             &     2             &   1
&     3             &     3             &   1      \\
\hline
\pagebreak[2]
\mrow{4}{0.12\h} &  \mrow{4}{6.0\h}  &   \mrow{4}{6}
&  \mrow{2}{Bare}   &  \phantom{0.0}--1.6
&     2             &     2             &   \dots
&     3             &     2             &   \dots
&     \dots            &     \dots            &   \dots     \\*
&                   &
&                   &  1.6--\phantom{0.0}
&     2             &     2             &   \dots
&     3             &     2             &   \dots
&     \dots            &     \dots            &   \dots     \\*
\cline{4-14}
&                   &
&  \mrow{2}{HYP}    &   \phantom{0.0}--3.5
&     2             &     2             &   1
&     3             &     2             &   1
&     \dots            &     \dots            &   \dots     \\*
&                   &
&                   &  3.5--\phantom{0.0}
&     2             &     2             &   \dots
&     3             &     2             &   1
&     \dots            &     \dots            &   \dots     \\
\hline
\pagebreak[2]
\mrow{5}{0.088}  &   \mrow{5}{6.3\h} &   \mrow{5}{8}
&  \mrow{3}{Bare}   &   \phantom{0.0}--1.0
&     3             &     1             &   2
&     4             &     2             &   2
&     5             &     3             &   2      \\*
&                   &
&                   &   1.0--3.5
&     3             &     2             &   2
&     4             &     3             &   2
&     5             &     3             &   2      \\*
&                   &
&                   &   3.5--\phantom{0.0}
&     3             &     2             &   1
&     4             &     3             &   1
&     5             &     3             &   1      \\*
\cline{4-14}
&                   &
&  \mrow{2}{HYP}    &   \phantom{0.0}--3.5
&     3             &     3             &   2
&     4             &     3             &   2
&     5             &     3             &   2      \\*
&                   &
&                   &   3.5--\phantom{0.0}
&     3             &     3             &   1
&     4             &     3             &   1
&     5             &     3             &   1      \\
\hline
\pagebreak[2]
\mrow{10}{0.057} &   \mrow{10}{6.72} &   \mrow{10}{10}
&  \mrow{5}{Bare}   &   \phantom{0.0}--1.6
&     4             &     3             &   2
&     5             &     3             &   2
&     6             &     4             &   2      \\*
&                   &
&                   &   1.6--1.9
&     5             &     3             &   2
&     6             &     4             &   2
&     7             &     4             &   2      \\*
&                   &
&                   &   1.9--3.0
&     5             &     3             &   2
&     6             &     4             &   2
&     7             &     5             &   2      \\*
&                   &
&                   &   3.0--3.5
&     5             &     3             &   2
&     6             &     4             &   2
&     7             &     5             &   3      \\*
&                   &
&                   &   3.5--\phantom{0.0}
&     5             &     3             &   1
&     6             &     4             &   2
&     7             &     5             &   3      \\*
\cline{4-14}
&                   &
&  \mrow{5}{HYP}    &   \phantom{0.0}--1.6
&     4             &     3             &   2
&     5             &     3             &   2
&     6             &     4             &   2      \\*
&                   &
&                   &   1.6--1.9
&     5             &     3             &   2
&     6             &     3             &   2
&     7             &     4             &   2      \\*
&                   &
&                   &   1.9--3.0
&     5             &     3             &   2
&     6             &     4             &   2
&     7             &     5             &   2      \\*
&                   &
&                   &   3.0--3.5
&     5             &     3             &   2
&     6             &     4             &   2
&     7             &     5             &   3      \\*
&                   &
&                   &   3.5--\phantom{0.0}
&     5             &     3             &   1
&     6             &     4             &   2
&     7             &     5             &   3      \\
\hline
\pagebreak[2]
\mrow{16}{0.042} &  \mrow{16}{7.0\h} &  \mrow{16}{20}
&  \mrow{9}{Bare}   &   \phantom{0.0}--1.0
&     5             &     3             &   2
&     6             &     3             &   2
&     7             &     4             &   2      \\*
&                   &
&                   &   1.0--2.4
&     6             &     3             &   2
&     7             &     4             &   2
&     8             &     5             &   2      \\*
&                   &
&                   &   2.4--2.7
&     6             &     3             &   2
&     8             &     4             &   2
&     9             &     5             &   2      \\*
&                   &
&                   &   2.7--3.0
&     7             &     4             &   2
&     8             &     5             &   2
&     9             &     6             &   2      \\*
&                   &
&                   &   3.0--3.5
&     7             &     4             &   2
&     8             &     5             &   2
&     9             &     6             &   3      \\*
&                   &
&                   &   3.5--3.8
&     7             &     4             &   1
&     8             &     5             &   2
&     9             &     6             &   3      \\*
&                   &
&                   &   3.8--6.0
&     7             &     5             &   1
&     8             &     6             &   2
&     9             &     7             &   3      \\*
&                   &
&                   &   6.0--9.0
&     7             &     5             &   2
&     8             &     6             &   3
&     9             &     7             &   4      \\*
&                   &
&                   &   9.0--\phantom{0.0}
&     7             &     5             &   3
&     8             &     6             &   4
&     9             &     7             &   5      \\*
\cline{4-14}
&                   &
&  \mrow{7}{HYP}    &   \phantom{0.0}--2.5
&     6             &     3             &   2
&     8             &     4             &   2
&     7             &     4             &   2      \\*
&                   &
&                   &   2.5--3.0
&     7             &     4             &   2
&     8             &     5             &   2
&     8             &     5             &   2      \\*
&                   &
&                   &   3.0--3.5
&     7             &     4             &   2
&     8             &     5             &   2
&     8             &     5             &   2      \\*
&                   &
&                   &   3.5--3.8
&     7             &     4             &   1
&     8             &     5             &   1
&     8             &     5             &   2      \\*
&                   &
&                   &   3.8--6.0
&     7             &     5             &   1
&     8             &     6             &   1
&     8             &     6             &   2      \\*
&                   &
&                   &   6.0--9.0
&     7             &     5             &   2
&     8             &     6             &   2
&     8             &     6             &   3      \\*
&                   &
&                   &   9.0--\phantom{0.0}
&     7             &     5             &   3
&     8             &     6             &   3
&     8             &     6             &   4      \\
\hline
\pagebreak[4]
\mrow{27}{0.032} &   \mrow{27}{7.28} &  \mrow{27}{28}
&  \mrow{14}{Bare}  &   \phantom{0.0}--1.0
&     7             &     3             &   2
&     8             &     4             &   2
&     9             &     5             &   2      \\*
&                   &
&                   &   1.0--1.6
&     7             &     4             &   2
&     8             &     5             &   2
&     9             &     6             &   2      \\*
&                   &
&                   &   1.6--2.5
&     8             &     4             &   2
&     9             &     5             &   2
&     10            &     6             &   2      \\*
&                   &
&                   &   2.5--2.8
&     9             &     4             &   2
&     9             &     5             &   2
&     10            &     6             &   2      \\*
&                   &
&                   &   2.8--2.9
&     9             &     4             &   2
&     10            &     5             &   2
&     11            &     6             &   2      \\*
&                   &
&                   &   2.9--3.0
&     9             &     5             &   2
&     10            &     6             &   2
&     11            &     7             &   2      \\*
&                   &
&                   &   3.0--3.5
&     9             &     5             &   2
&     10            &     6             &   2
&     11            &     7             &   3      \\*
&                   &
&                   &   3.5--4.3
&     9             &     5             &   1
&     10            &     6             &   2
&     11            &     7             &   3      \\*
&                   &
&                   &   4.3--5.8
&     9             &     6             &   1
&     10            &     7             &   2
&     11            &     8             &   3      \\*
&                   &
&                   &   5.8--6.0
&     9             &     7             &   1
&     10            &     8             &   2
&     11            &     9             &   3      \\*
&                   &
&                   &   6.0--9.0
&     9             &     7             &   2
&     10            &     8             &   3
&     11            &     9             &   4      \\*
&                   &
&                   &   9.0--12.0
&     9             &     7             &   3
&     10            &     8             &   4
&     11            &     9             &   5      \\*
&                   &
&                   &   12.0--15
&     9             &     7             &   4
&     10            &     8             &   5
&     11            &     9             &   6      \\*
&                   &
&                   &   15.0--\phantom{0.0}
&     9             &     7             &   5
&     10            &     8             &   6
&     11            &     9             &   7      \\*
\cline{4-14}
&                   &
&  \mrow{13}{HYP}   &   \phantom{0.0}--1.7
&     7             &     4             &   2
&     8             &     5             &   2
&     9             &     6             &   2      \\*
&                   &
&                   &   1.7--2.5
&     8             &     4             &   2
&     9             &     5             &   2
&     10            &     6             &   2      \\*
&                   &
&                   &   2.5--2.8
&     9             &     4             &   2
&     9             &     5             &   2
&     11            &     6             &   2      \\*
&                   &
&                   &   2.8--2.9
&     9             &     4             &   2
&     10            &     5             &   2
&     11            &     6             &   2      \\*
&                   &
&                   &   2.9--3.0
&     9             &     5             &   2
&     10            &     6             &   2
&     11            &     7             &   2      \\*
&                   &
&                   &   3.0--3.5
&     9             &     5             &   2
&     10            &     6             &   2
&     11            &     7             &   3      \\*
&                   &
&                   &   3.5--4.3
&     9             &     5             &   1
&     10            &     6             &   2
&     11            &     7             &   3      \\*
&                   &
&                   &   4.3--5.8
&     9             &     6             &   1
&     10            &     7             &   2
&     11            &     8             &   3      \\*
&                   &
&                   &   5.8--6.0
&     9             &     7             &   1
&     10            &     8             &   2
&     11            &     9             &   3      \\*
&                   &
&                   &   6.0--9.0
&     9             &     7             &   2
&     10            &     8             &   3
&     11            &     9             &   4      \\*
&                   &
&                   &   9.0--12.
&     9             &     7             &   3
&     10            &     8             &   4
&     11            &     9             &   5      \\*
&                   &
&                   &   12.--15.
&     9             &     7             &   4
&     10            &     8             &   5
&     11            &     9             &   6      \\*
&                   &
&                   &   15.--\phantom{0.0}
&     9             &     7             &   5
&     10            &     8             &   6
&     11            &     9             &   7      \\*
\hline
\hline

\end{longtable}
}

\FloatBarrier

\begin{figure}
    \centering
    \includegraphics[width=0.49\textwidth]{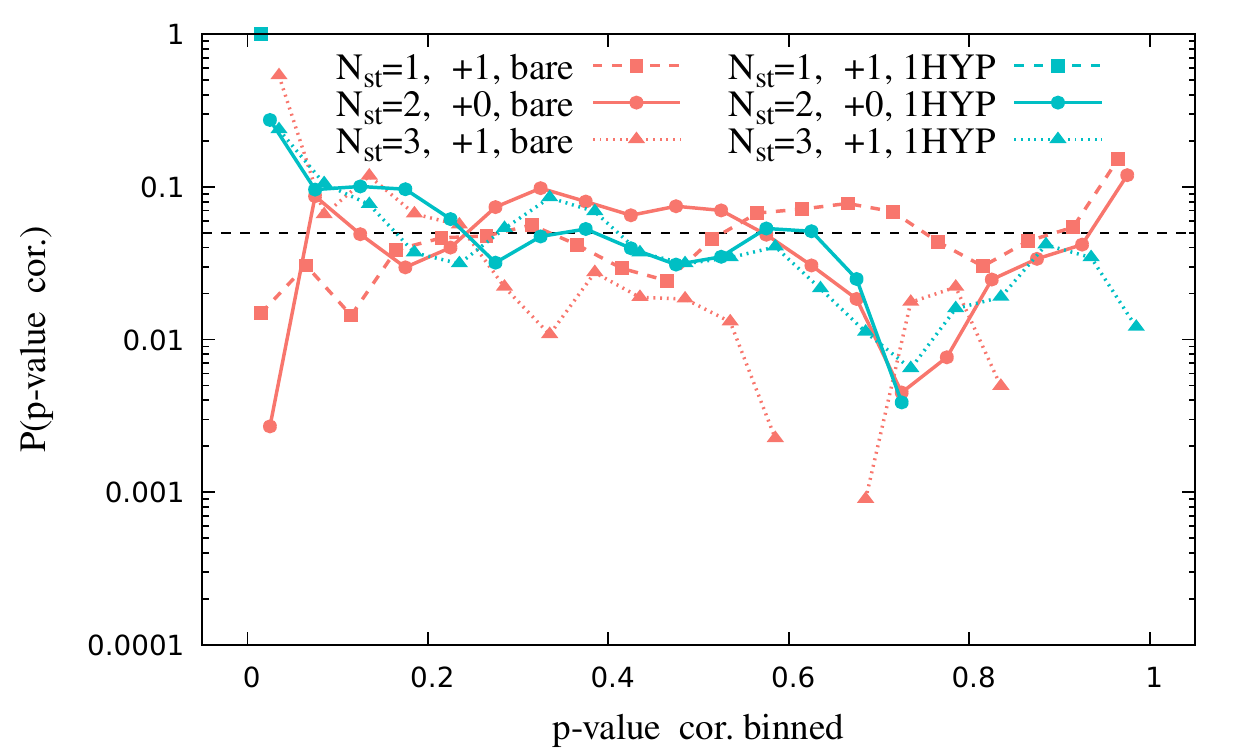}%
    \hfill%
    \includegraphics[width=0.49\textwidth]{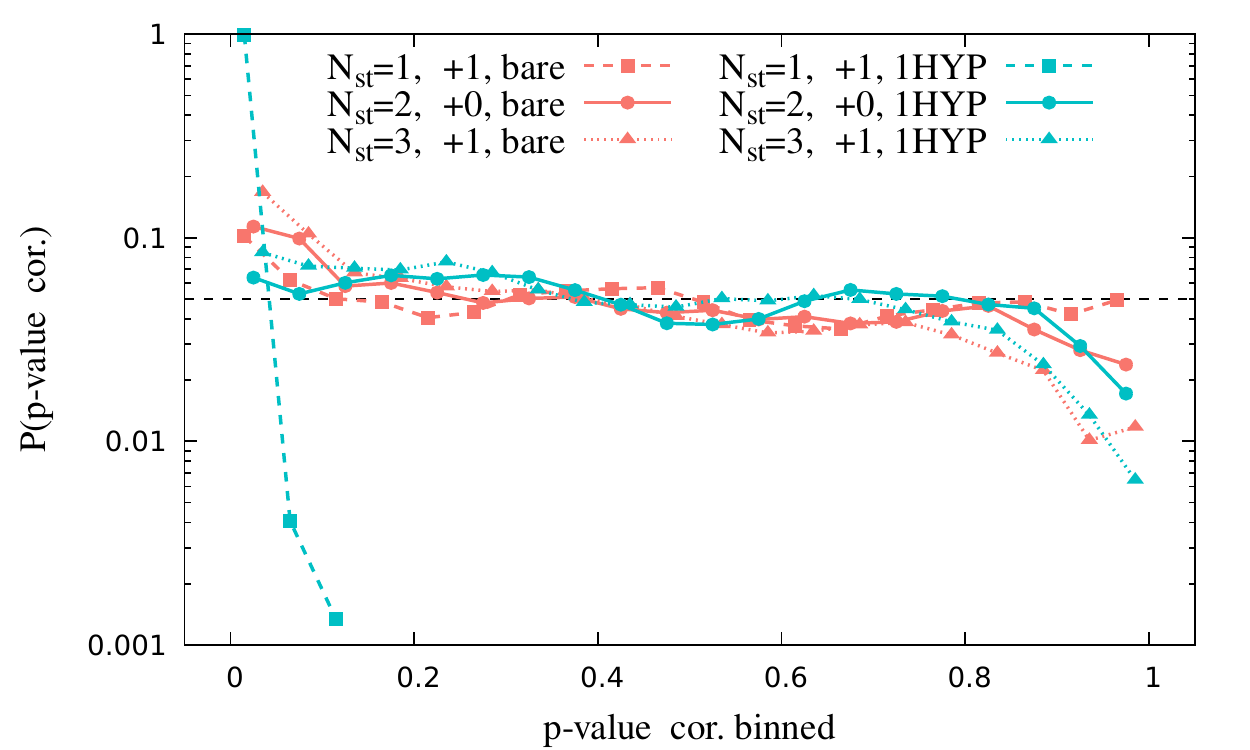}%
    \caption{\label{fig:p-values}%
    Distribution of $p$~values for fits to the correlator on the physical \ensemble{7.00}{i} ensemble.
    The horizontal dashed line corresponds to a flat distribution at $P(p~\text{value})=\sfrac{1}{20}$, while colored
    lines serve as guides to the eyes.
    We separately show results at small distances, i.e., $|\bm{r}| \le 0.2$~fm (left) or at large distances, i.e.,
    $|\bm{r}| > 0.2$~fm (right).
    The former constitutes a much smaller sample of fits (fewer combinations of $\sfrac{|\bm{r}|}{a}$).
    Left: at $|\bm{r}| \le 0.2$~fm, the $p$~value distribution is reasonably flat in the case of bare links and
    $\Nstates \le 2$, while fits with $\Nstates \ge 2$ in similar ranges fare similarly well for smeared links.
    Right: at $|\bm{r}| > 0.2$~fm, the $p$~value distribution with bare links is quite flat.
    Fits with $\Nstates = 1$ in similar intervals are disfavored for smeared links.}
\end{figure}

\begin{figure}
    \centering
    \includegraphics[width=0.49\textwidth]{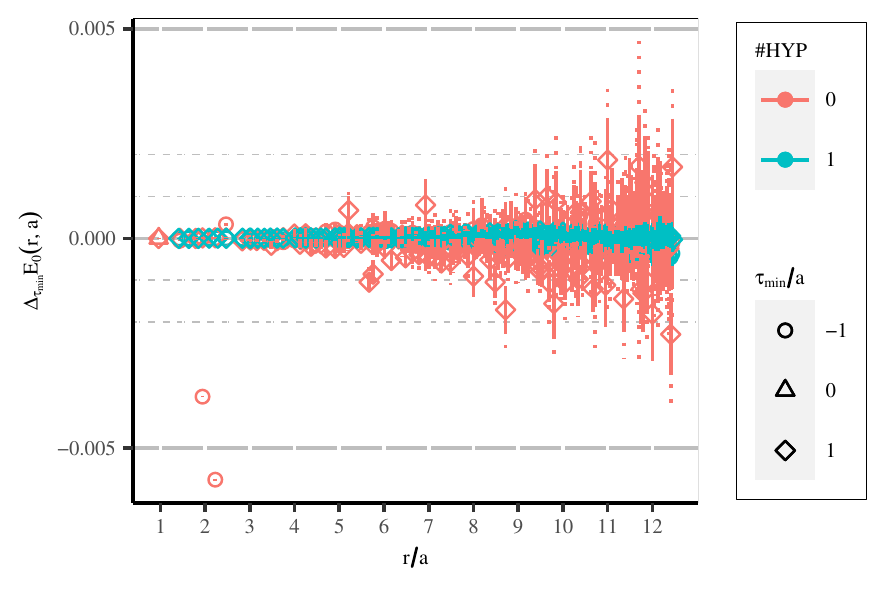}%
    \hfill%
    \includegraphics[width=0.49\textwidth]{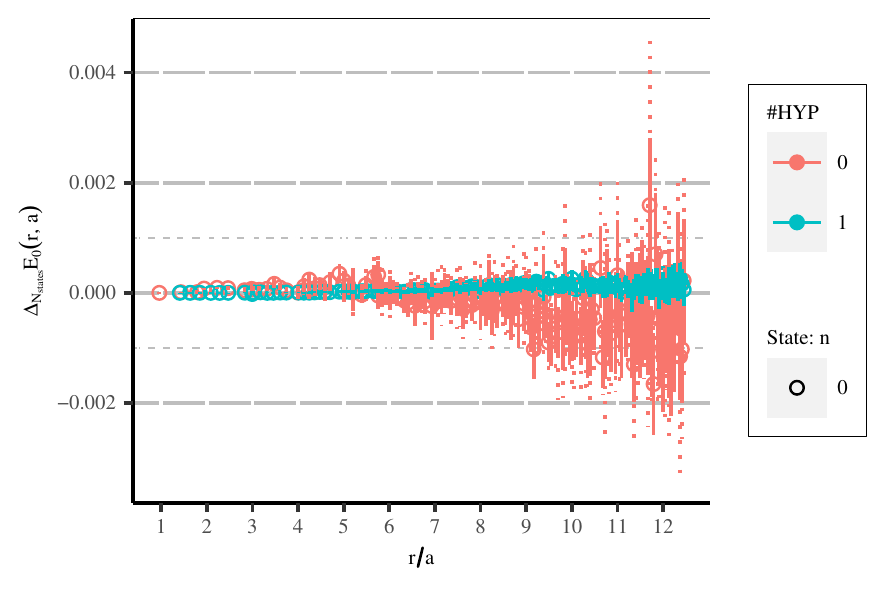}%
    \caption{\label{fig:stability_analysis}%
    Stability plots of the correlation function fits for the physical \ensemble{7.00}{i} ensemble.
    Differences between extracted ground state energy values $E_{0}(r,a)$ from different fits are usually covered by
    the statistical errors corresponding to our canonical choice of fits, i.e., with $\Nstates=2$ and
    $\sfrac{\tmin^{(2,0)}}{a}$.
    Dotted error bars represent Hessian errors, while solid error bars represent jackknife errors obtained from the
    distribution of the resamples.
    Left: for fits with $\Nstates=2$, we vary $\sfrac{\tmin}{a}$ by $\pm1$ against our canonical fit.
    We clearly see that the fit stability breaks down at a few small $\sfrac{\bm{r}}{a}$ values for the choice
    $\sfrac{\tmin^{(2,-)}}{a}$.
    Right: we compare the fits with $\Nstates=3$ and $\sfrac{\tmin^{(3,+)}}{a}$ to our canonical fits $\Nstates=2$ and
    $\sfrac{\tmin^{(2,0)}}{a}$.
    For $\sfrac{\tmin^{(3,0)}}{a}$ we already see that the fit stability breaks down in a few cases.}
\end{figure}

We show representative plots of $p$~value distributions for correlator fits with different numbers of states on the
physical \ensemble{7.00}{i} ensemble in Fig.~\ref{fig:p-values}.
We show stability plots for the ground state energy on the same ensemble under variation of the time range or of the
number of states in Fig.~\ref{fig:stability_analysis}.

\subsection{Tree-level corrections}
\label{app:tree-level corrections}

We collect the tree-level corrections for bare or smeared links in Table~\ref{tab:tree-level_corrections_both}, which
are part of an ongoing project aiming at a full one-loop calculation~\cite{lpt_paper} using the \texttt{HiPPy} software
package and the \texttt{HPsrc} software framework~\cite{Hart:2004bd, Hart:2009nr}.

\begin{table}
    \centering
    \caption{\label{tab:tree-level_corrections_both}%
    Spatial Euclidean distance $\sfrac{|\bm{r}|}{a}$ and tree-level improved distances $\sfrac{r_{I}}{a}$ for bare
    links or for links after one step of HYP-smearing in the fourth, fifth, and sixth column.
    Each row corresponds to any permutation of the three spatial coordinates $\sfrac{x_{i}}{a}$ since these belong to
    the same representation of the cubic group $W_{3}$.
    For bare links the improved distance is smaller than the corresponding Euclidean distance for on-axis vectors with
    $\sfrac{|\bm{r}|}{a} < 5$, and for a few off-axis vectors in the same range; the relative modification is at most
    $0.3\%$ for $\sfrac{|\bm{r}|}{a} > \sqrt{12}$.
    For smeared links the improved distance is larger than the corresponding Euclidean distance except for
    $\sfrac{\bm{r}}{a} = (2,2,1)$ or $(2,2,2)$; the relative modification is at most $0.3\%$ for $\sfrac{|\bm{r}|}{a} >
    \sqrt{10}$.}
    \begin{tabular}{ccc@{\quad}c@{\quad}c@{\quad}c}
        \hline\hline
$\sfrac{x_{1}}{a}$ & $\sfrac{x_{2}}{a}$ & $\sfrac{x_{3}}{a}$ & $\sfrac{|\bm{r}|}{a}$ &
$\sfrac{r_{I}}{a}$ (bare links) &
$\sfrac{r_{I}}{a}$ (smeared links) \\
\hline
1	&	0	&	0	&	1.0     	&	0.959904	$\pm$	0.000003	&	1.409072	$\pm$	0.000027	 \\
1	&	1	&	0	&	1.414214	&	1.433383	$\pm$	0.000013	&	1.634790	$\pm$	0.000047	 \\
1	&	1	&	1	&	1.732051	&	1.786648	$\pm$	0.000031	&	1.846760	$\pm$	0.000072	 \\
2	&	0	&	0	&	2.0     	&	1.940264	$\pm$	0.000049	&	2.086964	$\pm$	0.000109	 \\
2	&	1	&	0	&	2.236068	&	2.225023	$\pm$	0.000081	&	2.282411	$\pm$	0.000149	 \\
2	&	1	&	1	&	2.449490	&	2.465341	$\pm$	0.000119	&	2.469253	$\pm$	0.000197	 \\
2	&	2	&	0	&	2.828427	&	2.827621	$\pm$	0.000208	&	2.832540	$\pm$	0.000319	 \\
2	&	2	&	1	&	3.0     	&	3.012822	$\pm$	0.000264	&	2.996896	$\pm$	0.000390	 \\
3	&	0	&	0	&	3.0     	&	2.979371	$\pm$	0.000265	&	3.019798	$\pm$	0.000401	 \\
3	&	1	&	0	&	3.162278	&	3.151661	$\pm$	0.000327	&	3.174122	$\pm$	0.000479	 \\
3	&	1	&	1	&	3.316625	&	3.315550	$\pm$	0.000396	&	3.323076	$\pm$	0.000565	 \\
2	&	2	&	2	&	3.464102	&	3.477834	$\pm$	0.000467	&	3.457683	$\pm$	0.000649	 \\
3	&	2	&	0	&	3.605551	&	3.604281	$\pm$	0.000550	&	3.608266	$\pm$	0.000761	 \\
3	&	2	&	1	&	3.741657	&	3.746234	$\pm$	0.000637	&	3.742821	$\pm$	0.000868	 \\
4	&	0	&	0	&	4.0     	&	3.994281	$\pm$	0.000855	&	4.009312	$\pm$	0.001140	 \\
3	&	2	&	2	&	4.123106	&	4.131366	$\pm$	0.000932	&	4.123893	$\pm$	0.001236	 \\
4	&	1	&	0	&	4.123106	&	4.118891	$\pm$	0.000960	&	4.130599	$\pm$	0.001270	 \\
3	&	3	&	0	&	4.242641	&	4.244320	$\pm$	0.001049	&	4.245404	$\pm$	0.001382	 \\
4	&	1	&	1	&	4.242641	&	4.240985	$\pm$	0.001072	&	4.248958	$\pm$	0.001406	 \\
3	&	3	&	1	&	4.358899	&	4.363650	$\pm$	0.001165	&	4.361678	$\pm$	0.001525	 \\
4	&	2	&	0	&	4.472136	&	4.472231	$\pm$	0.001313	&	4.477483	$\pm$	0.001703	 \\
4	&	2	&	1	&	4.582576	&	4.584957	$\pm$	0.001442	&	4.587801	$\pm$	0.001860	 \\
3	&	3	&	2	&	4.690416	&	4.698346	$\pm$	0.001549	&	4.694392	$\pm$	0.001997	 \\
4	&	2	&	2	&	4.898979	&	4.904615	$\pm$	0.001863	&	4.904903	$\pm$	0.002375	 \\
4	&	3	&	0	&	5.0     	&	5.003469	$\pm$	0.002027	&	5.006342	$\pm$	0.002573	 \\
5	&	0	&	0	&	5.0     	&	5.000618	$\pm$	0.002127	&	5.010612	$\pm$	0.002666	 \\
4	&	3	&	1	&	5.099020	&	5.104096	$\pm$	0.002184	&	5.105849	$\pm$	0.002764	 \\
5	&	1	&	0	&	5.099020	&	5.100262	$\pm$	0.002286	&	5.109318	$\pm$	0.002860	 \\
3	&	3	&	3	&	5.196152	&	5.205202	$\pm$	0.002313	&	5.203278	$\pm$	0.002930	 \\
5	&	1	&	1	&	5.196152	&	5.198344	$\pm$	0.002452	&	5.206361	$\pm$	0.003060	 \\
4	&	3	&	2	&	5.385165	&	5.392954	$\pm$	0.002689	&	5.393750	$\pm$	0.003380	 \\
5	&	2	&	0	&	5.385165	&	5.388792	$\pm$	0.002802	&	5.395594	$\pm$	0.003484	 \\
5	&	2	&	1	&	5.477226	&	5.481937	$\pm$	0.002983	&	5.487994	$\pm$	0.003703	 \\
4	&	4	&	0	&	5.656854	&	5.663307	$\pm$	0.003299	&	5.667292	$\pm$	0.004106	 \\
4	&	4	&	1	&	5.744563	&	5.752137	$\pm$	0.003495	&	5.755734	$\pm$	0.004344	 \\
5	&	2	&	2	&	5.744563	&	5.751750	$\pm$	0.003562	&	5.756765	$\pm$	0.004404	 \\
4	&	3	&	3	&	5.830952	&	5.841093	$\pm$	0.003653	&	5.842952	$\pm$	0.004549	 \\
5	&	3	&	0	&	5.830952	&	5.837748	$\pm$	0.003783	&	5.843586	$\pm$	0.004667	 \\
5	&	3	&	1	&	5.916080	&	5.923874	$\pm$	0.003991	&	5.929386	$\pm$	0.004919	 \\
4	&	4	&	2	&	6.0     	&	6.009986	$\pm$	0.004115	&	6.013449	$\pm$	0.005097	 \\
6	&	0	&	0	&	6.0     	&	6.006756	$\pm$	0.004501	&	6.017224	$\pm$	0.005441	 \\
\hline\hline
    \end{tabular}
\end{table}

\subsection{Detailed definition of the scales and the string tension}
\label{app:scales and string tension}

\begin{figure}
    \centering
    \begin{minipage}[t]{0.33\textwidth}\vspace{0pt}%
        \includegraphics[width=\textwidth]{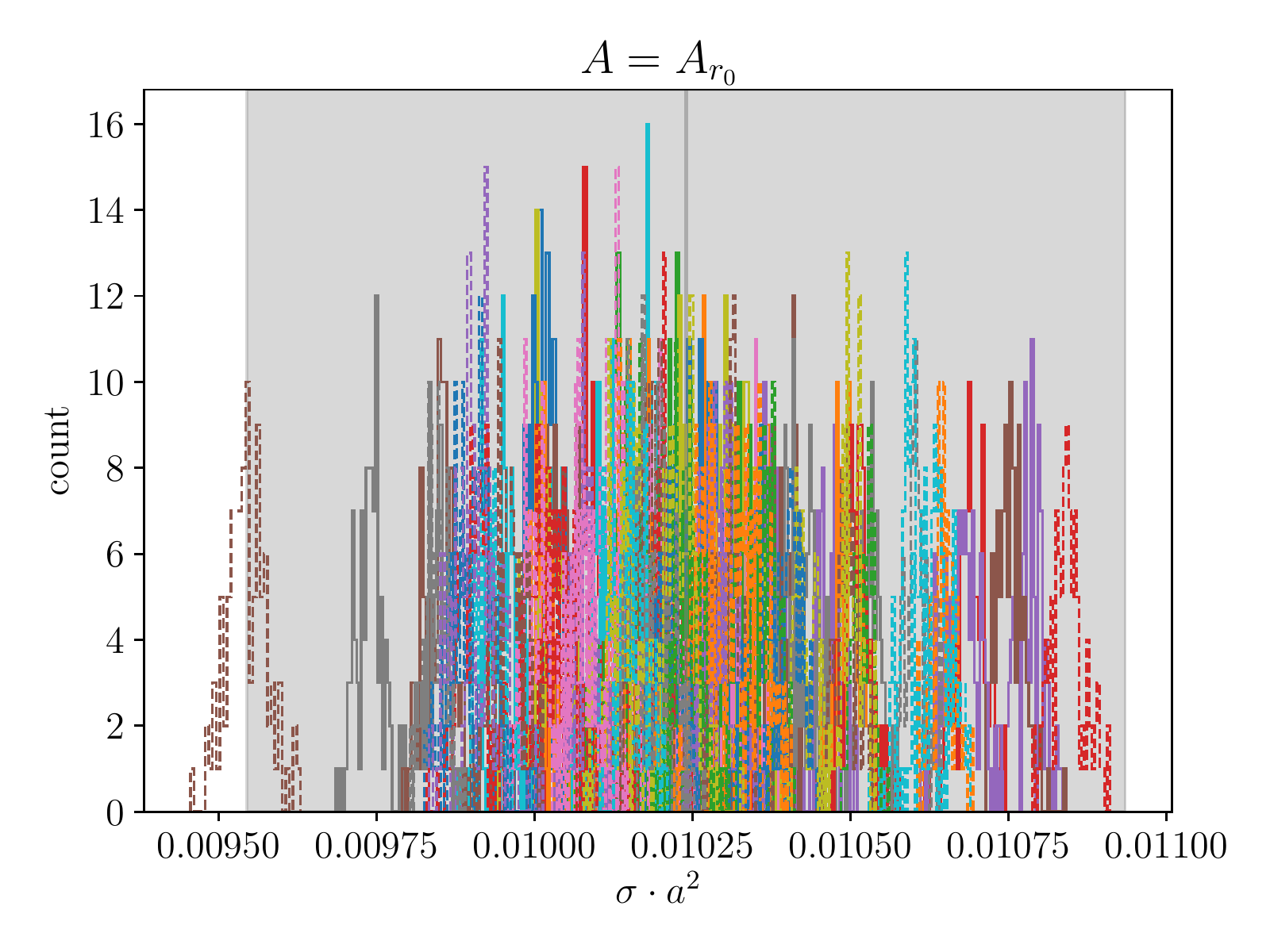}%
    \end{minipage}%
    \hfill%
    \begin{minipage}[t]{0.33\textwidth}\vspace{0pt}%
        \includegraphics[width=\textwidth]{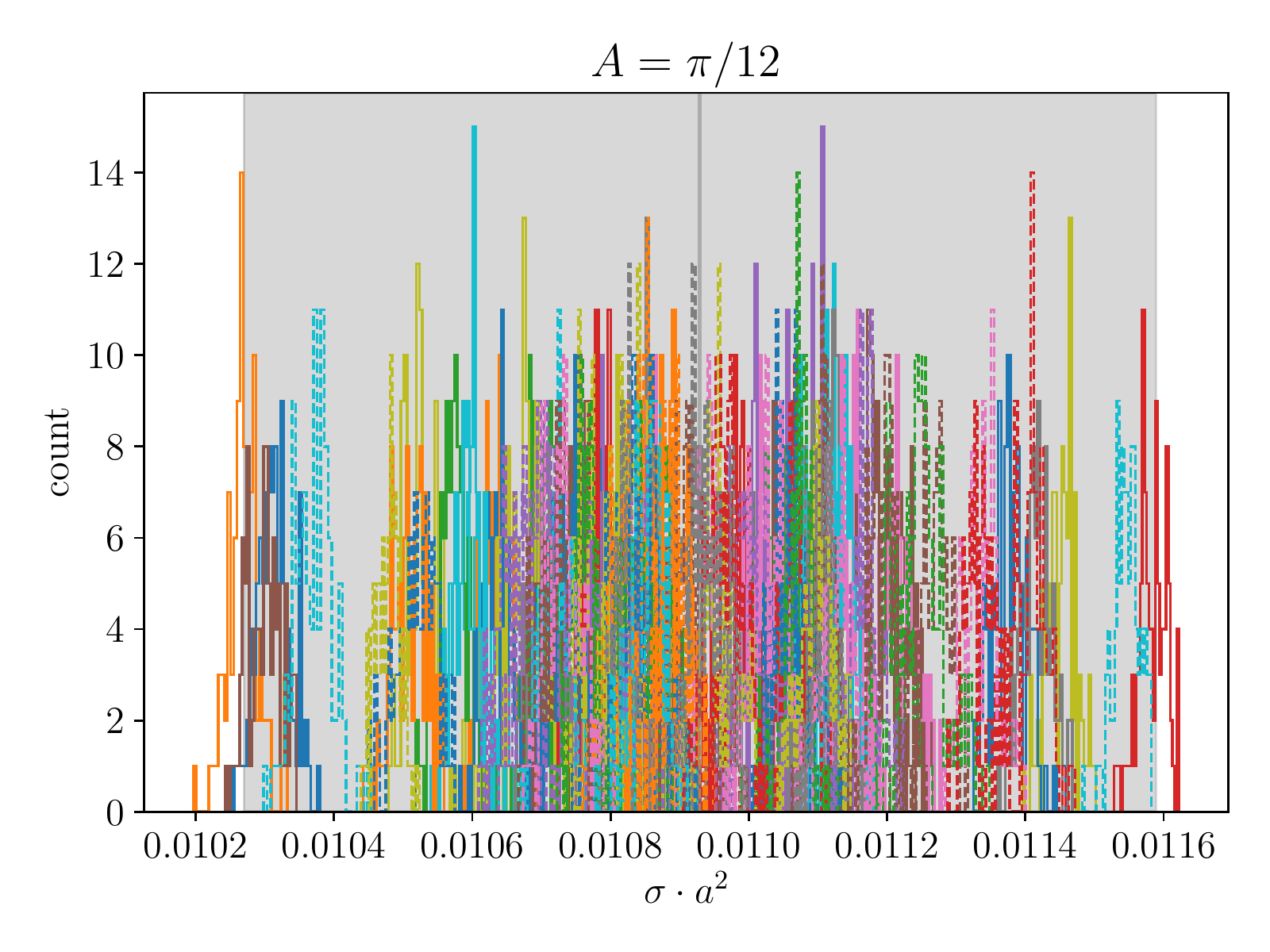}%
    \end{minipage}%
    \hfill
    \begin{minipage}[t]{0.33\textwidth}\vspace{0pt}%
        \caption{\label{fig:systematical_distribution_sigma}%
        The distribution of the results obtained on the $N_{J}$ jackknife pseudoensembles for each set of $N_{P}$
        random picks (designated by the color) is not dissimilar to a Gaussian distribution.
        However, the distribution of the jackknife means over the $N_{P}$ sets of random picks is usually not similar
        to a Gaussian distribution.
        The data are shown for the bare physical \ensemble{7.0}{i} ensemble.
        The gray vertical line and the gray band represent the corresponding mean value and error estimate in
        Table~\ref{tab:r_i_over_a}.}
    \end{minipage}
\end{figure}

\begin{figure}
    \centering
    \includegraphics[width=0.33\textwidth]{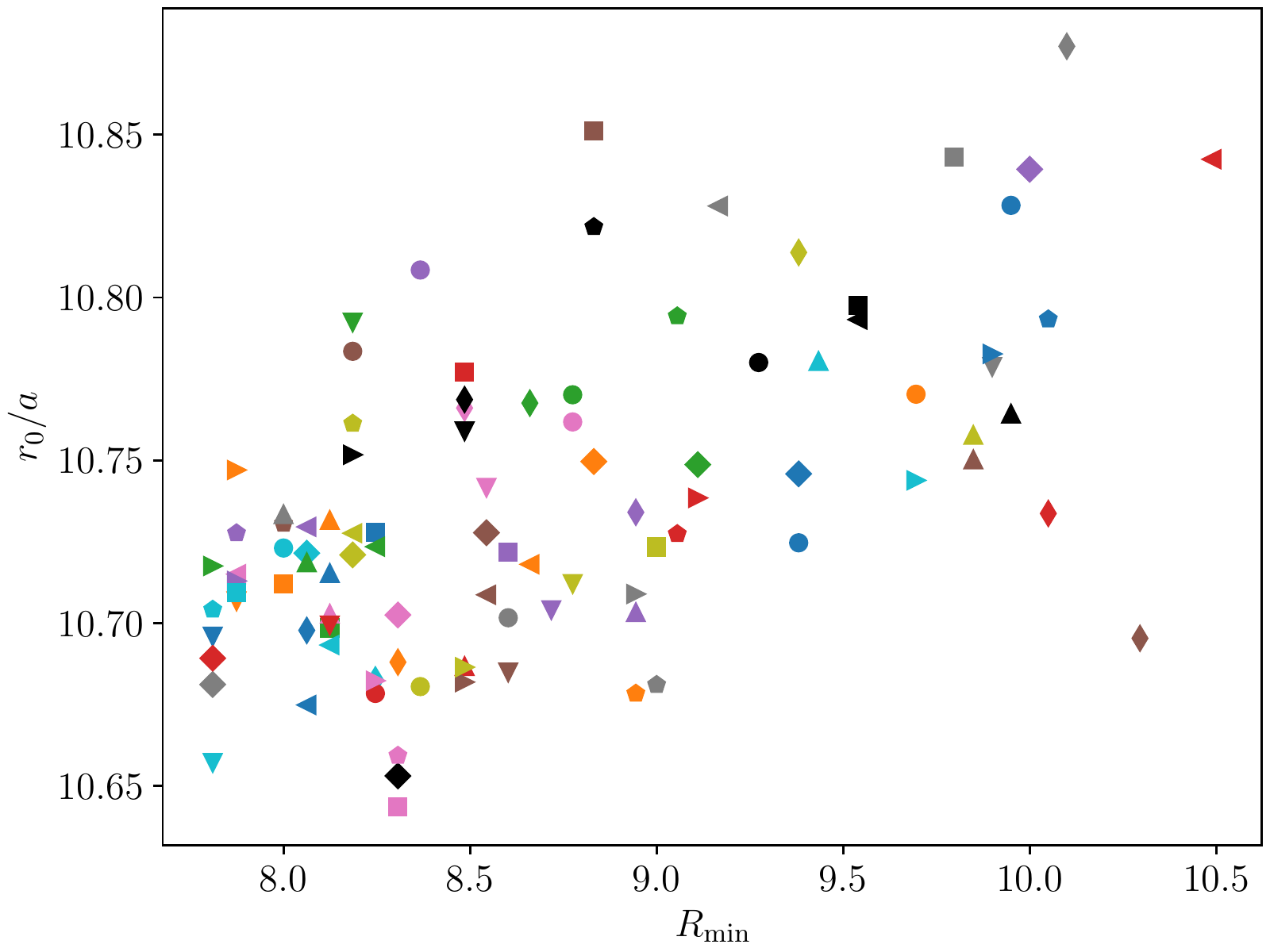}%
    \hfill%
    \includegraphics[width=0.33\textwidth]{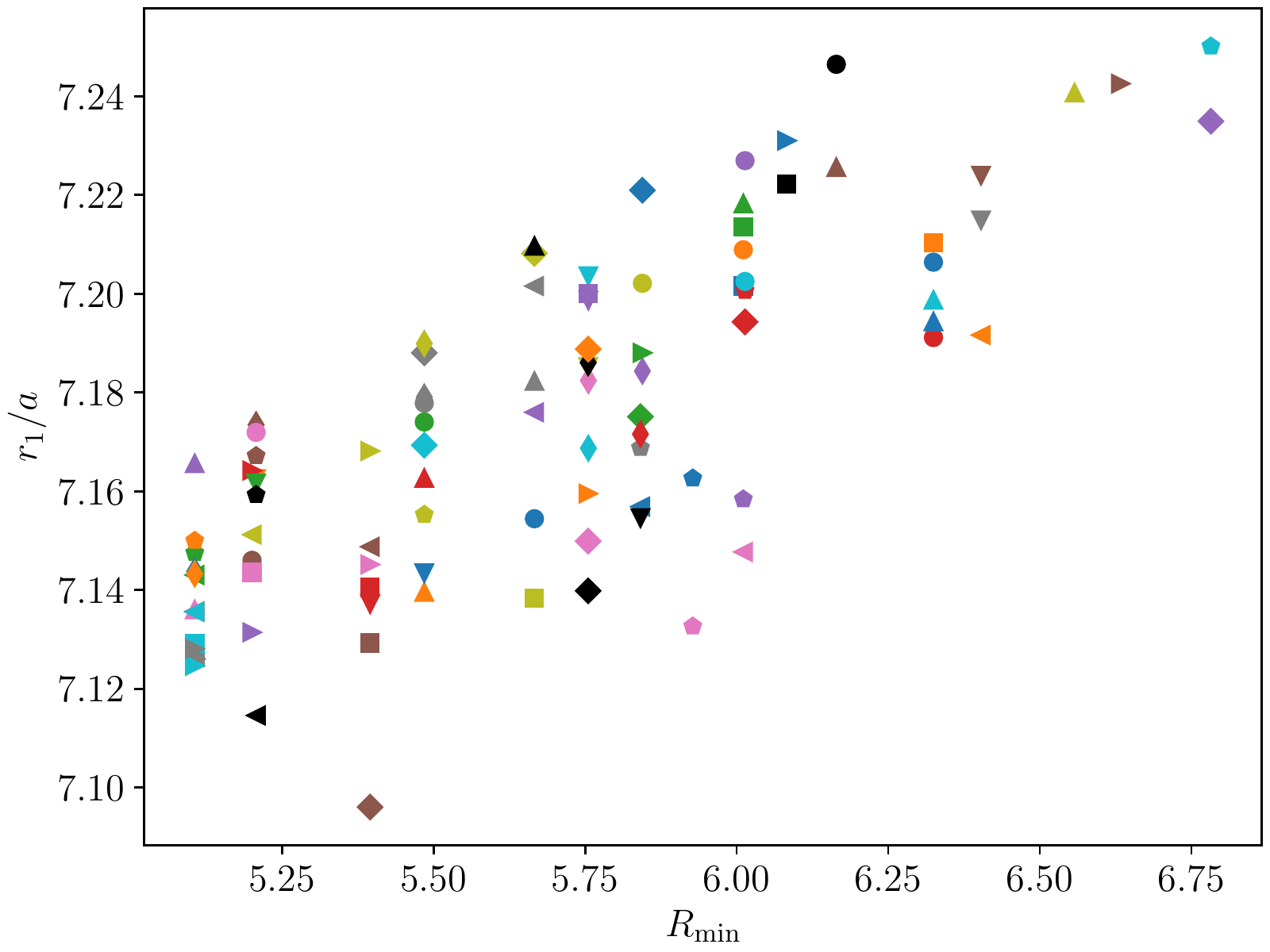}%
    \hfill%
    \includegraphics[width=0.33\textwidth]{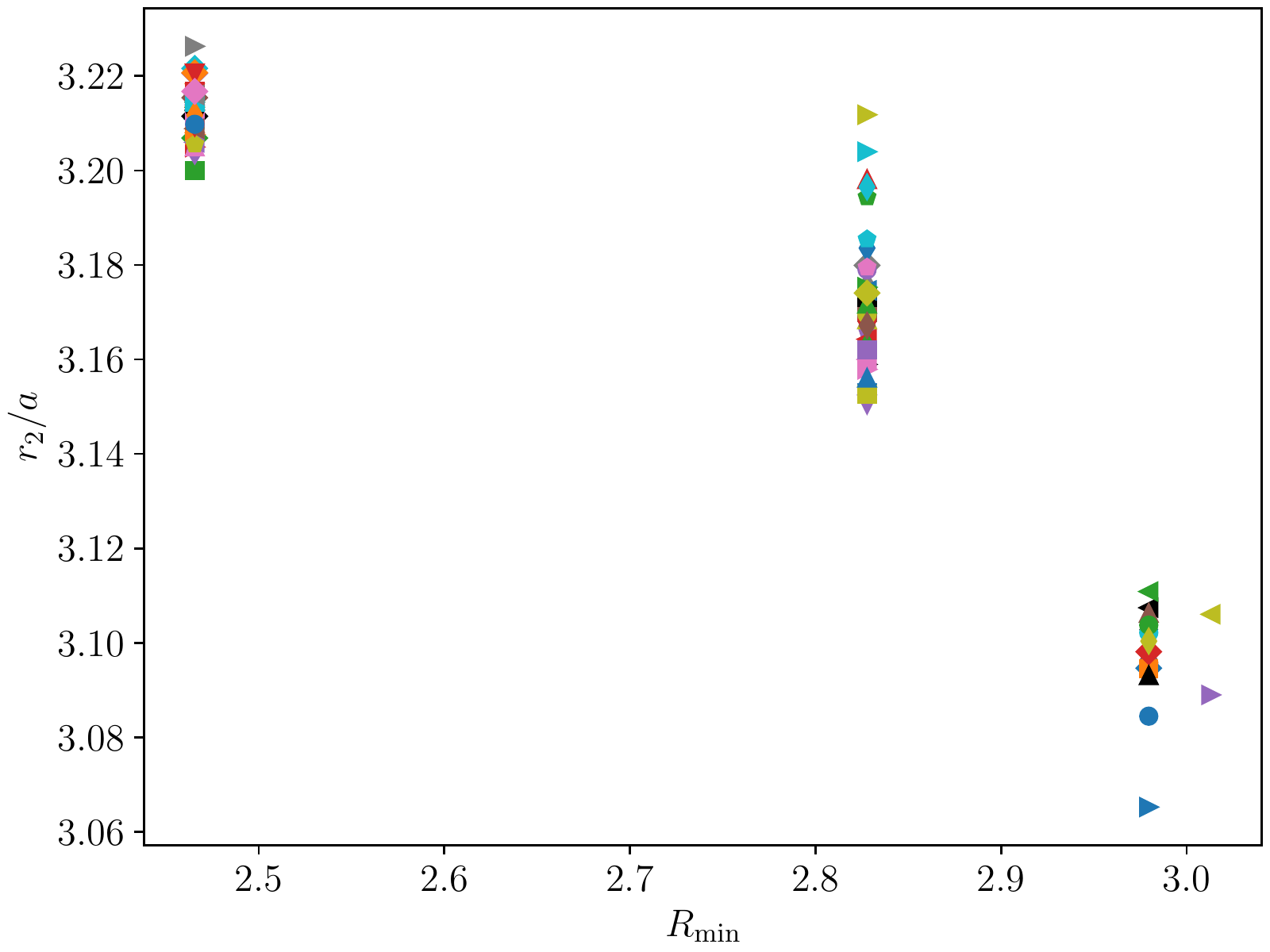}%
    \\%
    \begin{minipage}[t]{0.33\textwidth}\vspace{0pt}%
        \includegraphics[width=\textwidth]{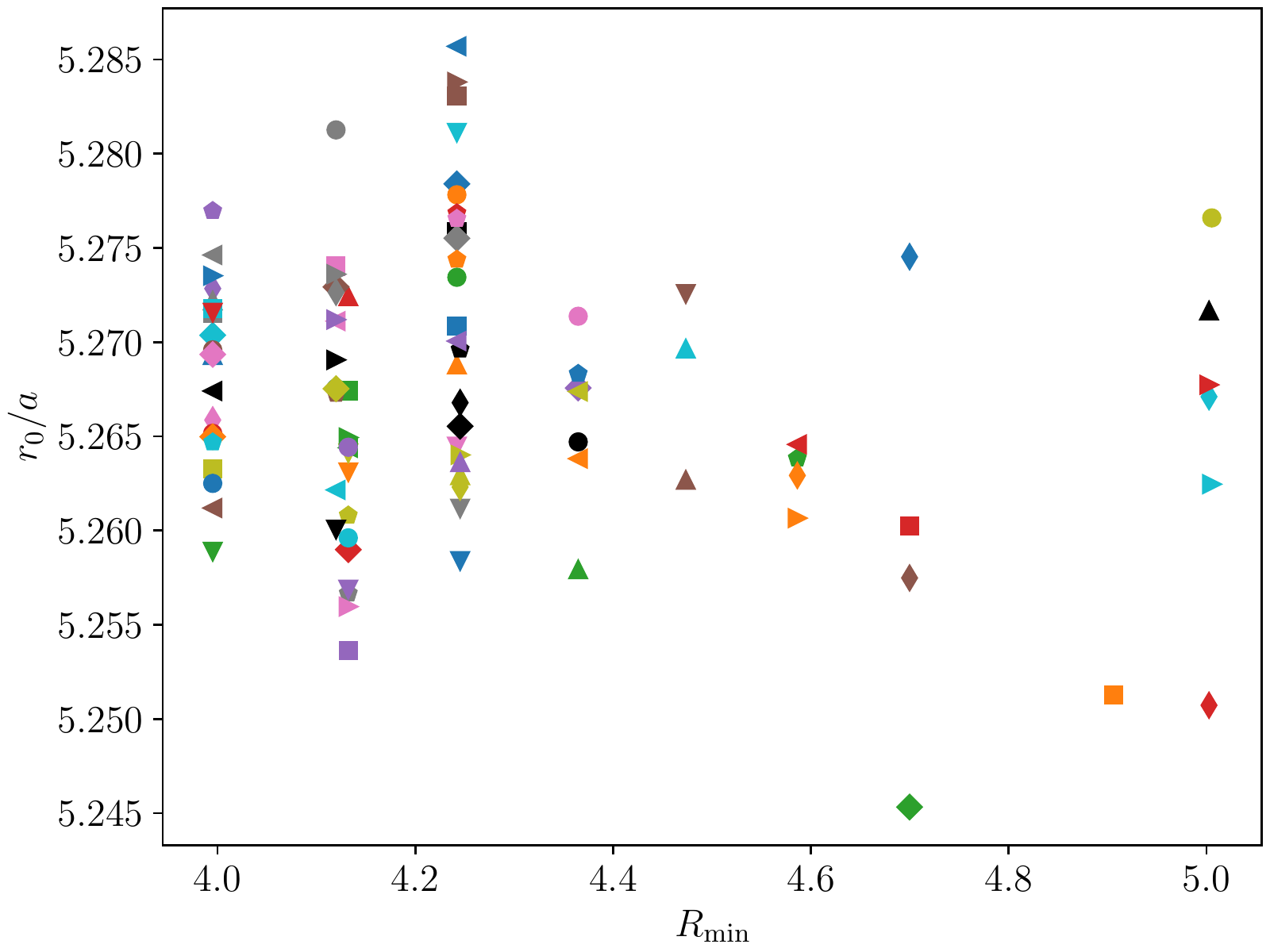}%
    \end{minipage}%
    \hfill%
    \begin{minipage}[t]{0.33\textwidth}\vspace{0pt}%
        \includegraphics[width=\textwidth]{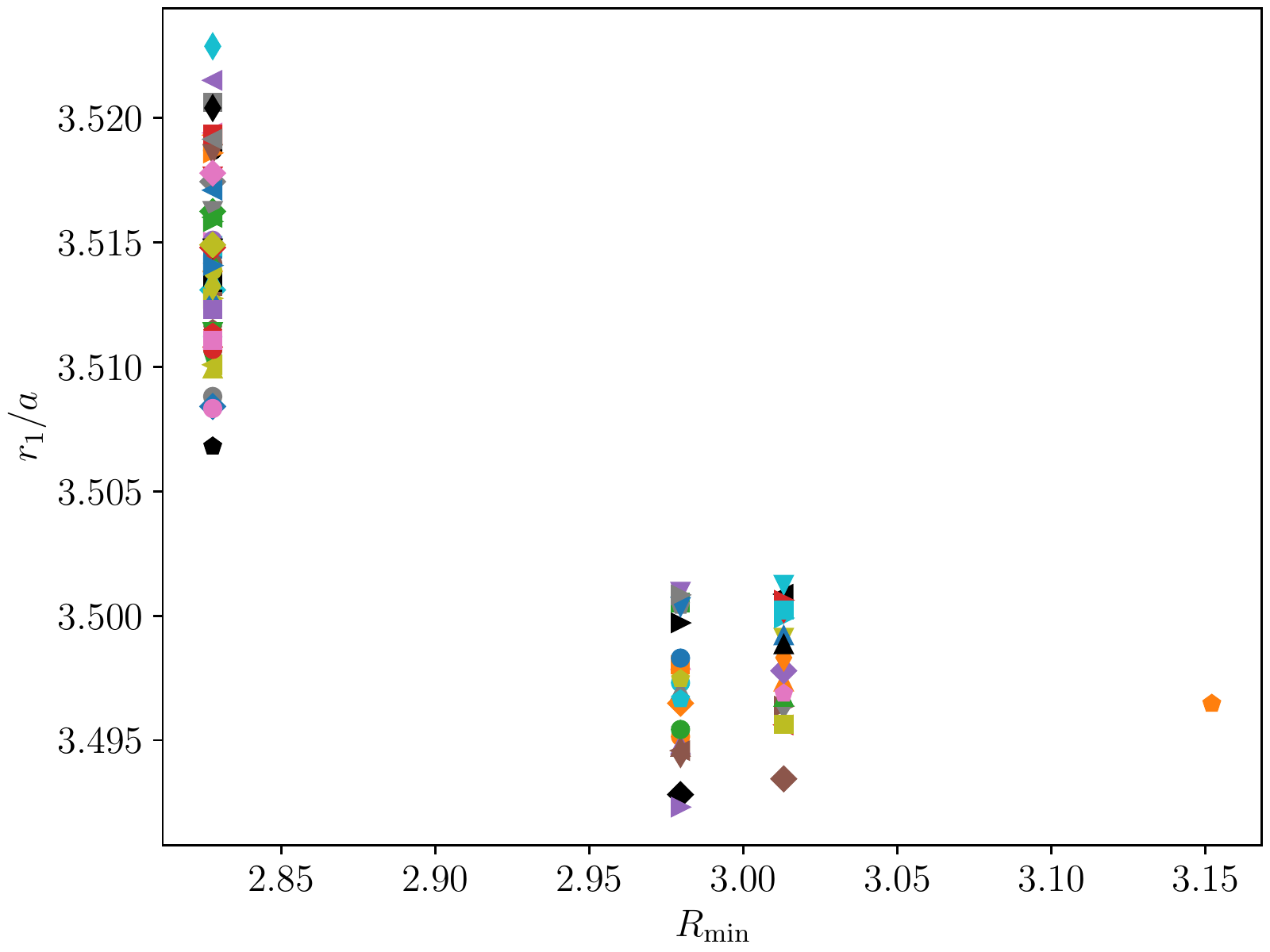}%
    \end{minipage}%
    \hfill%
    \begin{minipage}[t]{0.33\textwidth}\vspace{0pt}%
        \caption{\label{fig:random_picks_rmin}%
        Dependence of the fit results on the $R_{\text{min}}$ of the $N_{P}$ randomly selected data points for the
        first jackknife pseudoensemble of the bare physical \ensemble{7.00}{i} ensemble (top) and the bare physical
        \ensemble{6.30}{i} ensemble (bottom).
        There are correlations between the extracted $\sfrac{r_{i}}{a}$ and $R_{\text{min}}$ that are similar between
        the different ensembles but strongly vary with the considered $R$ range.}
    \end{minipage}
\end{figure}

\begin{figure}
    \centering
    \begin{minipage}[t]{0.33\textwidth}\vspace{0pt}%
        \includegraphics[width=\textwidth]{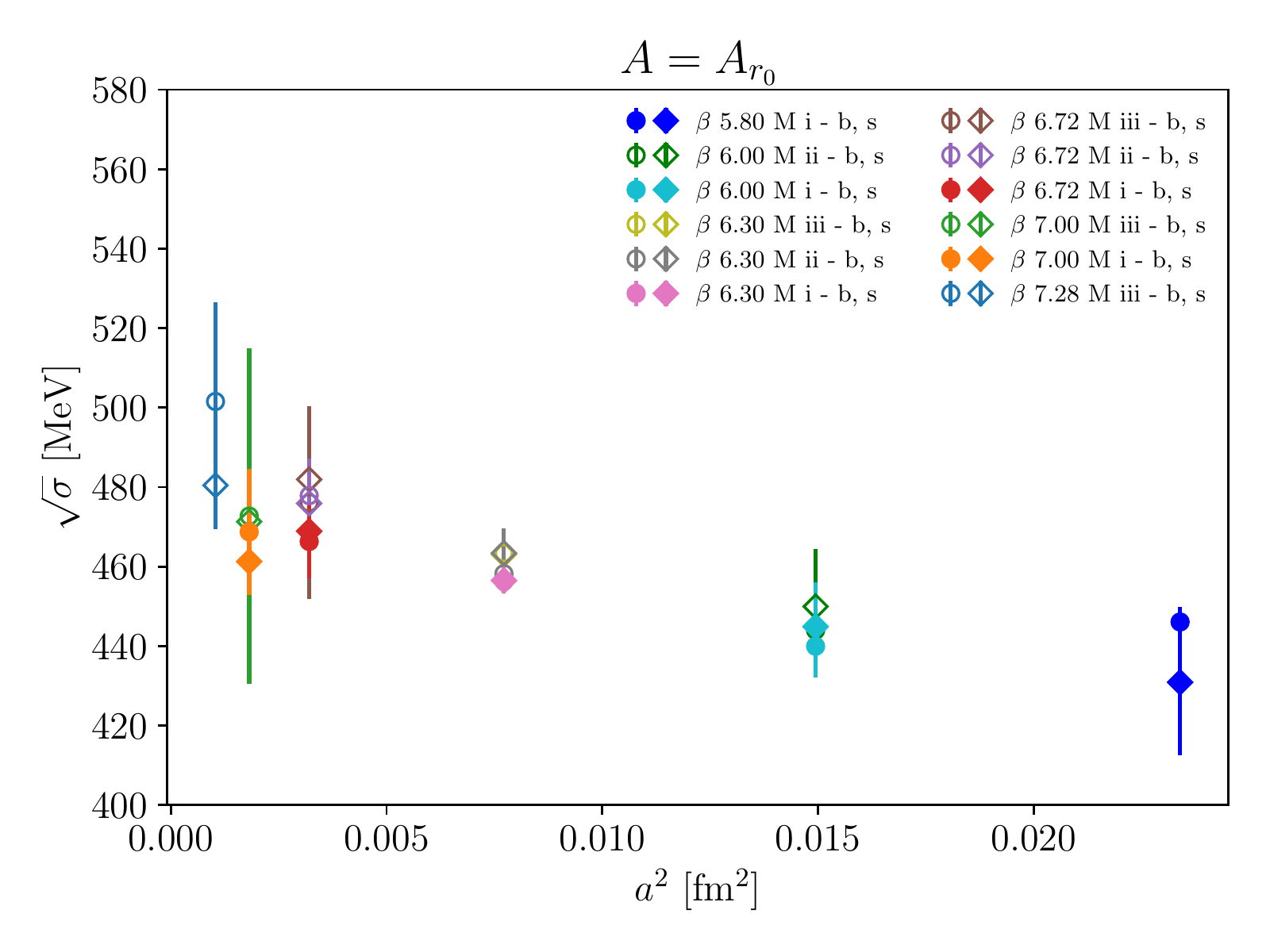}%
    \end{minipage}%
    \hfill
    \begin{minipage}[t]{0.33\textwidth}\vspace{0pt}%
        \includegraphics[width=\textwidth]{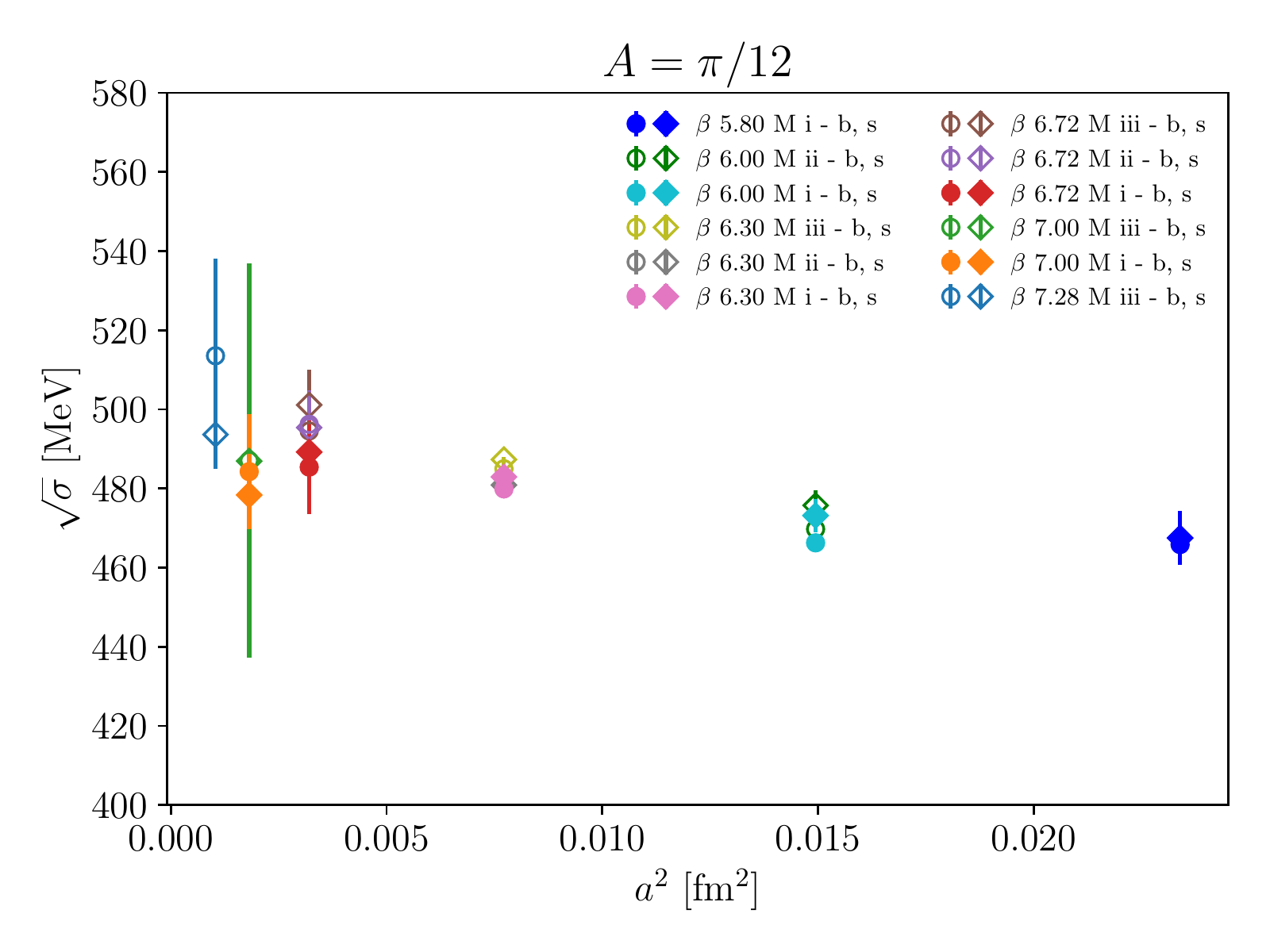}%
    \end{minipage}%
    \hfill
    \begin{minipage}[t]{0.33\textwidth}\vspace{0pt}%
        \caption{\label{fig:sigma_over_a}%
        The string tension $\sqrt{\sigma}$ for all ensembles (indicated by colors) and bare ($\circ$) and smeared
        ($\diamond$) gauge links and for the different choices of $A$.
        We use the lattice scale $a_{f_{p4s}}$ to convert to physical units and $a_{f_{p4s}}^{2} $ for the
        $x$-coordinate.
        Filled symbols correspond to physical light quark mass ensembles, while open symbols represent larger than
        physical quark masses.}
    \end{minipage}
\end{figure}

We show the fit range dependence of the extracted values of $a^{2}\sigma$ for the physical \ensemble{7.00}{i} ensemble
(with bare links) using two different fixed values of the Coulomb coefficient in
Fig.~\ref{fig:systematical_distribution_sigma}.
The distribution with the random picks is fairly Gaussian.
For the determinations of the scales $\sfrac{r_{i}}{a}$, the corresponding distributions are in
Fig.~\ref{fig:systematical_distribution} in Sec.~\ref{sec:interpolation}.
These distributions for $\sfrac{r_{i}}{a}$ clearly exhibit non-Gaussian characteristics and, in some cases,
correlations between $R_\text{min}$ and the obtained value of $\sfrac{r_{i}}{a}$, see Fig.~\ref{fig:random_picks_rmin}.
The string tension in physical units shows a fairly mild lattice spacing dependence, much smaller than the dependence
on assumptions about the Coulomb coefficient, see Fig.~\ref{fig:sigma_over_a}.

\FloatBarrier

\subsection{Relative scale setting}
\label{app:relative scale setting}

\begin{figure}[ht]
    \centering
    \includegraphics[width=1.0\textwidth]{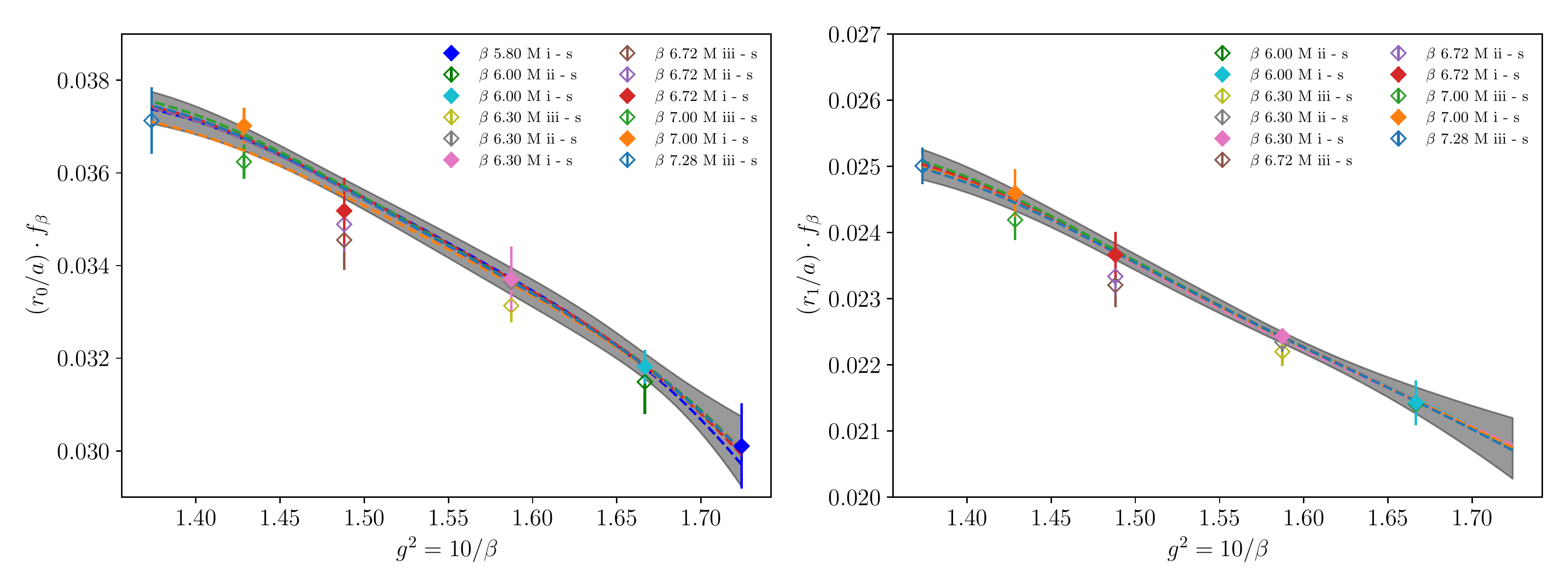}%
    \caption{\label{fig:r_i_over_a_Allton_summary_smeared}%
    The potential scales $\sfrac{r_{i}}{a}$, $i=0,1$ multiplied by the two-loop $\beta$-function, $f_{\beta}$ as in
    Eq.~\eqref{eq:two_loop_beta_function}, for all ensembles (indicated by colors) and smeared links.
    Any further details about the plot can be found in the caption of the corresponding figure for bare links in
    Fig.~\ref{fig:r_i_over_a_Allton_summary} in Sec.~\ref{sec:quark_mass_dependence}.}
\end{figure}

In this section, we collect additional material relevant for the relative scale setting, namely the scales
$\sfrac{r_{i}}{a}$, $i=0,1,2$, and the string tension $a^{2}\sigma$ from the direct fits in Table~\ref{tab:r_i_over_a},
two different parametrizations in terms of Allton fits in Table~\ref{tab:Allton_coefficients}, and the smoothened
scales $\sfrac{r_{i}}{a}$, $i=0,1$ and $\sigma r_{0}^{2}$, respectively, in Table~\ref{tab:r_i_over_a_Allton}.
Finally, in Fig.~\ref{fig:r_i_over_a_Allton_summary_smeared}, we show the Allton fits for the scales with smeared links.

\begin{table}[ht]
    \centering
    \caption{\label{tab:r_i_over_a}%
    $\sfrac{r_{i}}{a}$, $i=0,1,2$ and $a^{2}\sigma$ for all ensembles with the two different choices of $A$.
    The first of each row is for bare links, the second one for smeared links.
    The smeared-link values in brackets are replaced with the bare ones as explained in the text.}
    \begin{tabular*}{\textwidth}{l@{\extracolsep\fill}ccccc}
\hline
\hline
Ensemble & $\sfrac{r_{0}}{a}$ & $\sfrac{r_{1}}{a}$ & $\sfrac{r_{2}}{a}$ & $a^{2} \sigma$ ($A = A_{r_{0}}$) & $a^{2} \sigma$ ($A = \sfrac{\pi}{12}$) \\
\hline
\ensemble{5.80}{i} & $3.09640 \pm 0.01286$ & \dots & \dots & $0.11952 \pm 0.00206$ & $0.13035 \pm 0.00118$ \\
& $3.02320 \pm 0.09242$ & \dots & \dots & $0.11153 \pm 0.00949$ & $0.13128 \pm 0.00383$ \\[3pt]
\ensemble{6.00}{ii} & $3.83872 \pm 0.01950$ & $2.56673 \pm 0.03562$ & \dots & $0.07564 \pm 0.00258$ & $0.08468 \pm 0.00025$ \\
& $3.77465 \pm 0.08278$ & [$2.41291 \pm 0.07079$] & \dots & $0.07769 \pm 0.00499$ & $0.08685 \pm 0.00143$ \\[3pt]
\ensemble{6.00}{i} & $3.85113 \pm 0.01557$ & $2.56844 \pm 0.04089$ & \dots & $0.07428 \pm 0.00266$ & $0.08343 \pm 0.00011$ \\
& $3.81394 \pm 0.04370$ & [$2.42701 \pm 0.12536$] & \dots & $0.07595 \pm 0.00379$ & $0.08591 \pm 0.00151$ \\[3pt]
\ensemble{6.30}{iii} & $5.15895 \pm 0.03287$ & $3.47263 \pm 0.03436$ & \dots & $0.04258 \pm 0.00061$ & $0.04663 \pm 0.00041$ \\
& $5.18334 \pm 0.05663$ & [$3.42049 \pm 0.09021$] & \dots & $0.04251 \pm 0.00052$ & $0.04708 \pm 0.00012$ \\[3pt]
\ensemble{6.30}{ii} & $5.22042 \pm 0.03414$ & $3.49545 \pm 0.02301$ & \dots & $0.04163 \pm 0.00065$ & $0.04577 \pm 0.00023$ \\
& $5.26352 \pm 0.08370$ & [$3.52911 \pm 0.06433$] & \dots & $0.04257 \pm 0.00115$ & $0.04584 \pm 0.00013$ \\[3pt]
\ensemble{6.30}{i} & $5.26708 \pm 0.02218$ & $3.50938 \pm 0.01723$ & \dots & $0.04128 \pm 0.00039$ & $0.04565 \pm 0.00020$ \\
& $5.27320 \pm 0.10936$ & [$3.45729 \pm 0.11000$] & \dots & $0.04131 \pm 0.00059$ & $0.04623 \pm 0.00038$ \\[3pt]
\ensemble{6.72}{iii} & $7.88980 \pm 0.08546$ & $5.29637 \pm 0.03357$ & $2.45044 \pm 0.06290$ & $0.01866 \pm 0.00190$ & $0.02014 \pm 0.00126$ \\
& $7.85715 \pm 0.14605$ & $5.27694 \pm 0.07599$ & [$2.12384 \pm 0.19803$] & $0.01912 \pm 0.00025$ & $0.02067 \pm 0.00028$ \\[3pt]
\ensemble{6.72}{ii} & $7.92209 \pm 0.05545$ & $5.32775 \pm 0.02521$ & $2.45062 \pm 0.07633$ & $0.01880 \pm 0.00074$ & $0.02027 \pm 0.00070$ \\
& $7.93435 \pm 0.13460$ & $5.30683 \pm 0.06247$ & [$2.12747 \pm 0.16083$] & $0.01865 \pm 0.00026$ & $0.02020 \pm 0.00021$ \\[3pt]
\ensemble{6.72}{i} & $8.03302 \pm 0.07694$ & $5.38261 \pm 0.02038$ & $2.46825 \pm 0.07423$ & $0.01790 \pm 0.00071$ & $0.01940 \pm 0.00094$ \\
& $8.00053 \pm 0.16195$ & $5.38097 \pm 0.07864$ & [$2.14853 \pm 0.18989$] & $0.01810 \pm 0.00025$ & $0.01970 \pm 0.00022$ \\[3pt]
\ensemble{7.00}{iii} & $10.55417 \pm 0.38163$ & $7.06784 \pm 0.12644$ & $3.19097 \pm 0.13449$ & $0.01041 \pm 0.00186$ & $0.01106 \pm 0.00226$ \\
& $10.58371 \pm 0.10773$ & $7.06499 \pm 0.08959$ & $3.10774 \pm 0.03702$ & $0.01035 \pm 0.00025$ & $0.01105 \pm 0.00040$ \\[3pt]
\ensemble{7.00}{i} & $10.72634 \pm 0.15351$ & $7.16677 \pm 0.08142$ & $3.20894 \pm 0.14144$ & $0.01024 \pm 0.00070$ & $0.01093 \pm 0.00066$ \\
& $10.81068 \pm 0.11230$ & $7.18318 \pm 0.10637$ & $3.13717 \pm 0.06911$ & $0.00992 \pm 0.00022$ & $0.01066 \pm 0.00027$ \\[3pt]
\ensemble{7.28}{iii} & $13.58998 \pm 0.28745$ & $9.26165 \pm 0.12521$ & $4.21509 \pm 0.02629$ & $0.00668 \pm 0.00066$ & $0.00700 \pm 0.00067$ \\
& $13.93468 \pm 0.27129$ & $9.38512 \pm 0.10500$ & $4.21051 \pm 0.08218$ & $0.00613 \pm 0.00028$ & $0.00647 \pm 0.00023$ \\
\hline
\hline
\end{tabular*}
\end{table}

\begin{table}[ht]
    \centering
    \caption{\label{tab:Allton_coefficients}%
    Coefficients of the Allton fits, Eq.~\eqref{eq:Allton_fit}.
    For each of $\sfrac{r_{0}}{a}$ or $\sfrac{r_{1}}{a}$, the first of the two rows shows bare results, while the second
    one shows smeared results.}
    \begin{tabular*}{\textwidth}{c@{\extracolsep{\fill}}ccccc}
        \hline
        \hline
        \multicolumn{6}{c}{linear in $a\mtot$} \\
        & $C_{00}$ & $C_{01}$ & $\sfrac{C_{20}}{10^{5}}$ & $\sfrac{D_{2}}{10^{3}}$ & $\chi^{2}_{\text{red.}}$ \\
        \hline
        $\sfrac{r_{0}}{a}$
        & $21.94181 \pm 1.68077$ & $0.08587 \pm 0.02668$ & $1.08538 \pm 0.49989$ & $2.86204 \pm 1.48991$ & $\sfrac{6.03305}{8} = 0.75413$ \\
        & $20.50314 \pm 2.06333$ & $0.10875 \pm 0.03774$ & $0.36230 \pm 1.13404$ & $0.57839 \pm 3.37576$ & $\sfrac{1.30035}{8} = 0.16254$ \\[3pt]
        $\sfrac{r_{1}}{a}$
        & $34.72718 \pm 1.76851$ & $0.08556 \pm 0.02828$ & $3.13605 \pm 2.24815$ & $5.78128 \pm 4.79208$ & $\sfrac{1.24069}{7} = 0.17724$ \\
        & $32.21438 \pm 2.83572$ & $0.12787 \pm 0.05211$ & $3.04080 \pm 2.67043$ & $5.77559 \pm 5.62018$ & $\sfrac{0.43440}{7} = 0.06206$ \\
        \hline
        \multicolumn{6}{c}{quadratic in $a\mtot$} \\
        & $C_{00}$ & $C_{02}$ & $\sfrac{C_{20}}{10^{5}}$ & $\sfrac{D_{2}}{10^{3}}$ & $\chi^{2}_{\text{red.}}$ \\
        \hline
        $\sfrac{r_{0}}{a}$
        & $25.82044 \pm 0.53306$ & $0.15358 \pm 0.04626$ & $1.28496 \pm 0.54074$ & $5.02550 \pm 1.69661$ & $\sfrac{3.70085}{8} = 0.46261$ \\
        & $25.59835 \pm 0.46671$ & $0.18195 \pm 0.06659$ & $0.26569 \pm 1.27393$ & $2.03565 \pm 3.54790$ & $\sfrac{1.49736}{8} = 0.18717$ \\[3pt]
        $\sfrac{r_{1}}{a}$
        & $38.65447 \pm 0.74486$ & $0.15049 \pm 0.05064$ & $3.28647 \pm 2.38685$ & $7.10549 \pm 5.08679$ & $\sfrac{1.01596}{7} = 0.14514$ \\
        & $38.05123 \pm 0.77302$ & $0.22051 \pm 0.09517$ & $3.41874 \pm 2.93632$ & $8.03369 \pm 6.03171$ & $\sfrac{0.49725}{7} = 0.07104$ \\
        \hline
        \hline
    \end{tabular*}
\end{table}

\begin{table}[ht]
    \centering
    \caption{\label{tab:r_i_over_a_Allton}%
    $\sfrac{r_{i}}{a}$, $i=0,1$, and $\sqrt{r_{0}^{2} \sigma}$ for all ensembles with the different choices of $A$
    using the smoothened $\sfrac{r_{0}}{a}$ from the Allton fits.
    The values in brackets stem from the Allton fits, but we do not have a direct determination.
    The first of each row is for bare links, the second one for smeared links.}
    \begin{tabular*}{\textwidth}{c@{\extracolsep\fill}cccc}
\hline
\hline
Ensemble & $\sfrac{r_{0}}{a}$ & $\sfrac{r_{1}}{a}$ & $\sqrt{\sigma r_{0}^{2}}$ ($A = A_{r_{0}}$) & $\sqrt{\sigma r_{0}^{2}}$ ($A = \sfrac{\pi}{12}$) \\
\hline
l3248f211b580m00235m0647m831 & $3.09673 \pm 0.01204$ & \dots & $1.07061 \pm 0.01013$ & $1.11805 \pm 0.00666$ \\
 & $3.01140 \pm 0.07614$ & \dots & $1.00569 \pm 0.04979$ & $1.09112 \pm 0.03185$ \\[3pt]
l3264f211b600m00507m0507m628 & $3.83628 \pm 0.00856$ & $2.56244 \pm 0.02403$ & $1.05509 \pm 0.01812$ & $1.11634 \pm 0.00299$ \\
 & $3.79648 \pm 0.03130$ & $2.56593 \pm 0.02591$ & $1.05822 \pm 0.03509$ & $1.11882 \pm 0.01302$ \\[3pt]
l4864f211b600m00184m0507m628 & $3.84897 \pm 0.00986$ & $2.56725 \pm 0.02304$ & $1.04899 \pm 0.01895$ & $1.11173 \pm 0.00293$ \\
 & $3.81537 \pm 0.03320$ & $2.57260 \pm 0.02423$ & $1.05150 \pm 0.02777$ & $1.11832 \pm 0.01383$ \\[3pt]
l3296f211b630m0074m037m440 & $5.19001 \pm 0.01869$ & $3.48020 \pm 0.01224$ & $1.07091 \pm 0.00853$ & $1.12073 \pm 0.00641$ \\
 & $5.17709 \pm 0.03704$ & $3.47017 \pm 0.01736$ & $1.06741 \pm 0.01007$ & $1.12328 \pm 0.00816$ \\[3pt]
l4896f211b630m00363m0363m430 & $5.24553 \pm 0.01325$ & $3.50231 \pm 0.01153$ & $1.07022 \pm 0.00878$ & $1.12227 \pm 0.00396$ \\
 & $5.25407 \pm 0.04198$ & $3.50118 \pm 0.01247$ & $1.08402 \pm 0.01702$ & $1.12492 \pm 0.00914$ \\[3pt]
l6496f211b630m0012m0363m432 & $5.25416 \pm 0.01396$ & $3.50572 \pm 0.01203$ & $1.06745 \pm 0.00573$ & $1.12265 \pm 0.00389$ \\
 & $5.26608 \pm 0.04551$ & $3.50598 \pm 0.01362$ & $1.07037 \pm 0.01202$ & $1.13231 \pm 0.01081$ \\[3pt]
l48144f211b672m0048m024m286 & $7.86707 \pm 0.04293$ & $5.30732 \pm 0.02156$ & $1.07467 \pm 0.05516$ & $1.11633 \pm 0.03541$ \\
 & $7.86948 \pm 0.08455$ & $5.28136 \pm 0.04660$ & $1.08820 \pm 0.01372$ & $1.13151 \pm 0.01430$ \\[3pt]
l64144f211b672m0024m024m286 & $7.89244 \pm 0.03821$ & $5.31806 \pm 0.01868$ & $1.08215 \pm 0.02194$ & $1.12375 \pm 0.02018$ \\
 & $7.90203 \pm 0.07380$ & $5.29661 \pm 0.04068$ & $1.07905 \pm 0.01252$ & $1.12321 \pm 0.01195$ \\[3pt]
l96192f211b672m0008m022m260 & $8.05163 \pm 0.03896$ & $5.38486 \pm 0.01662$ & $1.07731 \pm 0.02207$ & $1.12140 \pm 0.02783$ \\
 & $8.10762 \pm 0.04793$ & $5.39212 \pm 0.02820$ & $1.09084 \pm 0.00982$ & $1.13805 \pm 0.00929$ \\[3pt]
l64192f211b700m00316m0158m188 & $10.57092 \pm 0.07717$ & $7.08912 \pm 0.03450$ & $1.07880 \pm 0.09674$ & $1.11155 \pm 0.11400$ \\
 & $10.63795 \pm 0.06414$ & $7.10727 \pm 0.04002$ & $1.08239 \pm 0.01456$ & $1.11827 \pm 0.02148$ \\[3pt]
l144288f211b700m000569m01555m1827 & $10.64095 \pm 0.09157$ & $7.11895 \pm 0.03886$ & $1.07678 \pm 0.03771$ & $1.11245 \pm 0.03490$ \\
 & $10.72618 \pm 0.07495$ & $7.15068 \pm 0.04681$ & $1.06813 \pm 0.01416$ & $1.10769 \pm 0.01581$ \\[3pt]
l96288f211b728m00223m01115m1316 & $13.90017 \pm 0.15482$ & $9.31454 \pm 0.08267$ & $1.13621 \pm 0.05777$ & $1.16337 \pm 0.05728$ \\
 & $14.00094 \pm 0.12505$ & $9.37267 \pm 0.08271$ & $1.09633 \pm 0.02682$ & $1.12639 \pm 0.02226$ \\
\hline
\hline
\end{tabular*}
\end{table}

\FloatBarrier

\subsection{Continuum extrapolations}
\label{app:continuum extrapolations}

\begin{figure}
    \centering
    \includegraphics[width=0.49\textwidth]{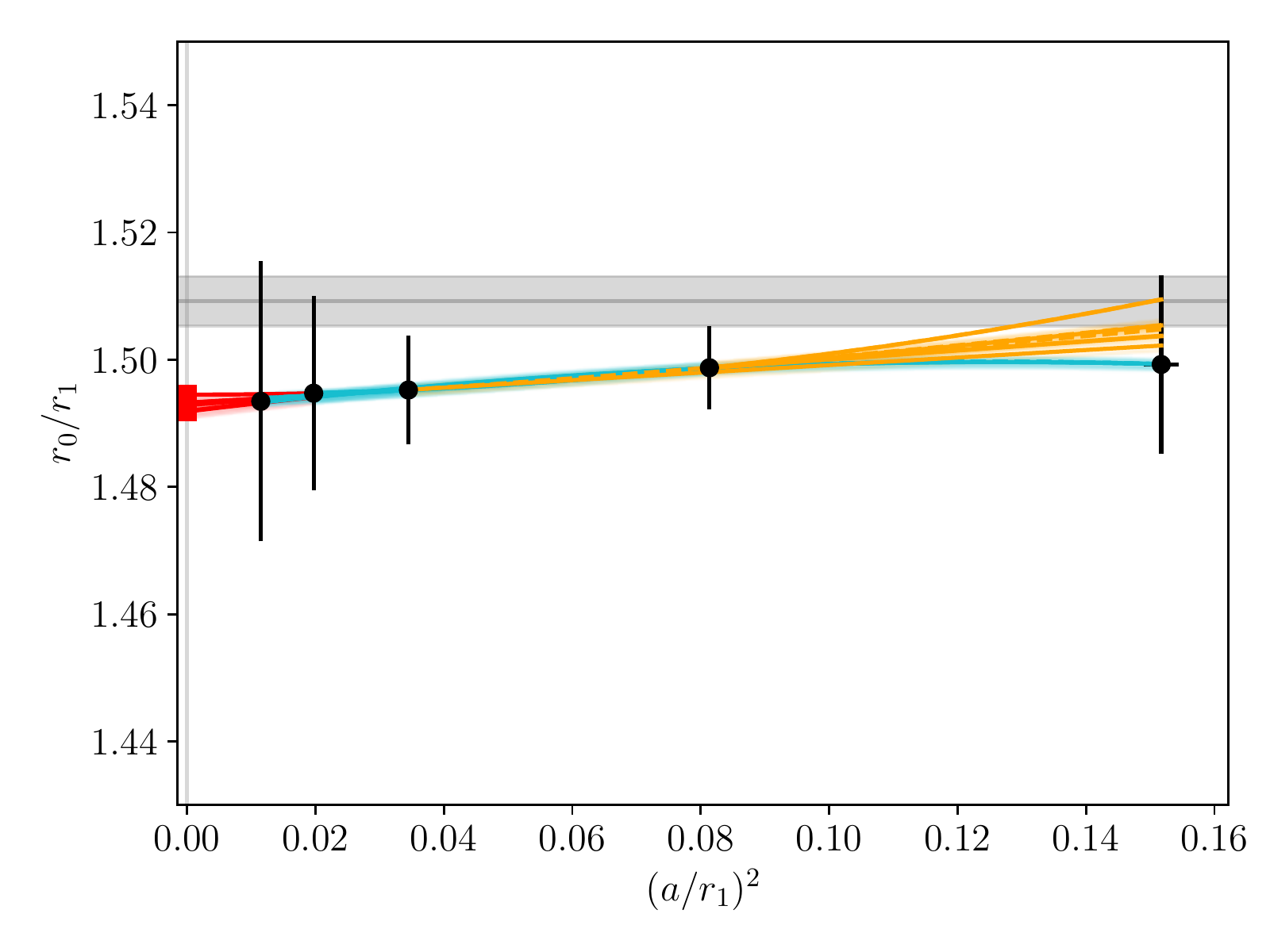}
    \includegraphics[width=0.49\textwidth]{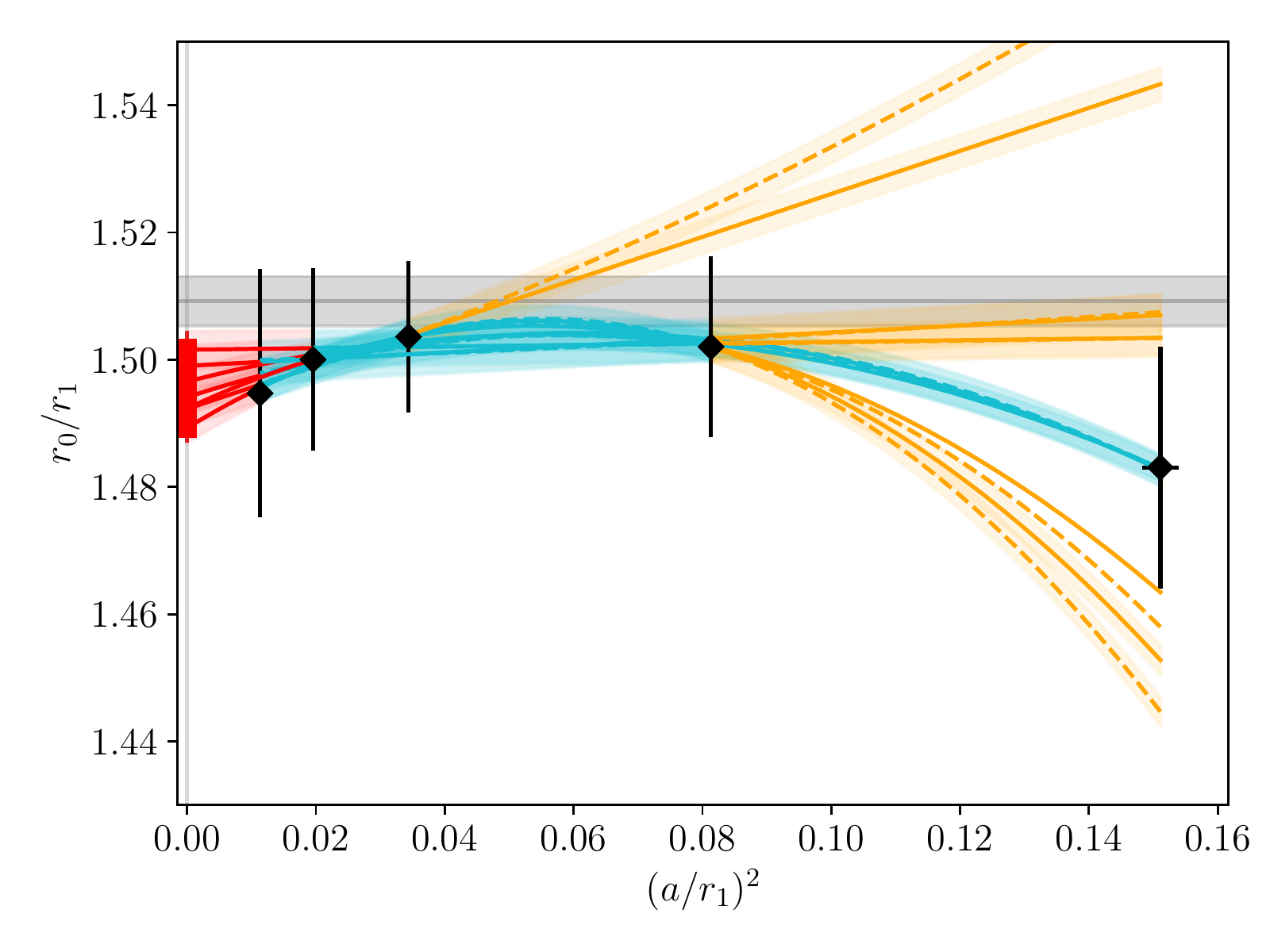}
    \caption{\label{fig:continuum_ri_rj_I_r1}%
    Continuum extrapolation of the parametrization of $\sfrac{r_{0}}{r_{1}}$ evaluated at the physical
    $\sfrac{\ml}{\ms}$-ratio for $6.0\le\beta\le7.28$ as a function of $(\sfrac{a}{r_{1}})^{2}$.
    The black points show the bare-link data (left) and the smeared-link data (right) with the corresponding continuum
    results shown in red.
    The lines and bands show the fit curves and errors, within the fit range in cyan, as extrapolations towards the
    continuum or coarser lattices in red/orange, respectively.
    The gray solid line and band indicate the HotQCD result in ($2+1$)-flavor QCD~\cite{HotQCD:2014kol}.
    Note that the weighted average is absent in this figure since it is the same that is already shown in the
    corresponding panels for extrapolation as a function of $(\sfrac{a}{r_{0}})^{2}$ in Fig.~\ref{fig:continuum_ri_rj_I}
    in Sec.~\ref{sec:ratios}.}
\end{figure}

\begin{figure}[ht]
    \centering
    \includegraphics[width=\textwidth]{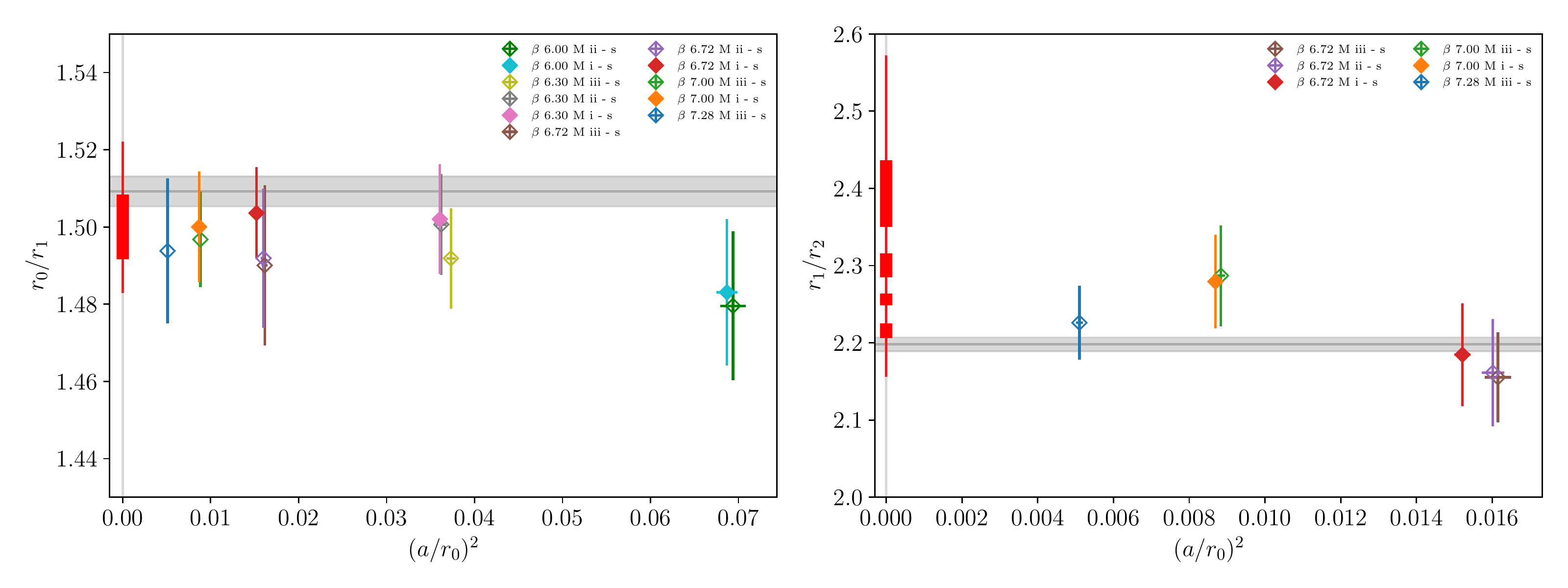}
    \caption{\label{fig:continuum_ri_rj_III_smeared}%
    Continuum results (red) and smoothened data (other colors) for the ratios $\sfrac{r_{0}}{r_{1}}$ or
    $\sfrac{r_{1}}{r_{2}}$ are shown in the left or right columns, respectively, as functions of
    $(\sfrac{a}{r_{0}})^{2}$, using smeared-link data.
    The gray solid line and band show the ($2+1$)-flavor QCD reference values~\cite{HotQCD:2014kol, Bazavov:2017dsy}.
    The corresponding plot for bare links is shown in Fig.~\ref{fig:continuum_ri_rj_III} in Sec.~\ref{sec:cont-limit}.}
\end{figure}

\begin{figure}[ht]
    \centering
    \includegraphics[width=1.0\textwidth]{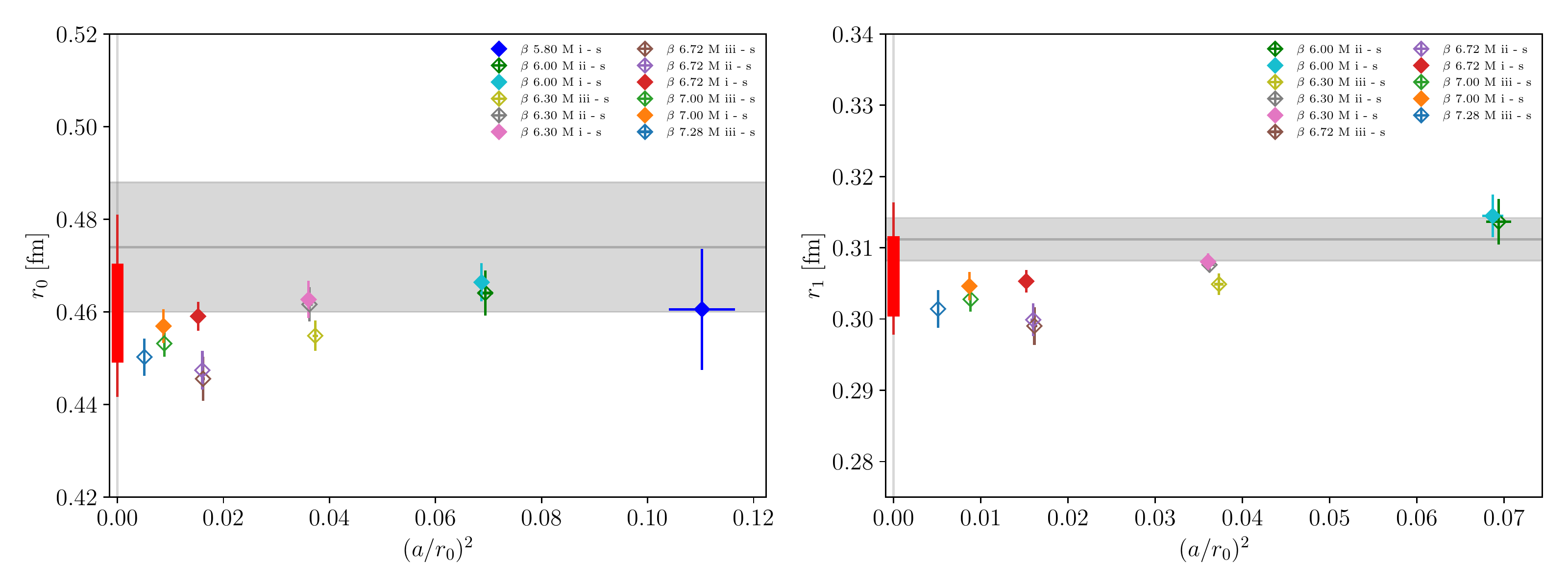}%
    \caption{\label{fig:continuum_r0_r1_smeared}%
    Continuum results (red) and smoothened data (other colors) for $a_{f_{p4s}}\sfrac{r_{0,1}}{a}$ are shown in the
    left or right columns, respectively, as functions of $(\sfrac{a}{r_{0}})^{2}$, using bare links.
    The gray solid line and band show the published ($2+1+1$)-flavor QCD values~\cite{EuropeanTwistedMass:2014osg,
    Dowdall:2013rya} for $r_{0}$ and $r_{1}$, respectively.
    The corresponding plots for bare links are shown in Fig.~\ref{fig:continuum_r0_r1} in Sec.~\ref{sec:cont-limit}.}
\end{figure}

\begin{figure}[ht]
    \centering
    \includegraphics[width=\textwidth]{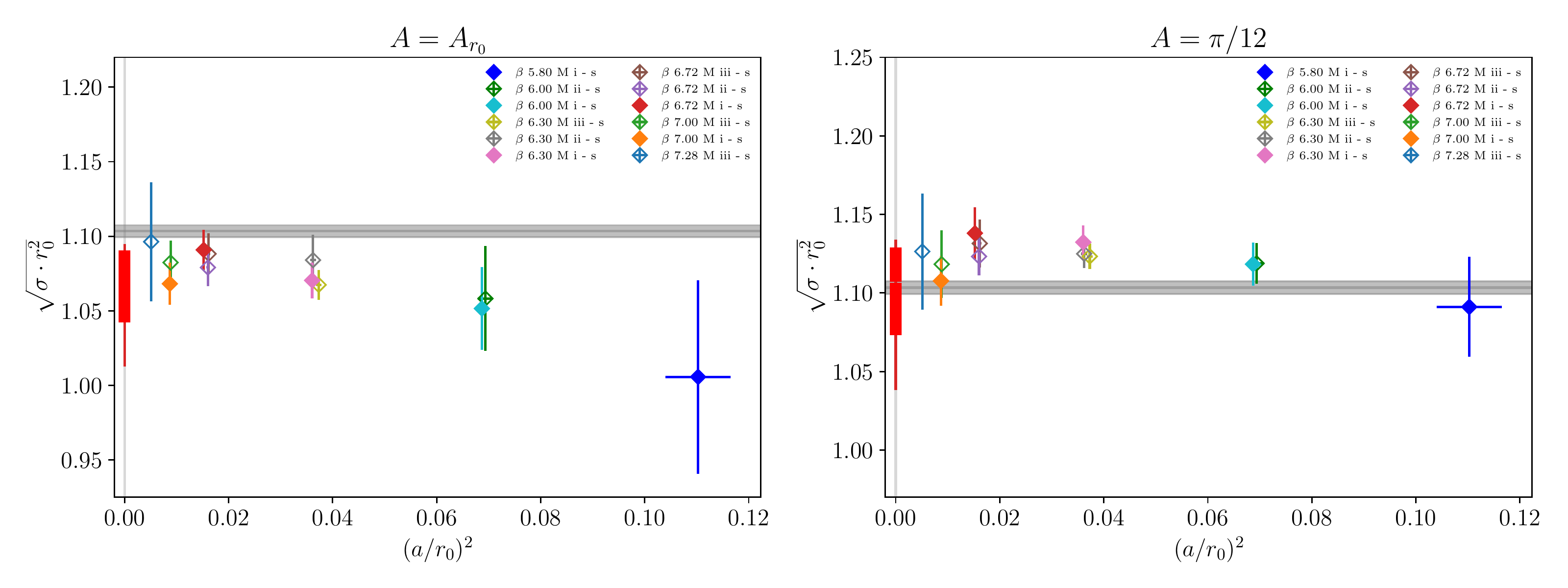}
    \caption{\label{fig:continuum_sqrt_sigma_r0_sq_smeared}%
    Continuum results (red) and smoothened data (other colors) for $\sqrt{\sigma r_{0}^{2}}$ assuming two different
    coefficients $A$ for the Coulomb term are shown in the left or right columns, respectively, as functions of
    $(\sfrac{a}{r_{0}})^{2}$, using smeared links.
    The gray solid line and band show the published ($2+1$)-flavor QCD value~\cite{Cheng:2007jq}.
    The corresponding plots for bare links are shown in Fig.~\ref{fig:continuum_sqrt_sigma_r0_sq} in
    Sec.~\ref{sec:cont-limit}.}
\end{figure}

In this section, we collect additional material relevant for the continuum extrapolations.
The continuum extrapolations of bare- or smeared-link data for $\sfrac{r_{0}}{r_{1}}$ as a function of
$(\sfrac{a}{r_{1}})^{2}$ are shown in Fig.~\ref{fig:continuum_ri_rj_I_r1}.
We show the continuum results and the approach to the continuum limit for smeared-link data, in particular, for
$\sfrac{r_{0}}{r_{1}}$ and $\sfrac{r_{1}}{r_{2}}$ in Fig.~\ref{fig:continuum_ri_rj_III_smeared}, for
$a_{f_{p4s}}\sfrac{r_{0}}{a}$ and $a_{f_{p4s}}\sfrac{r_{1}}{a}$ in Fig.~\ref{fig:continuum_r0_r1_smeared}, or for
$\sqrt{\sigma r_{0}^{2}}$ with two different coefficients for the Coulomb term in
Fig.~\ref{fig:continuum_sqrt_sigma_r0_sq_smeared}.


\begin{figure}
    \centering
    \includegraphics[width=\textwidth]{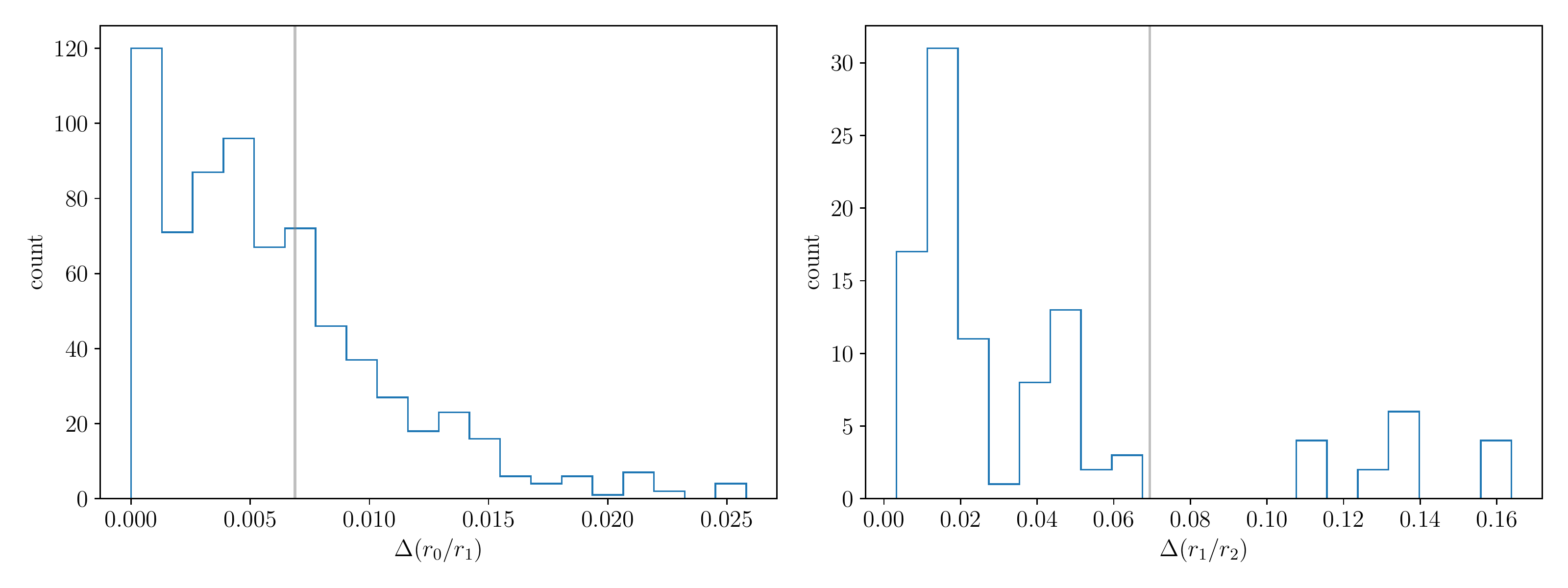}%
    \\%
    \includegraphics[width=\textwidth]{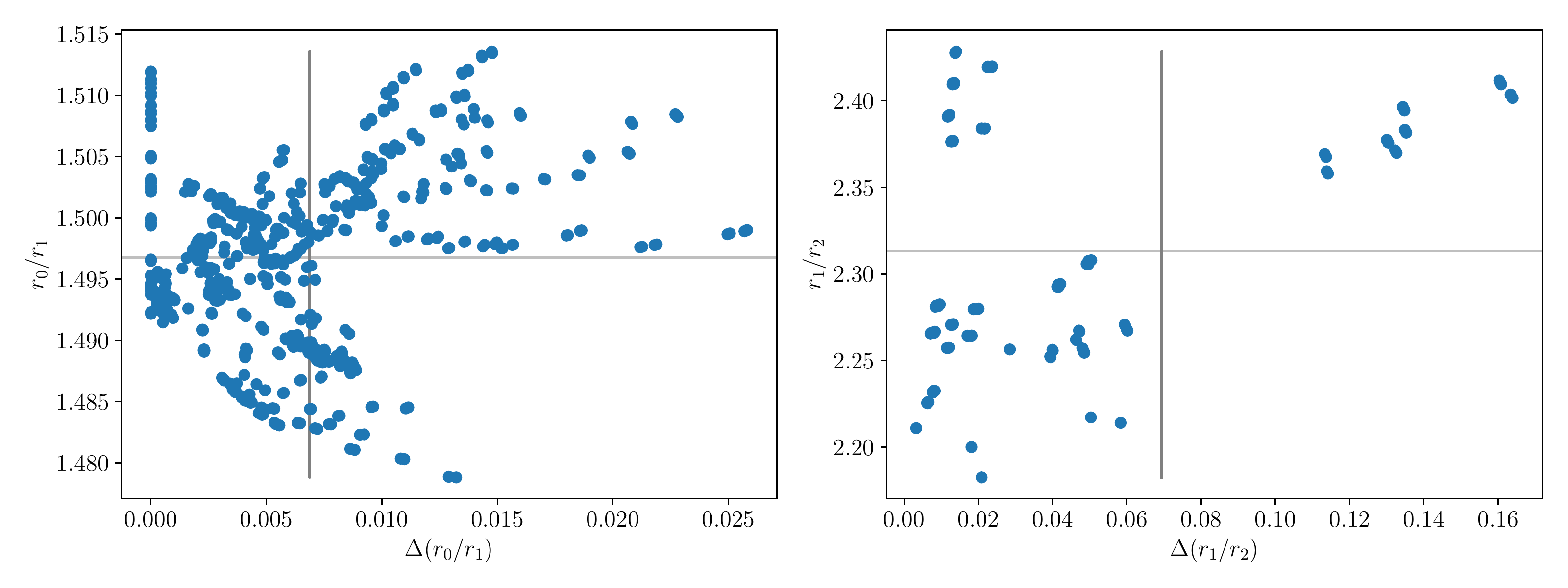}%
    \caption{\label{fig:errors_ratio}%
    Distribution of the errors (top) and correlation plots between central values and regression errors (bottom) for
    the ratios $\sfrac{r_{0}}{r_{1}}$ and $\sfrac{r_{1}}{r_{2}}$, respectively.
    The gray lines indicate our error and central value also shown in the histogram
    Fig.~\ref{fig:continuum_ri_rj_hist}.}
\end{figure}

\begin{figure}
    \centering
    \includegraphics[width=1.0\textwidth]{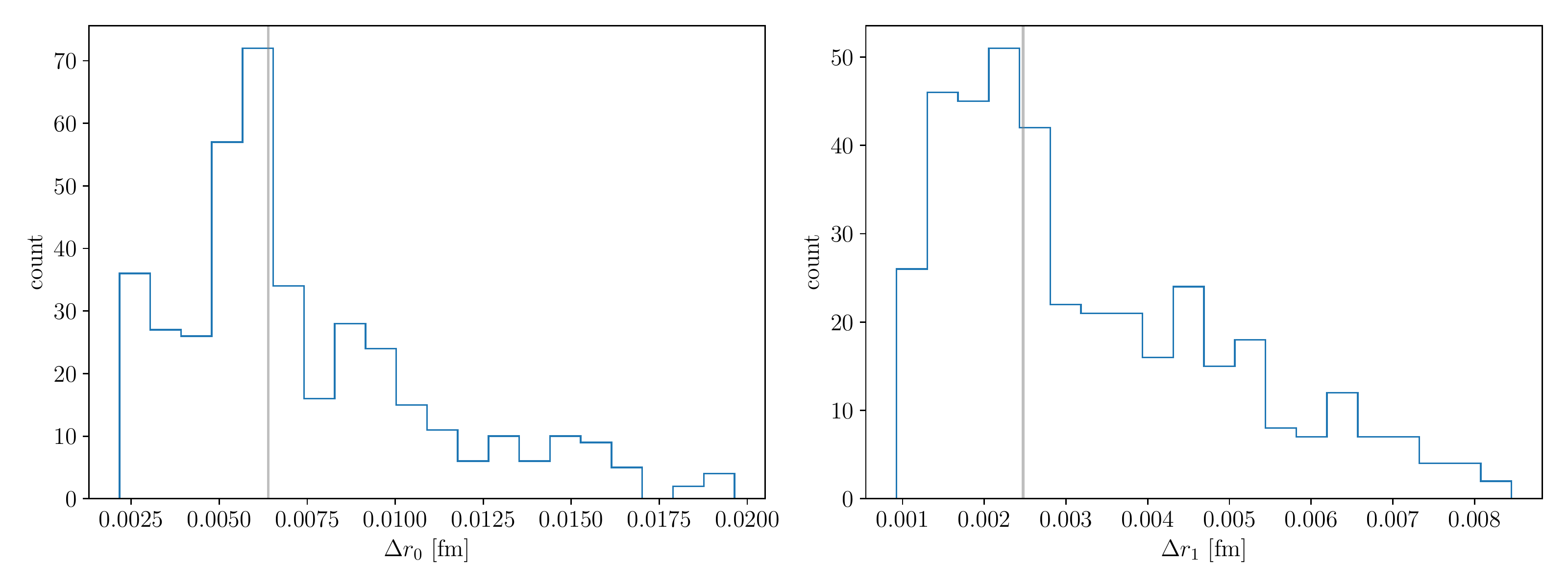}%
    \\%
    \includegraphics[width=1.0\textwidth]{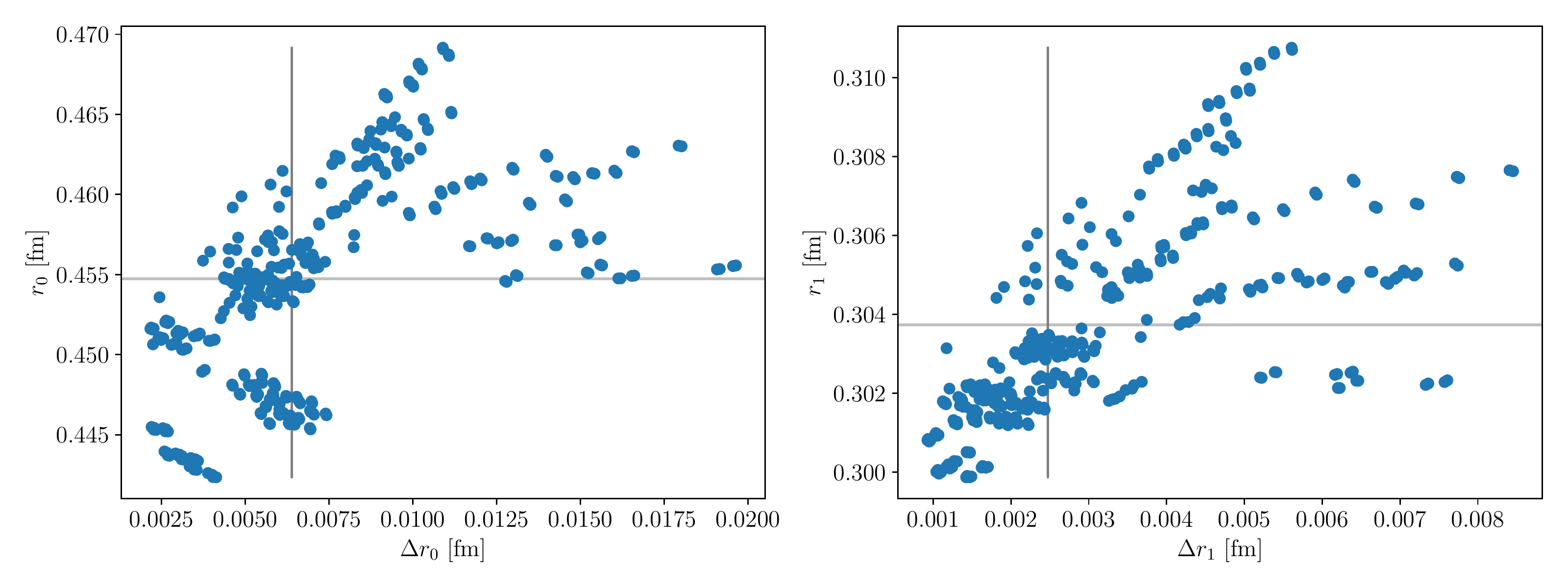}%
    \caption{\label{fig:errors_ri}%
    Distribution of the errors (top) and correlation plots between central values and regression errors (bottom) for
    the scales $r_{0}$ and $r_{1}$, respectively.
    The gray lines indicate our error and central value also shown in the histogram
    Fig.~\ref{fig:continuum_r0_r1_hist}.}
\end{figure}

\begin{figure}
    \centering
    \includegraphics[width=0.49\textwidth]{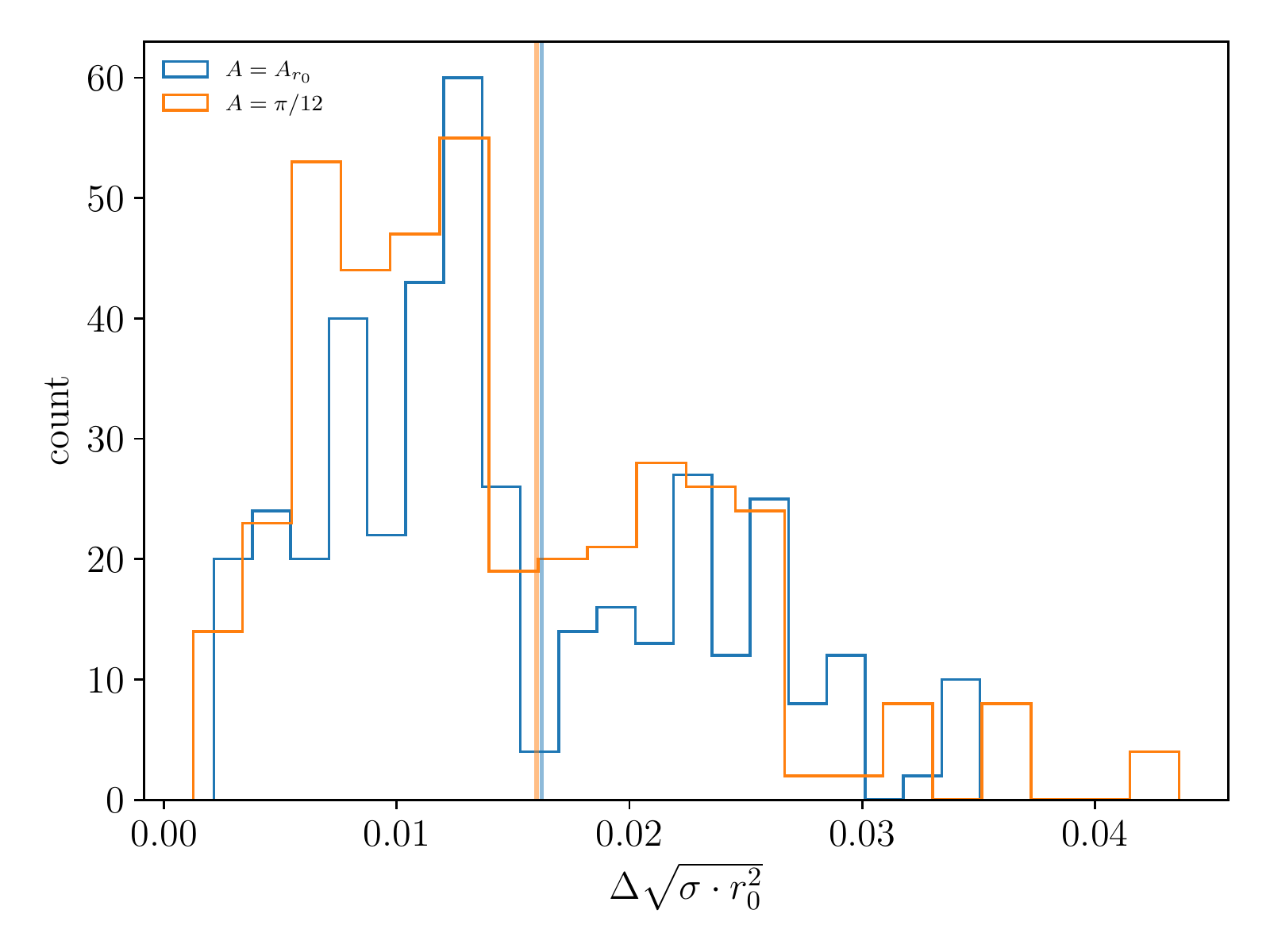}%
    \hfill%
    \includegraphics[width=0.49\textwidth]{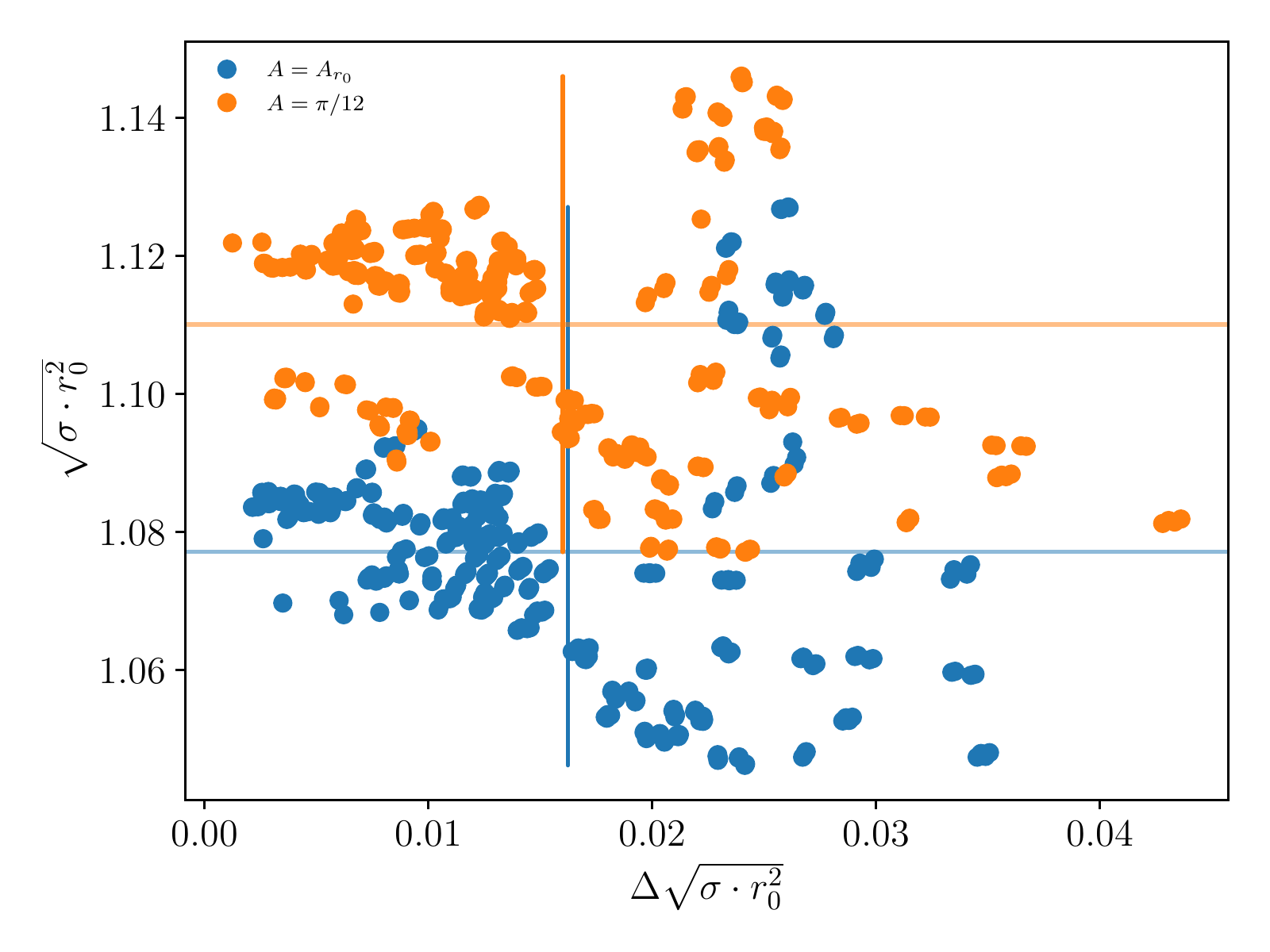}%
    \caption{\label{fig:errors_string}%
    Distribution of the errors (left) and correlation plots between central values and regression errors (right) for
    the string tension $\sqrt{\sigma r_{0}^{2}}$ for the two different choices of the Coulomb parameter, respectively.
    The lines indicate our errors and central values, respectively, also shown in the histogram
    Fig.~\ref{fig:continuum_sqrt_sigma_r0_sq_hist}.}
\end{figure}

We provide further information regarding the distribution of errors from the different continuum extrapolations, in
particular, for $\sfrac{r_{0}}{r_{1}}$ and $\sfrac{r_{1}}{r_{2}}$ in Fig.~\ref{fig:errors_ratio}, for
$a_{f_{p4s}}\sfrac{r_{0}}{a}$ and $a_{f_{p4s}}\sfrac{r_{1}}{a}$ in Fig.~\ref{fig:errors_ri}, or for $\sqrt{\sigma
r_{0}^{2}}$ with two different coefficients for the Coulomb term in Fig.~\ref{fig:errors_string}.
We show the histogram of the Hessian regression errors together with the error estimate from the width of the central
values' histogram and, in further panels correlation plots between the central values and the regression errors.
In the case of $\sfrac{r_{0}}{r_{1}}$, 64 instances in the left-most bin are due to fits with zero degree of freedom
(where we use zero for the error).
In all three cases among $\sfrac{r_{0}}{r_{1}}$, $a_{f_{p4s}}\sfrac{r_{0}}{a}$, and $a_{f_{p4s}}\sfrac{r_{1}}{a}$ we see
a fairly sharp drop of the error distribution beyond the bin containing our error estimate.
While there is no obvious correlation between central values and regression errors for $\sfrac{r_{0}}{r_{1}}$, there
are correlations between larger regression errors and larger central values for $a_{f_{p4s}}\sfrac{r_{0}}{a}$ and
$a_{f_{p4s}}\sfrac{r_{1}}{a}$.
For $\sfrac{r_{1}}{r_{2}}$, this is quite different.
Our error estimate is significantly larger than the majority of regression errors.
Although all regressions with large errors correspond to high central values, not all regressions with central values
entail large errors.
Lastly, for the string tension our error estimate is marginally larger than the bulk of the distribution of errors, and
there is no clear pattern of correlation between the central value and the regression error.

\FloatBarrier
\section{Perturbative QCD formulas}
\label{app:pQCD}

In this Appendix, we collect various formulas obtained in perturbative QCD that are used in Sec.~\ref{sec:charm}.
Appendix~\ref{app:coefficients} contains the perturbative expansion coefficients for the force in the case of massless
quarks, Appendix~\ref{app:charm} provides the one- and two-loop corrections due to a massive charm quark, and in
Appendix~\ref{app:functions} one may find the definitions of some special functions showing up in
Appendix~\ref{app:charm}.

\subsection{Coefficients in the force and coupling at two loops}
\label{app:coefficients}

The one- and two-loop coefficients, $a_{i}$, appearing in Eqs.~\eqref{eq:statfor} and~\eqref{eq:statfor_resummed} have
been computed in Refs.~\cite{Fischler:1977yf, Billoire:1979ih, Peter:1996ig, Peter:1997me, Schroder:1998vy}, and the
three-loop coefficients, which go beyond our accuracy, in Refs.~\cite{Smirnov:2008pn, Anzai:2009tm, Smirnov:2009fh,
Lee:2016cgz}:
\begin{align}
a_{1}^{(\Nf)} =& \frac{31\CA}{9} - \frac{10}{9}\Nf , \\
a_{2}^{(\Nf)} =& \left(\frac{4343}{162} + 4\pi^{2} - \frac{\pi^{4}}{4} + \frac{22}{3} \zeta_{3}\right) \CA^{2}
        - \left(\frac{899}{81} + \frac{28}{3} \zeta_{3}\right) \CA\Nf \nonumber \\
    &- \left(\frac{55}{6} - 8\zeta_{3}\right) \CF\Nf + \frac{100}{81} \Nf^{2} , \\
a_{3}^{\text{L}} =& \frac{16\pi^{2}}{3} \CA^{3} ,
\end{align}
where
$\CA = \Nc = 3$, and $\CF = \sfrac{(\Nc^{2}-1)}{(2\Nc)} = \sfrac{4}{3}$.
Note that $\zeta_{n} \equiv \zeta(n) = \displaystyle \sum_{i=1}^{\infty} \sfrac{1}{i^{n}}$ is the Riemann zeta function.
The logarithmic terms affecting the static force and potential are most conveniently extracted in an effective field
theory framework~\cite{Brambilla:1999qa, Brambilla:1999xf, Brambilla:2004jw}, as discussed in
Sec.~\ref{sec:charm}, and have been computed and resummed to all orders at N$^{2}$LL and N$^{3}$LL accuracy in
Refs.~\cite{Brambilla:1999qa, Kniehl:1999ud, Pineda:2000gza, Brambilla:2006wp, Brambilla:2009bi}.

The running of the strong coupling $\alsNf(\mu)$ is determined by the $\beta$-function.
The first two coefficients of the $\beta$-function, $\beta_{1,2}$, are scheme independent and given by
\begin{align}
\beta_{0}^{(\Nf)} =& \frac{11}{3} \CA - \frac{2}{3} \Nf , \\
\beta_{1}^{(\Nf)} =& \frac{34}{3} \CA^{2} - \frac{10}{3} \CA\Nf - 2\CF\Nf .
\end{align}
The coupling $\alsNf(\mu)$ with $\Nf$ massless flavors is related to $\alsNfm(\mu)$ with $\Nf-1$ massless flavors via
\begin{equation}
\label{eq:alschange}
\alsNfp(\mu) = \alsNf(\mu) \left\{1 + \sum\limits_{n=1}^{\infty} \left[\alsNf(\mu)\right]^{n}
    \left[\sum\limits_{l=0}^{n} c_{nl} \ln^{l}\left(\frac{\mu^{2}}{m^{2}}\right)\right]\right\} .
\end{equation}
Following~\cite{Recksiegel:2001xq} (see Refs.~\cite{Chetyrkin:2005ia, Schroder:2005hy} for the four-loop decoupling),
we have for the first terms
\begin{align}
& c_{11} = \frac{1}{6\pi} , \quad c_{10} = 0 , \\
& c_{22} = \frac{1}{36\pi^{2}} , \quad c_{21} = \frac{19}{24\pi^{2}} , \quad c_{20} = -\frac{11}{72\pi^{2}} ,
\end{align}
when $m$ is the $\MS$ mass renormalized at the $\MS$ mass scale: $m= m^{\MS}(m^{\MS})$.\footnote{%
In the PDG~\cite{ParticleDataGroup:2020ssz} and in Refs.~\cite{Chetyrkin:2005ia, Schroder:2005hy} the coefficient
$c_{21}$ reads $\sfrac{11}{(24\pi^{2})}$ because there the $\MS$ mass $m$ in the decoupling
relation~\eqref{eq:alschange} is taken at the renormalization scale $\mu$.
We follow Ref.~\cite{Recksiegel:2001xq} and understand the $\MS$ mass $m$ in Eq.~\eqref{eq:alschange} as computed at
the $\MS$ mass scale.}

\subsection{Finite-mass corrections}
\label{app:charm}

Adding the effect of a quark of mass $m$ to $\Nf$ massless flavors modifies the order $\als^{N}$ term in the static
potential from $V^{(\Nf),[N]}$ into
\begin{equation}
    \label{eq:full_potential}
    V_{m}^{(\Nf),[N]} = V^{(\Nf),[N]} + \delta V_{m}^{(\Nf),[N]}.
\end{equation}
The correction due to the quark of finite mass $m$ is known up to two-loop accuracy.
The $\order(\als^{2})$ corrections have been computed in Refs.~\cite{Eiras:1999xx, Melles:2000dq, Eiras:2000rh,
Melles:2000ey, Hoang:2000fm}.
At $\order(\als^{3})$, the correction was computed first in momentum space in \cite{Melles:2000dq}.
The Fourier transform was performed (the coordinate-space potential was studied using different integral
representations in \cite{Melles:2000dq}) and $\delta V_{m}^{(\Nf),[3]}(r)$ was obtained also in one-parameter integral
form in~\cite{Hoang:2000fm, Melles:2000ey}.
Many of the original references contain misprints; corrected formulas can be found in Ref.~\cite{Recksiegel:2001xq}.

At one-loop accuracy, the finite mass correction to the static potential reads
\begin{equation}
    \label{eq:deltaV2}
    \delta V_{m}^{(\Nf),[2]}(r,\nu) = - \frac{\CF \alsNf(\nu)}{r} \frac{\alsNf(\nu)}{3\pi}
        \int\limits_{1}^{\infty} \d x \, f(x) \, \e{-2mrx} ,
\end{equation}
where $f(x) = \sfrac{\sqrt{x^{2}-1}}{x^{2}} \left(1 + \sfrac{1}{2x^{2}}\right)$.
At two-loop accuracy, the finite mass correction to the static potential is given by
\begin{align}
\label{eq:deltaV3}
\delta V_{m}^{(\Nf),[3]}(r,\nu) =& - \frac{\CF \alsNf}{r} \left(\frac{\alsNf}{3\pi}\right)^{2} \times \Bigg\{
    \left[9\pi^{2} \big(c_{20} - 2\ln(mr) (c_{21} - 2 c_{22} \ln(mr))\big) - \ln^{2}(mr) + \frac{57}{4} \ln(mr) +
        \frac{11}{8}\right] \nonumber \\
    &+ \frac{57}{4} \Big[f_{1} \Gamma(0,2f_{2} mr) + b_{1} \Gamma(0,2b_{2} mr)\Big] + \left[-2\ln(mr) - \frac{5}{3}\Nf
        + \frac{83}{6}\right] \int\limits_{1}^{\infty} \d x\, f(x) \, \e{-2mrx} \nonumber \\
    &+ \left(\frac{33}{2} - \Nf\right) \int\limits_{1}^{\infty} \d x \, f(x) \, \left(\e{-2mrx} \, \Ei(2mrx) + \e{2mrx}
        \, \Ei(-2mrx) - 2\ln(2mrx)\right) \nonumber \\
    &- \int\limits_{1}^{\infty} \d x \, f(x) \, \e{-2mrx} \left(\frac{1}{x^{2}} + 2\ln(2x) + 8mrx + f(x) \, x \,
        \ln\frac{x - \sqrt{x^{2} - 1}}{x + \sqrt{x^{2} - 1}}\right) \Bigg\} ,
\end{align}
where\footnote{%
This parametrization matches the one from Ref.~\cite{Hoang:2000fm} when renaming $f \to c$ and $b \to d$.}
\begin{align}
f_{1} &= \frac{\ln A - \ln b_{2}}{\ln f_{2} - \ln b_{2}} ,
    &&b_{1} = \frac{\ln A - \ln f_{2}}{\ln b_{2} - \ln f_{2}} , \nonumber \\
f_{2} &= 0.470 \pm 0.005 , &&b_{2} = 1.120 \pm 0.010 , \\
\ln A &= \sfrac{161}{228} + \sfrac{13\zeta_{3}}{19} -\ln 2 . && \nonumber
\end{align}
In Eq.~\eqref{eq:deltaV3}, $\Ei$ denotes the exponential-integral function and $\Gamma$ with two arguments denotes the
incomplete gamma function.
Their definitions and some useful properties can be found in Appendix~\ref{app:functions}.
The mass $m$ in the above formulas is the $\MS$ mass renormalized at the $\MS$ mass scale: $m =
m^{\MS}(m^{\MS})$.\footnote{%
For the numerical evaluation of the above integrals, it is convenient to introduce the coordinate transformation $x \to
\sfrac{1}{\sqrt{1 - v^{2}}}$, $\d x \to v (1 - v^{2})^{-\sfrac{3}{2}} \d v$, that transforms the integral boundaries
from $(1, \infty)$ to $(0, 1)$.}

\begin{figure}
    \centering
    \includegraphics[width=0.49\textwidth]{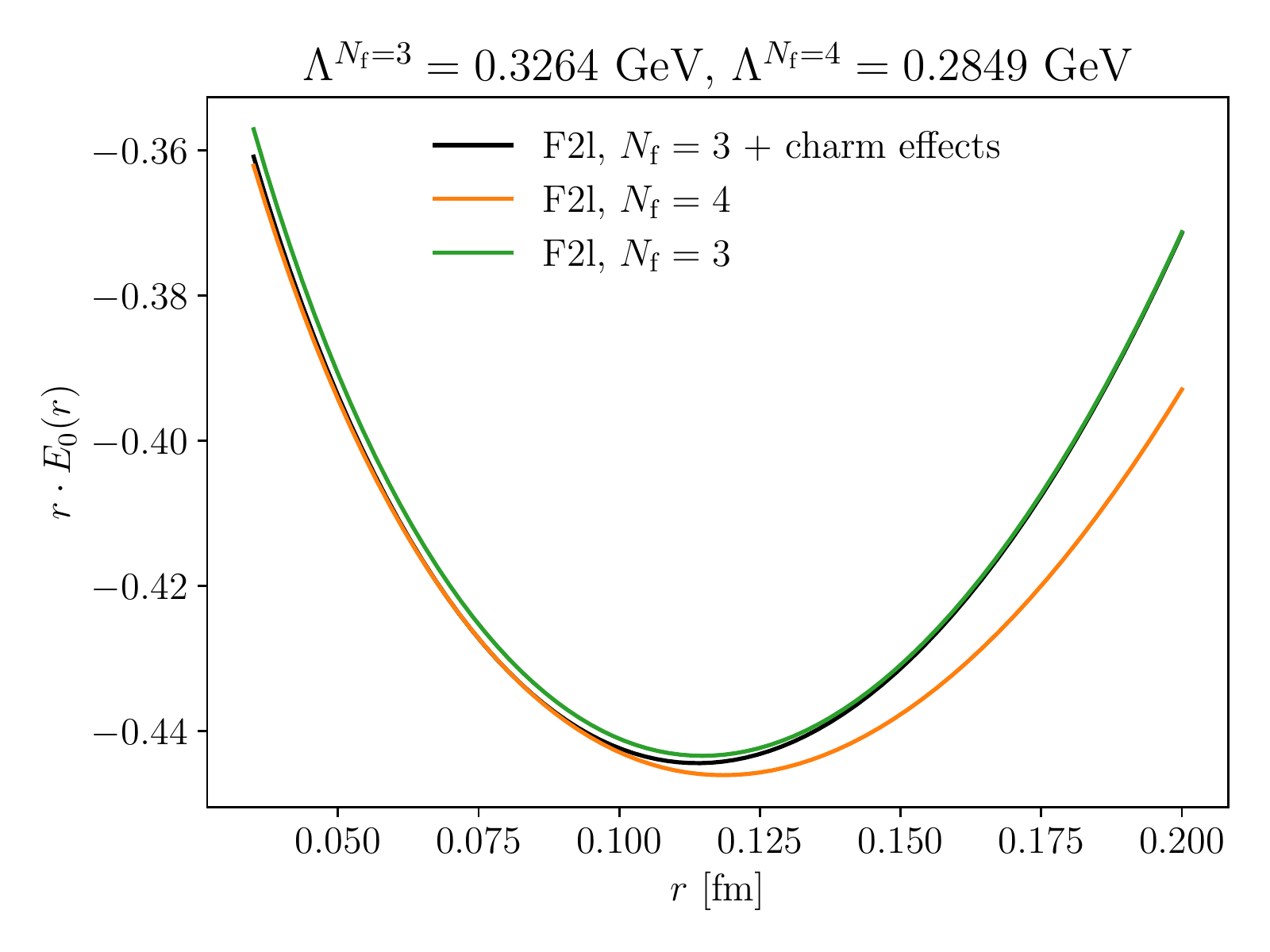}%
    \hfill%
    \includegraphics[width=0.49\textwidth]{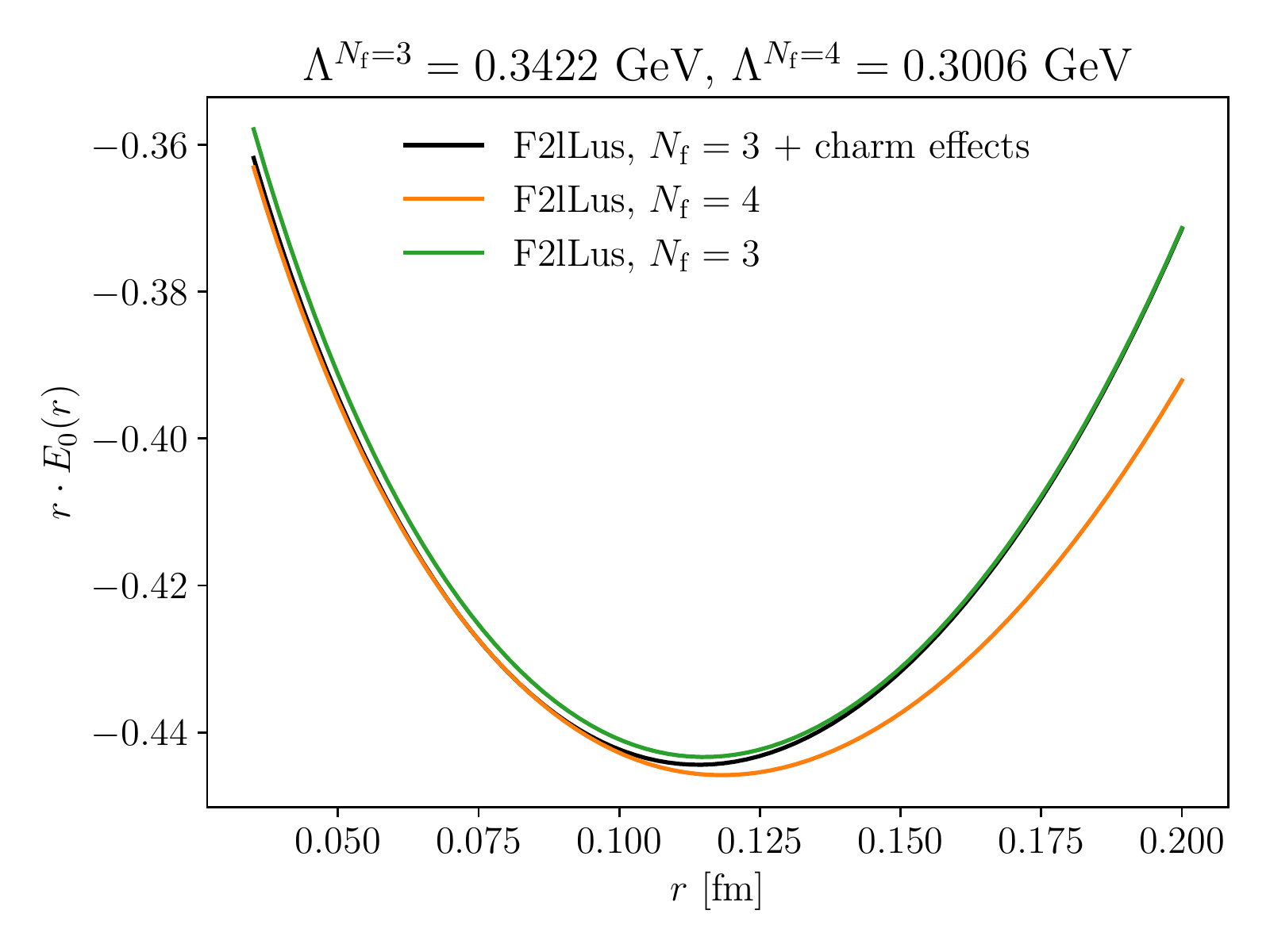}%
    \caption{\label{fig:theory_curves}%
    Left: static energy obtained from integrating the static force at two loops, Eq.~\eqref{eq:statfor}, for $\Nf=4$
    (orange curve), $\Nf=3$ (green curve), and for $\Nf=3$ plus the two-loop charm-mass correction using
    $\mc^{\MS}(\mc^{\MS}) = 1.28$~GeV (black curve); all curves using three-loop running of $\als$.
    The $\Nf=4$ curve has been matched to the $\Nf=3$ plus the two-loop charm-mass correction curve at $0.08$~fm by
    shifting the static energy by a constant.
    Right: like left but with the static force computed at N$^{2}$LL accuracy, Eq.~\eqref{eq:statfor_resummed}.}
\end{figure}

Decoupling requires that
\begin{equation}
    V_{m}^{(\Nf),[N]}(r,\nu) \to V^{(\Nf),[N]}(r,\nu) 
        , \qquad \text{for} \quad m \to \infty ,
    \label{Vmoo}
\end{equation}
and
\begin{equation}
    V_{m}^{(\Nf),[N]}(r,\nu) \to V^{(\Nf+1),[N]}(r,\nu) + \order((\alsNf)^{N+1}) , \qquad \text{for} \quad m \to 0 .
    \label{Vm0}
\end{equation}
One can verify analytically from the above expressions that the expected decoupling conditions hold in the limits $m
\to \infty$ and $m \to 0$ at one ($N=2$) and two ($N=3$) loops.
For a numerical verification at two loops see Fig.~\ref{fig:theory_curves}.
Since we use expressions with $\Nf$ flavors in the right-hand side of Eq.~\eqref{eq:full_potential}, the decoupling
\eqref{Vmoo} is exact in the $m \to \infty$ limit.
Hence, in Fig.~\ref{fig:theory_curves} the $\Nf=3$ green curve overlaps exactly at large distances with the black curve
obtained from the $\Nf=3$ static energy plus charm-mass corrections.
In contrast, the decoupling in Eq.~\eqref{Vm0} gets higher-order corrections when expressing the three flavor coupling
in terms of the four flavor one.
In order to account for these higher-order corrections, in Fig.~\ref{fig:theory_curves} we have matched the $\Nf=4$
orange curve with the black one at 0.08~fm by adding a small constant to the static energy.
The curves show the expected behavior, i.e., the curve with charm-mass effects interpolates smoothly between the
$\Nf=4$ one at the short distances ($\mc \ll \sfrac{1}{r}$) and the $\Nf=3$ one at large distances ($\mc \gg
\sfrac{1}{r}$).

\subsection{Special functions}
\label{app:functions}

The exponential-integral function is given by
\begin{equation}
    \Ei(x) = - \int\limits_{-x}^{\infty} \d t \, \frac{\e{-t}}{t} = \int\limits_{-\infty}^{x} \d t \, \frac{\e{t}}{t} ,
\end{equation}
fulfilling (for $x>0$) the relation
\begin{equation}
    \Ei(-x) = -\Ei(1,x) ,
\end{equation}
where
\begin{equation}
    \Ei(1,x) = \int\limits_{x}^{\infty} \d t \, \frac{\e{-t}}{t} = \int\limits_{1}^{\infty} \d t \, \frac{\e{-tx}}{t} =
        \int\limits_{0}^{1} \d t \, \frac{\e{-\sfrac{x}{t}}}{t} = -\gammaE - \ln(x) + \int\limits_{0}^{x} \d t \,
            \frac{1-\e{-t}}{t} \underset{x \to 0}{\longrightarrow} -\gammaE - \ln(x) .
\end{equation}
Note that $\Ei(1,x)$ is the $n=1$ case of the general $\text{E}_{n}$-function defined via
\begin{equation}
    \text{E}_{n}(x) \equiv \Ei(n,x) = \int\limits_{1}^{\infty} \d t \, \frac{\e{-xt}}{t^{n}} = x^{n-1} \Gamma(1-n,x) .
\end{equation}
$\Gamma$ with two arguments is the (upper) incomplete gamma function,
\begin{equation}
    \Gamma(a,x) = \int\limits_{x}^{\infty} \d t \, t^{a-1} \e{-t} \underset{a \to 0}{\longrightarrow} \Ei(1,x) .
\end{equation}
It can be expressed in terms of the regular gamma function and the lower incomplete gamma function as
\begin{equation}
    \Gamma(a,x) = \Gamma(a) - \gamma(a,x) ,
\end{equation}
where
\begin{equation}
    \gamma(a,x) = \frac{x^{a}}{a} + \sum\limits_{k=1}^{\infty} \frac{(-1)^{k}}{k! (a+k)} x^{a+k} ,
\end{equation}
such that
\begin{equation}
    \Gamma(a,x) = \Gamma(a) - \frac{x^{a}}{a} - \sum\limits_{k=1}^{\infty} \frac{(-1)^{k}}{k! (a+k)} x^{a+k}
        \underset{x \to 0}{\longrightarrow} - \frac{x^{a}}{a} , \quad \text{for } a < 0 .
\end{equation}

\bibliography{notes}

\end{document}